\documentclass[12pt,fleqn,a4paper]{article} 
\usepackage{epsfig,graphics,graphicx}
\usepackage{clshan-math,clshan-dm-dd}
\def \lsim {\:\raisebox{-0.7ex}{$\stackrel{\textstyle<}{\sim}$}\:}
\def \gsim {\:\raisebox{-0.7ex}{$\stackrel{\textstyle>}{\sim}$}\:}
\begin{document}
\thispagestyle{empty}
\begin{flushright}
 March 2013
\end{flushright}
\begin{center}
{\Large\bf
 Model--Independent Identification of Inelastic WIMPs \\ \vspace{0.2cm}
 from Direct Dark Matter Detection Experiments}       \\
\vspace*{0.7cm}
 {\sc Sen Miao}$^{1, \ddagger}$,
 {\sc Chung-Lin Shan}$^{2, \S}$,
  and {\sc Yu-Feng Zhou}$^{1, \P}$ \\~\\
\vspace{0.5cm}
 ${}^1$
 {\it State Key Laboratory of Theoretical Physics,   \\
      Kavli Institute for Theoretical Physics China, \\
      Institute of Theoretical Physics,
      Chinese Academy of Sciences                    \\
      Beijing 100190, China}                         \\~\\
 ${}^2$
 {\it Physics Division,
      National Center for Theoretical Sciences       \\
      No.~101, Sec.~2, Kuang-Fu Road,
      Hsinchu City 30013, Taiwan, R.O.C.}            \\~\\~\\
 ${}^{\ddagger}$
 {\it E-mail:} {\tt miaosen@itp.ac.cn}               \\
\vspace{0.1cm}
 ${}^{\S}$
 {\it E-mail:} {\tt clshan@phys.nthu.edu.tw}         \\
\vspace{0.1cm}
 ${}^{\P}$
 {\it E-mail:} {\tt yfzhou@itp.ac.cn}                \\~\\
\vspace{0.1cm}
\end{center}
\vspace{1cm}
\begin{abstract}
 In this paper,
 we introduce model--independent data analysis procedures
 for identifying inelastic WIMP--nucleus scattering
 as well as for reconstructing
 the mass and the mass splitting of inelastic WIMPs
 simultaneously and separately.
 Our simulations show that,
 with ${\cal O}(50)$ observed WIMP signals
 from one experiment,
 one could already distinguish the inelastic WIMP scattering scenarios
 from the elastic one.
 By combining two or more data sets with positive signals,
 the WIMP mass and the mass splitting could even be reconstructed
 with statistical uncertainties of
 less than a factor of two.
\end{abstract}
\clearpage
\section{Introduction}
 Different astronomical observations and measurements
 indicate that
 more than 80\% of all matter in the Universe is dark
 (i.e.~interacts at most very weakly
  with electromagnetic radiation and ordinary matter).
 The dominant component of this cosmological Dark Matter (DM)
 must be due to some yet to be discovered, non--baryonic particles.
 Weakly Interacting Massive Particles (WIMPs) $\chi$
 arising in several extensions of
 the Standard Model of electroweak interactions
 are one of the leading DM candidates.
 Currently,
 the most promising method to detect different WIMP candidates
 is the direct detection of the recoil energy
 deposited in a low--background underground detector
 by
 scattering of ambient WIMPs off target nuclei
 (for reviews,
  see Refs.~\cite{SUSYDM96, Bertone05,
                  Cerdeno10, Schnee11, Rau11, Freese12}).

 The first positive signal of WIMPs
 was reported by the DAMA Collaboration
 with observation of the annual modulation of
 the event rate of possible DM--target interaction
 by the DAMA/NaI detector
 \cite{DAMA} 
 (for the latest updated results,
  see Refs.~\cite{Bernabei13}).
 The CoGeNT Collaboration announced also their result with
 a modulated component of unknown origin
 \cite{Aalseth10, Aalseth11,
       CoGeNT-IDM2012}.
 The CRESST Collaboration published in 2011 their newest result
 with more observed events than expected backgrounds
 in the CRESST-II experiment
 \cite{Angloher11}.
 However,
 other direct DM detection experiments
 can so far mostly observe only very few candidate events
 and a large part ($\sim$ 40\% to 60\%)
 of these events would be unrejected backgrounds
 \cite{Ahmed09b,
       Armengaud11,
       Aprile11c, Aprile12,
       Akimov11b,
       SCKim12}.

 Many theoretical scenarios have been proposed
 to find an explanation
 for reconciling all these results
 from different experiments with various target nuclei.
 Among them,
 one of the mostly discussed and experimentally constrained frameworks
 is inelastic Dark Matter (iDM) models
 \cite{TuckerSmith,
       Chang08}.
 The basic assumption of iDM models is that
 the incident WIMPs scatter {\em inelastically} off target nuclei
 and then transit to a slightly heavier (exciting) state
 with a tiny mass (energy) splitting $\delta$,
 while the elastic scattering channel is suppressed or forbidden.
 Due to the kinematics of inelastic scattering,
 the targets with heavier nuclei have greater sensitivities
 than those with lighter nuclei.
 This suggests that
 the signals reported by DAMA
 which utilises relatively heavy iodine nuclei
 may not be detected by other experiments
 with light target nuclei
 such as germanium and silicon.
 Comparisons between exclusion limits
 on the cross section versus mass (splitting)
 ($\sigma - \mchi (\delta)$) planes
 in the iDM scenarios
 by using various experimental data
 have
 been done
 \cite{MarchRussell08, SchmidtHoberg09}.
 In Ref.~\cite{SchmidtHoberg09}
 the authors considered also the case of low--mass WIMPs
 (\mbox{$\mchi \sim$ 5 GeV})
 with a tiny mass splitting (\mbox{$\delta \sim$ 10 keV}).
 Meanwhile,
 a method for reconstructing the iDM parameters,
 most importantly the WIMP mass $\mchi$ and the mass splitting $\delta$,
 based on likelihood analysis
 has been suggested
 \cite{Finkbeiner09}.

 In the recent years,
 several experimental collaborations
 have (re)analyzed their (null) observation results
 and found severe constraints on
 the parameter space of iDM,
 i.e.~on the mass splitting $\delta$ versus the WIMP mass $\mchi$ plane.
 From these analyses,
 exclusion limits for mass splitting up to 250 keV
 in the mass range between 40 GeV and 1 TeV
 with different detector materials
 have been given
 \cite{Ahmed09b,
       ZEPLIN-iDM,
       XENON-iDM,
       Chang10,
       Ahmed10b,
       SCKim12}.
 However,
 in all these works,
 data analyses have been done basically by
 estimations of the (differential) event rate,
 which is strongly halo--model dependent.
 In addition,
 types and (relative) strengths of different DM--nucleus (quark) interactions
 are important factors in direct and indirect DM detection experiments
 \cite{SUSYDM96, Bertone05}.

 Hence,
 in this paper,
 as complementarity and extension of our earlier work
 on developing methods for reconstructing WIMP properties
 in elastic scattering framework
 as model--independently as possible
 \cite{DMDDf1v, DMDDmchi, DMDDsigma} 
 we introduce new model--independent approaches
 for identifying inelastic WIMP--nucleus scattering scenarios
 as well as for reconstructing the most important properties of inelastic WIMPs:
 the mass $\mchi$ and the mass splitting $\delta$
 simultaneously and separately.
 Our method is based on an estimation of a characteristic energy $\Qvthre$
 corresponding to a threshold (minimal required) (one--dimensional) velocity of
 incident (inelastic) WIMPs,
 $\vthre$.%
\footnote{
 Note that,
 in conventional elastic scattering framework,
 all incident WIMPs with non--zero velocity could scatter off target nuclei
 and deposit recoil energies $Q$.
 This means that
 $\Qvthre = 0$ for elastic WIMPs and
 $\Qvthre > 0$ for inelastic WIMPs.
 In fact,
 later we will show that
 the characteristic energy $\Qvthre$ is
 proportional to the mass splitting $\delta$.
}
 This can be done by determining the maximum of the integral
 over the one--dimensional velocity distribution function
 of incident WIMPs.
 Once this characteristic energy can be solved
 by using data sets of positive (inelastic) WIMP signals
 with different target nuclei,
 one can then determine the (degenerate) WIMP mass
 and the tiny mass splitting straightforwardly.

 The remainder of this paper is organized as follows.
 In Sec.~2,
 we develop the formalism of the methods
 for reconstructing the one--dimensional WIMP velocity distribution function,
 determining the WIMP mass and the mass splitting
 as well as estimating the characteristic energy
 and in turn identifying the inelastic WIMP--nucleus scattering scenarios.
 In Sec.~3,
 we demonstrate the ability and shortcomings of
 our model--independent methods
 by presenting numerical results
 based on Monte--Carlo simulations.
 The possibility of distinguishing the inelastic WIMP scenarios
 from the elastic one will be particularly discussed.
 We conclude
 in Sec.~4.
 Some technical details for our analysis
 will be given in the appendices.
\section{Formalism}
 In this section,
 we develop the formulae needed in our model--independent
 reconstructions of different properties of halo WIMPs
 in iDM scenarios,
 both of an approximated analytic method
 and an iterative numerical procedure
 will be considered.

 We start with
 the basic expression for the differential event rate
 for (elastic) WIMP--nucleus scattering given by \cite{SUSYDM96}:
\beq
   \dRdQ
 = \calA \FQ \int_{\vmin}^{\vmax} \bfrac{f_1(v)}{v} dv
\~.
\label{eqn:dRdQ}
\eeq
 Here $R$ is the direct detection event rate,
 i.e.~the number of events
 per unit time and unit mass of detector material,
 $Q$ is the energy deposited in the detector,
 $F(Q)$ is the (elastic) nuclear form factor,
 $f_1(v)$ is the one--dimensional velocity distribution function
 of the WIMPs impinging on the detector,
 $v$ is the absolute value of the WIMP velocity
 in the laboratory frame.
 The constant coefficient $\calA$ is defined as
\beq
        \calA
 \equiv \frac{\rho_0 \sigma_0}{2 \mchi \mrN^2}
\~,
\label{eqn:calA}
\eeq
 where $\rho_0$ is the WIMP density near the Earth
 and $\sigma_0$ is the total cross section
 ignoring the form factor suppression.
 The reduced mass $\mrN$ is defined by
\(
        \mrN
 \equiv \mchi \mN / \abrac{\mchi + \mN}
\),
 where $\mchi$ is the WIMP mass and
 $\mN$ that of the target nucleus.
 Finally,
 $\vmin$ is the minimal incoming velocity of incident WIMPs
 that can deposit the energy $Q$ in the detector:
\beq
   \vmin
 = \frac{1}{\sqrt{2 \mN Q}} \bbrac{\afrac{\mN}{\mrN} Q + \delta}
 = \alpha \sqrt{Q} + \frac{\alpha_{\delta}}{\sqrt{Q}}
\~,
\label{eqn:vmin_in}
\eeq
 with the transformation constants
\beq
        \alpha
 \equiv \sfrac{\mN}{2 \mrN^2}
\~,
\label{eqn:alpha}
\eeq
 and
\beq
        \alpha_{\delta}
 \equiv \frac{\delta}{\sqrt{2 \mN}}
\~;
\label{eqn:alpha_delta}
\eeq
 $\vmax$ is the maximal WIMP velocity
 in the Earth's reference frame,
 which is related to
 the escape velocity from our Galaxy
 at the position of the Solar system,
 $\vesc~\gsim~600$ km/s.
\subsection{Reconstruction of the velocity distribution \boldmath $f_1(v)$}
 Following the process in Ref.~\cite{DMDDf1v},
 we define
\beq
   \Dd{F_1(v)}{v}
 = \frac{f_1(v)}{v}
\~,
\label{eqn:dF1v_dv}
\eeq
 Eq.~(\ref{eqn:dRdQ}) can then be rewritten as%
\footnote{
 For simplicity,
 we set at first here the maximal cut--off of
 the one--dimensional WIMP velocity distribution
 as infinity.
 Later we will discuss correction of the formulae given here
 in the practical use with real (generated) data events.
}
\beq
   \frac{1}{\calA \FQ} \adRdQ
 = \int_{\vmin}^{\vmax \to \infty} \bfrac{f_1(v)}{v} dv
 = F_1(v = \vmax \to \infty) - F_1(\vmin)
\~.
\label{eqn:F1vmin}
\eeq
 Since WIMPs in today's Universe move quite slow,
 $f_1(v)$ must vanish as $v$ approaches infinity:
\(
     f_1(v \to \infty)
 \to 0
\).
 Thus
 $d F_1(v) / dv \big|_{v \to \infty} \to 0$
 and
 $F_1(v \to \infty)$ in turn approaches a finite value.
 Differentiating both sides of Eq.~(\ref{eqn:F1vmin}) with respect to $\vmin$
 and using Eq.~(\ref{eqn:vmin_in}),
 we can find
\beq
     \Dd{F_1(\vmin)}{\vmin}
  =  \frac{1}{\calA}
     \cbrac{\abrac{  \alpha \sqrt{Q}
                   - \frac{\alpha_{\delta}}{\sqrt{Q}} }^{-1}
            \cbrac{- 2 Q \cdot
                     \dd{Q} \bbrac{\frac{1}{\FQ} \adRdQ} } }
            _{Q = Q(\vmin)}
\~.
\label{eqn:dF1vmin_dvmin_in}
\eeq
 Here,
 from the definition (\ref{eqn:vmin_in}) of $\vmin$,
 we can firstly find that
\beq
   \Dd{\vmin}{Q}
 = \frac{1}{2 Q}
   \abrac{  \alpha \sqrt{Q}
          - \frac{\alpha_{\delta}}{\sqrt{Q}} }
\~,
\label{eqn:dvmin_dQ_in}
\eeq
 and an analytic expression of $Q(\vmin)$
 can be solved from the definition (\ref{eqn:vmin_in})
 directly as:
\beq
   Q(\vmin)
 = \frac{    \vmin^2
         -   2 \alpha \alpha_{\delta}
         \pm \vmin \sqrt{\vmin^2 - 4 \alpha \alpha_{\delta}} }
        {2 \alpha^2}
\~.
\label{eqn:Q_vmin_in}
\eeq
 Note that,
 corresponding to one specific value of $\vmin$,
 there are two possible values of $Q(\vmin)$,
 except of
\beq
   \Qvthre
 = Q\abrac{\vmin = \vthre = 2 \sqrt{\alpha \alpha_{\delta}}}
 = \frac{\alpha_{\delta}}{\alpha}
 = \afrac{\mchi}{\mchi + \mN} \delta
\~.
\label{eqn:Qvthre}
\eeq
 Here $\vthre$ is the {\em threshold} (minimal required) velocity
 of incident {\em inelastic} WIMPs,
 which can produce recoil energy at all.

 Since the expression (\ref{eqn:dF1vmin_dvmin_in}) of $d F_1(\vmin) / d \vmin$
 holds for arbitrary $\vmin$,
 we can write down the following result directly:
\beq
   \frac{f_1(v)}{v}
 = \Dd{F_1(v)}{v}
 = \frac{1}{\calA}
   \cbrac{\abrac{  \alpha \sqrt{Q}
                 - \frac{\alpha_{\delta}}{\sqrt{Q}} }^{-1}
          \cbrac{- 2 Q \cdot
                   \dd{Q} \bbrac{\frac{1}{\FQ} \adRdQ} } }
          _{Q = Q(v)}
\~,
\label{eqn:f1v_v_in}
\eeq
 with
\beq
   Q(v)
 = \frac{    v^2
         -   2 \alpha \alpha_{\delta}
         \pm v \sqrt{v^2 - 4 \alpha \alpha_{\delta}} }
        {2 \alpha^2}
\~.
\label{eqn:Q_v_in}
\eeq
 Although the right--hand side of this expression
 depends on the as yet unknown constant $\calA$,
 $f_1(v)$ is the {\em normalized} velocity distribution,
 i.e.~it is defined to satisfy:
\(
   \intz f_1(v) \~ dv
 = 1
\).
 Therefore,
 the normalized one--dimensional velocity distribution function
 of inelastic WIMPs
 can be given by
\beq
   f_1(v)
 = \calN
   \cbrac{\abrac{  \alpha \sqrt{Q}
                 + \frac{\alpha_{\delta}}{\sqrt{Q}} }
          \abrac{  \alpha \sqrt{Q}
                 - \frac{\alpha_{\delta}}{\sqrt{Q}} }^{-1}
          \cbrac{- 2 Q \cdot
                   \dd{Q} \bbrac{\frac{1}{\FQ} \adRdQ} } }
          _{Q = Q(v)}
\~,
\label{eqn:f1v_in}
\eeq
 with the normalization constant $\calN$:
\beq
     \calN
 =   2
     \cbrac{\intz \bbrac{\frac{1}{Q}
                         \abrac{  \alpha \sqrt{Q}
                                - \frac{\alpha_{\delta}}{\sqrt{Q}} } }
                  \bbrac{\frac{1}{\FQ} \adRdQ} dQ }^{-1}
\~.
\label{eqn:calN_in}
\eeq
 Note that,
 for the case of elastic WIMP--nucleus scattering,
 $\alpha_{\delta} = 0$,
 the expressions (\ref{eqn:f1v_in}) and (\ref{eqn:calN_in})
 can then be reduced to the simple, analytic forms
 given in Eqs.~(12) and (13) of Ref.~\cite{DMDDf1v}.

\subsection{Determinations of the WIMP mass \boldmath $\mchi$
            and the mass splitting $\delta$}
 The expression for reconstructing
 the one--dimensional velocity distribution of inelastic WIMPs
 given in Eqs.~(\ref{eqn:f1v_in}) and (\ref{eqn:calN_in})
 can unfortunately not be used directly,
 since at first there are {\em two unknowns},
 i.e.~the WIMP mass $\mchi$ (involved in $\alpha$)
 and the mass splitting $\delta$ (involved in $\alpha_{\delta}$).
 Moreover,
 the typical {\em peaky} shape of the recoil spectrum of inelastic WIMPs
 (see e.g.~Figs.~\ref{fig:idRdQ-Ge76-100-025-050} to \ref{fig:idRdQ-100-050-050})
 makes its reconstruction
 more complicated
 and therefore a similar development of
 procedures introduced in Refs.~\cite{DMDDf1v, DMDDmchi}
 is basically impossible.

 Hence,
 in this subsection,
 we introduce a new approach
 for determining the WIMP mass and the mass splitting
 based on the estimation of the characteristic energy $\Qvthre$,
 which requires the reconstruction of
 the peaky inelastic WIMP recoil spectrum.

\subsubsection{Determinations of \boldmath $\mchi$ and $\delta$}
 From the definition (\ref{eqn:vmin_in}) of $\vmin$,
 one have
\beq
   \left.\Dd{\vmin}{Q}\right|_{Q = \Qvthre}
 = \frac{1}{2 \Qvthre}
   \abrac{  \alpha \sqrt{\Qvthre}
          - \frac{\alpha_{\delta}}{\sqrt{\Qvthre}} }
 = 0
\~.
\label{eqn:dvmin_dQ}
\eeq
 Then the characteristic energy
 corresponding to the {\em minimal} value of $\vmin$, $\vthre$,
 can easily be solved as
\cheqnref{eqn:Qvthre}
\beq
   \Qvthre
 = \frac{\alpha_{\delta}}{\alpha}
 = \afrac{\mchi}{\mchi + \mN} \delta
\~,
\eeq
\cheqnN{-1}
 which is proportional to the mass splitting $\delta$
 and the proportionality constant is simply
 a function of the WIMP mass $\mchi$.
 Hence,
 by combining two experimental data sets
 with different target nuclei, $X$ and $Y$,
 one can derive analytic expressions for determining $\mchi$ and $\delta$
 as functions of the characteristic energy $\QvthreX$ and $\QvthreY$:
\beq
   \mchi
 = \frac{\QvthreY \mY - \QvthreX \mX}{\QvthreX - \QvthreY}
\~,
\label{eqn:mchi_in}
\eeq
 and
\beq
   \delta
 = \frac{\QvthreX \QvthreY \abrac{\mY - \mX}}{\QvthreY \mY - \QvthreX \mX}
\~.
\label{eqn:delta_in}
\eeq
 Then,
 since $Q_{{\rm thre}, (X, Y)}$ are two independent variables,
 by using the standard Gaussian error propagation,
 the statistical uncertainties on the reconstructed $\mchi$ and $\delta$
 can be given as
\beqn
     \sigma\abrac{\mchi}
 \=  \bbrac{  \aPp{\mchi}{\QvthreX}^2 \sigma^2\abrac{\QvthreX}
            + \aPp{\mchi}{\QvthreY}^2 \sigma^2\abrac{\QvthreY} }^{1 / 2}
     \non\\
 \=  \frac{\vbrac{\mX - \mY} \QvthreX \QvthreY}
          {\abrac{\QvthreX - \QvthreY}^2}
     \bbrac{  \frac{\sigma^2\abrac{\QvthreX}}{\QvthreX^2}
            + \frac{\sigma^2\abrac{\QvthreY}}{\QvthreY^2} }^{1 / 2}
\~,
\label{eqn:sigma_mchi_in}
\eeqn
 and
\beqn
     \sigma\abrac{\delta}
 \=  \bbrac{  \aPp{\delta}{\QvthreX}^2 \sigma^2\abrac{\QvthreX}
            + \aPp{\delta}{\QvthreY}^2 \sigma^2\abrac{\QvthreY} }^{1 / 2}
     \non\\
 \=  \frac{\vbrac{\mX - \mY} \mX \mY \QvthreX^2 \QvthreY^2}
          {\abrac{\QvthreY \mY - \QvthreX \mX}^2}
     \bbrac{  \frac{\sigma^2\abrac{\QvthreX}}{\mX^2 \QvthreX^4}
            + \frac{\sigma^2\abrac{\QvthreY}}{\mY^2 \QvthreY^4} }^{1 / 2}
\~.
\label{eqn:sigma_delta_in}
\eeqn
\subsubsection{Ansatz for reconstructing the inelastic--scattering recoil spectrum}
 Note that,
 for the use of Eqs.~(\ref{eqn:mchi_in}) and (\ref{eqn:delta_in}),
 one needs to estimate the characteristic energy $\Qvthre$
 corresponding to the threshold (minimal required) velocity
 of incident inelastic WIMPs $\vthre$,
 which could produce recoil energy at all.
 This means that
 $v = \vthre$ is the lowest bound of the velocity of inelastic WIMPs,
 which could contribute to
 the integral over the one--dimensional velocity distribution function $f_1(v)$
 on the right--hand side of Eq.~(\ref{eqn:dRdQ}).
 This means in turn that
 the integral over $f_1(v)$,
 or, equivalently,
 a ``reduced'' differential event rate
 (i.e.~the differential event rate
  divided by the squared nuclear form factor):
\beq
         \frac{1}{\FQ} \adRdQ
 \propto \int_{\vmin}^{\vmax} \bfrac{f_1(v)}{v} dv
\label{eqn:dRdQ_FQ}
\eeq
 should be maximal once $Q = \Qvthre$.

 On the other hand,
 for the simplest isothermal spherical halo model,
 the normalized one--dimensional velocity distribution function
 can be expressed as
 \cite{DMDDf1v}:
\beq
   f_{1,\Gau}(v)
 = \frac{4}{\sqrt{\pi}} \afrac{v^2}{v_0^3} e^{-v^2 / v_0^2}
\~,
\label{eqn:f1v_Gau}
\eeq
 where $v_0 \approx 220$ km/s is the Sun's orbital speed
 around the Galactic center.
 More realistically,
 by taking into account
 the orbital motion of the Solar system around the Galaxy,
 the more frequently used shifted Maxwellian velocity distribution
 has been given by
 \cite{DMDDf1v}:
\beq
   f_{1, \sh}(v)
 = \frac{1}{\sqrt{\pi}} \afrac{v}{\ve v_0}
   \bbigg{ e^{-(v - \ve)^2 / v_0^2} - e^{-(v + \ve)^2 / v_0^2} }
\~,
\label{eqn:f1v_sh}
\eeq
 where
 $\ve$ is the {\em time--dependent} Earth's velocity
 in the Galactic frame:
\beq
   \ve(t)
 = v_0 \bbrac{1.05 + 0.07 \cos\afrac{2 \pi (t - t_{\rm p})}{1~{\rm yr}}}
\~,
\label{eqn:ve}
\eeq
 where $t_p \simeq$ June 2nd is the date
 on which the Earth's velocity relative to the WIMP halo is maximal.

 Substituting these two expressions into Eq.~(\ref{eqn:dRdQ})
 and using Eq.~(\ref{eqn:vmin_in}),
 one can easily obtain that%
\footnote{
 As in Sec.~2.1,
 we assume here that
 the cut--off on $f_1(v)$, $\vmax$, as well as
 the experimental maximal cut--off energy $\Qmax$
 are large enough
 and the integral in the higher velocity/energy range
 can be neglected.
}
\cheqna
\beq 
         \frac{1}{\FQ} \adRdQ_{\rm in,~\Gau}
 \propto e^{- \abig{\alpha \sqrt{Q} + \alpha_{\delta}   / \sqrt{Q}}^2 / v_0^2}
 \propto e^{- \abig{\alpha^2     Q  + \alpha_{\delta}^2 /       Q }   / v_0^2}
\~,
\label{eqn:dRdQ_in_Gau}
\eeq
\cheqnb
 and
\beq
          \frac{1}{\FQ} \adRdQ_{\rm in,~\sh}
 \propto  \erf{\T \afrac{\alpha \sqrt{Q} + \alpha_{\delta} / \sqrt{Q} + \ve}{v_0}}
        - \erf{\T \afrac{\alpha \sqrt{Q} + \alpha_{\delta} / \sqrt{Q} - \ve}{v_0}}
\~.
\label{eqn:dRdQ_in_sh}
\eeq
\cheqn
 Then,
 similar to the use of the exponential approximation for reconstructing
 the recoil spectrum of elastic WIMP--nucleus scattering
 \cite{DMDDf1v},
 in order to approximate
 the measured recoil spectrum
 and take into account the extra contribution
 predicted in inelastic WIMP scattering scenarios,
 we introduce {\em empirically} here
\beq
   \adRdQ_{\rm in,~expt}
 = r_0 \~ e^{-k Q - k' / Q}
\~,
\label{eqn:dRdQ_in}
\eeq
 where
\beq
   r_0
 = \frac{N_{\rm tot}}
        {\D \calE \int_{\Qmin}^{\Qmax} e^{-k Q - k' / Q} \~ dQ}
\~,
\label{eqn:r0_in}
\eeq
 with the total WIMP signal
 events
 in our data set $N_{\rm tot}$
 and the experimental exposure $\calE$;
 $k$ and $k'$ are two fitting parameters
 which we want to estimate by using the measured recoil energies directly.
 By using the ansatz (\ref{eqn:dRdQ_in}),
 the position of the peak of the inelastic recoil spectrum
 can be solved easily and then estimated by
 (mathematical details are given in Appendix A.2)
\beq
   \Qpk
 = \frac{\D \Expv{Q^{-1 / 2}}_{\rm inf}}{\D \Expv{Q^{-3 / 2}}_{\rm inf}}
\~,
\label{eqn:Qpk}
\eeq
 with the statistical uncertainty given by
\beqn
    \sigma(\Qpk)
 \= \Qpk
    \bbrac{  \frac{\D \sigma^2\abrac{\Expv{Q^{-1 / 2}}_{\rm inf} } }
                  {\D \Expv{Q^{-1 / 2}}_{\rm inf}^2 }
           + \frac{\D \sigma^2\abrac{\Expv{Q^{-3 / 2}}_{\rm inf} } }
                  {\D \Expv{Q^{-3 / 2}}_{\rm inf}^2 }
           + \frac{\D 2 {\rm cov}\abrac{\Expv{Q^{-1 / 2}}_{\rm inf}, \Expv{Q^{-3 / 2}}_{\rm inf}} }
                  {\D \Expv{Q^{-1 / 2}}_{\rm inf} \Expv{Q^{-3 / 2}}_{\rm inf}} }^{1/2}
\~,
    \non\\
\label{eqn:sigma_Qpk}
\eeqn
 where the $\lambda-$th momentum of the recoil energy spctrum
 can be estimated by
\beq
     \Expv{Q^{\lambda}}_{\rm inf}
 \equiv
     \frac{\D \intz Q^{\lambda} \aBig{dR/dQ}_{\rm in,~expt} \~ dQ}
          {\D \intz             \aBig{dR/dQ}_{\rm in,~expt} \~ dQ}
 \to
     \frac{1}{N_{\rm tot}} \sum_{a} Q_a^{\lambda}
\~.
\label{eqn:expv_Qlambda}
\eeq
 Note that
 we assumed here (unrealistically) that
 the minimal experimental cut--off energy $\Qmin$ is negligible
 and the maximal one $\Qmax$ is infinity
 (a numerical correction will be discussed
  in the next subsection).
 Finally,
 two fitting parameters in the ansatz (\ref{eqn:dRdQ_in})
 can be estimated separately as
\cheqna
\beq
   k_{\rm ana}
 = \frac{1}{2}
   \afrac{\D \Expv{Q^{-1 / 2}}_{\rm inf} \Expv{Q^{-3 / 2}}_{\rm inf}}
         {\D \Expv{Q^{ 1 / 2}}_{\rm inf} \Expv{Q^{-3 / 2}}_{\rm inf} - \Expv{Q^{-1 / 2}}_{\rm inf}^2}
\~,
\label{eqn:k_in_ana}
\eeq
 and
\cheqnb
\beq
   k'_{\rm ana}
 = \frac{1}{2}
   \afrac{\D \Expv{Q^{-1 / 2}}_{\rm inf} \Expv{Q^{-3 / 2}}_{\rm inf}}
         {\D \Expv{Q^{-1 / 2}}_{\rm inf} \Expv{Q^{-5 / 2}}_{\rm inf} - \Expv{Q^{-3 / 2}}_{\rm inf}^2}
\~.
\label{eqn:kp_in_ana}
\eeq
\cheqn

 By using the approximation (\ref{eqn:dRdQ_in}),
 the expression (\ref{eqn:f1v_in})
 for reconstructing the one--dimensional WIMP velocity distribution
 can be rewritten as
\beqn
    f_1(v)
 \= \calN
    \cBiggl{\abrac{  \alpha \sqrt{Q}
                   + \frac{\alpha_{\delta}}{\sqrt{Q}} }
            \abrac{  \alpha \sqrt{Q}
                   - \frac{\alpha_{\delta}}{\sqrt{Q}} }  }^{-1}
    \non\\
 \conti ~~~~~~~~~~~~~~~~ \times 
    \cBiggr{\cbrac{  \frac{2 Q}{\FQ}
                     \bbrac{\dd{Q} \ln \FQ + \abrac{k \~ - \~ \frac{k'}{Q^2}} }
                     \adRdQ} }
          _{Q = Q(v)}
\~,
\label{eqn:f1v_in_rec}
\eeqn
 with the normalization constant $\calN$ given in Eq.~(\ref{eqn:calN_in}).
 Note that,
 for reconstructing $f_1(v)$
 by using Eqs.~(\ref{eqn:f1v_in_rec}) and (\ref{eqn:calN_in}),
 the constant $r_0$ in Eq.~(\ref{eqn:r0_in}) can be cancelled out,
 since this appears in both of the expression (\ref{eqn:f1v_in}) or (\ref{eqn:f1v_in_rec})
 and the expression (\ref{eqn:calN_in}).

 On the other hand,
 as discussed at the beginning of this subsection,
 $\vthre$ is the lowest bound of the integral
 in Eqs.~(\ref{eqn:dRdQ}) or (\ref{eqn:dRdQ_FQ}),
 which gives a maximal value of the (reduced) event rate
 and thus have to satisfy the following condition:
\beq
     \dd{Q}\bbrac{\frac{1}{\FQ} \adRdQ}
 =  0
\~.
\eeq
 By using the ansatz (\ref{eqn:dRdQ_in})
 for reconstructing the inelastic scattering spectrum,
 one can find that
\beq
     \abrac{k - \frac{k'}{\Qvthre^2}}
   + \frac{2}{F(\Qvthre)} \aDd{F}{Q}_{Q = \Qvthre}
 =   0
\~.
\label{eqn:Qvthre_sol}
\eeq
 $\Qvthre$ can then be solved
 numerically
 and the statistical uncertainty on $\Qvthre$
 can be estimated by
\beqn
     \sigma\abrac{\Qvthre}
 \=  \cleft{  \sum_{\lambda, \rho = -3}^{0}
              \bbrac{  \aPp{\Qvthre}{k}
                       \aPp{k }{\D \Expv{Q^{\lambda + 1 / 2}}}
                     + \aPp{\Qvthre}{k'}
                       \aPp{k'}{\D \Expv{Q^{\lambda + 1 / 2}}} } }
     \non\\
 \conti ~~~~~~~~~~~~ \times 
              \bbrac{  \aPp{\Qvthre}{k}
                       \aPp{k }{\D \Expv{Q^{\rho    + 1 / 2}}}
                     + \aPp{\Qvthre}{k'}
                       \aPp{k'}{\D \Expv{Q^{\rho    + 1 / 2}}} }
     \non\\
 \conti ~~~~~~~~~~~~~~~~~~ \times 
     \cBiggr{  {\rm cov}
               \abrac{\Expv{Q^{\lambda + 1 / 2}},
                      \Expv{Q^{\rho    + 1 / 2}} } }^{1 / 2}
     \non\\
 \=  \frac{1}{2}
     \vbrac{  \frac{k'}{Q^3}
           + \dd{Q}\bbrac{\frac{1}{F(Q)} \aDd{F}{Q}} }_{Q = \Qvthre}^{-1}
     \non\\
 \conti ~~~~~~ \times 
     \cleft{  \sum_{\lambda, \rho = -3}^{0}
              \bbrac{  \aPp{k }{\D \Expv{Q^{\lambda + 1 / 2}}}
                     - \frac{1}{\Qvthre^2}
                       \aPp{k'}{\D \Expv{Q^{\lambda + 1 / 2}}} } }
     \non\\
 \conti ~~~~~~~~~~~~~~~~~~~~~~~~ \times 
              \bbrac{  \aPp{k }{\D \Expv{Q^{\rho    + 1 / 2}}}
                     - \frac{1}{\Qvthre^2}
                       \aPp{k'}{\D \Expv{Q^{\rho    + 1 / 2}}} }
     \non\\
 \conti ~~~~~~~~~~~~~~~~~~~~~~~~~~~~~~ \times 
     \cBiggr{  {\rm cov}
               \abrac{\Expv{Q^{\lambda + 1 / 2}},
                      \Expv{Q^{\rho    + 1 / 2}} } }^{1 / 2}
\~.
\label{eqn:sigma_Qvthre}
\eeqn
 Here we have
\beq
   {\rm cov}\abrac{\Expv{Q^{\lambda + 1 / 2}}, \Expv{Q^{\rho + 1 / 2}}}
 = \frac{1}{N_{\rm tot} - 1}
   \bbigg{  \Expv{Q^{\lambda + \rho + 1}}
          - \Expv{Q^{\lambda + 1 / 2}} \Expv{Q^{\rho + 1 / 2}} }
\~,
\label{eqn:cov_expvQ}
\eeq
 and
 $\p k^{(\prime)} / \p \Expv{Q^{\lambda + 1 / 2}}$
 for $\lambda = -3,~-2,~-1,~0$
 are given in Appendix C.
 Note that,
 for the analytic estimates of $k$ and $k'$
 given by Eqs.~(\ref{eqn:k_in_ana}) and (\ref{eqn:kp_in_ana}),
 one has
\beq
   \Pp{k _{\rm ana}}{\D \Expv{Q^{-5 / 2}}_{\rm inf}}
 = \Pp{k'_{\rm ana}}{\D \Expv{Q^{ 1 / 2}}_{\rm inf}}
 = 0
\~.
\eeq
\subsection{A numerical correction}
 By using Eqs.~(\ref{eqn:expv_Qlambda}),
 (\ref{eqn:k_in_ana}) and (\ref{eqn:kp_in_ana})
 to estimate the spectrum fitting parameters $k$ and $k'$ analytically,
 one has to assume that
 the (experimental) minimal cut--off energy should be negligibly small
 (\mbox{$\Qmin \simeq 0$})
 and the maximal one large enough
 (\mbox{$\Qmax~\gsim$ a few} hundred keV).
 This would still be a challenge for next--generation detectors.
 Moreover,
 due to the local escape velocity of halo WIMPs $\vesc$,
 or, equivalently,
 the maximal cut--off on its one--dimensional velocity distribution $\vmax$,
 there are not only a kinematic maximal but also a minimal cut--off energies,
 below and above which
 the incident WIMPs could produce recoil energies in our detector
 (see e.g.~Figs.~\ref{fig:idRdQ-100-050-050}).
 From Eq.~(\ref{eqn:Q_v_in}),
 one can get
\cheqna
\beq
   Q_{\rm min, kin}
 = \frac{    \vmax^2
         -   2 \alpha \alpha_{\delta}
         -   \vmax \sqrt{\vmax^2 - 4 \alpha \alpha_{\delta}} }
        {2 \alpha^2}
\~,
\label{eqn:Qmin_kin_in}
\eeq
 and
\cheqnb
\beq
   Q_{\rm max, kin}
 = \frac{    \vmax^2
         -   2 \alpha \alpha_{\delta}
         +   \vmax \sqrt{\vmax^2 - 4 \alpha \alpha_{\delta}} }
        {2 \alpha^2}
\~.
\label{eqn:Qmax_kin_in}
\eeq
\cheqn
 As shown later in Sec.~3,
 this could be a serious issue
 once the mass splitting \mbox{$\delta~\gsim$ 30 keV},
 especially for light target nuclei,
 i.e.~$\rmXA{Si}{28}$ and $\rmXA{Ar}{40}$
 (see Figs.~\ref{fig:idRdQ-100-050-050}).

 Moreover,
 as we will show in Sec.~3.3,
 for larger WIMP masses (\mbox{$\mchi~\gsim$ 200 GeV}),
 the reconstructed WIMP mass could be strongly underestimated:
 the larger the mass splitting,
 the worse the mass reconstruction.
 Hence,
 we introduce in this subsection an iterative process
 for correcting the estimation of the fitting parameters,
 $k$ and $k'$,
 in the hope that
 this correction could reduce the systematic deviation of the solved $\Qvthre$
 and thereby alleviate the underestimation of the reconstructed WIMP mass.

 We start from Eq.~(\ref{eqn:expv_Qlambda}).
 Define
\beq
   \Expv{Q^{\lambda}}(k, k')
 = \frac{\D \Expv{x^{\lambda}}(k, k'; x = \Qmax) - \Expv{x^{\lambda}}(k, k'; x = \Qmin)}
        {\D \int_{\Qmin}^{\Qmax} e^{-k^{\ast} x - k'^{\ast} / x} \~ dx}
\~,
\label{eqn:expv_Qlambda_k_kp}
\eeq
 where $\Expv{x^{\lambda}}(k, k'; x)$
 are given in Eqs.~(\ref{eqn:expv_x_m1_2}) to (\ref{eqn:expv_x_m5_2})
 in Appendix A.2.
 In stead of $\Expv{Q^{\lambda}}_{\rm inf}$
 defined on the left--hand side of Eq.~(\ref{eqn:expv_Qlambda}),
 $\Expv{Q^{\lambda}}(k, k')$ defined above
 should be more suitable to be estimated by
 the average value of the summation over all measured recoil energies
 in the $\lambda$th power
 given on the right--hand side of Eq.~(\ref{eqn:expv_Qlambda}).
 Thus one has
\beq
   \Expv{x^{\lambda}}(k, k'; x)\bigg|_{\Qmin}^{\Qmax}
 = \abrac{\int_{\Qmin}^{\Qmax} e^{-k^{\ast} x - k'^{\ast} / x} \~ dx}
   \abrac{\frac{1}{N_{\rm tot}} \sum_{a} Q_a^{\lambda}}
\~.
\label{eqn:k_kp_in_num}
\eeq
 By setting $k^{(\prime) \ast} = k^{(\prime)}_{\rm ana}$
 estimated by Eqs.~(\ref{eqn:k_in_ana}) and (\ref{eqn:kp_in_ana}),
 one can solve $k^{(\prime)}$ numerically,
 denoted by $k^{(\prime)}_{\rm num}$,
 by using the expressions of $\Expv{x^{\lambda}}(k, k'; x)\Big|_{\Qmin}^{\Qmax}$,
 $\lambda = -1/2$ and $-3/2$,
 simultaneously%
\footnote{
 The detailed description about solving $k_{\rm num}$ and $k'_{\rm num}$
 from $\Expv{x^{-1/2}}(k, k'; x)\Big|_{\Qmin}^{\Qmax}$
 and  $\Expv{x^{-3/2}}(k, k'; x)\Big|_{\Qmin}^{\Qmax}$ simultaneously
 will be given in Appendix B.
}.

 Note that
 expressions (\ref{eqn:f1v_in_rec}) and (\ref{eqn:Qvthre_sol})
 for reconstructing the one--dimensional WIMP velocity distribution
 and solving the characteristic energy $\Qvthre$
 can be used by substituting
 both the analytic and numerical estimations
 of the fitting parameters $k$ and $k'$,
 whereas
 for using Eq.~(\ref{eqn:sigma_Qvthre})
 to estimate the statistical uncertainty on $\Qvthre$
 solved with the {\em numerically} estimated $k_{\rm num}$ and $k'_{\rm num}$,
 the summation runs only for $\lambda, \rho = -2$ and $-1$.

\section{Numerical results}
 In this section,
 we present numerical results
 of the reconstruction of the recoil spectrum,
 the identification of the positivity of $\Qvthre$
 (as the check of the iDM scenarios)
 as well as
 the reconstruction of the WIMP mass $\mchi$
 and the mass splitting $\delta$
 based on Monte--Carlo simulations.
 The special case of zero mass splitting $\delta = 0$,
 i.e.~the case of elastic WIMP scattering,
 will be particularly discussed
 at the end of this section
 as a demonstration of the usefulness of
 our model--independent approach
 for distinguishing the inelastic WIMP scenarios
 from the elastic one.

 For generating WIMP--induced signals in our simulations,
 we use the shifted Maxwellian velocity distribution
 given in Eq.~(\ref{eqn:f1v_sh})
 with the Sun's Galactic orbital velocity $v_0 = 220$ km/s;
 the time dependence of the Earth's velocity in the Galactic frame
 has been ignored,
 i.e.~\mbox{$\ve = 1.05 \~ v_0$} is used.
 Moreover,
 the maximal cut--off of the one--dimensional WIMP velocity distribution
 has been set as $\vmax = 700$ km/s.
 Meanwhile,
 the WIMP--nucleon cross section has been assumed
 to be only spin--independent (SI),
 \mbox{$\sigmapSI = 10^{-6}$ pb},
 and
 the commonly used analytic form
 for the elastic nuclear form factor
\beq
   F_{\rm SI}^2(Q)
 = \bfrac{3 j_1(q R_1)}{q R_1}^2 e^{-(q s)^2}
\label{eqn:FQ_WS}
\eeq
 has been adopted.
 Here $Q$ is the recoil energy
 transferred from the incident WIMP to the target nucleus,
 $j_1(x)$ is a spherical Bessel function,
\(
   q
 = \sqrt{2 m_{\rm N} Q}
\)
 is the transferred 3-momentum,
 for the effective nuclear radius we use
\(
   R_1
 = \sqrt{R_A^2 - 5 s^2}
\)
 with
\(
        R_A
 \simeq 1.2 \~ A^{1/3}~{\rm fm}
\)
 and a nuclear skin thickness
\(
        s
 \simeq 1~{\rm fm}
\).
 In addition,
 the experimental threshold energies
 have been assumed to be negligible ($\Qmin = 0$)
 and the maximal cut--off energies
 are set as \mbox{$\Qmax = 150$ keV}
 for all target nuclei%
\footnote{
 Note that,
 although we assume here (unrealistically) that
 one could experimentally measure recoil energies
 between $\Qmin = 0$ and \mbox{$\Qmax = 150$ keV},
 since the effect of
 the kinematic maximal and minimal cut--off energies
 has been taken into account in our simulations,
 we only generate events
 between $Q_{\rm min, kin}$ and
 the smaller one between $\Qmax$ and $Q_{\rm max, kin}$
 and then analyze the data sets numerically
 with these two cut--offs.
}.
 5,000 experiments with 50 total events on average
 in one experiment have been simulated.

\subsection{Reconstructing the recoil spectrum}

 In this subsection,
 we consider at first
 the reconstruction of the recoil spectrum,
 which is approximated by Eq.~(\ref{eqn:dRdQ_in})
 with the fitting parameters $k$ and $k'$,
 as well as the estimation of
 the characteristic energy $\Qvthre$.

\begin{figure}[t!]
\begin{center}
\hspace*{-1.6cm}
\includegraphics[width=8.5cm]{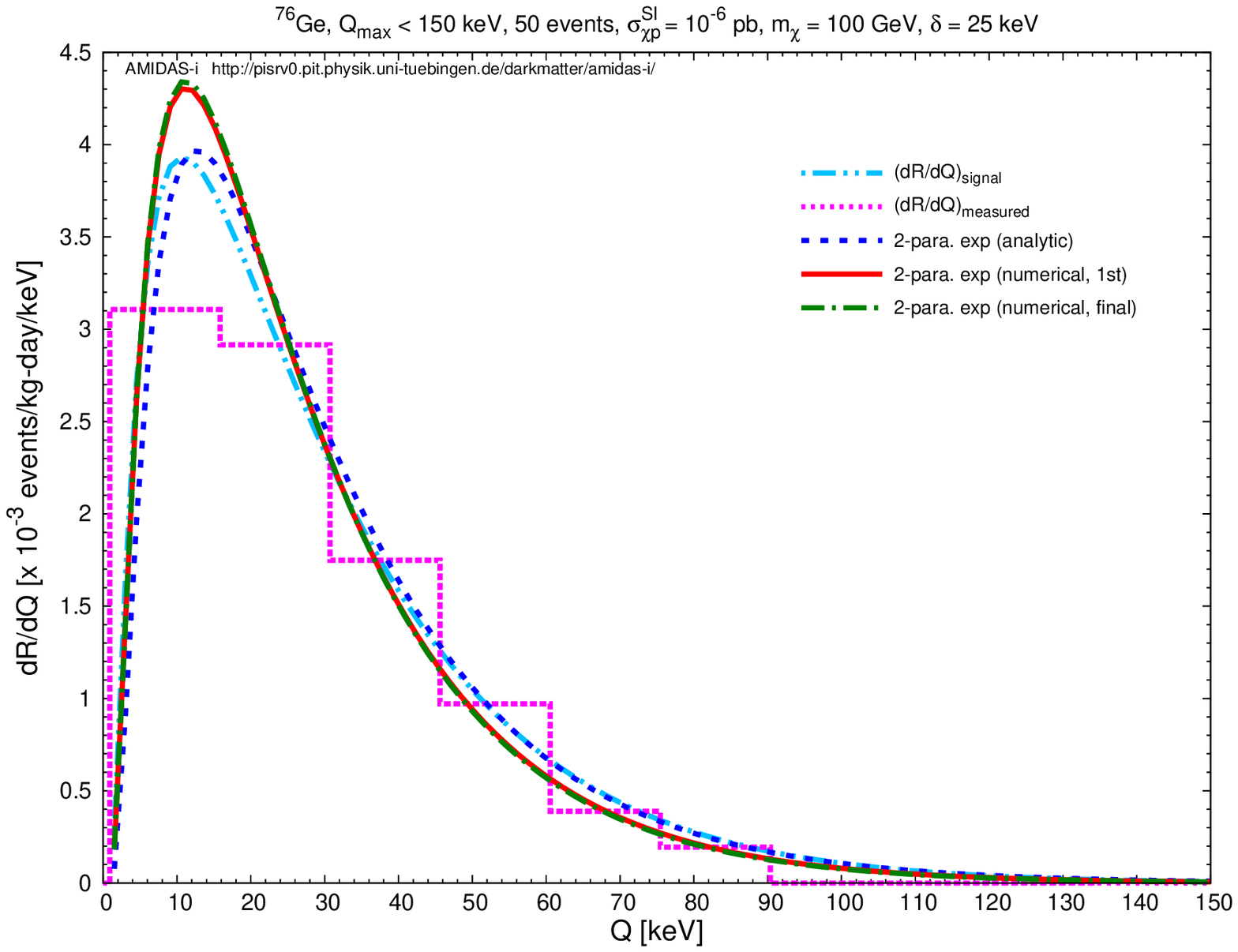}
\includegraphics[width=8.5cm]{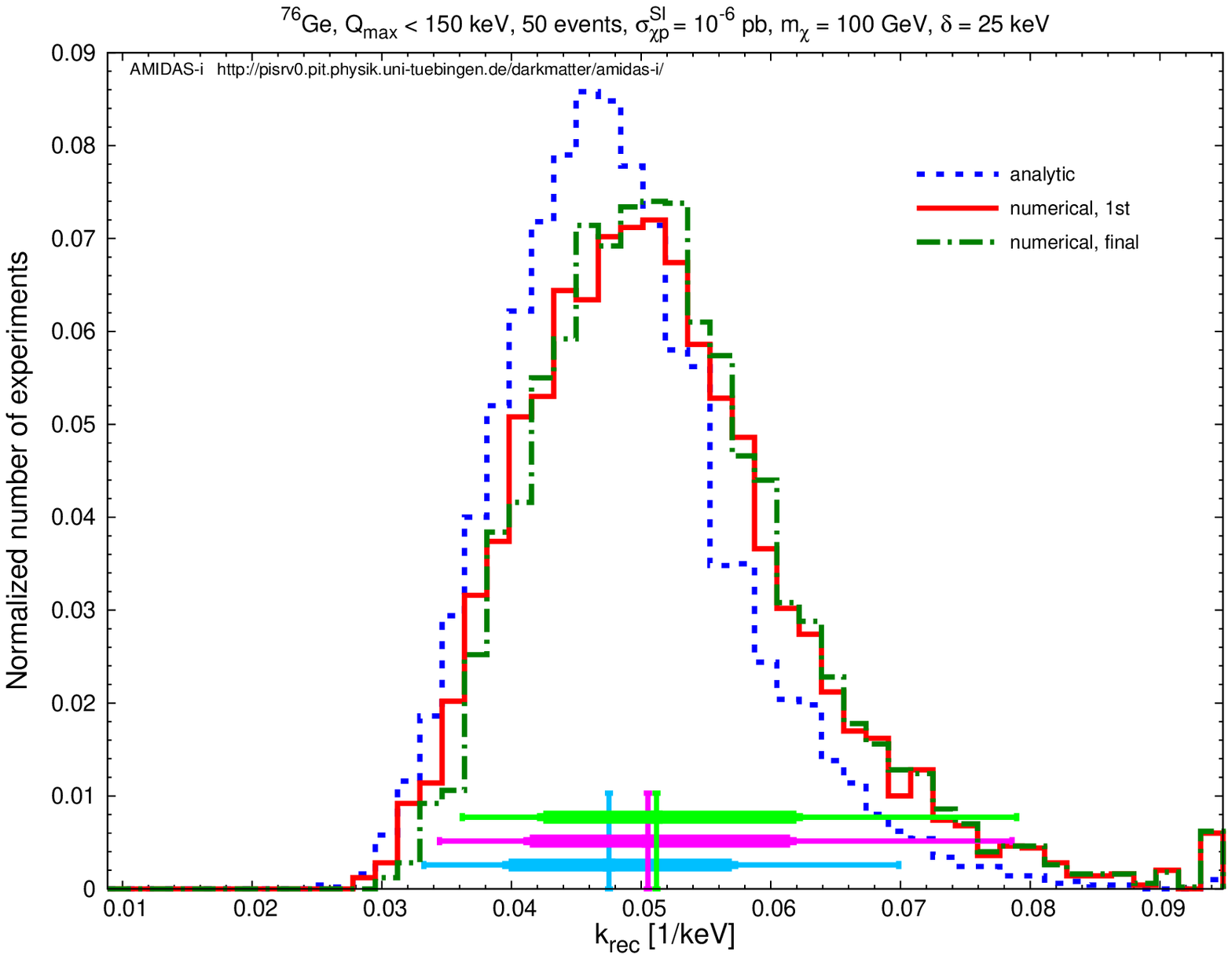}        \hspace*{-1.6cm} \\ \vspace{0.5cm}
\hspace*{-1.6cm}
\includegraphics[width=8.5cm]{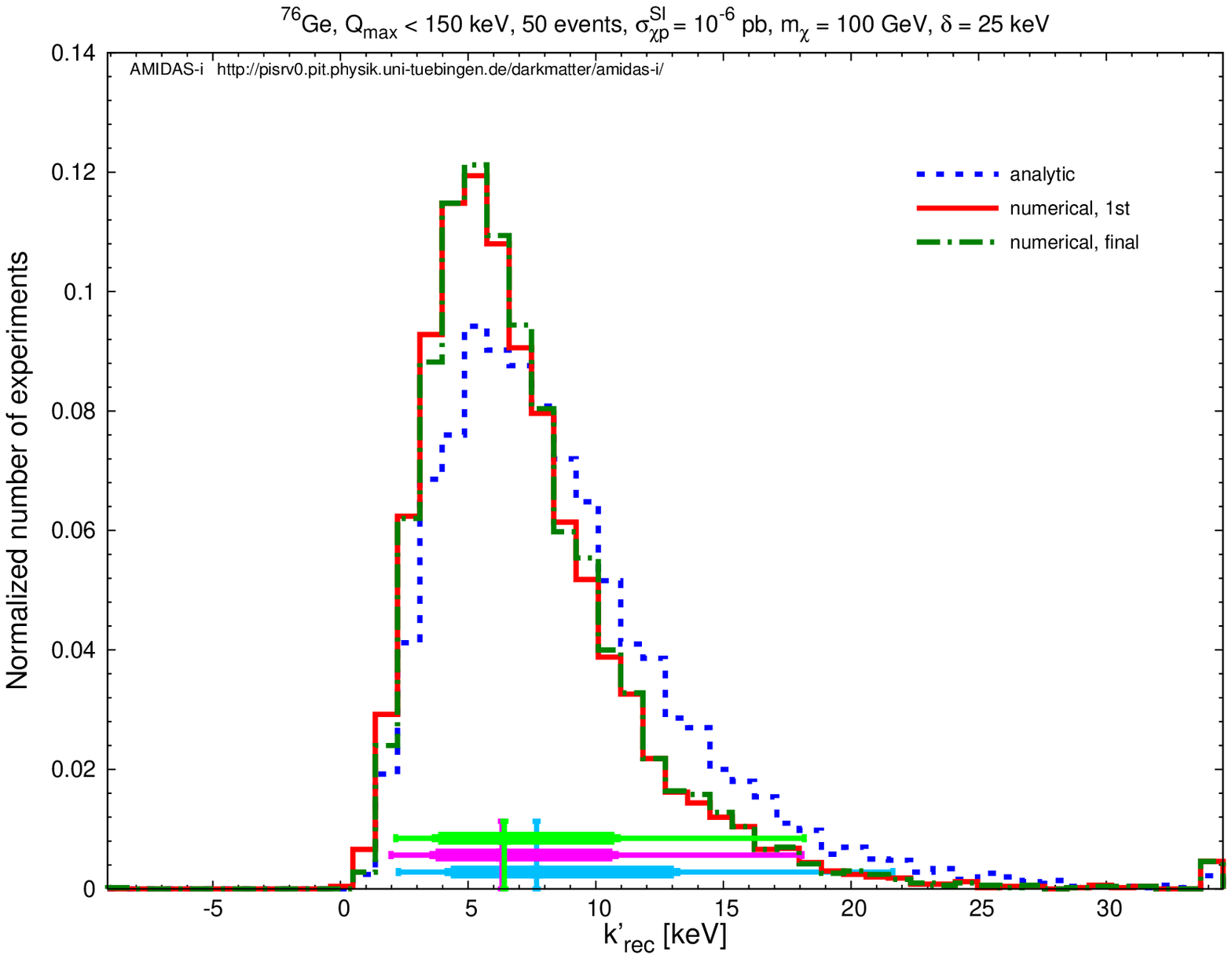}
\includegraphics[width=8.5cm]{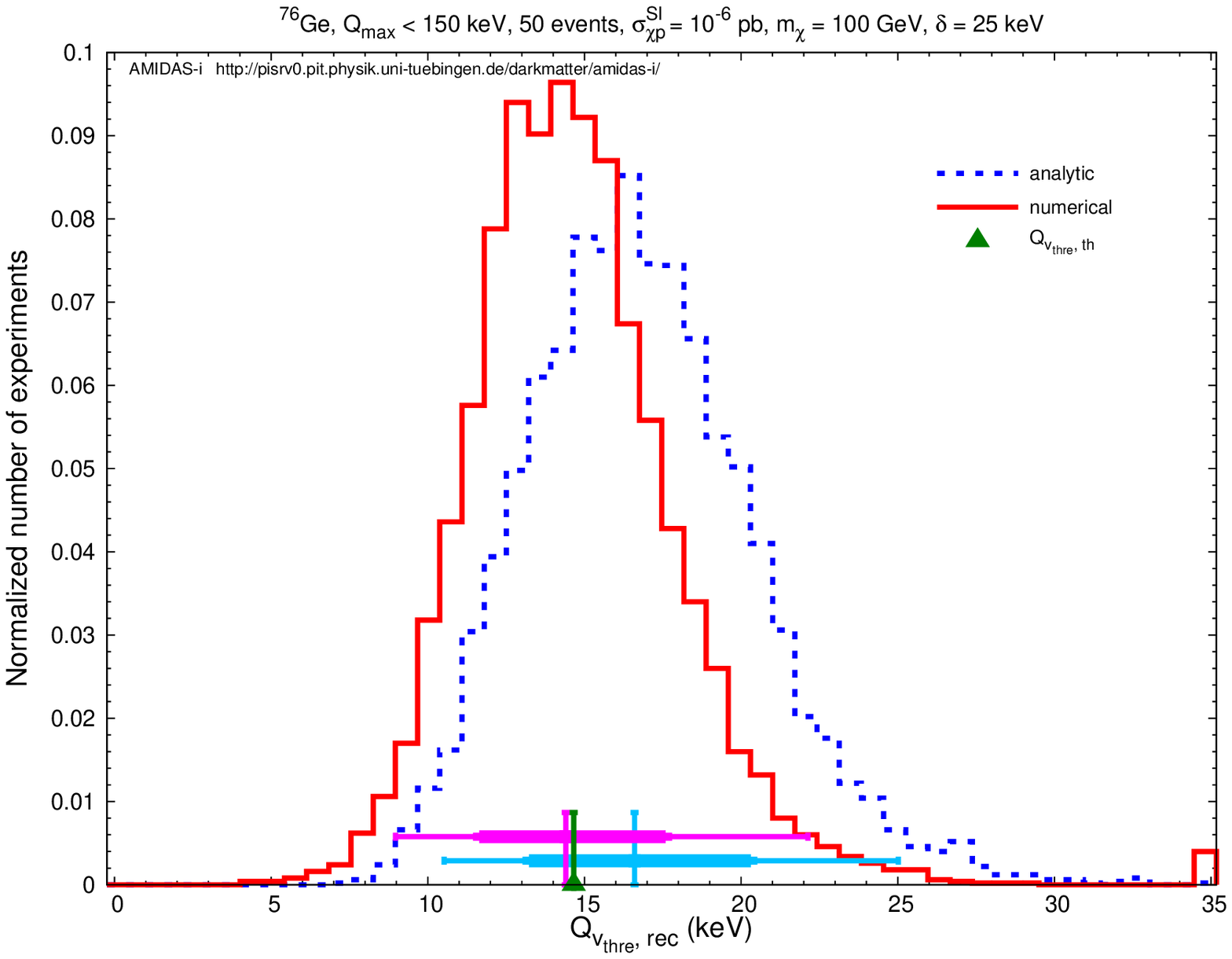}   \hspace*{-1.6cm} \\
\vspace{-0.25cm}
\end{center}
\caption{
 The measured recoil energy spectrum (dotted magenta histogram) for
 a $\rmXA{Ge}{76}$ target (top--left)
 as well as
 the distributions of the reconstructed fitting parameters
 $k$ (top--right),
 $k'$ (bottom--left)
 and the characteristic energy $\Qvthre$ (bottom--right).
 The dash--double--dotted cyan curve is
 the {\em inelastic} WIMP--nucleus scattering spectrum
 used for generating recoil events.
 The input WIMP mass and mass splitting
 are set as \mbox{$\mchi = 100$ GeV} and \mbox{$\delta = 25$ keV}.
 While the dashed blue curve is
 the 2-parameter exponential spectrum (\ref{eqn:dRdQ_in})
 with the parameters $k$ and $k'$ estimated analytically
 by Eqs.~(\ref{eqn:k_in_ana}) and (\ref{eqn:kp_in_ana})
 (top--left)
 or the distributions of the analytically estimated $k$, $k'$ and $\Qvthre$,
 the solid red and dash--dotted green curves are given
 with or for the parameters $k$ and $k'$ estimated numerically;
 the former and later show the results obtained
 from the first and final rounds of the iterative process.
 Meanwhile,
 the cyan (magenta, light green) vertical lines indicate
 the median values of the simulated results
 (corresponding to the blue, red and green histograms),
 whereas
 the horizontal thick (thin) bars show
 the 1$\~$(2)$\~\sigma$ ranges of the results.
 The green vertical line given additionally (bottom-right)
 indicates the theoretical value of $\Qvthre$
 estimated by Eq.~(\ref{eqn:Qvthre}),
 $Q_{v_{\rm thre}, {\rm th}}$.
 See the text for further details.
}
\label{fig:idRdQ-Ge76-100-025-050}
\end{figure}

 The top--left frame of Figs.~\ref{fig:idRdQ-Ge76-100-025-050} shows
 the measured recoil energy spectrum (dotted magenta histogram) for
 a $\rmXA{Ge}{76}$ target.
 The dash--double--dotted cyan curve is
 the {\em inelastic} WIMP--nucleus scattering spectrum
 used for generating recoil events.
 The input WIMP mass and mass splitting
 are set as \mbox{$\mchi = 100$ GeV} and \mbox{$\delta = 25$ keV}.
 We show here also the reconstructed recoil spectra:
 while the dashed blue curve is
 the 2-parameter exponential spectrum (\ref{eqn:dRdQ_in}) with
 the parameters $k$ and $k'$ estimated analytically
 by Eqs.~(\ref{eqn:k_in_ana}) and (\ref{eqn:kp_in_ana})%
\footnote{
 For the analytically reconstructed spectrum shown here,
 the constant $r_0$ in Eq.~(\ref{eqn:dRdQ_in})
 has been estimated by Eq.~(\ref{eqn:r0_in}),
 in which the lower and upper bounds of the integral in the denominator
 have to be set as $\Qmin = 0$ and $\Qmax = \infty$.
 Then Eq.~(\ref{eqn:r0_in}) can be rewritten as
\beq
   r_0
 = \frac{N_{\rm tot}}{\calE}
        \bbrac{2 \~ \sfrac{k'}{k} \~ K_1\abrac{2 \sqrt{k} \sqrt{k'}}}^{-1}
\~,
\label{eqn:r0_in_inf}
\eeq
 where $K_1(x)$ is the modified Bessel function of the second kind of order 1.
 Detailed derivation can be found in Appendix A.1.
},
 the solid red and dash--dotted green curves are given
 with the parameters $k$ and $k'$ estimated
 by the numerical method introduced in Sec.~2.3;
 the former and later show the results obtained
 from the first and final rounds of the iterative process%
\footnote{
 Note that,
 as we will discuss later,
 the results given by our iterative numerical procedure might diverge
 (deviate larger and larger from the theoretical values) round by round.
 Thus,
 once the results given by the $n$th run
 is too far away from the $(n - 1)$th run,
 our program will take the results of $(n - 1)$th run
 as the final results.
 This means that
 in some of the 5,000 simulated experiments
 the results of the final round are just
 those of the first or the second rounds.

 Note also that,
 because of the same divergency problem
 and of that
 the results given by the later round might not be
 much better than those of the first round,
 for the reconstructions of $\Qvthre$
 as well as the WIMP mass and the mass splitting
 shown later,
 we use only the results ($k_{\rm num}$ and $k'_{\rm num}$)
 given by the first round.
}.

\begin{figure}[t!]
\begin{center}
\hspace*{-1.6cm}
\includegraphics[width=4.8cm]{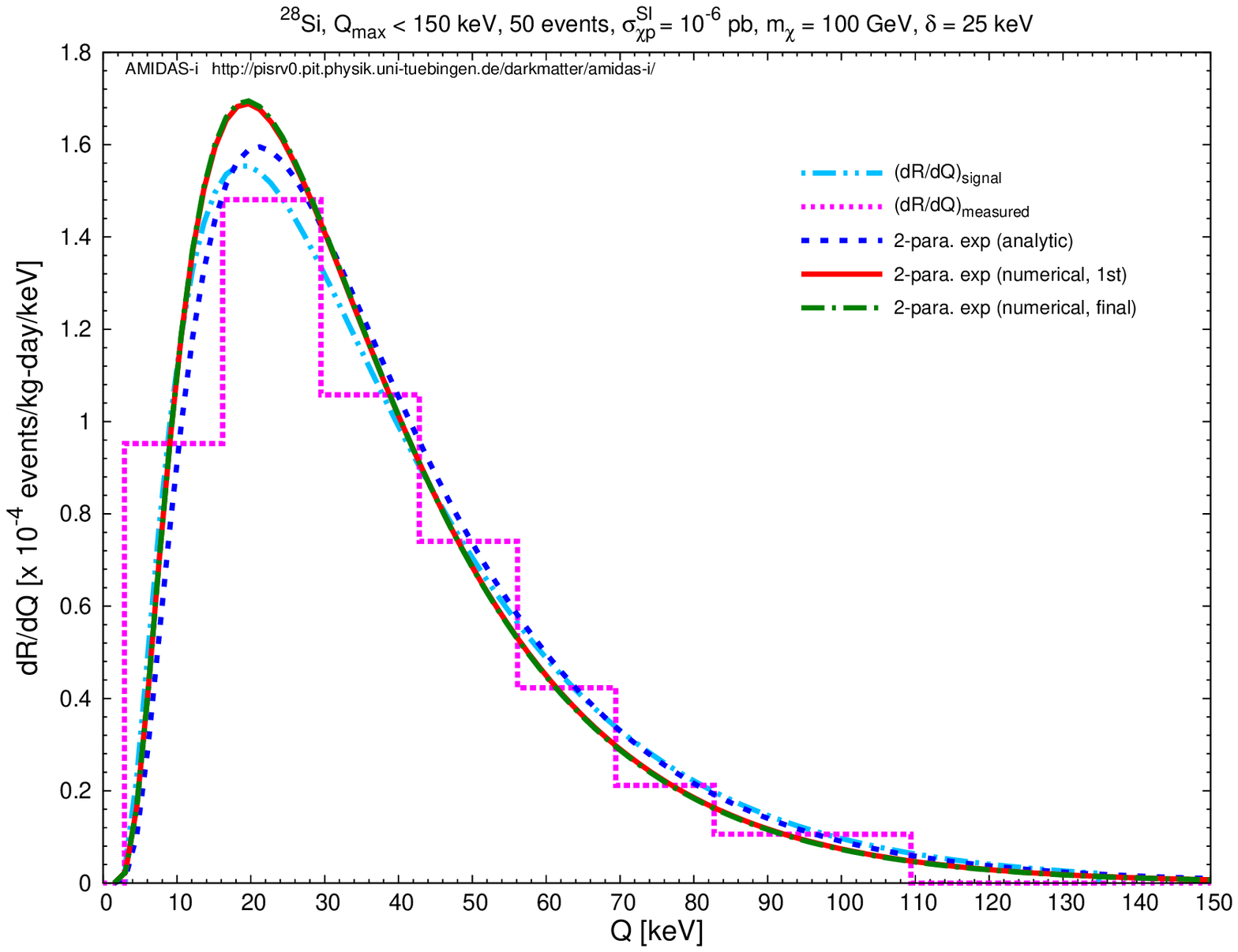}      \hspace*{-0.5cm}
\includegraphics[width=4.8cm]{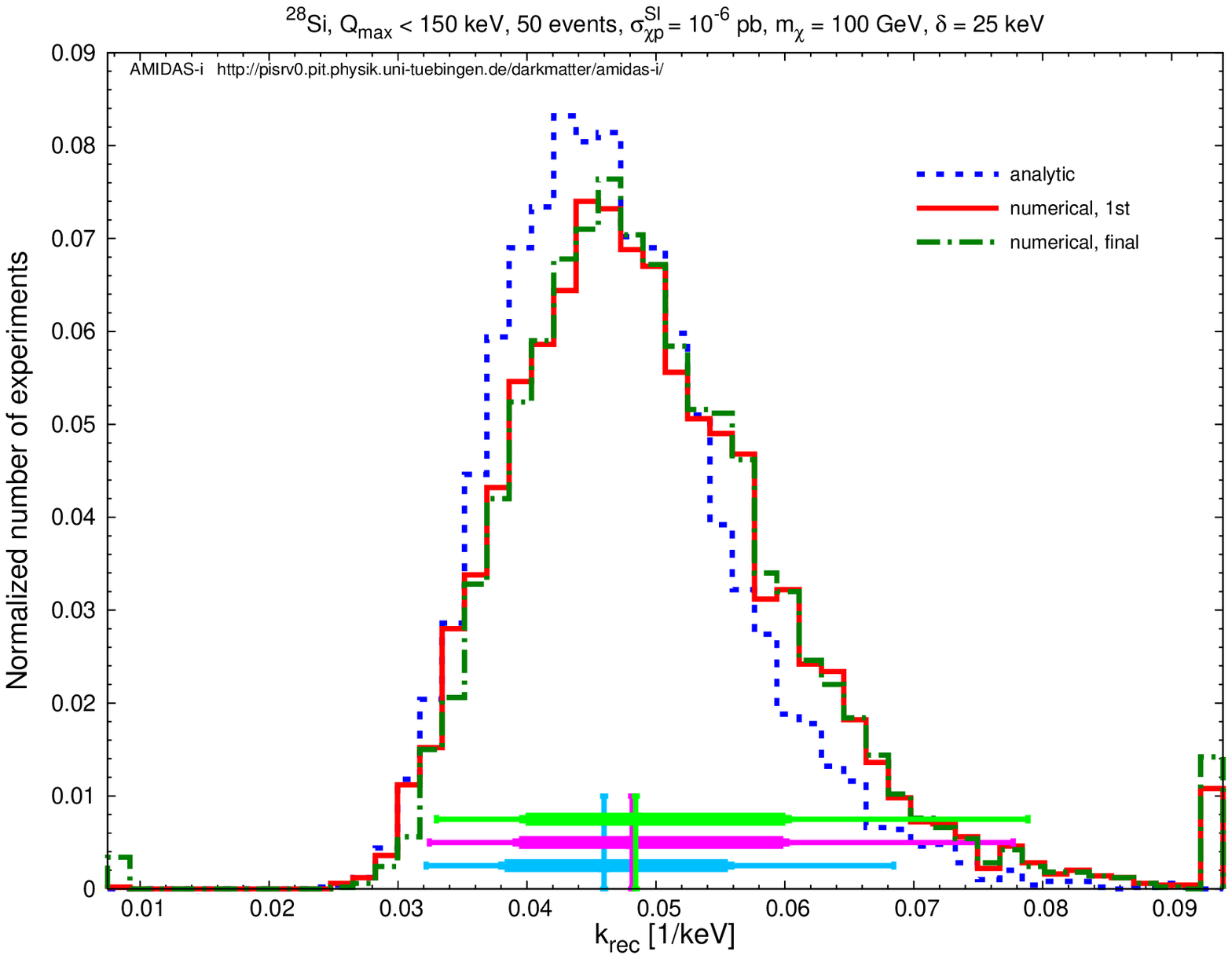}      \hspace*{-0.5cm}
\includegraphics[width=4.8cm]{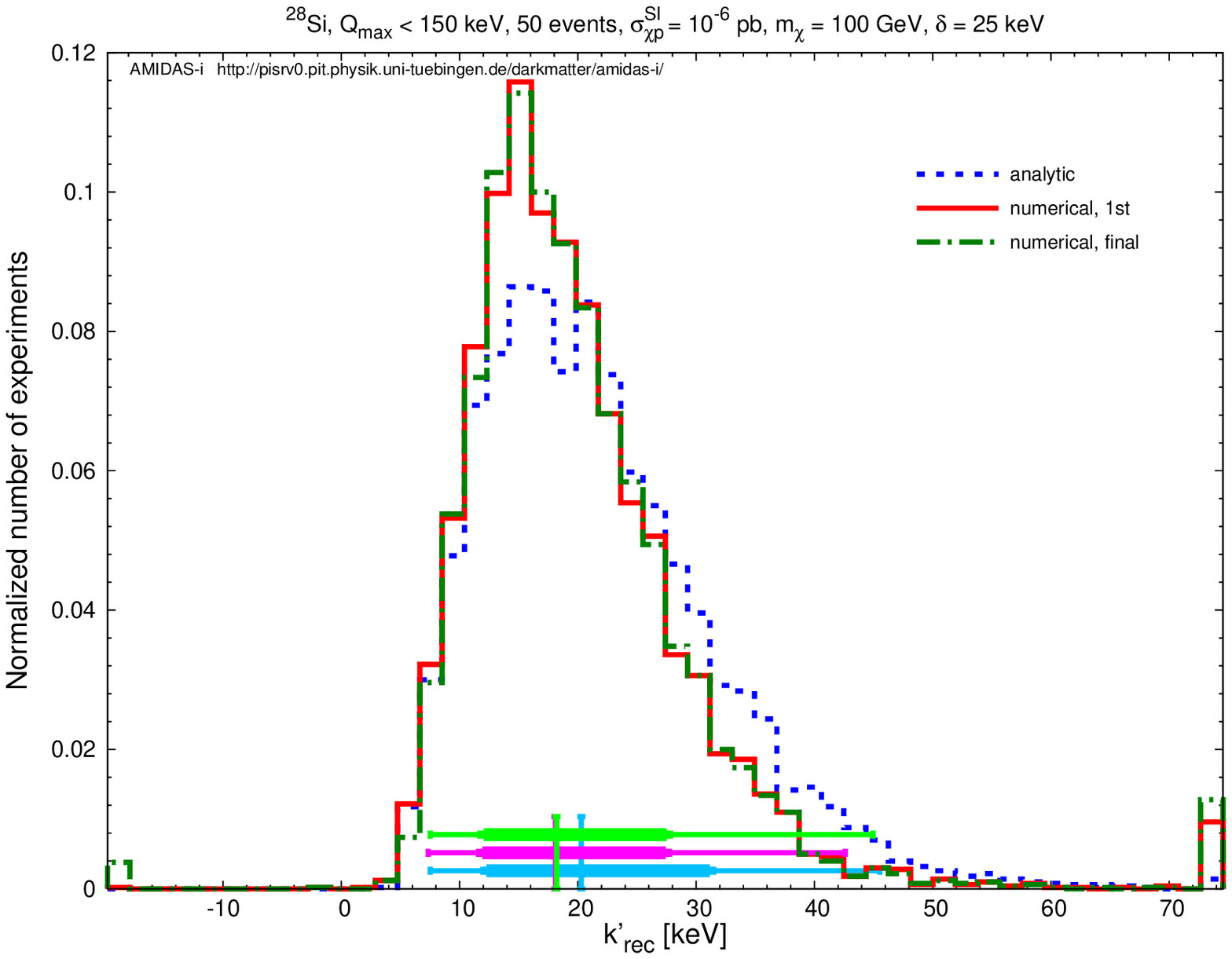}     \hspace*{-0.5cm}
\includegraphics[width=4.8cm]{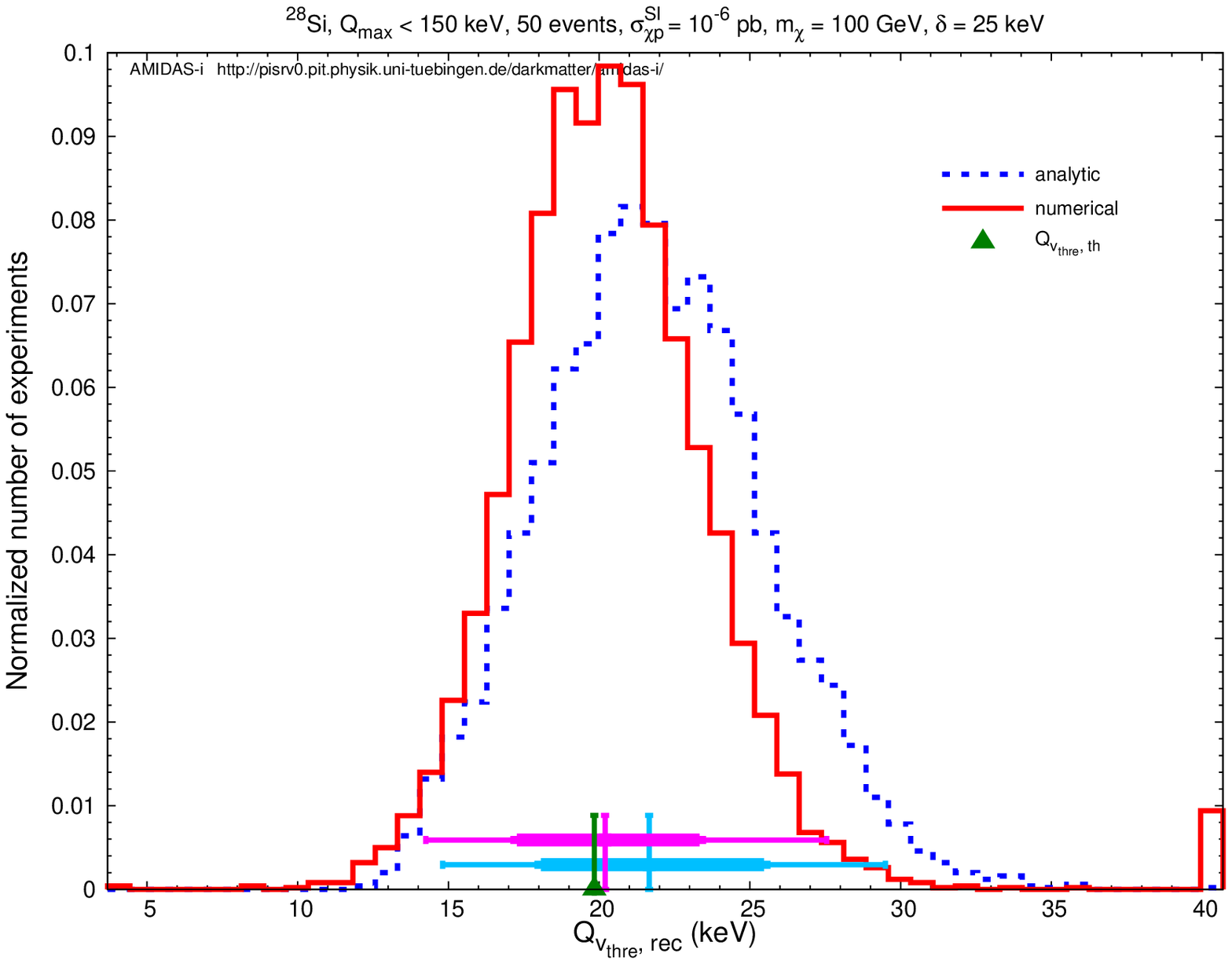} \hspace*{-1.6cm} \\
 ($\rmXA{Si}{28}$) \\
\vspace{0.75cm}
\hspace*{-1.6cm}
\includegraphics[width=4.8cm]{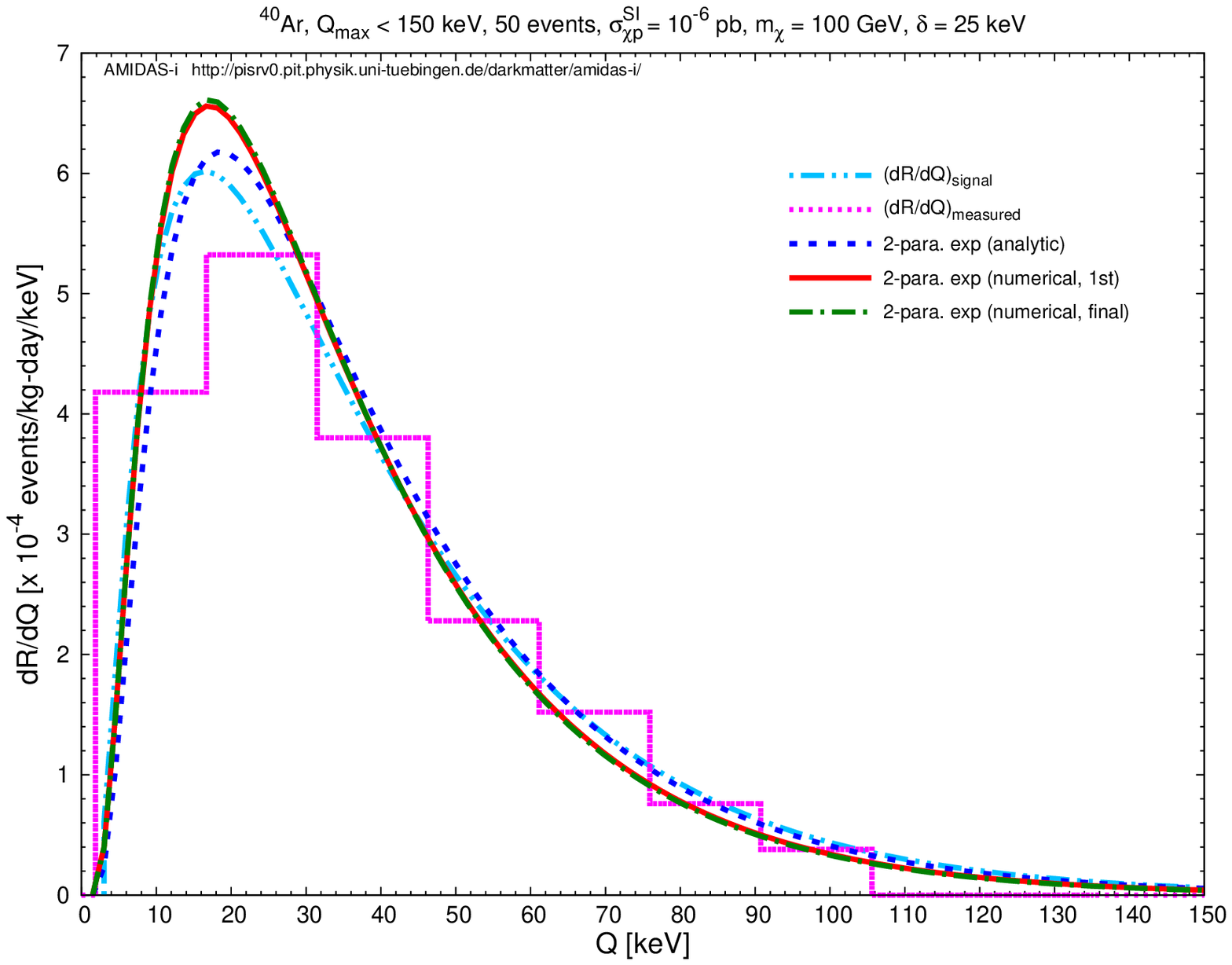}      \hspace*{-0.5cm}
\includegraphics[width=4.8cm]{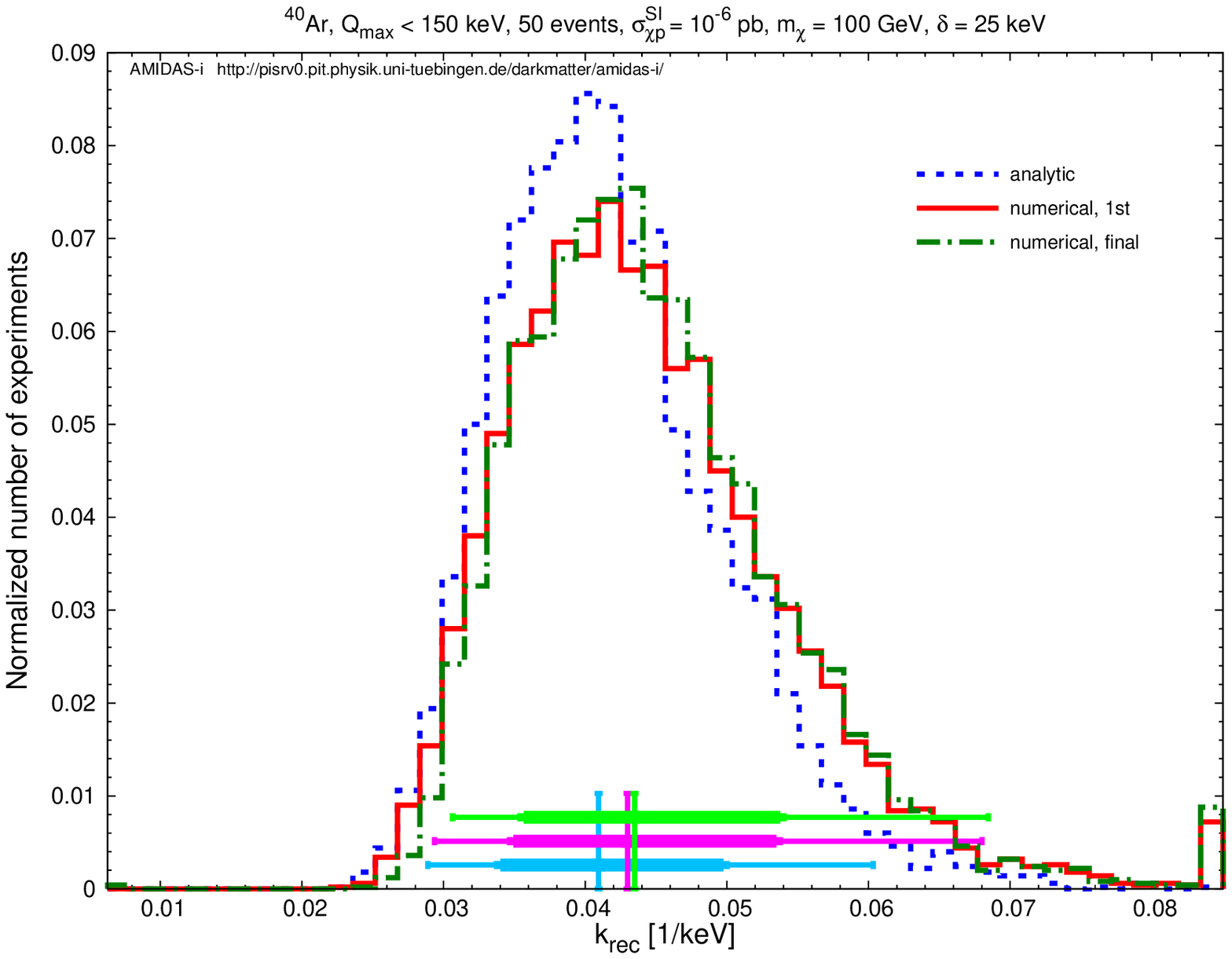}      \hspace*{-0.5cm}
\includegraphics[width=4.8cm]{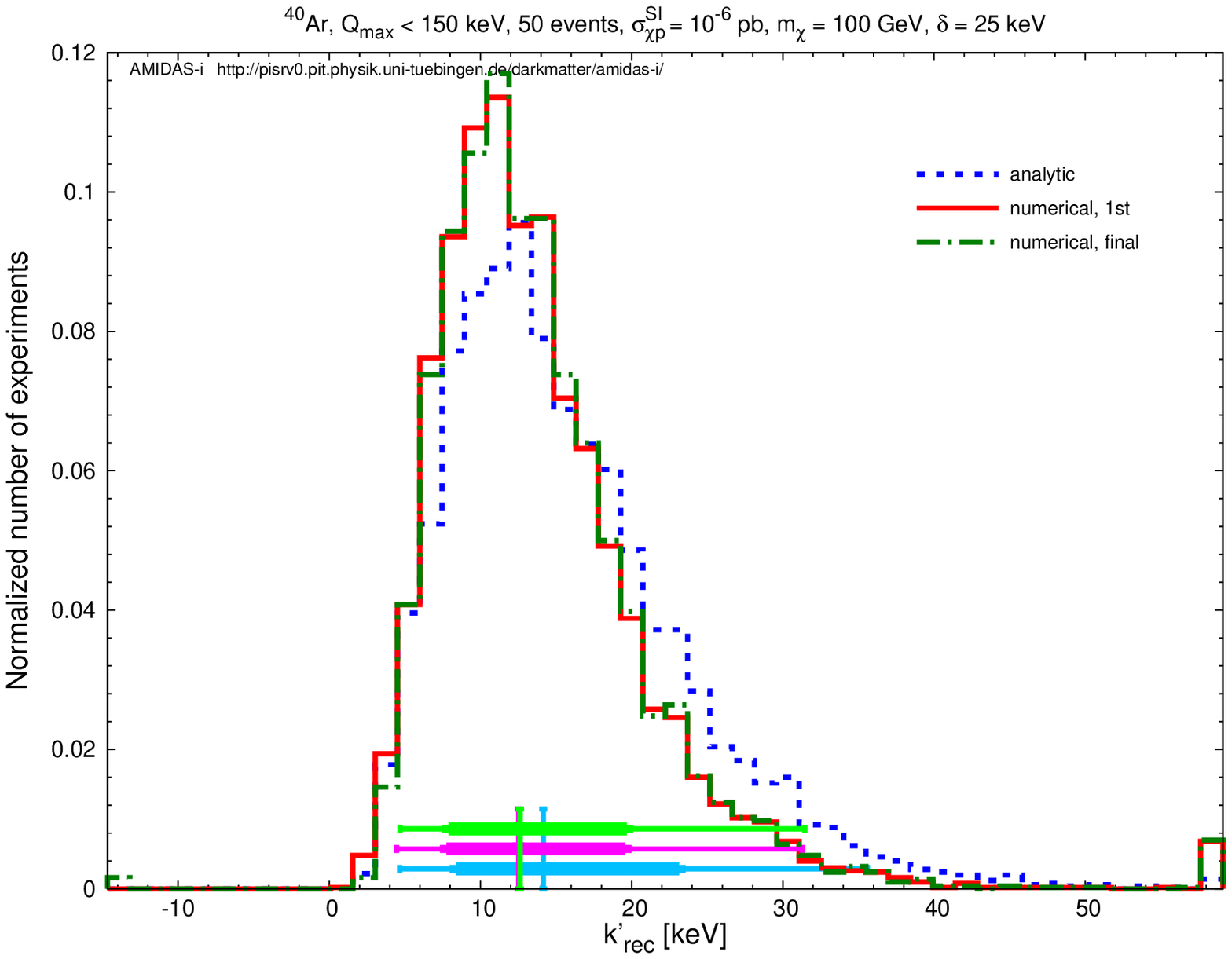}     \hspace*{-0.5cm}
\includegraphics[width=4.8cm]{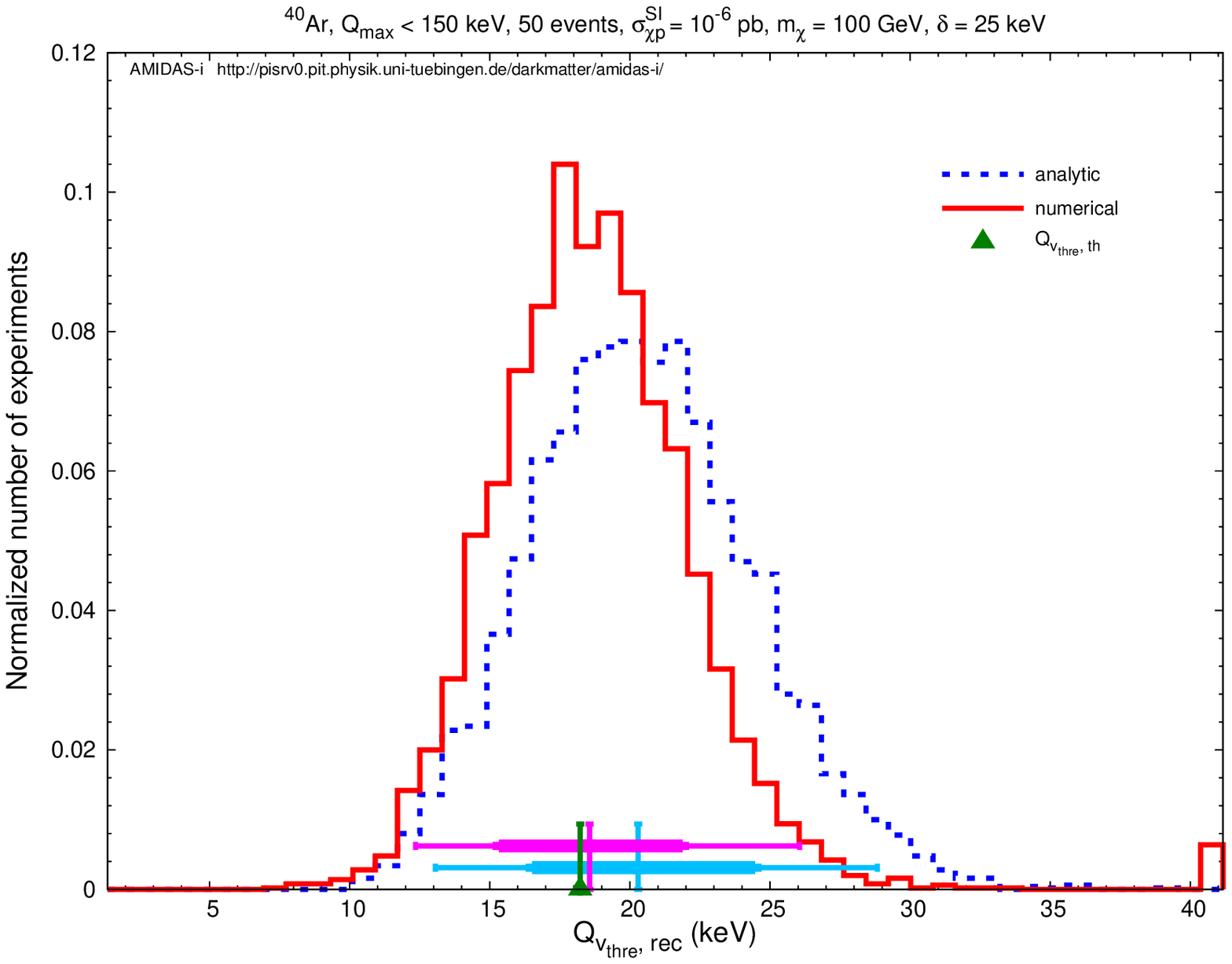} \hspace*{-1.6cm} \\
 ($\rmXA{Ar}{40}$) \\
\vspace{0.75cm}
\hspace*{-1.6cm}
\includegraphics[width=4.8cm]{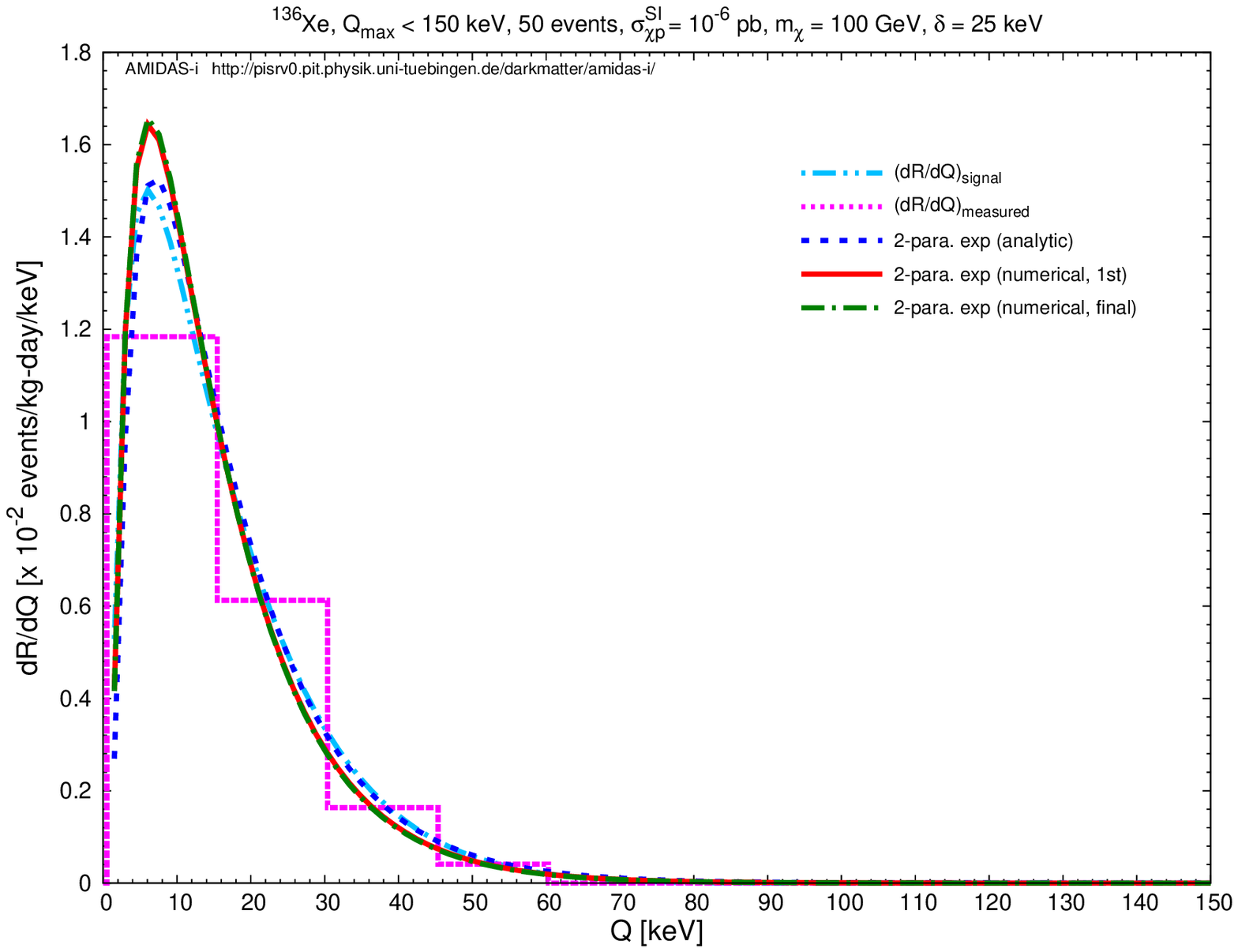}      \hspace*{-0.5cm}
\includegraphics[width=4.8cm]{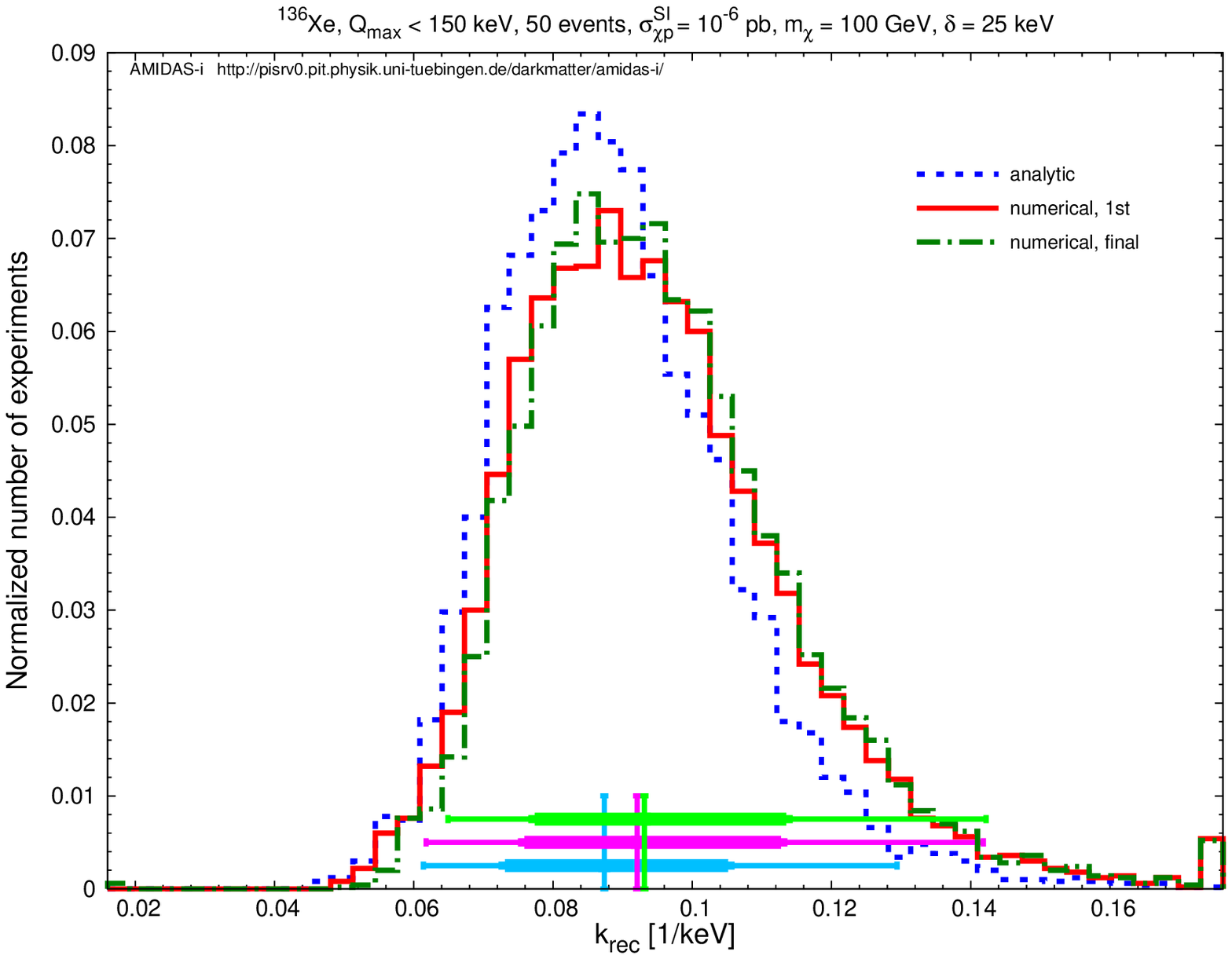}      \hspace*{-0.5cm}
\includegraphics[width=4.8cm]{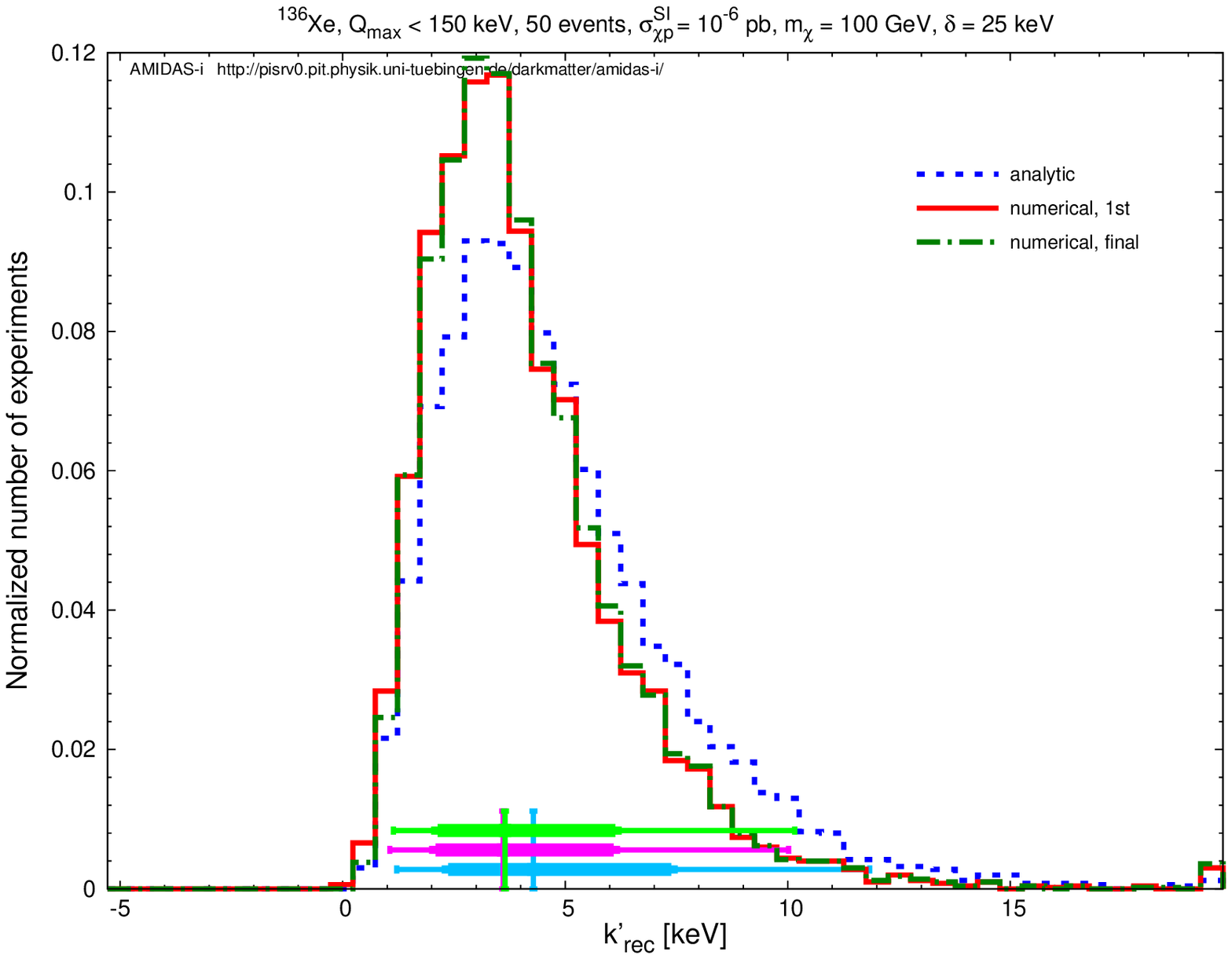}     \hspace*{-0.5cm}
\includegraphics[width=4.8cm]{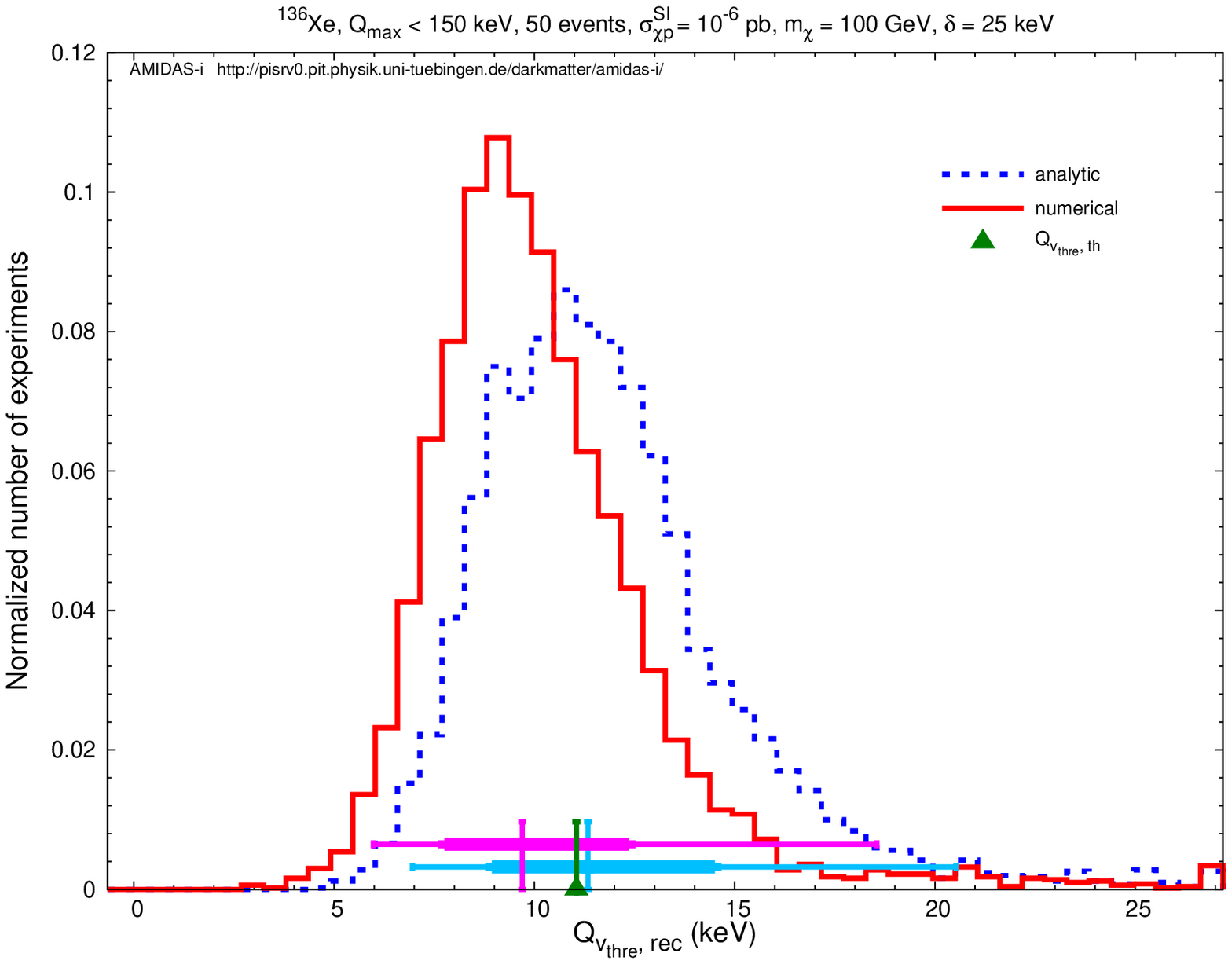} \hspace*{-1.6cm} \\
 ($\rmXA{Xe}{136}$) \\
\vspace{-0.25cm}
\end{center}
\caption{
 As in Figs.~\ref{fig:idRdQ-Ge76-100-025-050},
 except that
 $\rmXA{Si}{28}$  (top),
 $\rmXA{Ar}{40}$  (middle) and
 $\rmXA{Xe}{136}$ (bottom)
 have been used as target nuclei.
}
\label{fig:idRdQ-100-025-050}
\end{figure}
\begin{figure}[t!]
\begin{center}
\hspace*{-1.6cm}
\includegraphics[width=4.8cm]{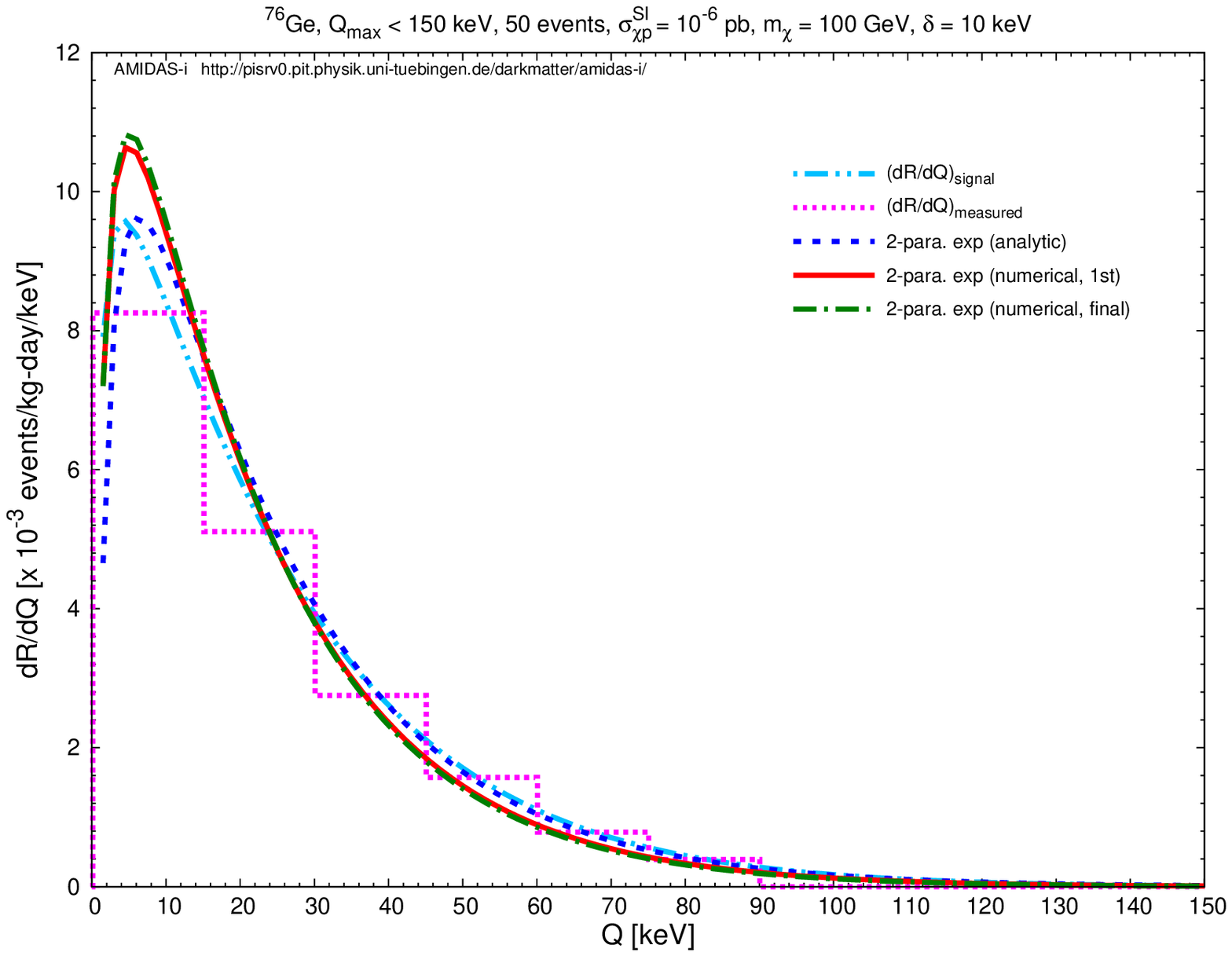}      \hspace*{-0.5cm}
\includegraphics[width=4.8cm]{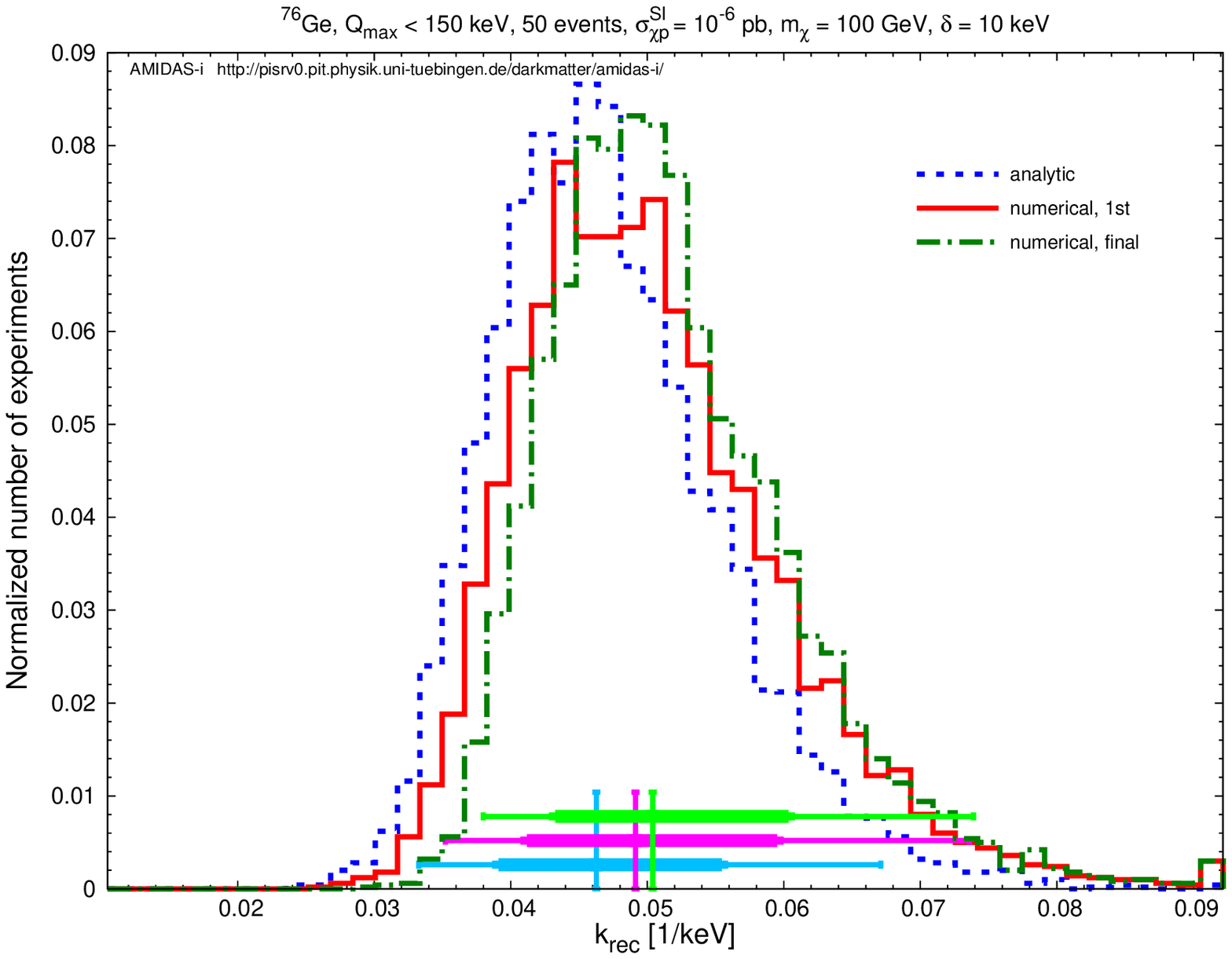}      \hspace*{-0.5cm}
\includegraphics[width=4.8cm]{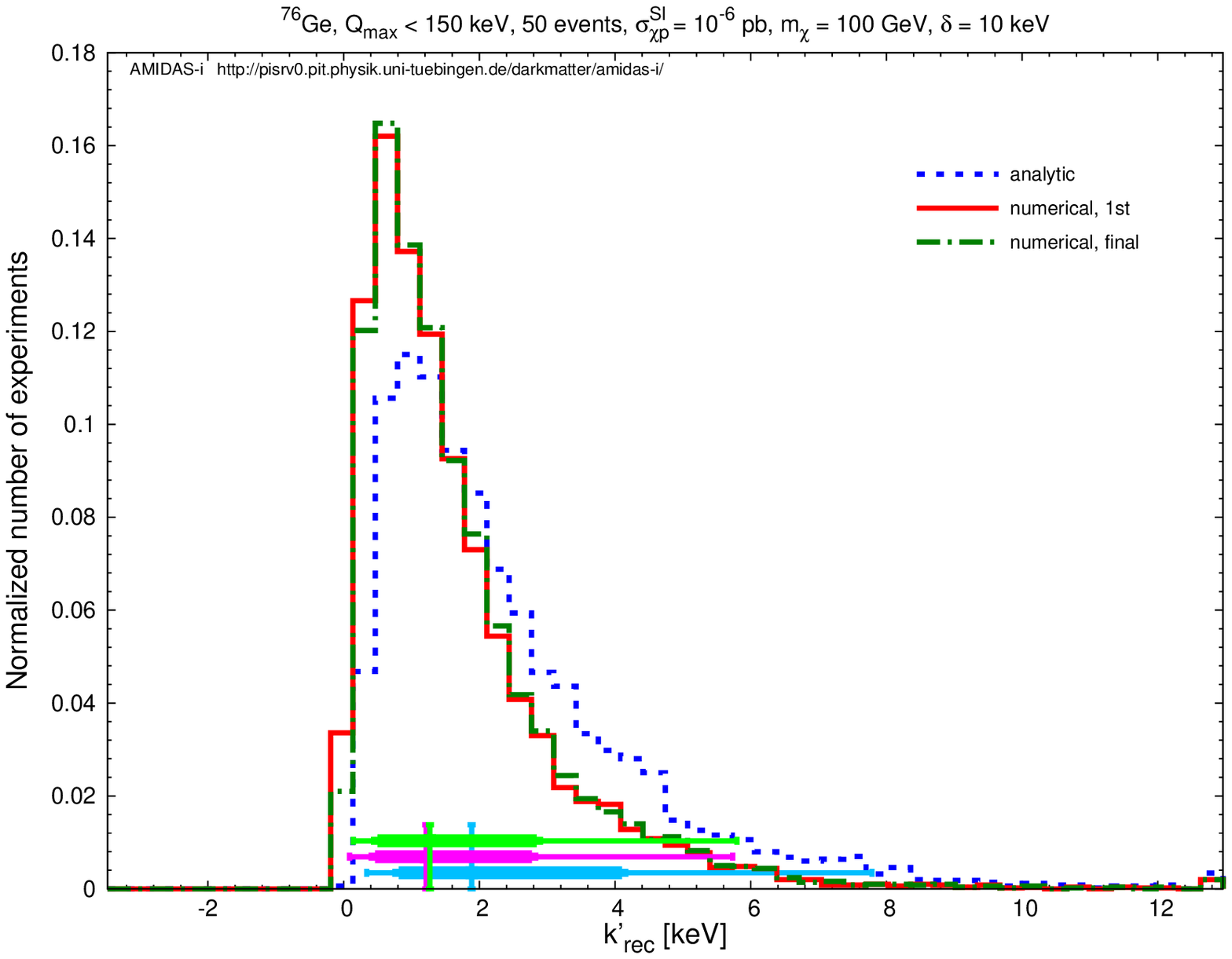}     \hspace*{-0.5cm}
\includegraphics[width=4.8cm]{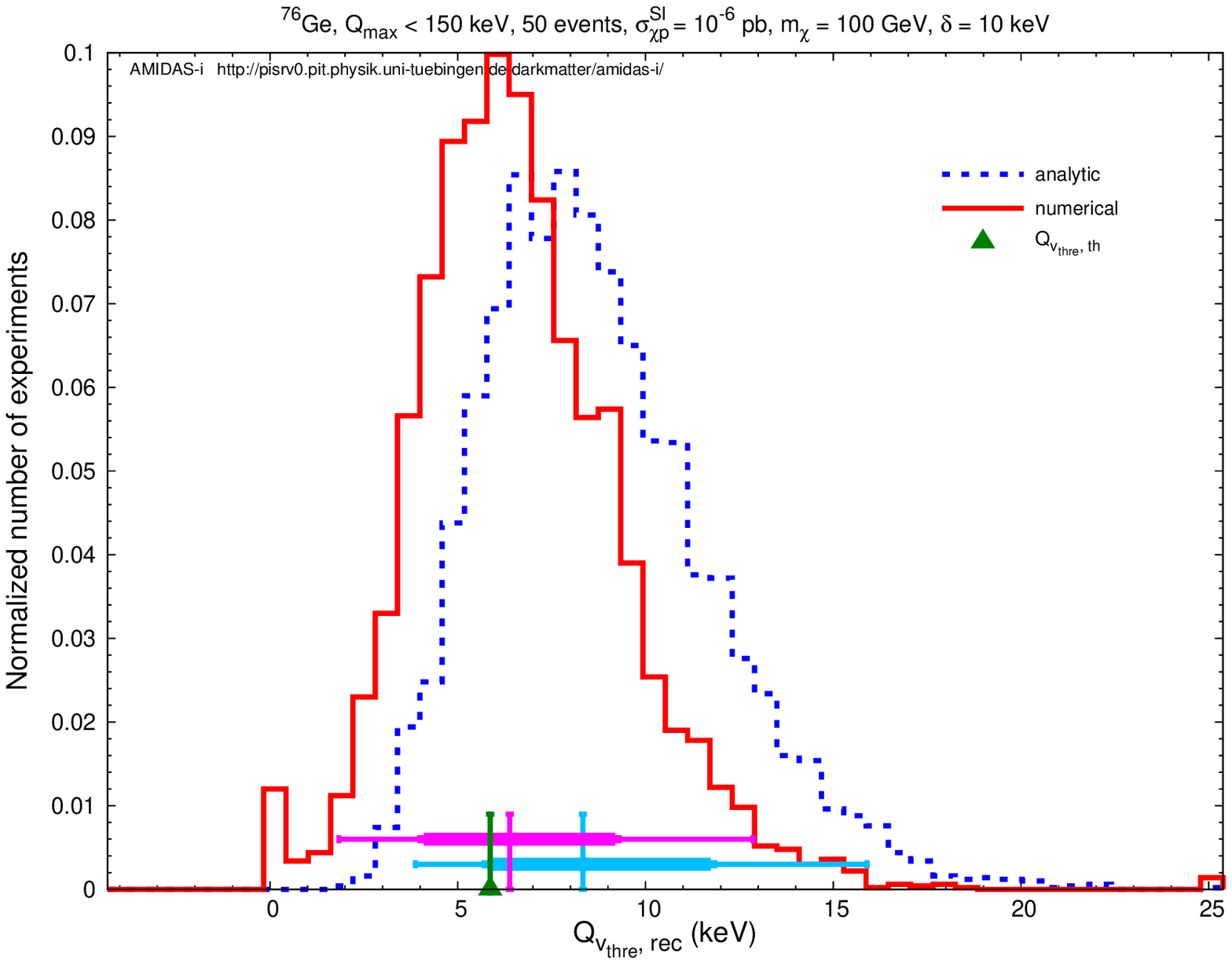} \hspace*{-1.6cm} \\
 ($\rmXA{Ge}{76}$) \\
\vspace{0.75cm}
\hspace*{-1.6cm}
\includegraphics[width=4.8cm]{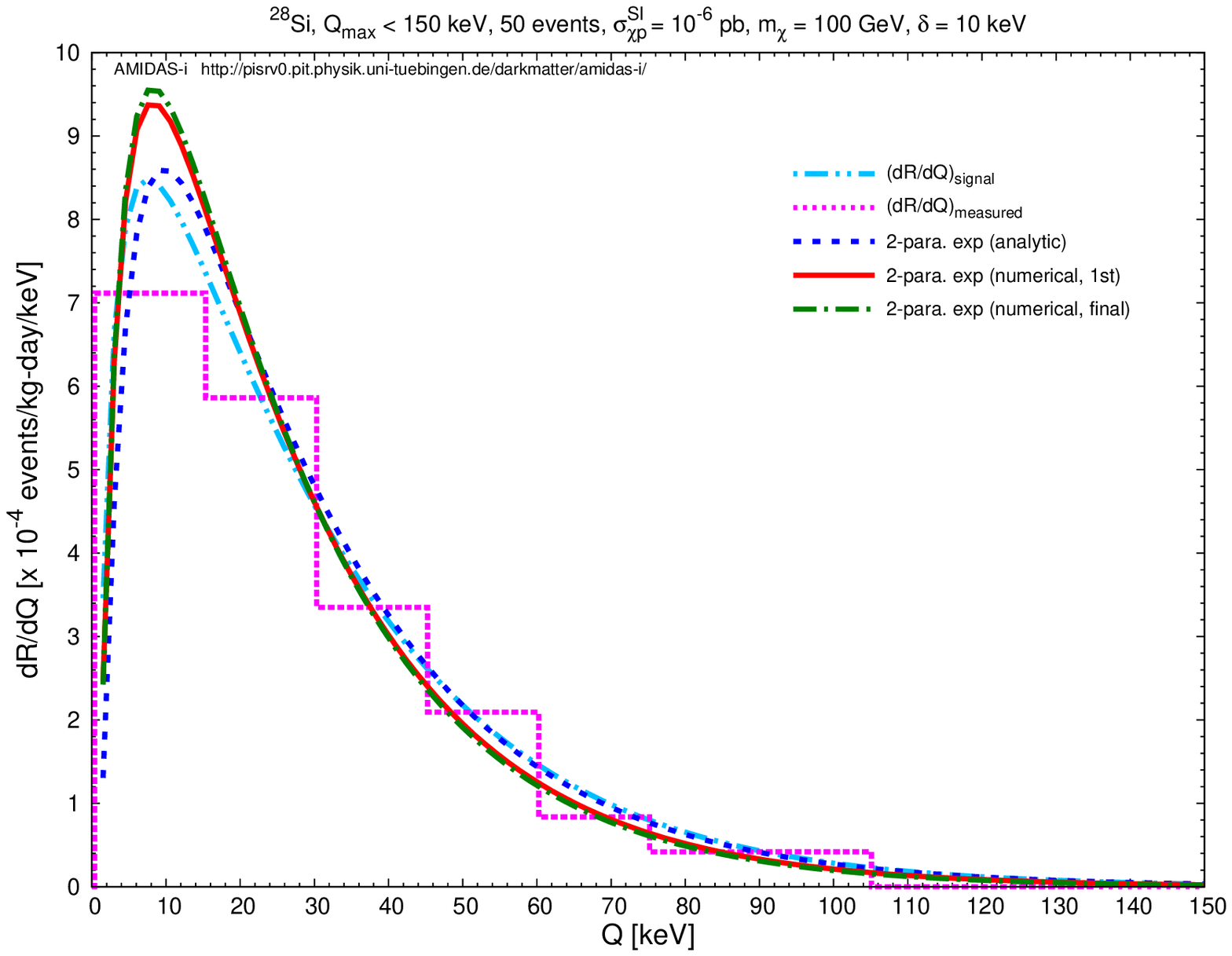}      \hspace*{-0.5cm}
\includegraphics[width=4.8cm]{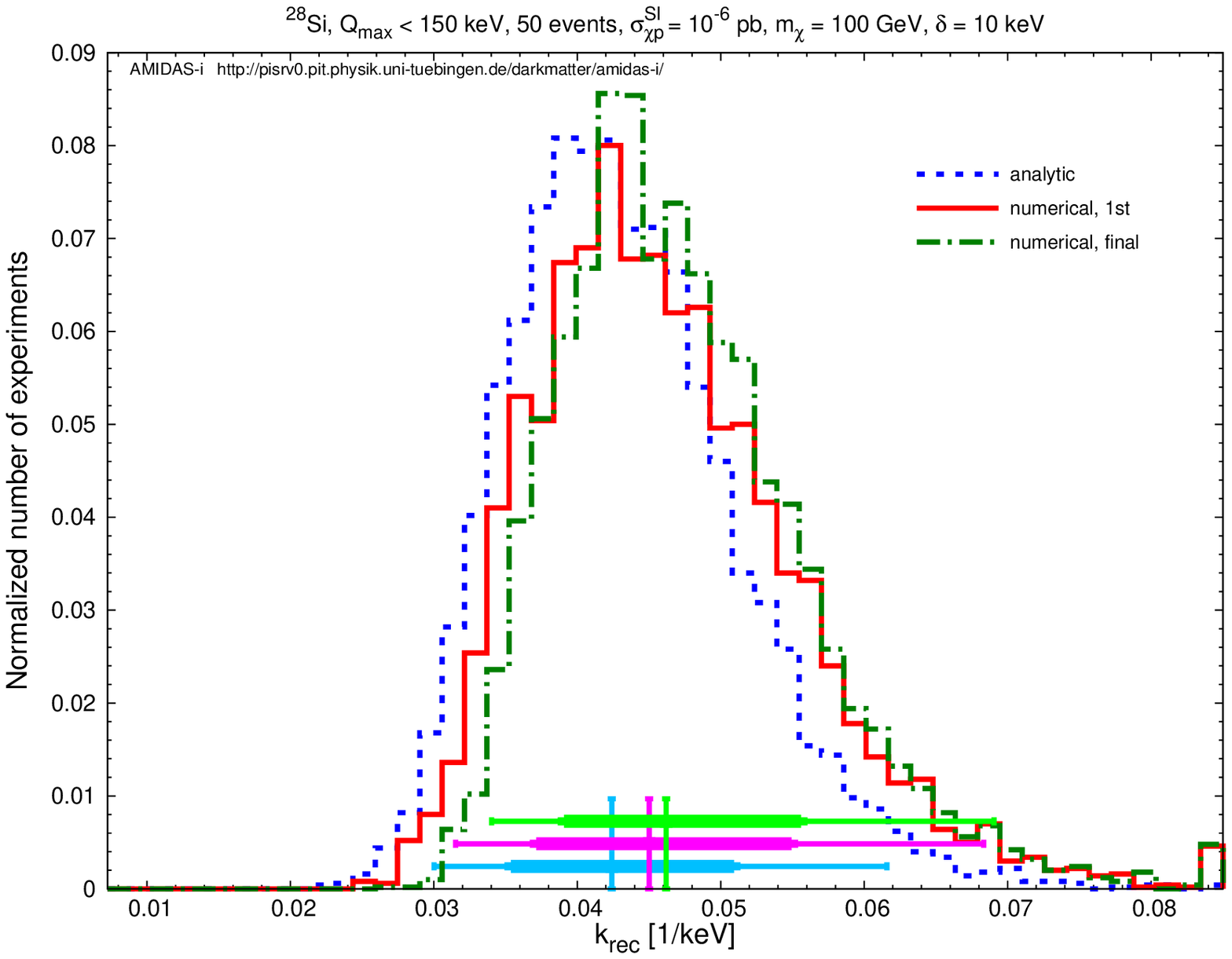}      \hspace*{-0.5cm}
\includegraphics[width=4.8cm]{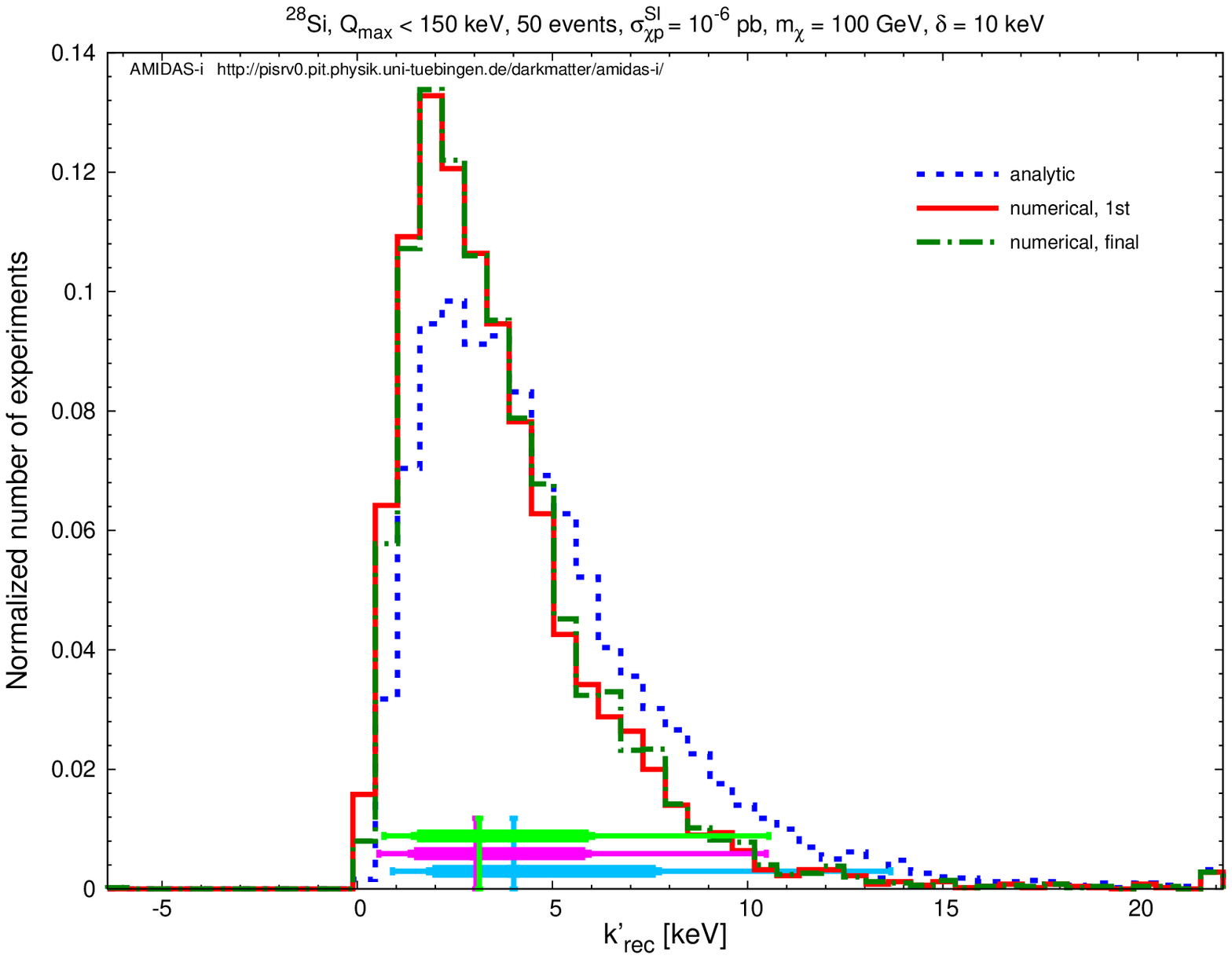}     \hspace*{-0.5cm}
\includegraphics[width=4.8cm]{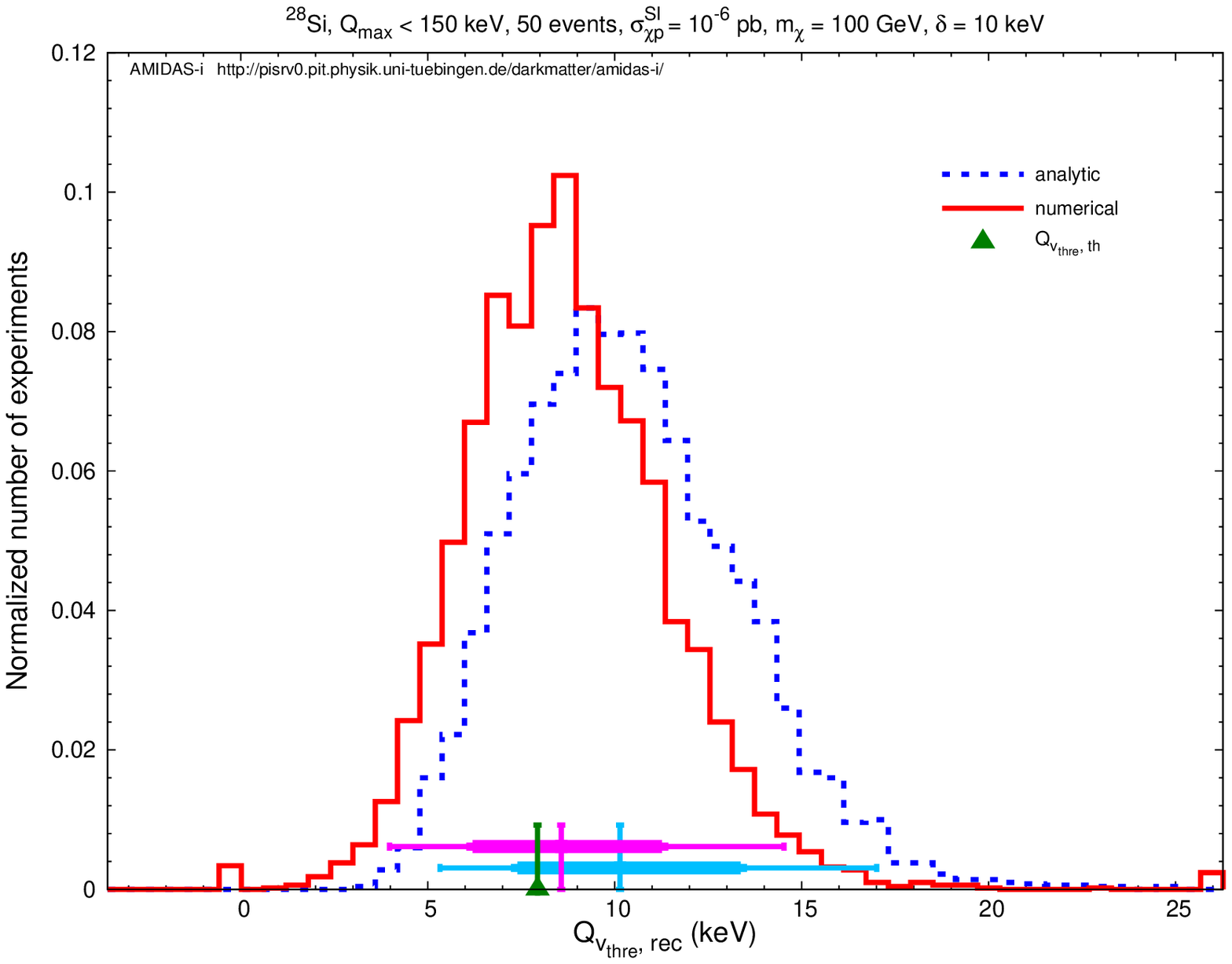} \hspace*{-1.6cm} \\
 ($\rmXA{Si}{28}$) \\
\vspace{0.75cm}
\hspace*{-1.6cm}
\includegraphics[width=4.8cm]{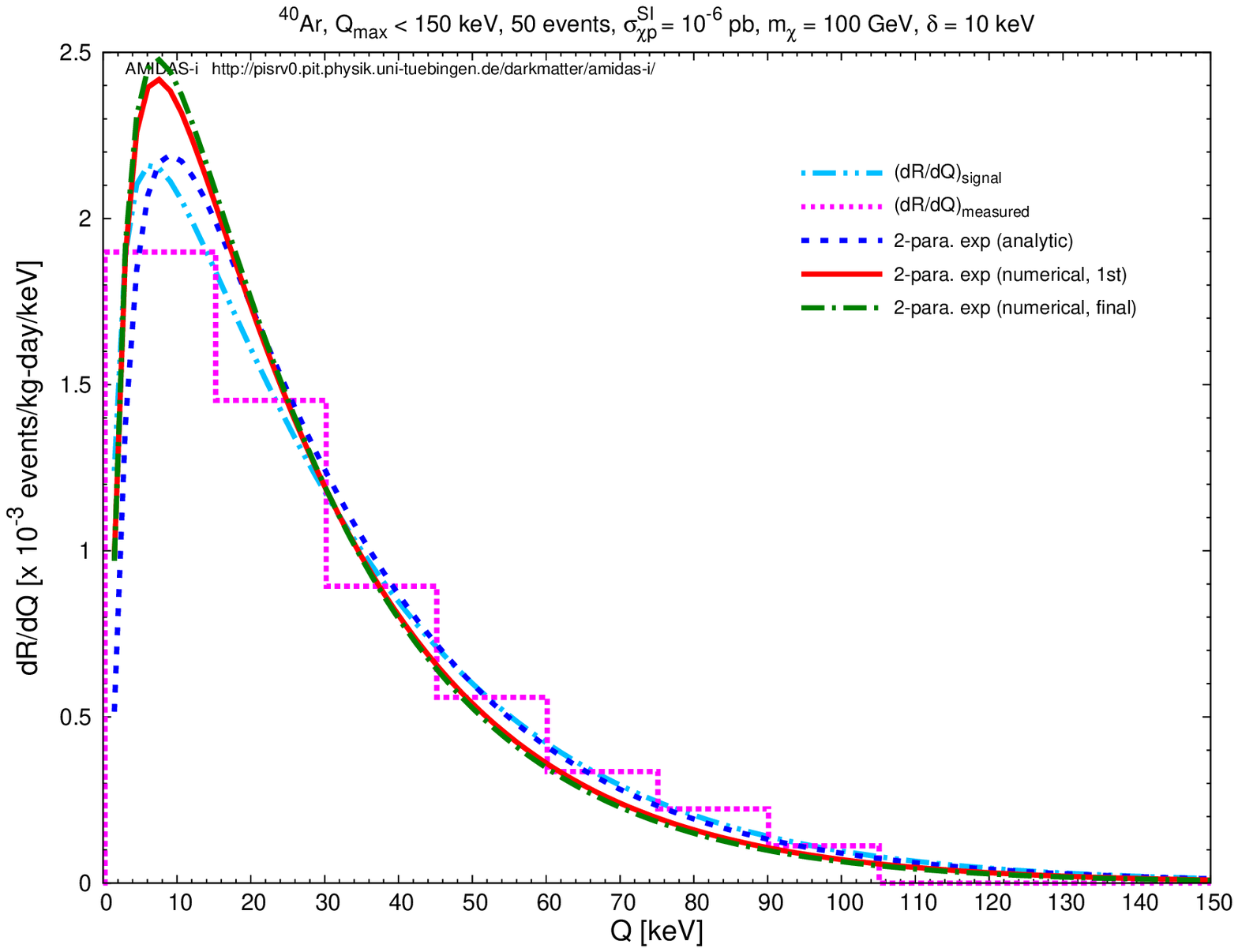}      \hspace*{-0.5cm}
\includegraphics[width=4.8cm]{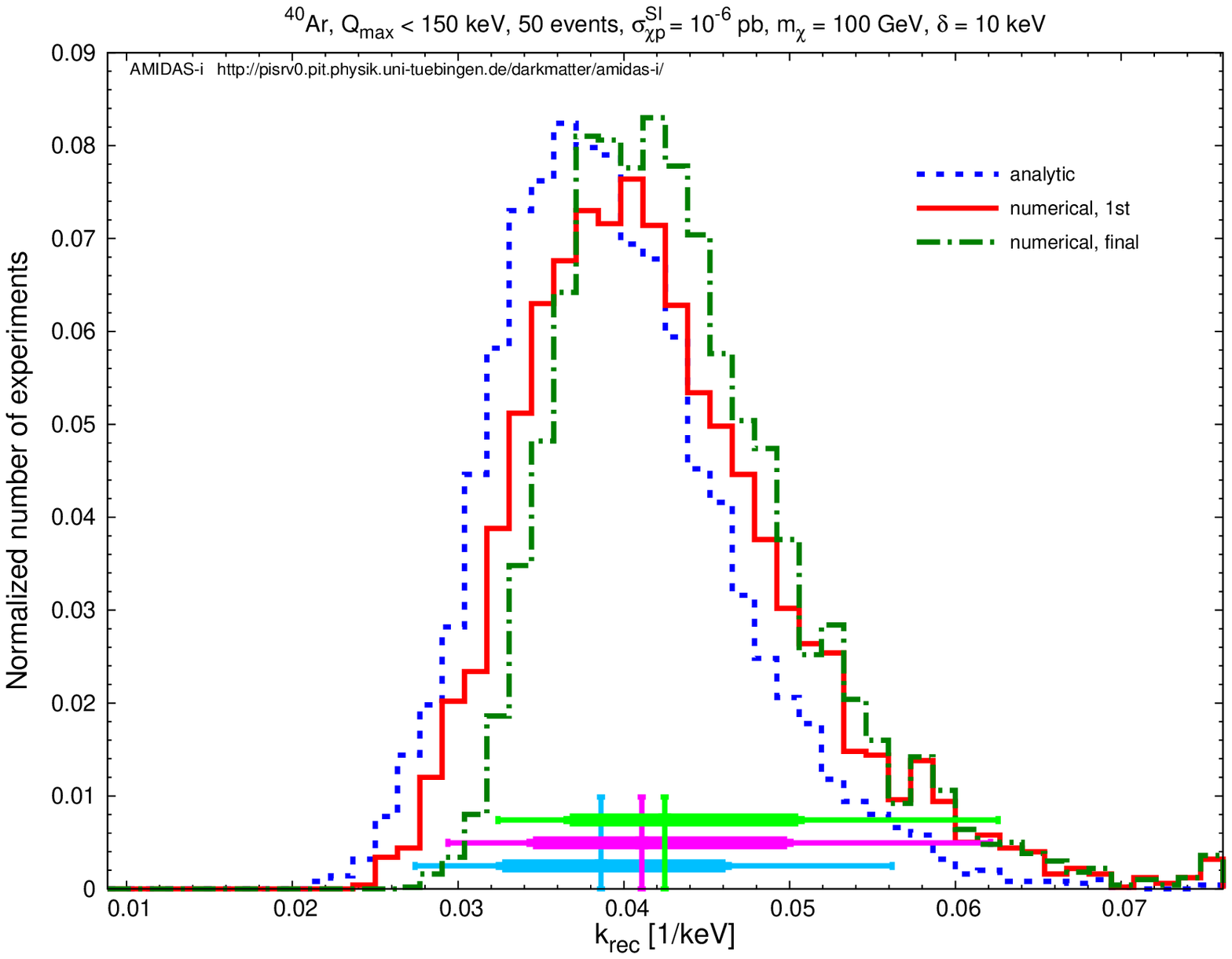}      \hspace*{-0.5cm}
\includegraphics[width=4.8cm]{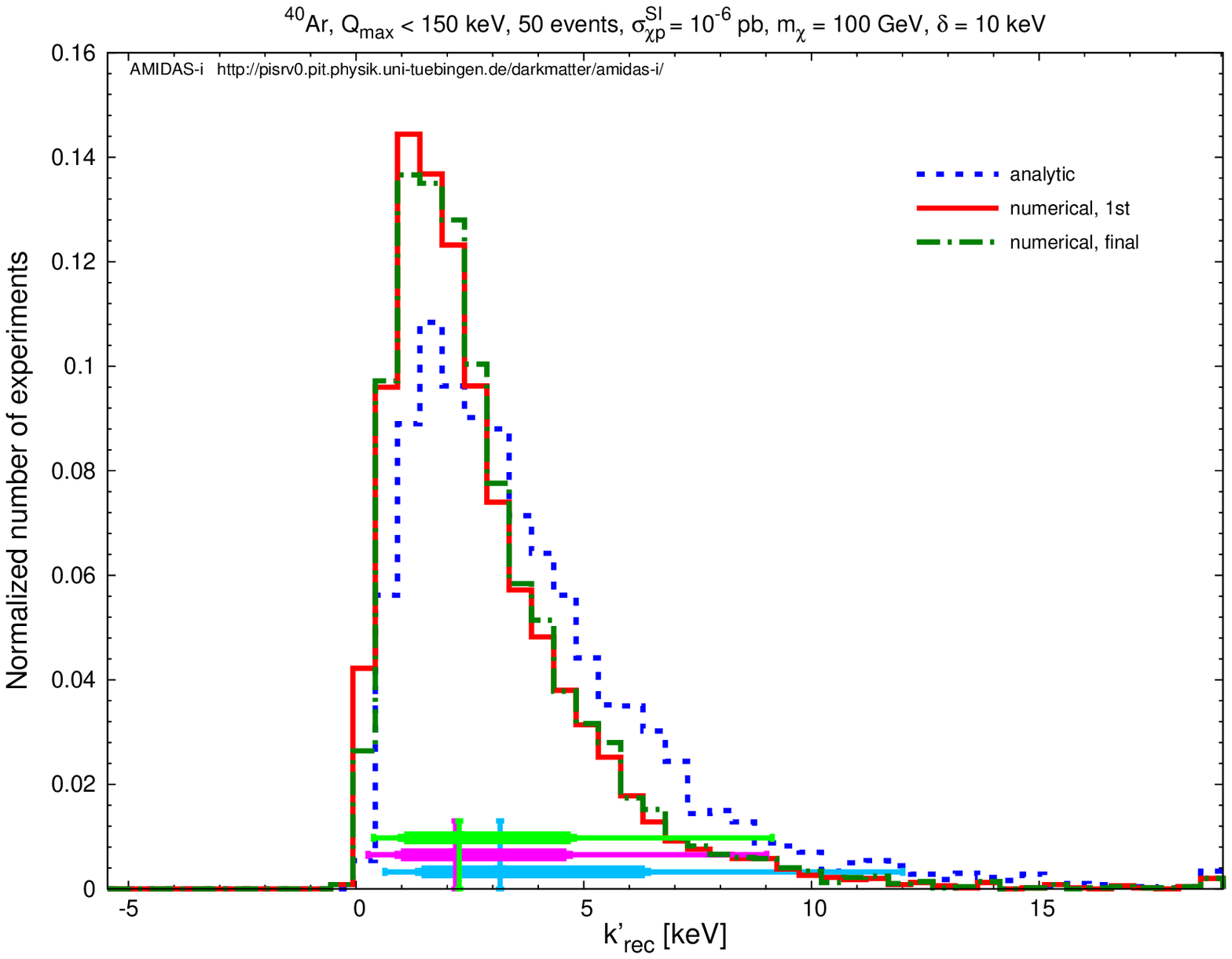}     \hspace*{-0.5cm}
\includegraphics[width=4.8cm]{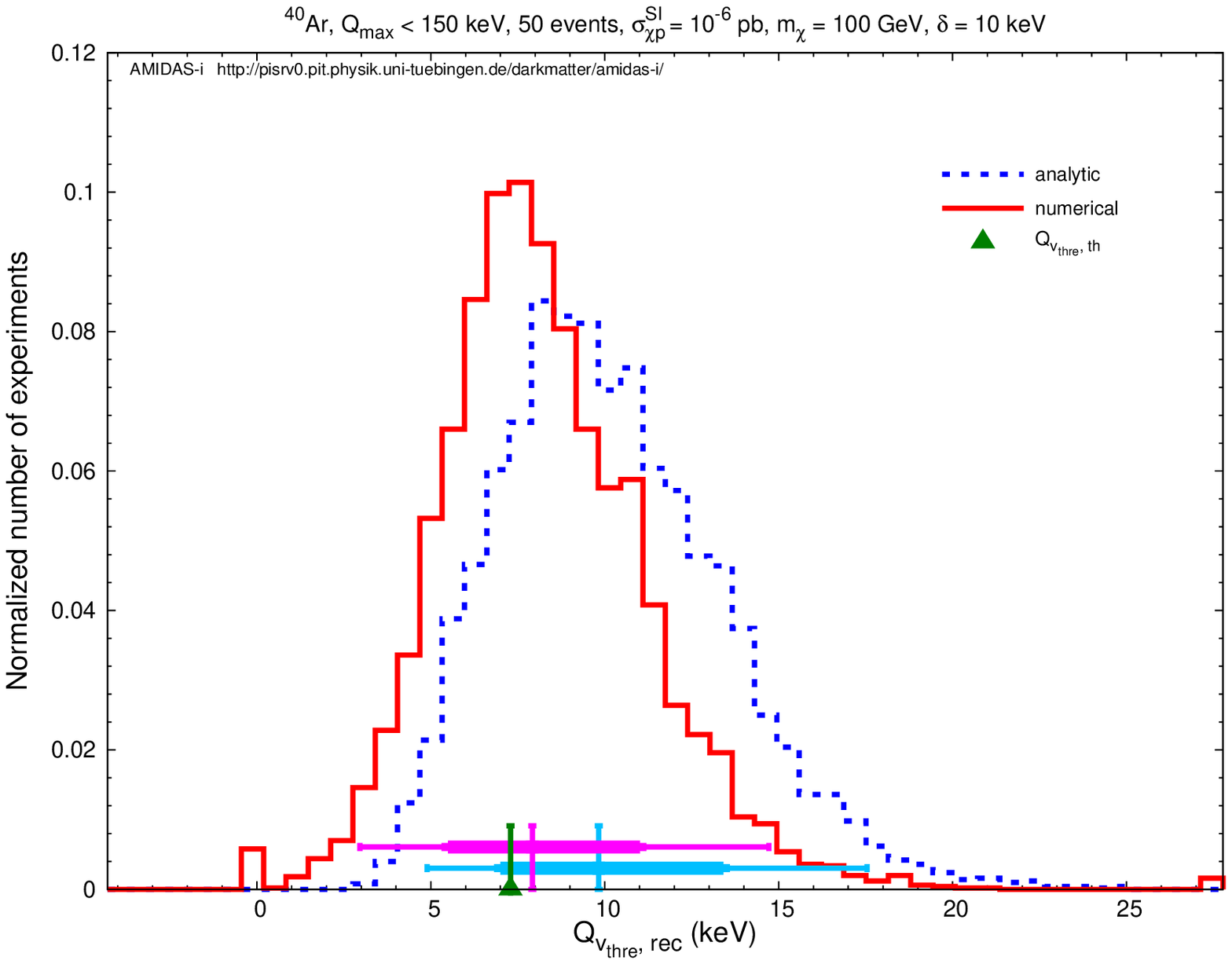} \hspace*{-1.6cm} \\
 ($\rmXA{Ar}{40}$) \\
\vspace{0.75cm}
\hspace*{-1.6cm}
\includegraphics[width=4.8cm]{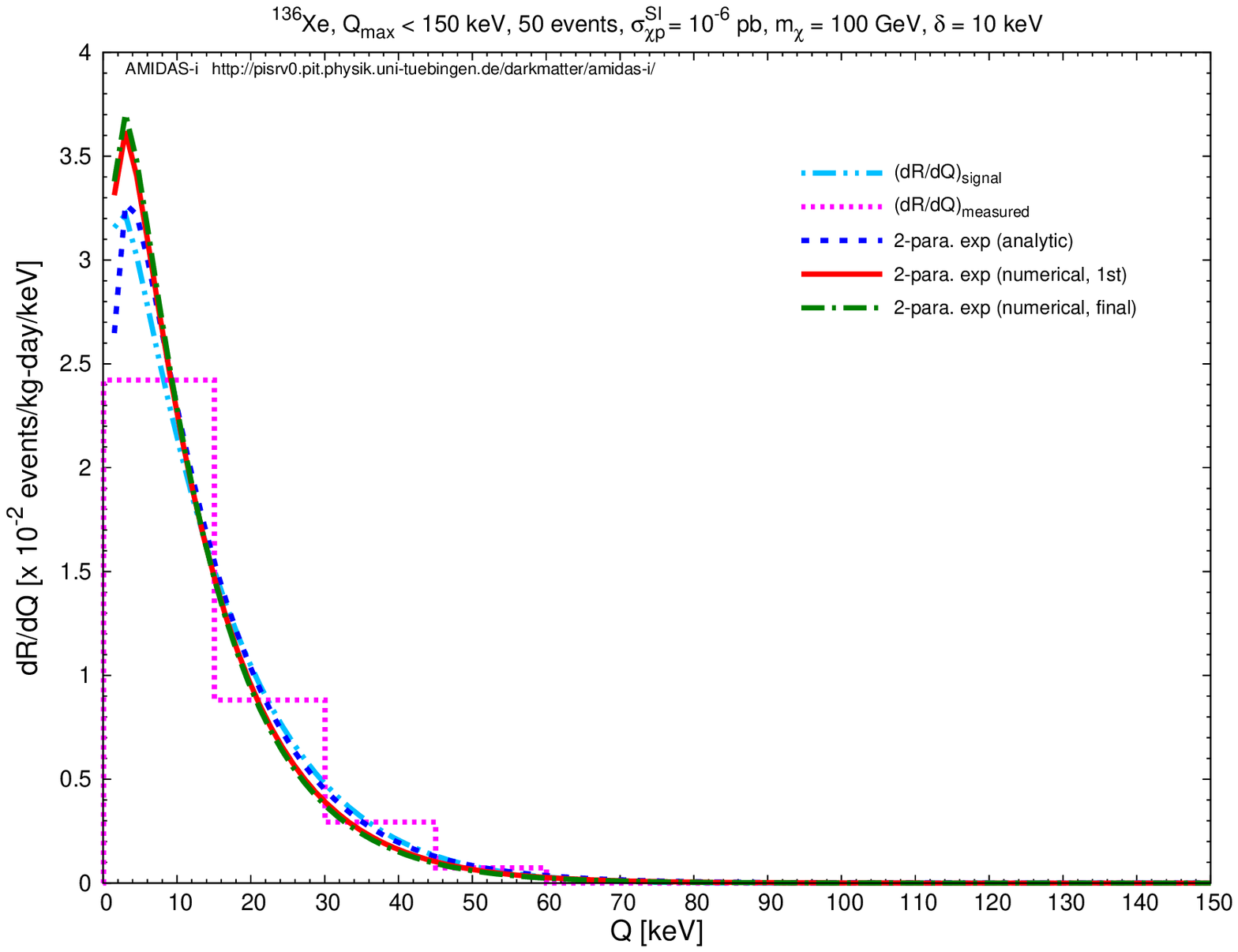}      \hspace*{-0.5cm}
\includegraphics[width=4.8cm]{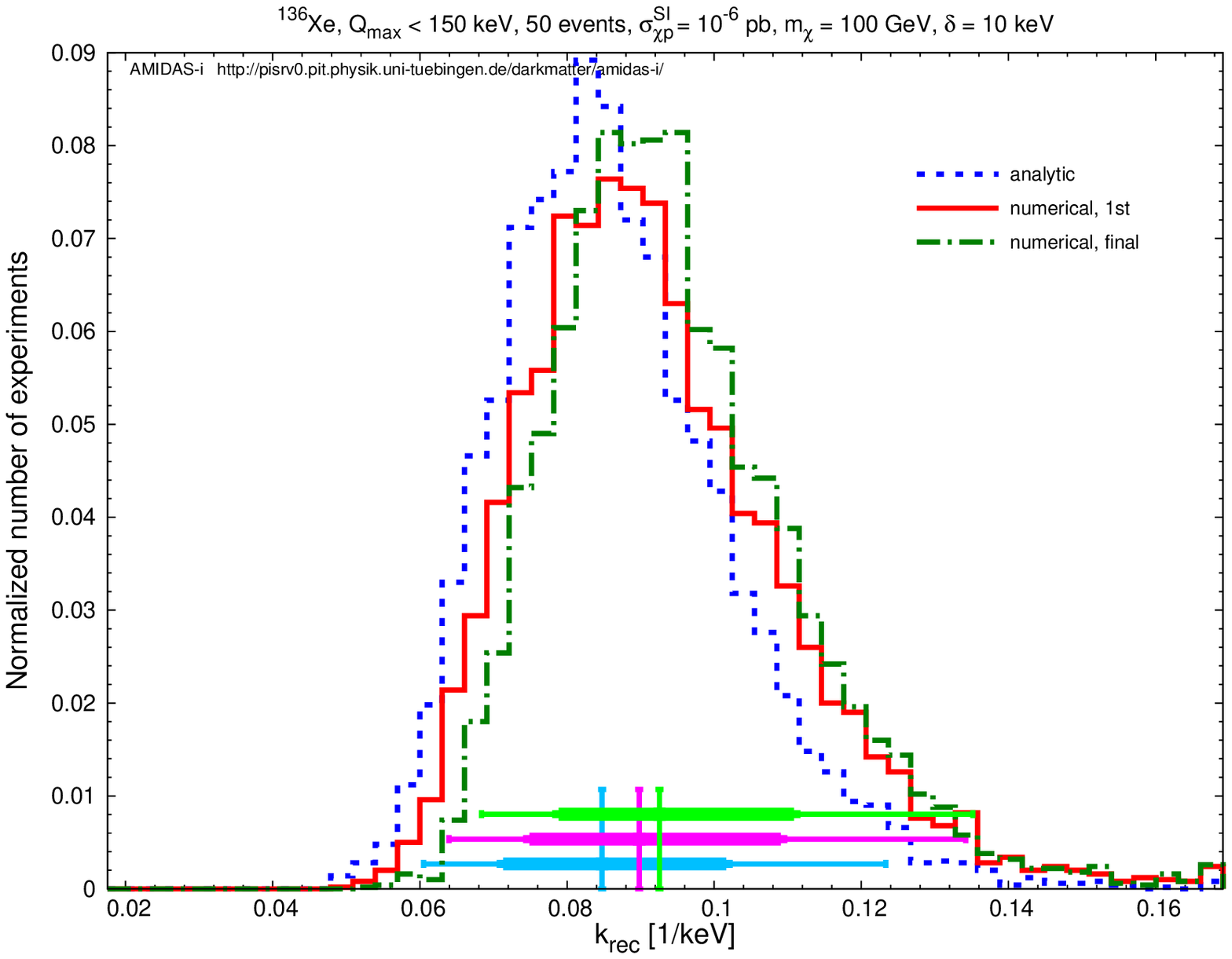}      \hspace*{-0.5cm}
\includegraphics[width=4.8cm]{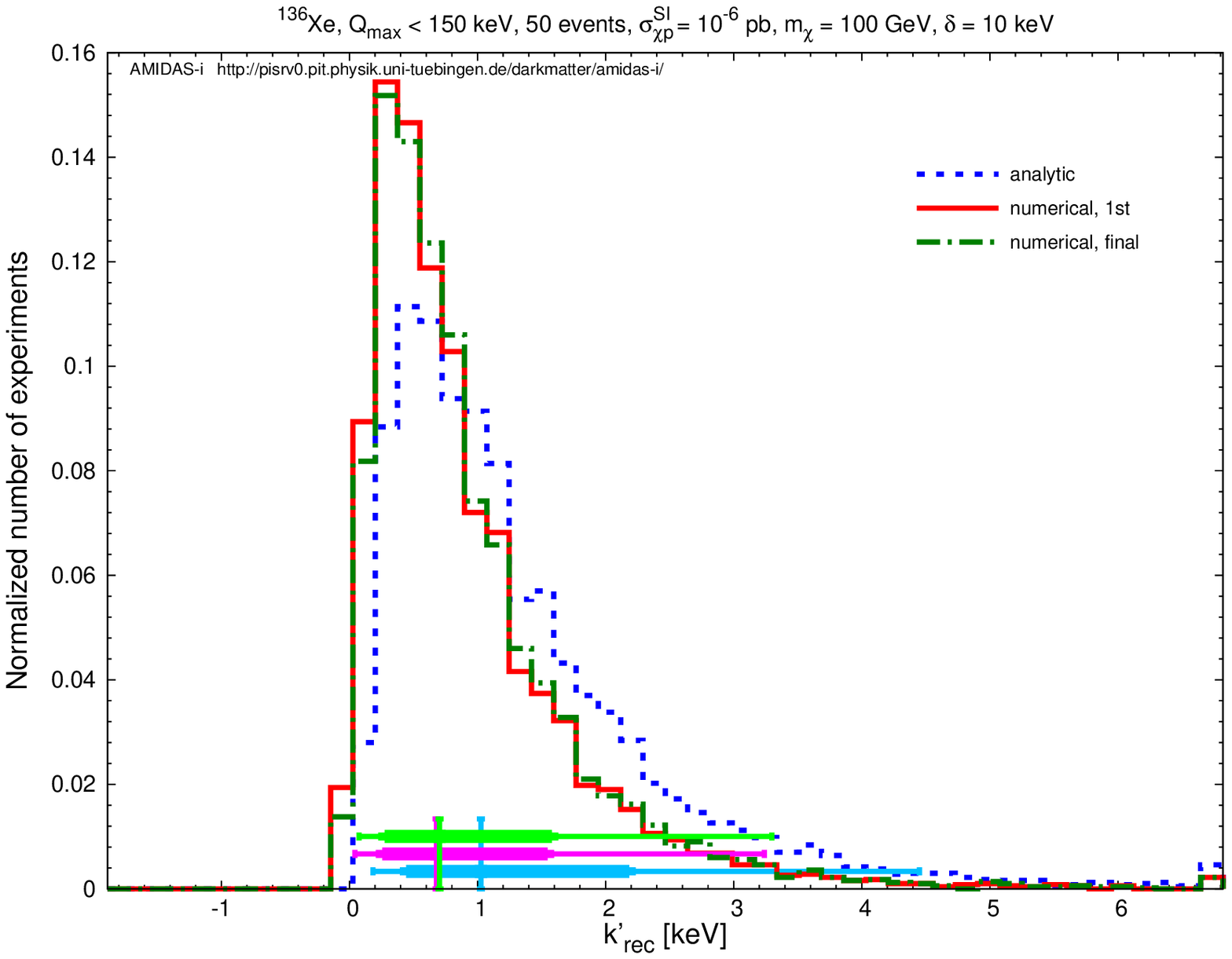}     \hspace*{-0.5cm}
\includegraphics[width=4.8cm]{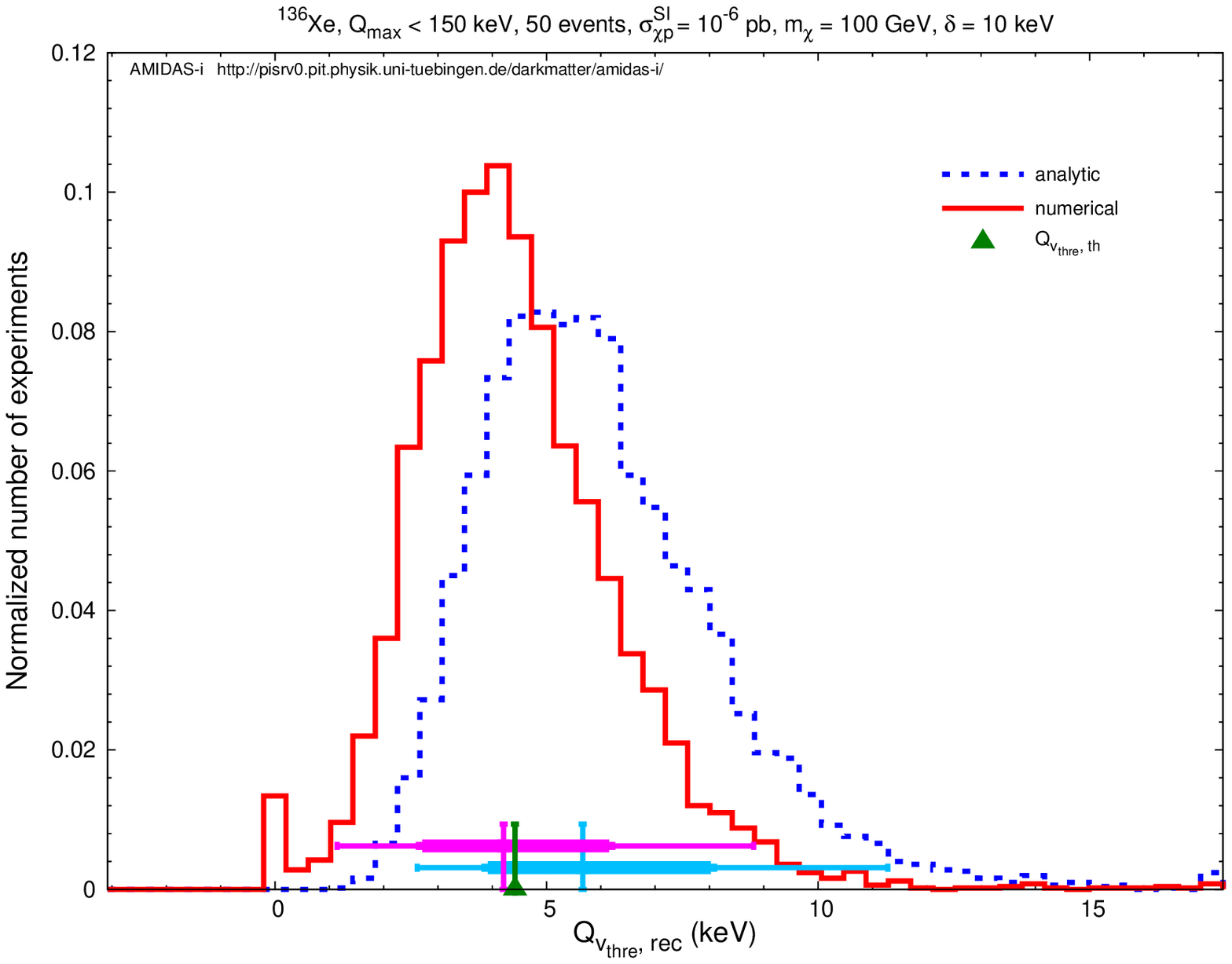} \hspace*{-1.6cm} \\
 ($\rmXA{Xe}{136}$) \\
\vspace{-0.25cm}
\end{center}
\caption{
 As in Figs.~\ref{fig:idRdQ-Ge76-100-025-050}
 and \ref{fig:idRdQ-100-025-050},
 except that
 a smaller mass splitting \mbox{$\delta = 10$ keV} has been used.
}
\label{fig:idRdQ-100-010-050}
\end{figure}
\begin{figure}[t!]
\begin{center}
\hspace*{-1.6cm}
\includegraphics[width=4.8cm]{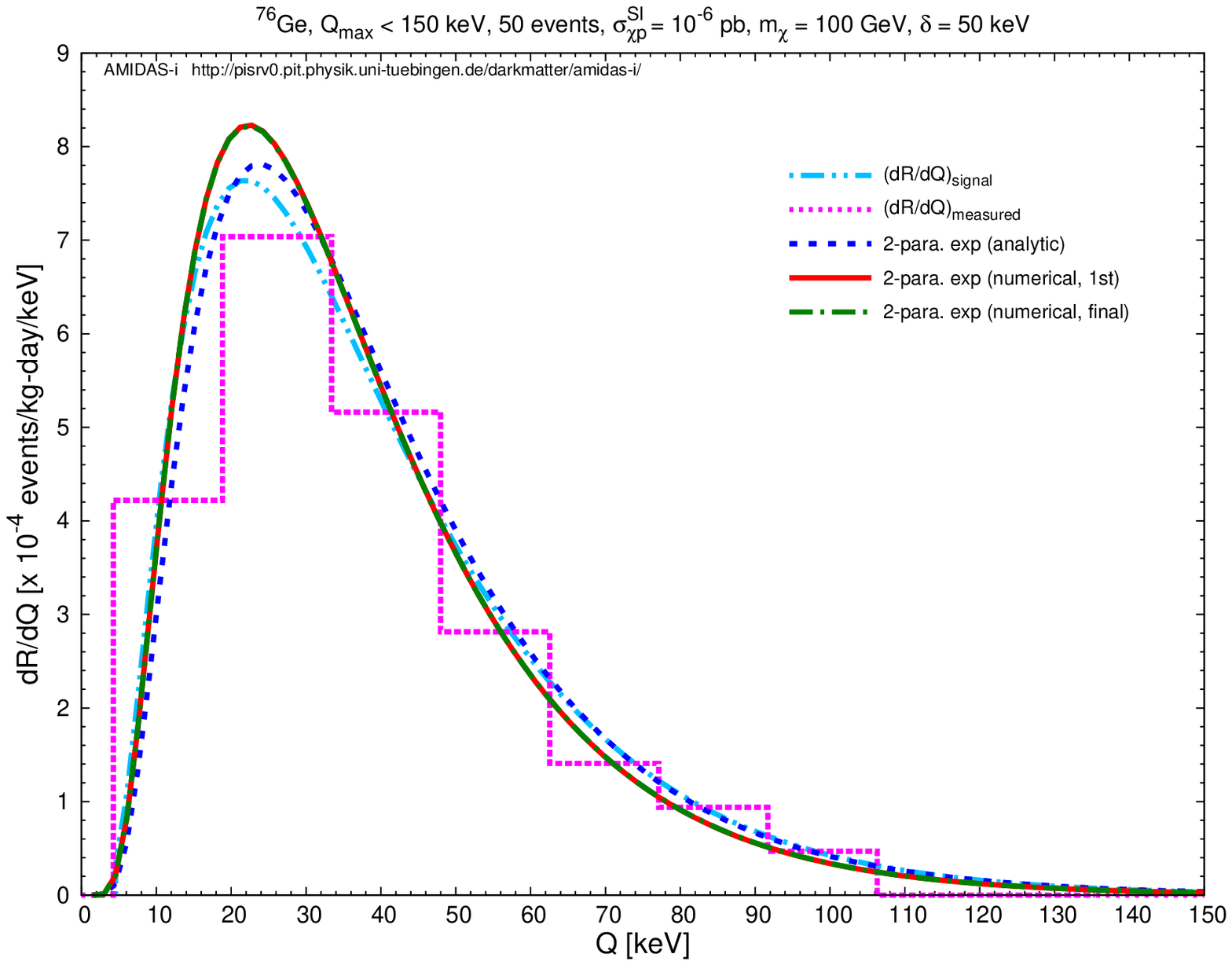}      \hspace*{-0.5cm}
\includegraphics[width=4.8cm]{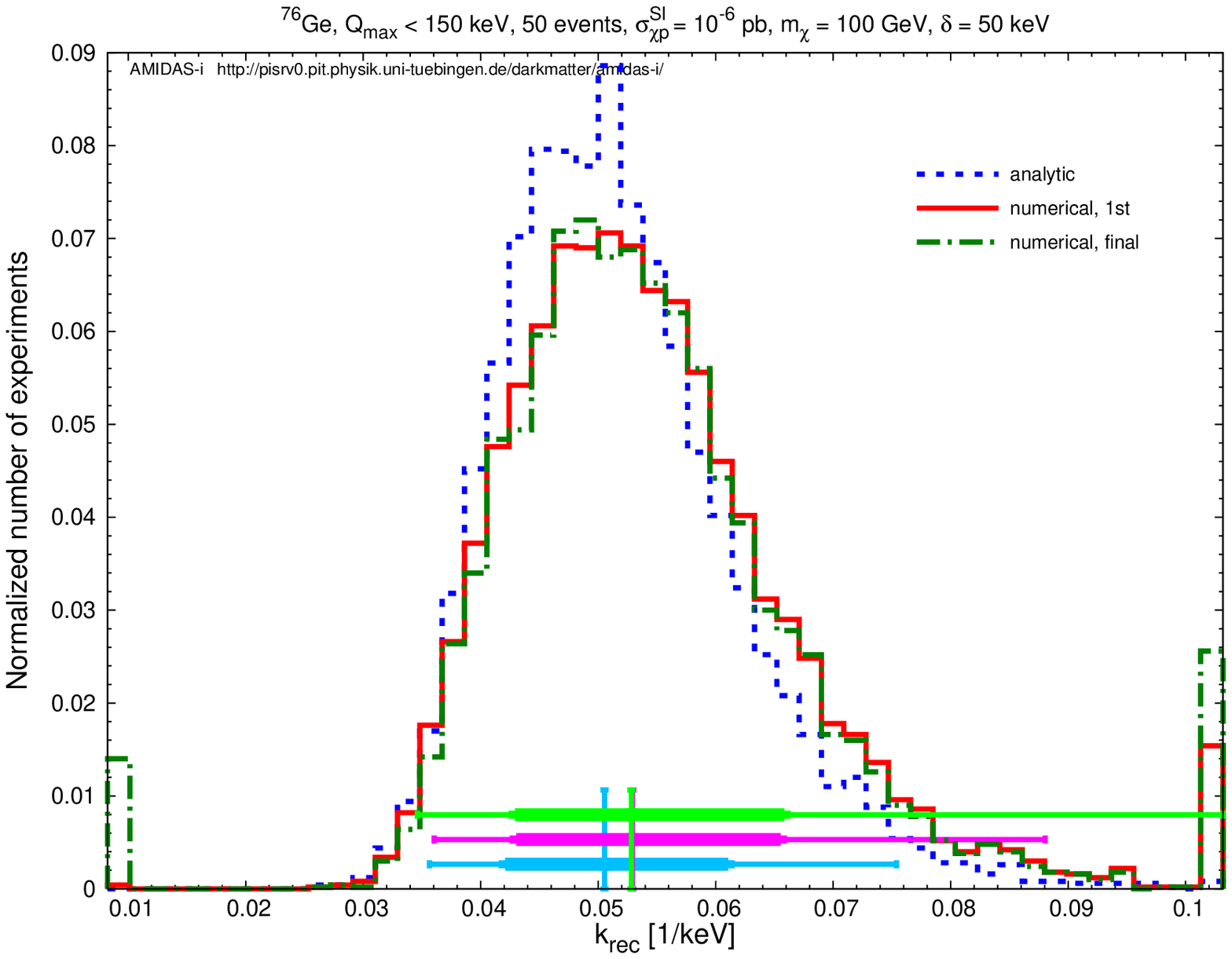}      \hspace*{-0.5cm}
\includegraphics[width=4.8cm]{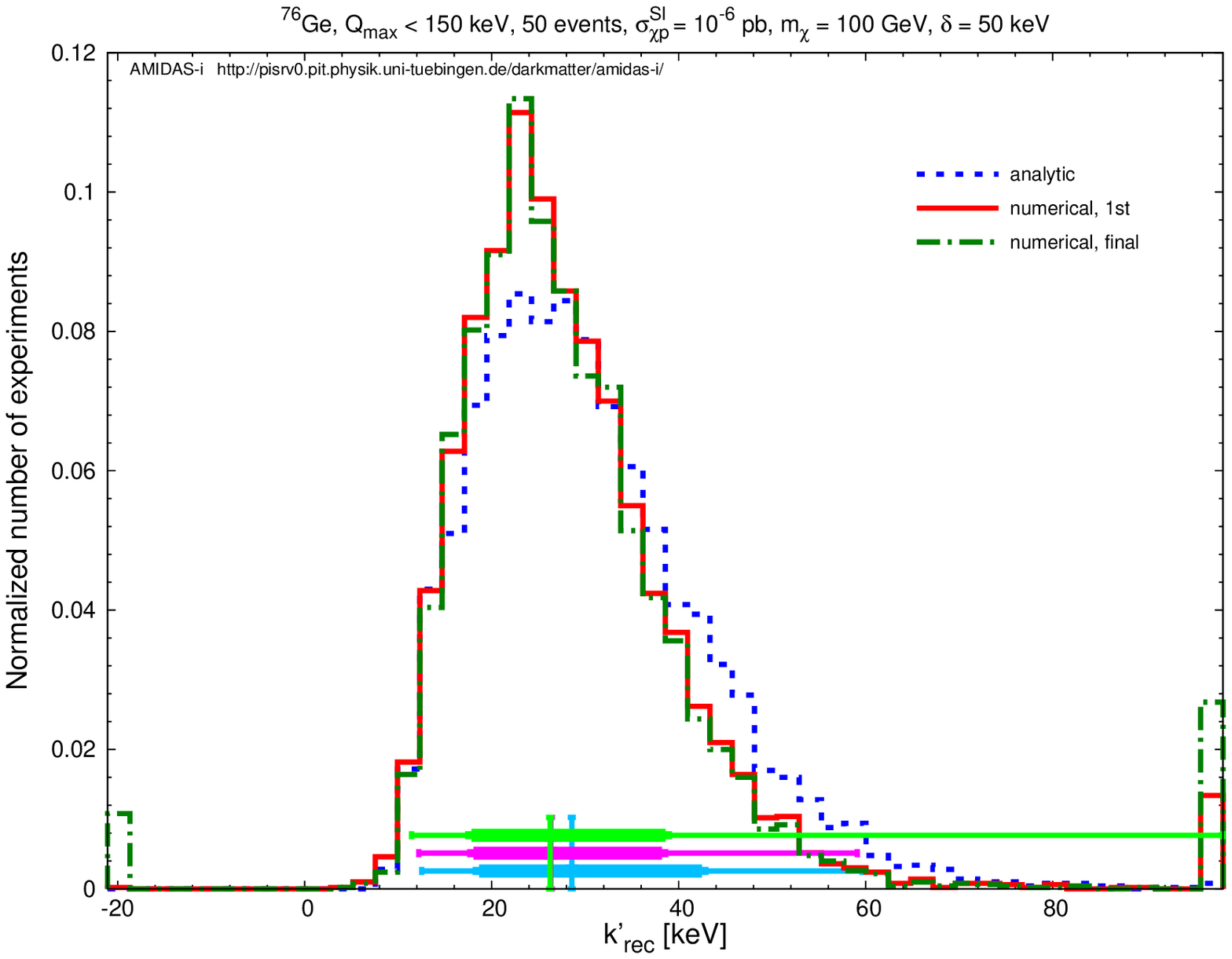}     \hspace*{-0.5cm}
\includegraphics[width=4.8cm]{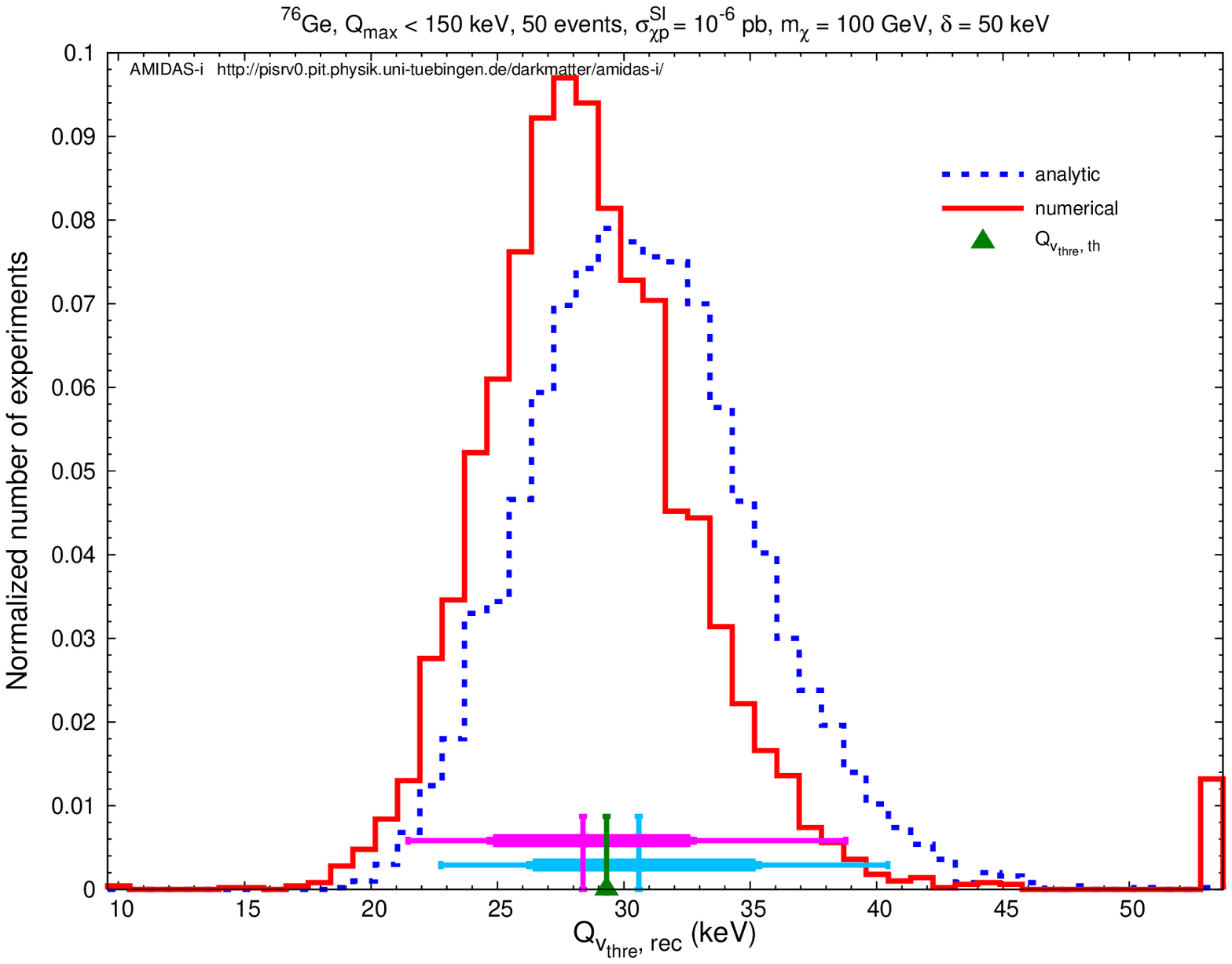} \hspace*{-1.6cm} \\
 ($\rmXA{Ge}{76}$) \\
\vspace{0.75cm}
\hspace*{-1.6cm}
\includegraphics[width=4.8cm]{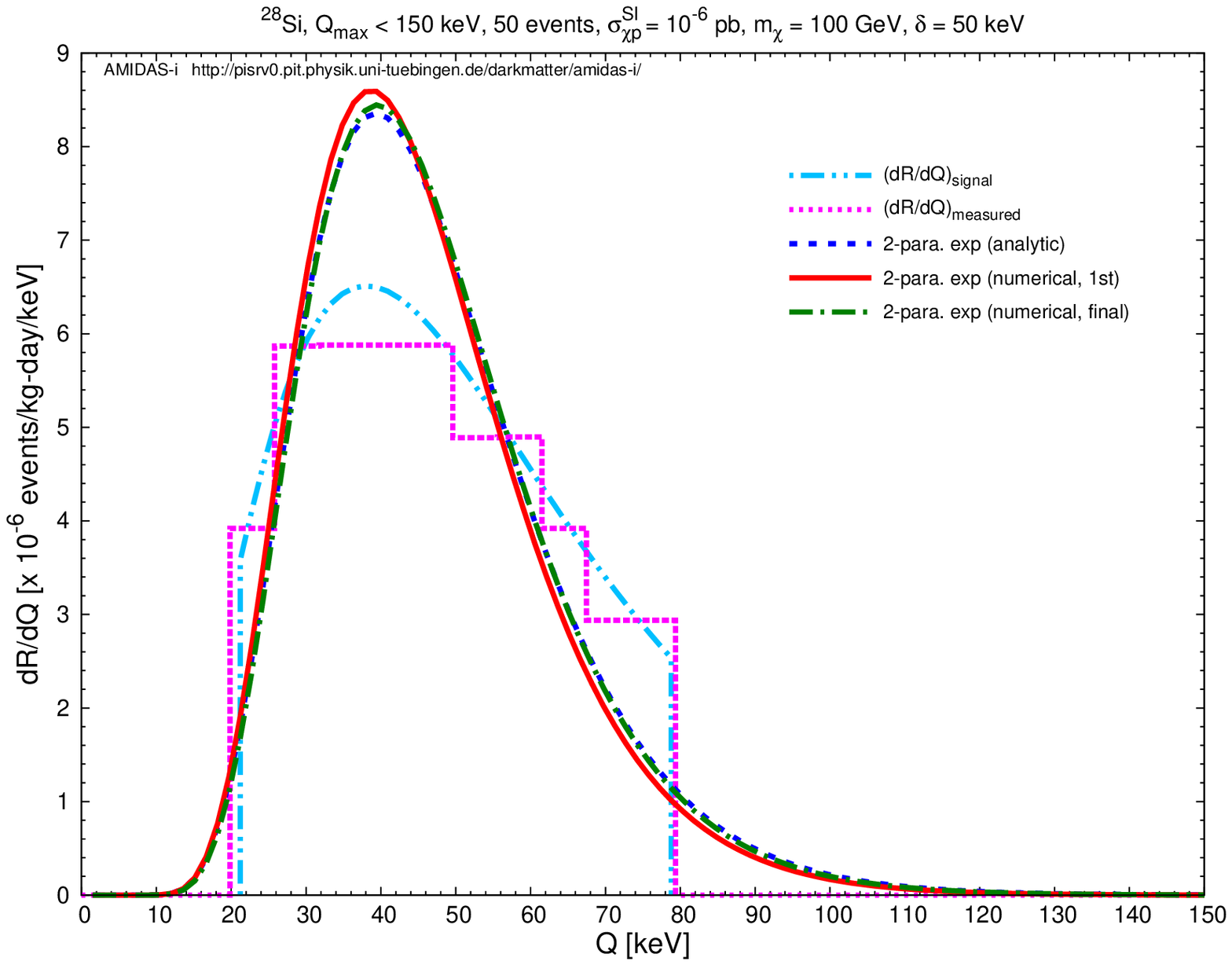}      \hspace*{-0.5cm}
\includegraphics[width=4.8cm]{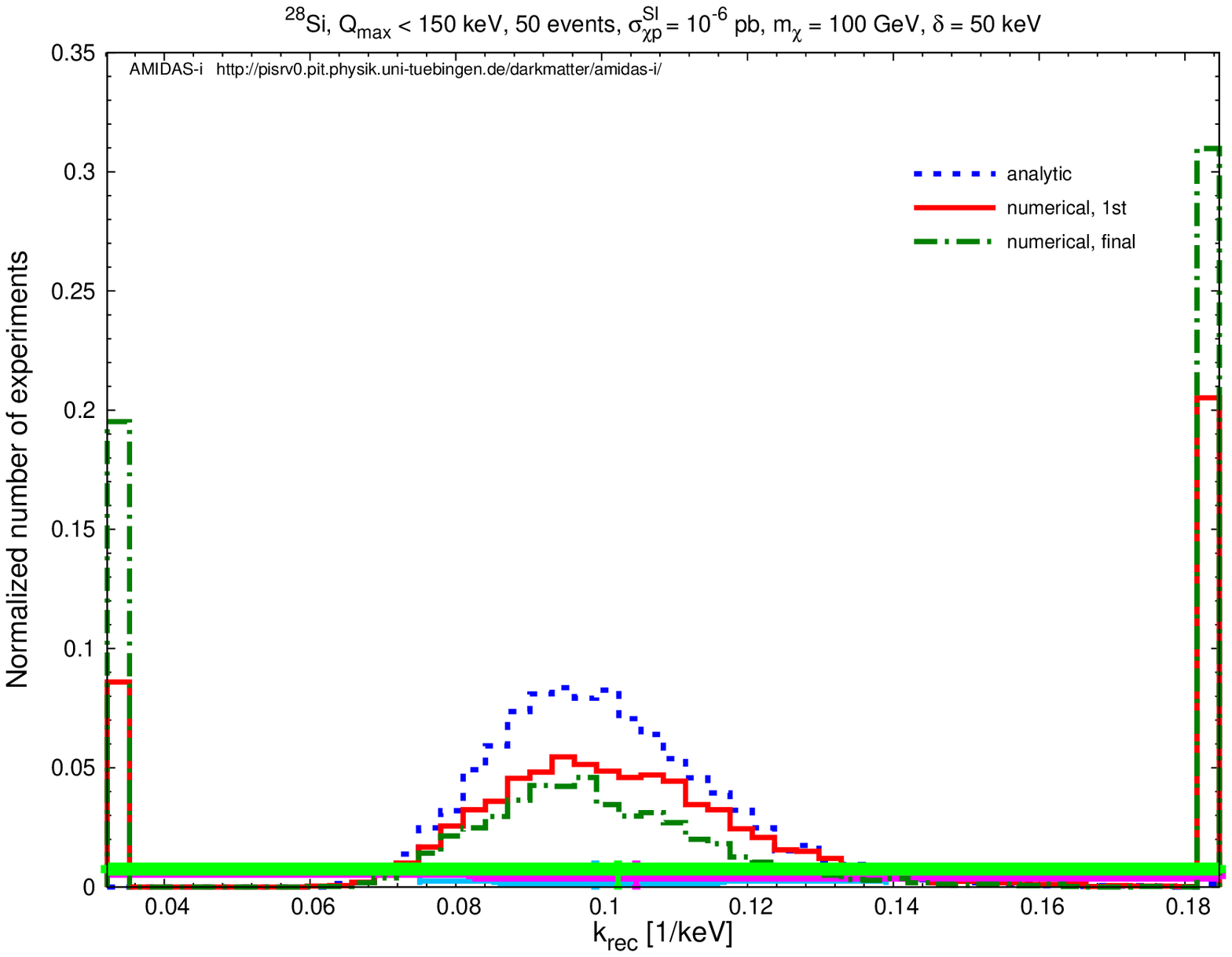}      \hspace*{-0.5cm}
\includegraphics[width=4.8cm]{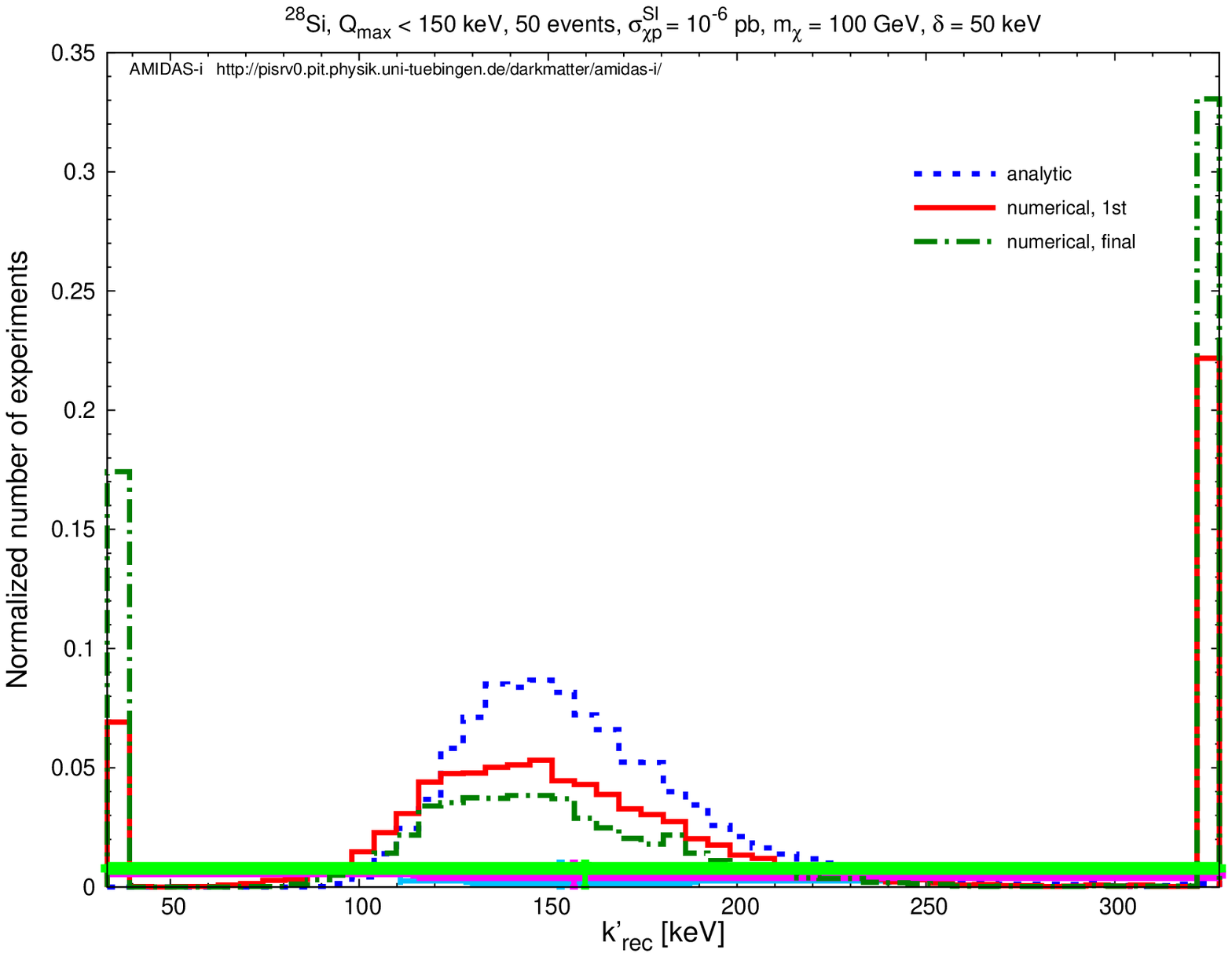}     \hspace*{-0.5cm}
\includegraphics[width=4.8cm]{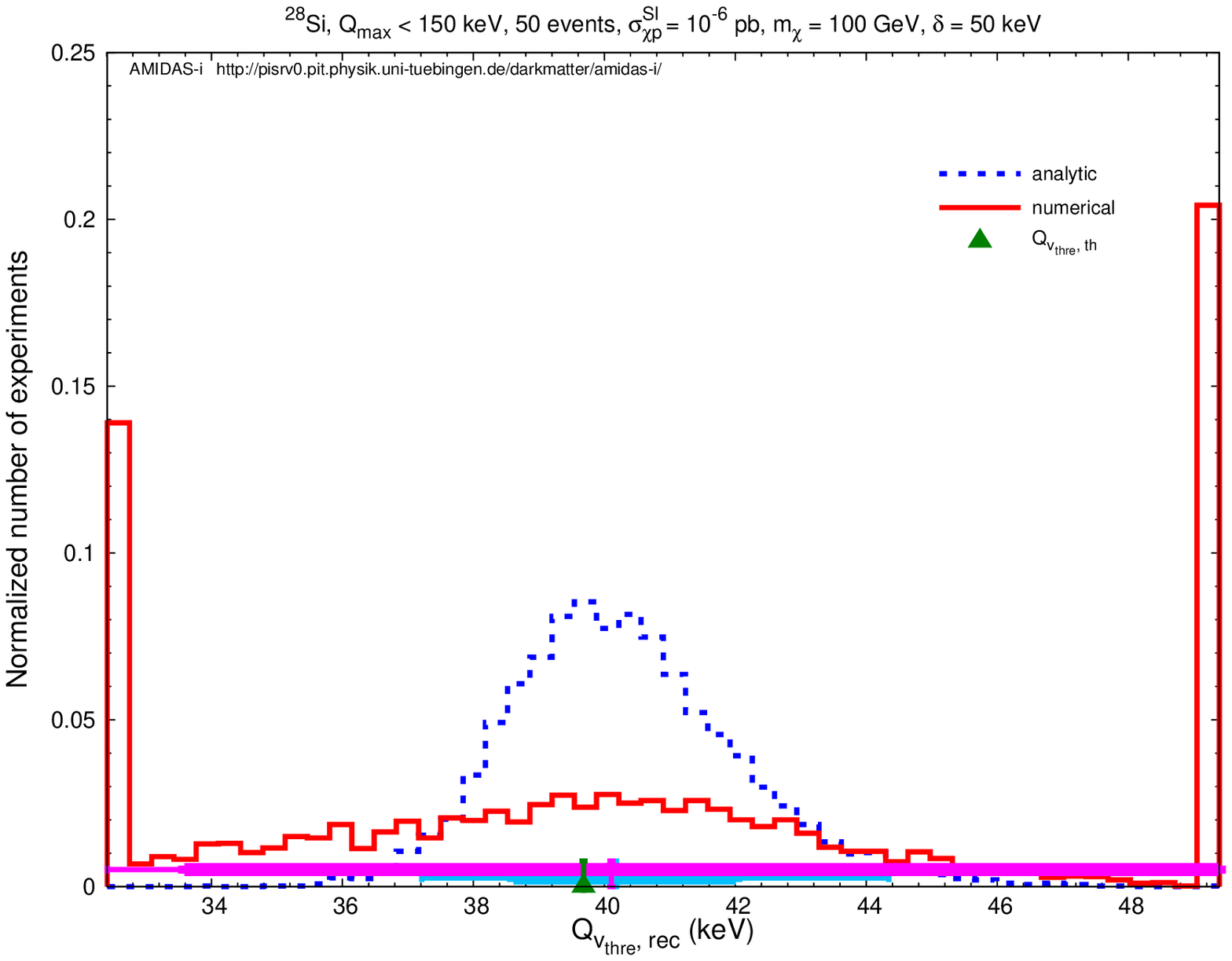} \hspace*{-1.6cm} \\
 ($\rmXA{Si}{28}$) \\
\vspace{0.75cm}
\hspace*{-1.6cm}
\includegraphics[width=4.8cm]{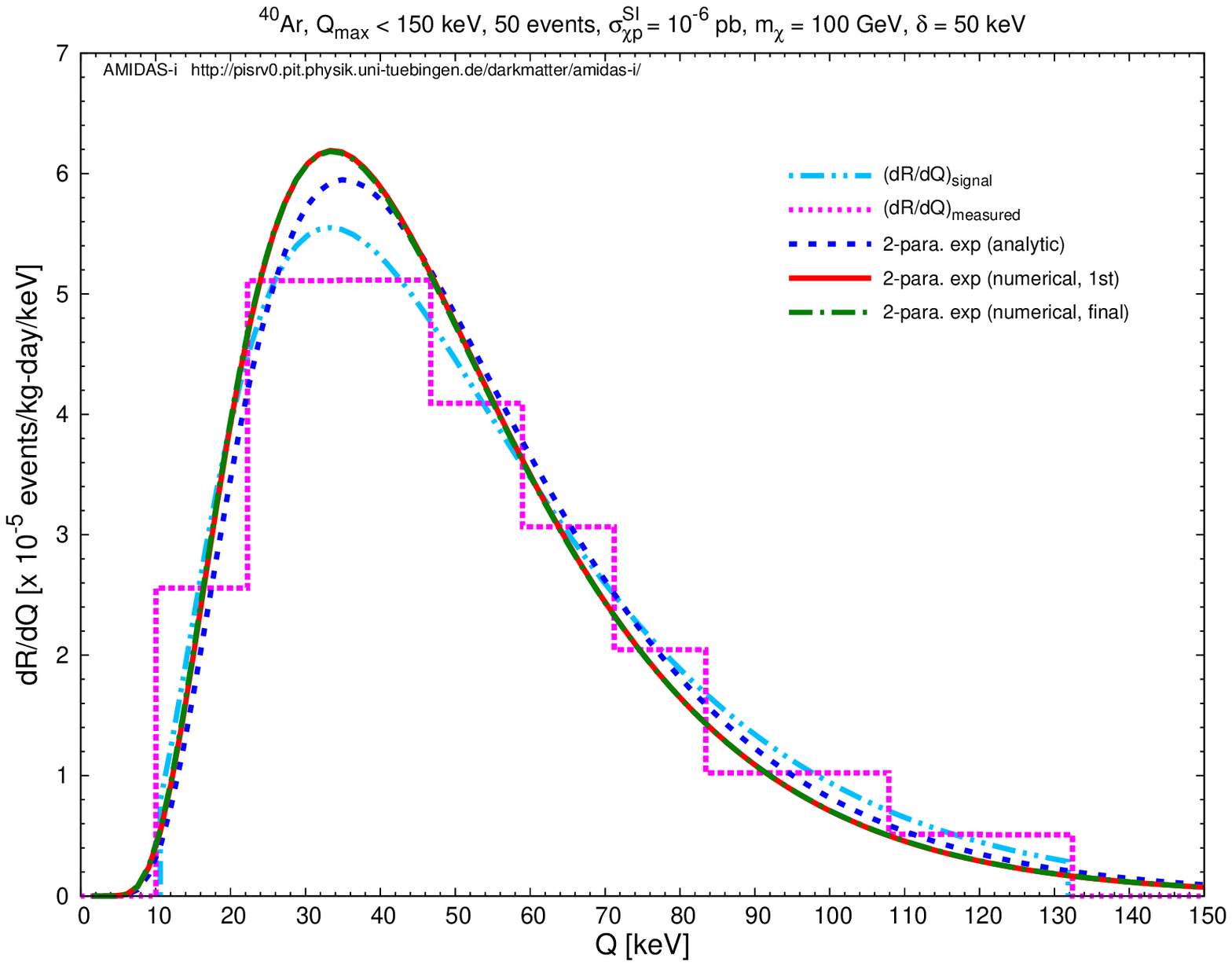}      \hspace*{-0.5cm}
\includegraphics[width=4.8cm]{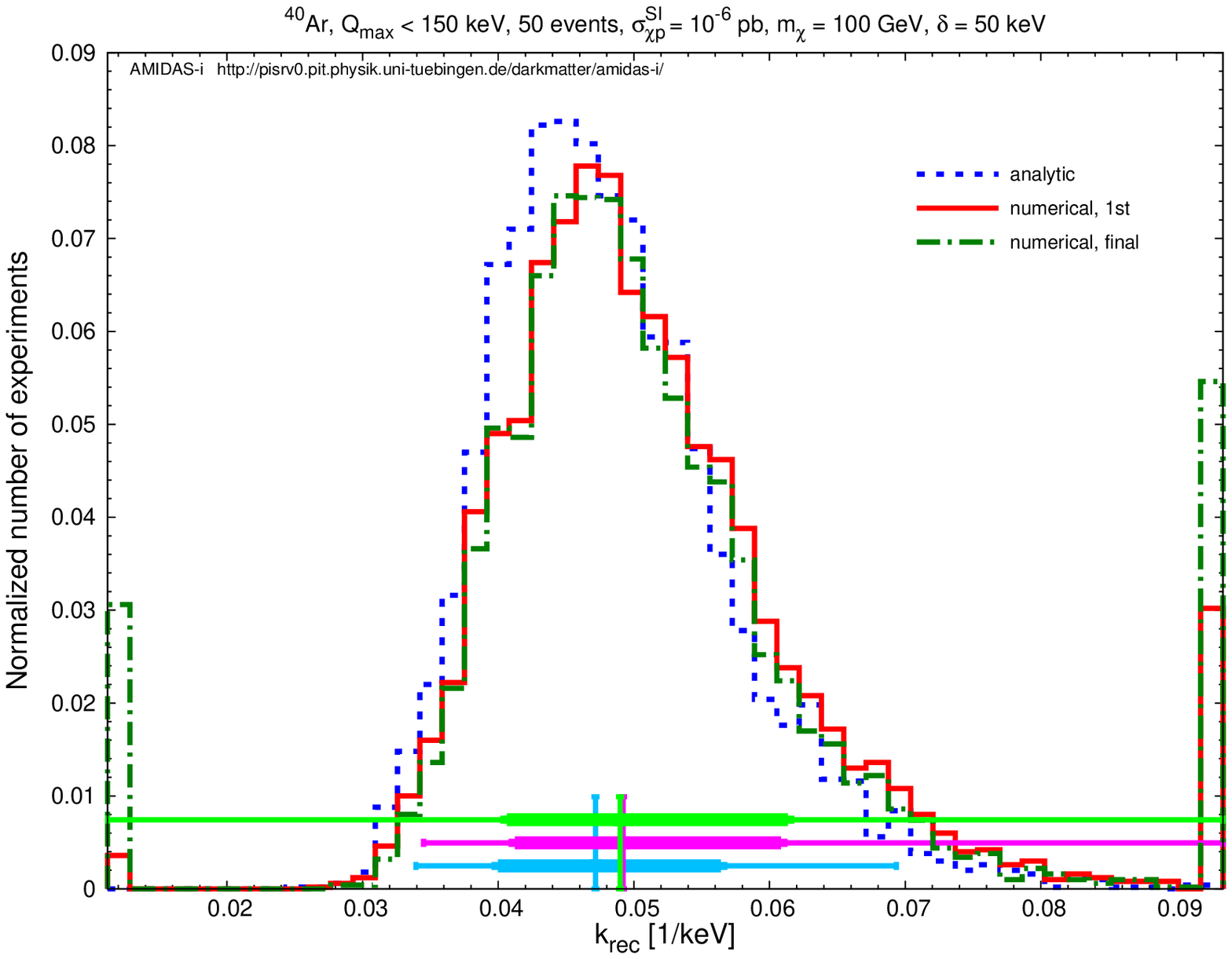}      \hspace*{-0.5cm}
\includegraphics[width=4.8cm]{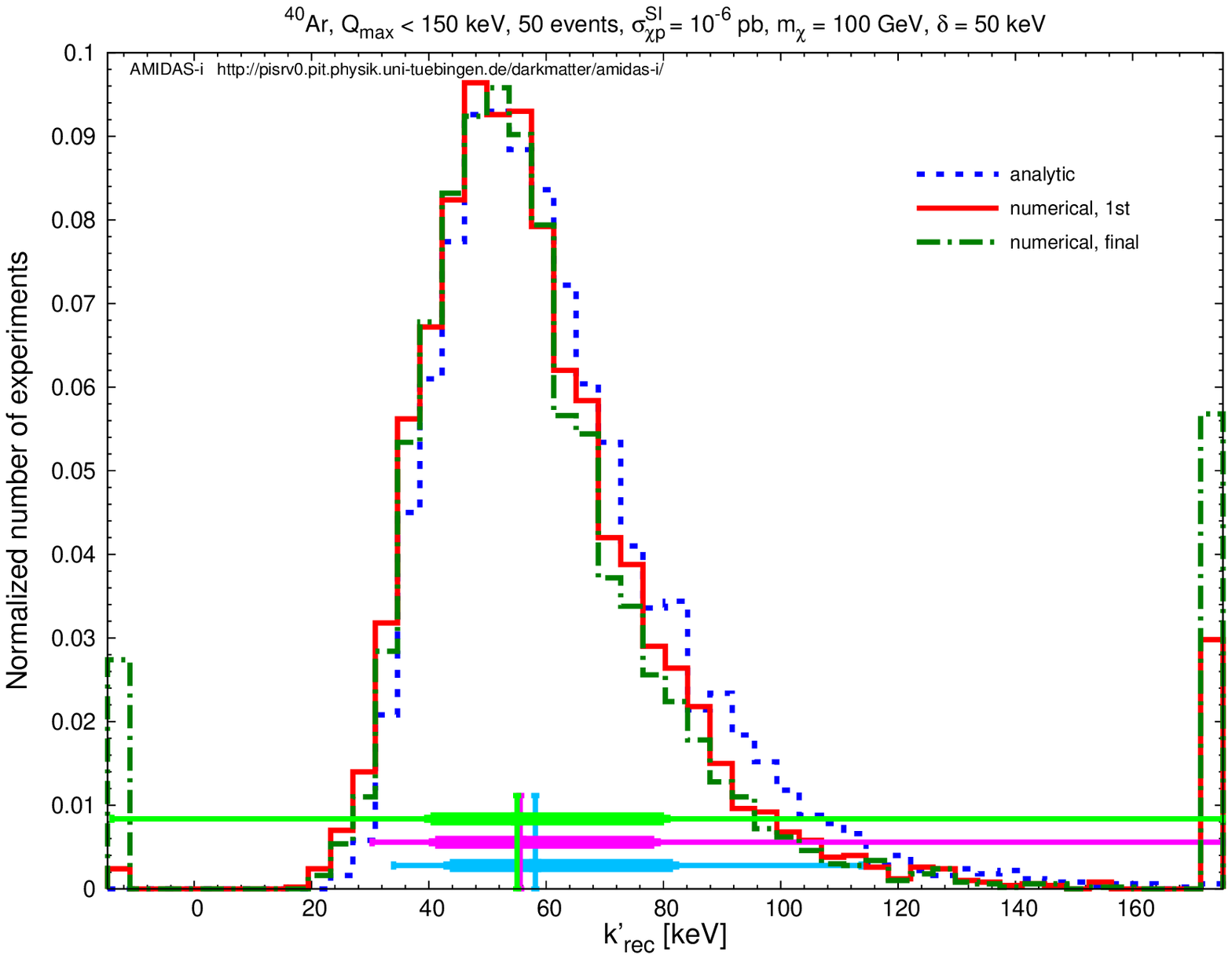}     \hspace*{-0.5cm}
\includegraphics[width=4.8cm]{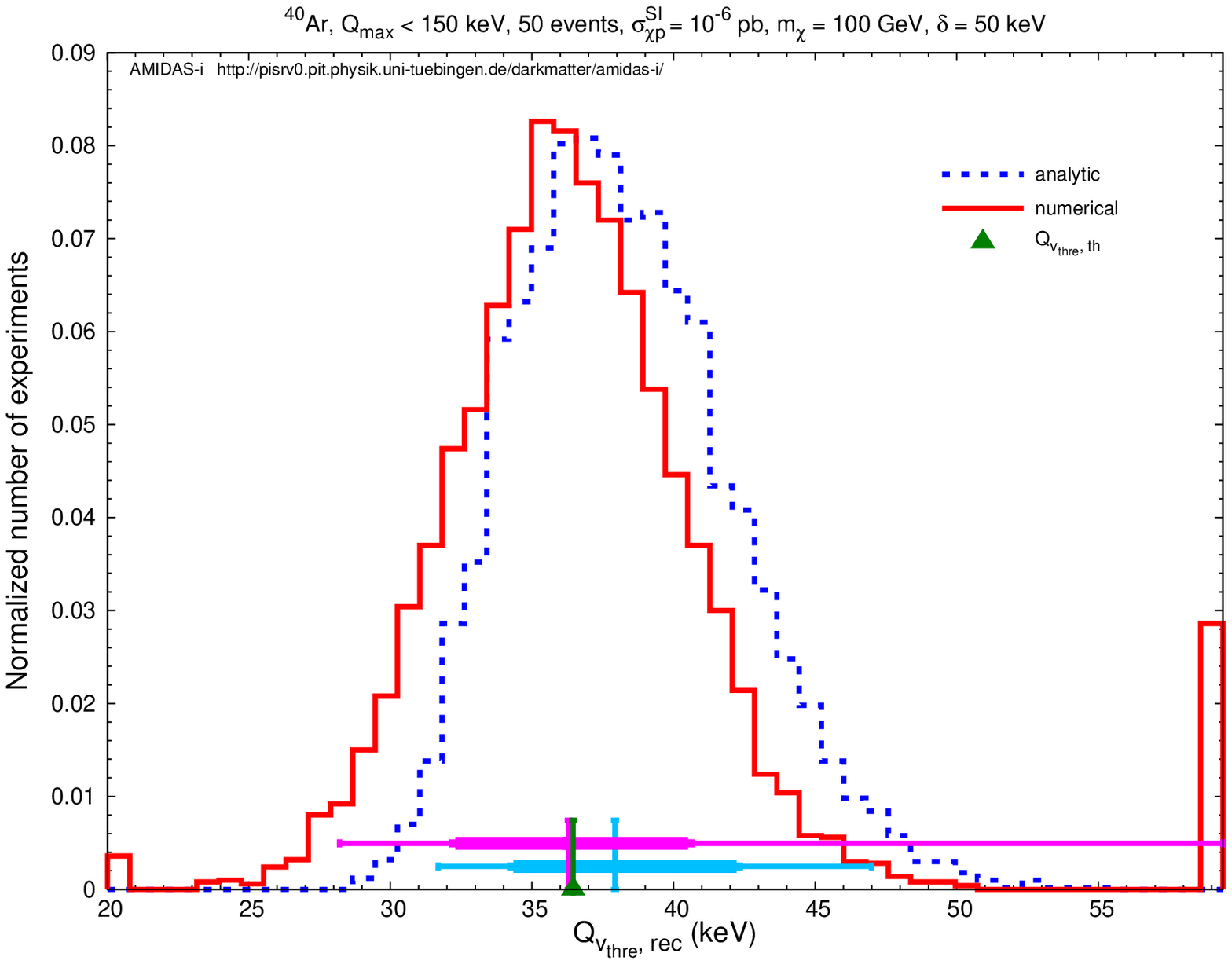} \hspace*{-1.6cm} \\
 ($\rmXA{Ar}{40}$) \\
\vspace{0.75cm}
\hspace*{-1.6cm}
\includegraphics[width=4.8cm]{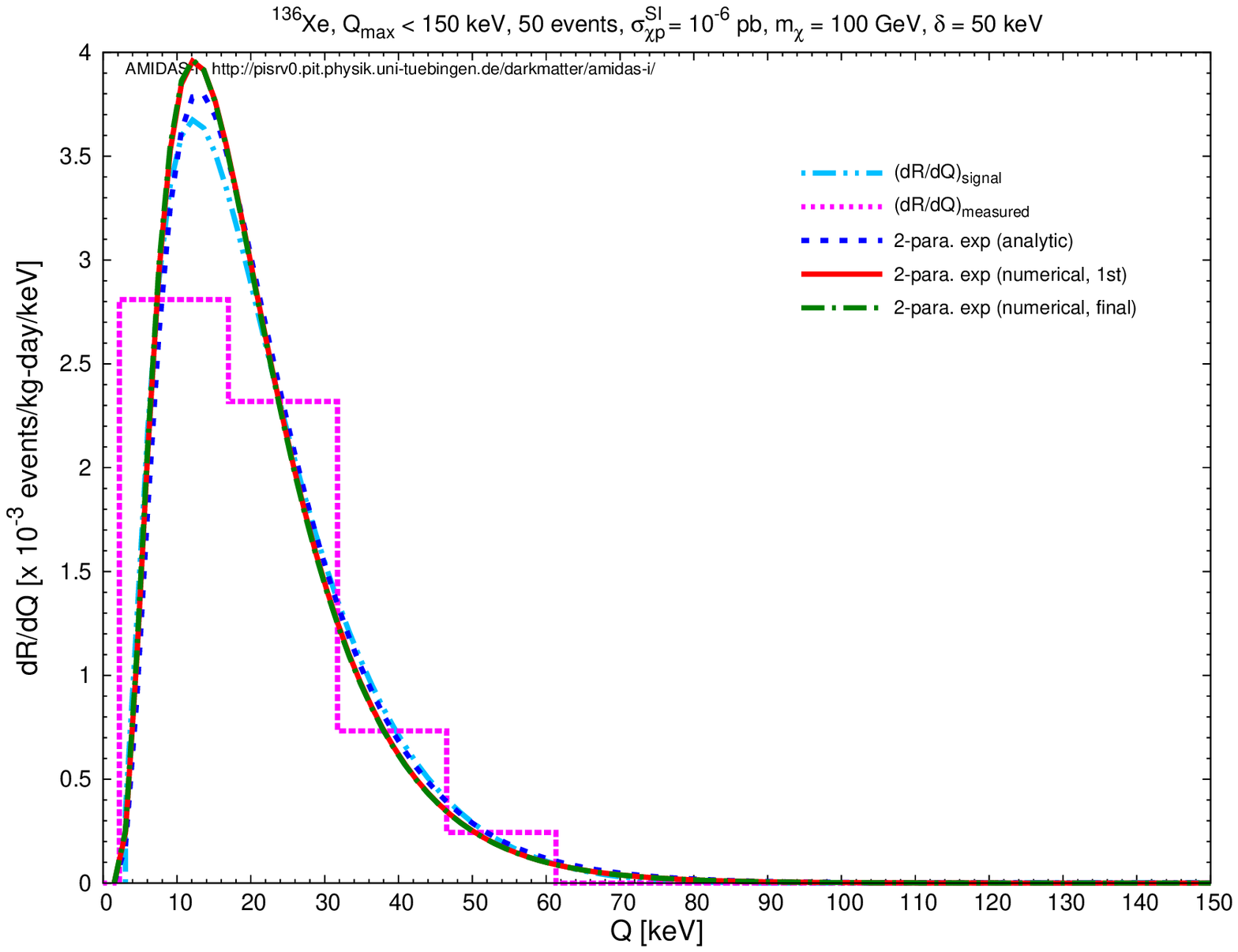}      \hspace*{-0.5cm}
\includegraphics[width=4.8cm]{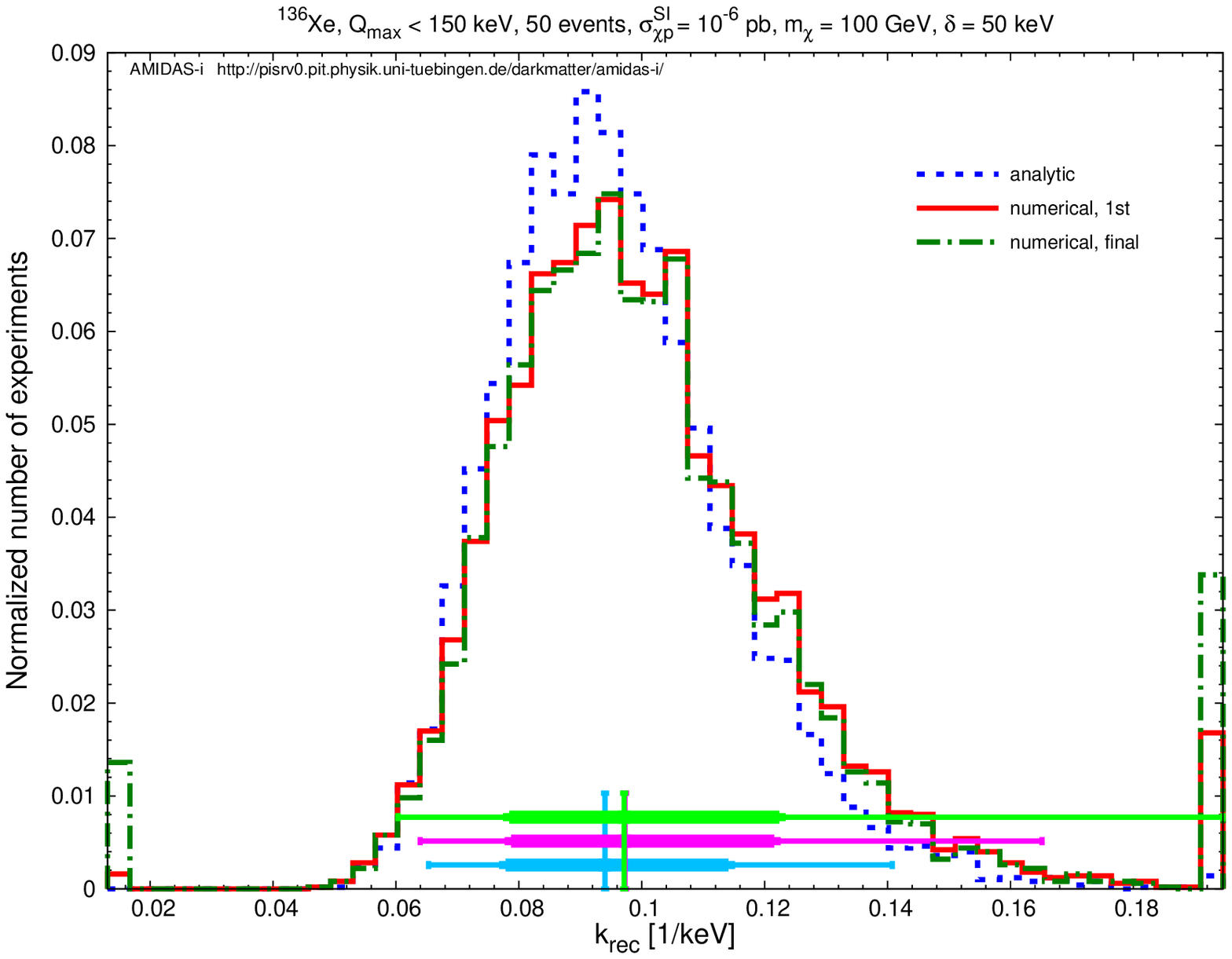}      \hspace*{-0.5cm}
\includegraphics[width=4.8cm]{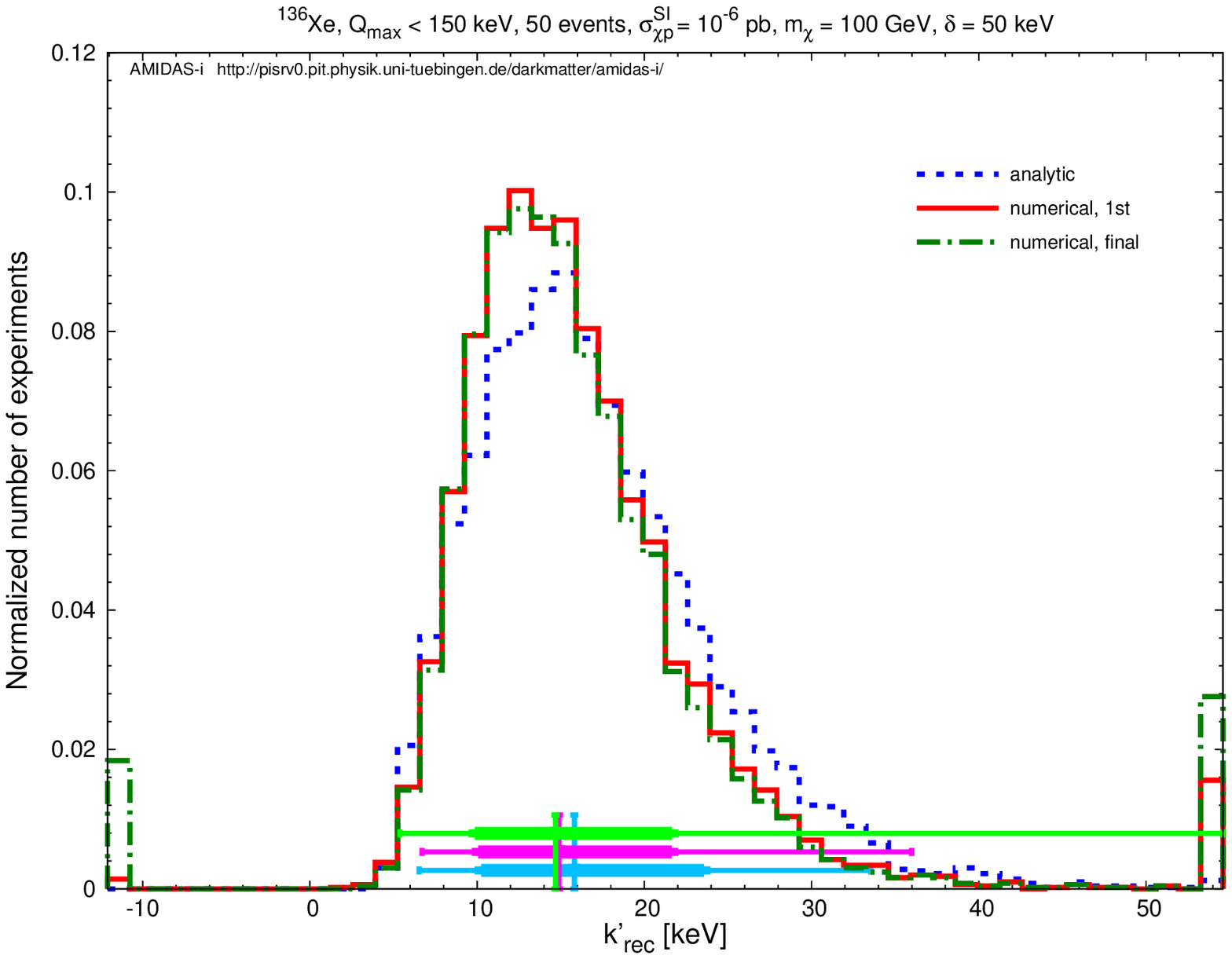}     \hspace*{-0.5cm}
\includegraphics[width=4.8cm]{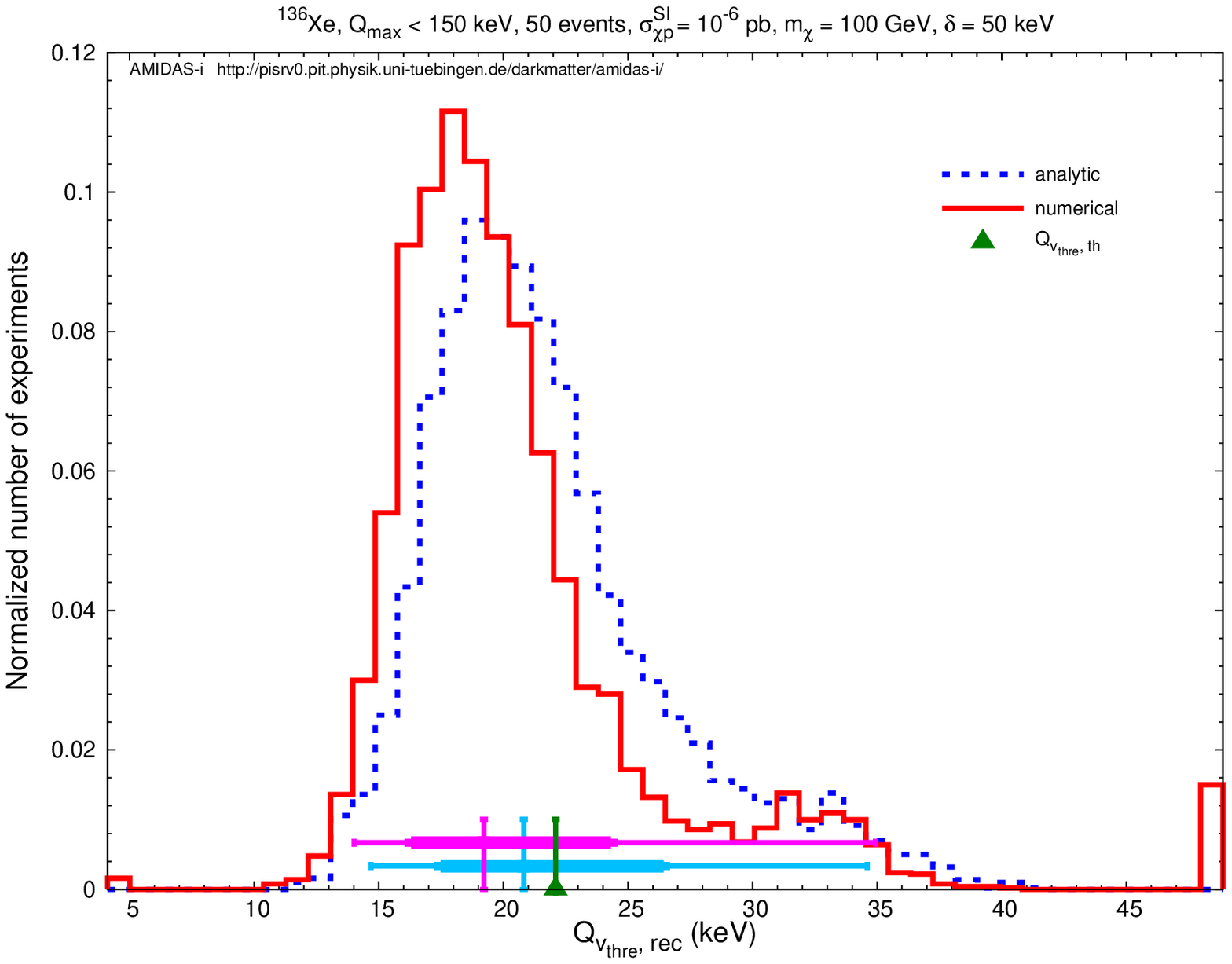} \hspace*{-1.6cm} \\
 ($\rmXA{Xe}{136}$) \\
\vspace{-0.25cm}
\end{center}
\caption{
 As in Figs.~\ref{fig:idRdQ-Ge76-100-025-050}
 and \ref{fig:idRdQ-100-025-050},
 except that
 a larger mass splitting \mbox{$\delta = 50$ keV} has been used.
}
\label{fig:idRdQ-100-050-050}
\end{figure}

 It can be found that,
 firstly,
 the recoil spectra reconstructed numerically
 could really be the better approximations
 of the original theoretical spectrum,
 at least in the sense that
 the peaks of these spectra coincide very well.
 Secondly, and not really as expected,
 although the numerically reconstructed recoil spectra
 could be the better results than the analytically reconstructed one,
 the iterative procedure could not improve the correction further:
 only in a small part of the simulated experiments
 (shown here and later with different initial setup)
 the results coming from the second (and also the final) rounds
 could be clearly better than those coming from the first round;
 in most part of simulations
 there would no (significant) difference
 between the results given by different rounds,
 or those from the later round could even be {\em worse}...

 Moreover,
 in the top--right and bottom--left frames of
 Figs.~\ref{fig:idRdQ-Ge76-100-025-050},
 we show the distributions of the reconstructed fitting parameters $k$ and $k'$,
 respectively:
 while the dashed blue curves show
 the distributions of $k$ and $k'$ estimated analytically
 by Eqs.~(\ref{eqn:k_in_ana}) and (\ref{eqn:kp_in_ana}),
 the solid red and dash--dotted green curves are
 those of the numerically estimated $k$ and $k'$,
 the former and later show the results given
 by the first and final rounds of the iterative process.
 Meanwhile,
 the cyan (magenta, light green) vertical lines indicate
 the median values of the simulated results
 (corresponding to the blue, red and green histograms),
 whereas
 the horizontal thick (thin) bars show
 the 1$\~$(2)$\~\sigma$ ranges of the results%
\footnote{
 Here the 1$\~$(2)$\~\sigma$ ranges mean that
 68.27\% (95.45\%) of the reconstructed values in the simulated experiments
 are in these ranges (central interval).
}.

 The top--right frame of Figs.~\ref{fig:idRdQ-Ge76-100-025-050}
 shows that
 the distributions of the reconstructed fitting parameter $k$
 by both of the analytic and numerical methods
 are indeed basically Gaussian
 (with small tails in the high--$k$ range).
 Meanwhile,
 in the bottom--left frame
 the distributions of the reconstructed $k'$
 by both methods
 would be Gaussian distributions
 with a $k' = 0$ cut--off.
 Note that,
 $k'$ is basically proportional to the square of the mass splitting, $\delta^2$,
 (see Eq.~(\ref{eqn:dRdQ_in_Gau}) and then Eq.~(\ref{eqn:alpha_delta})).
 The observation of the positively reconstructed $k'$
 indicates that
 our model--independent reconstruction proposed here
 should be useful and sensitive for identifying
 the inelastic WIMP--nucleus scattering scenarios.

 Moreover,
 these two plots show also that
 there are neither significant differences
 between the median values of $k$ and $k'$
 (and thus the reconstructed recoil spectra)
 estimated in the first and final rounds of the numerical procedure,
 nor those between the distributions of them
 in the simulated experiments.
 In contrast,
 the differences between
 the numerical results and the analytic ones
 are clear.
 Hence,
 later we show only results
 given by the first round of the iterative process
 as the numerical reconstruction.

 Finally,
 in the bottom--right frame of
 Figs.~\ref{fig:idRdQ-Ge76-100-025-050},
 we show the distributions of the characteristic energy $\Qvthre$
 solved by Eq.~(\ref{eqn:Qvthre_sol})
 with $k$ and $k'$ estimated analytically (dashed blue)
 and numerically (solid red),
 respectively.
 The green vertical line given additionally here
 indicates the theoretical value of $\Qvthre$
 estimated by Eq.~(\ref{eqn:Qvthre}),
 $Q_{v_{\rm thre}, {\rm th}}$.
 This plot shows that,
 firstly,
 both of the distributions of $\Qvthre$
 solved analytically and numerically
 are still basically Gaussian.
 Secondly,
 the median value of the numerically solved $\Qvthre$
 is indeed much closer to the theoretical estimate
 with a smaller 1$\~$(2)$\~\sigma$ upper and lower statistical uncertainties
 (and however a little bit longer tail
  in the high--energy range).

 Considering the currently running and built/planned
 next--generation detectors,
 in Figs.~\ref{fig:idRdQ-100-025-050}
 we present the results with
 several different target nuclei:
 $\rmXA{Si}{28}$  (top),
 $\rmXA{Ar}{40}$  (middle) and
 $\rmXA{Xe}{136}$ (bottom).
 Basically,
 all observations discussed above hold,
 except that
 the median value of the analytically solved $\Qvthre$
 with the $\rmXA{Xe}{136}$ target
 is now closer to the theoretical estimate.

 In Figs.~\ref{fig:idRdQ-100-010-050}
 we consider a smaller mass splitting
 \mbox{$\delta = 10$ keV}.
 For this extreme case%
\footnote{
 The special case of $\delta = 0$ (elastic WIMP--nucleus scattering)
 will be discussed particularly
 at the end of this section.
},
 firstly,
 the tails of the distributions of $k$, $k'$ and $\Qvthre$
 solved analytically and numerically
 found in Figs.~\ref{fig:idRdQ-Ge76-100-025-050} and \ref{fig:idRdQ-100-025-050}
 are reduced significantly.
 This indicates that,
 for smaller mass splittings,
 the statistical fluctuations of $k$, $k'$ and $\Qvthre$
 given by both of the analytic and numerical methods
 would in principle be smaller.
 However,
 while
 there is still no significant difference
 between the numerically estimated $k'$
 from different rounds of the iterative procedure
 (the third column),
 the estimations of the fitting parameter $k$
 (the second column)
 become clearly worse (and worse).
 This causes in turn
 larger deviations of the reconstructed recoil spectrum
 (see the first column).

 Moreover,
 in all four plots of the distribution of the solved $\Qvthre$
 (forth column),
 an excess around $\Qvthre \sim 0$
 can be seen clearly.
 This is caused by our setup of a lower bound
 used in solving $\Qvthre$
 and means thus that
 there is a small (but non--zero) possibility
 of obtaining non--physically negative $\Qvthre$,
 once the mass splitting is pretty small.

 Here
 we would like to remind that,
 as shown in the first and forth columns of
 Figs.~\ref{fig:idRdQ-100-010-050},
 since $\Qvthre$ is proportional to the mass splitting $\delta$
 (for a fixed WIMP mass $\mchi$, see Eq.~(\ref{eqn:Qvthre})),
 once the mass splitting is quite small ($\delta\~\lsim$ 10 keV),
 the position of the peak of the recoil spectrum, $\Qpk$
 (a little bit smaller than $\Qvthre$),
 would also be pretty small,
 probably
 smaller than the experimental minimal
 (software or even hardware) cut--off energy%
\footnote{
 For very light target nuclei,
 e.g.~$\rmXA{F}{19}$,
 $\Qvthre$ is a little bit {\em higher}.
 Thus
 detectors with such materials
 would be more suitable for identifying low--$\delta$ inelastic WIMPs
 \cite{Daw,
       Felizardo11b,
       Santos11b,
       Archambault12,
       Behnke12}.
}.
 In this case,
 the measured WIMP (inelastic) scattering spectrum
 (above the analysis threshold energy)
 would be {\em monotonically decreased} with increasing recoil energy
 and thus (very) difficult to be distinguished from
 the exponential--like elastic scattering spectrum
 (see e.g.~the bottom--left frame of Figs.~\ref{fig:idRdQ-100-025-050}
  and the left column of Figs.~\ref{fig:idRdQ-100-010-050}).
 Hence,
 a reduction of the threshold energy to be small enough/negligible
 would be necessary%
\footnote{
 So far,
 the minimal cut--off energies
 in most currently running experiments
 with semiconductor or liquid nobel gas detectors
 have been reduced to between 5 and \mbox{20 keV}
 \cite{Ahmed10b,
       Armengaud11,
       Angloher11,
       Akimov11b,
       Aprile12,
       Bernabei13,
       Agnese13a, Agnese13b,
       Akerib13},
 whereas
 the threshold energy of
 the CoGeNT p--type point contact Ge detector
 can be down to as low as \mbox{$\sim$ 2 keV}
 (\mbox{$\sim$ 0.5 $\keVee$})
 \cite{Aalseth11}.
 Meanwhile,
 the CDMS Collaboration developed new low--threshold technique,
 which can lower the analysis thresholds of their Ge and Si detectors
 to be smaller than \mbox{2 keV}
 \cite{CDMS-lowQthre}.
}.
 Once this requirement can be achieved,
 our simulations shown here and in the next subsections
 indicate the possibility of
 a model--independent identification of inelastic WIMPs
 for a small mass splitting of \mbox{$\lsim~10$ keV}.

 As comparison,
 in Figs.~\ref{fig:idRdQ-100-050-050}
 a larger mass splitting \mbox{$\delta = 50$ keV}
 is used.
 For this (relatively) high--$\delta$ case,
 the effect of the kinematic minimal and maximal cut--off energies
 (given in Eqs.~(\ref{eqn:Qmin_kin_in}) and (\ref{eqn:Qmax_kin_in}))
 becomes serious,
 especially for lighter target nuclei,
 e.g.~$\rmXA{Si}{28}$
 (second row).
 In this case
 the {\em analytic} results seem to be more reliable;
 the tails of the distributions of
 the reconstructed $k$, $k'$ and $\Qvthre$
 offered by the numerical iterative procedure
 in high-- (and even low--)value ranges
 become much longer and
 the divergency of the results mentioned earlier
 becomes now more problematic.
 Only numerical results offered from experiments
 with heavy target nuclei,
 e.g.~$\rmXA{Ge}{76}$ and $\rmXA{Xe}{136}$,
 could be used as auxiliary.
 Note here that
 the small bump of the distribution of $\Qvthre$
 reconstructed with the $\rmXA{Xe}{136}$ nucleus
 (forth frame in the bottom row)
 between 30 and 35 keV
 is caused by the (first) zero pont of the used nuclear form factor
 given in Eq.~(\ref{eqn:FQ_WS}).

\subsection{Identifying the positivity of \boldmath $\Qvthre$}
\begin{figure}[t!]
\begin{center}
\hspace*{-1.6cm}
\includegraphics[width=8.5cm]{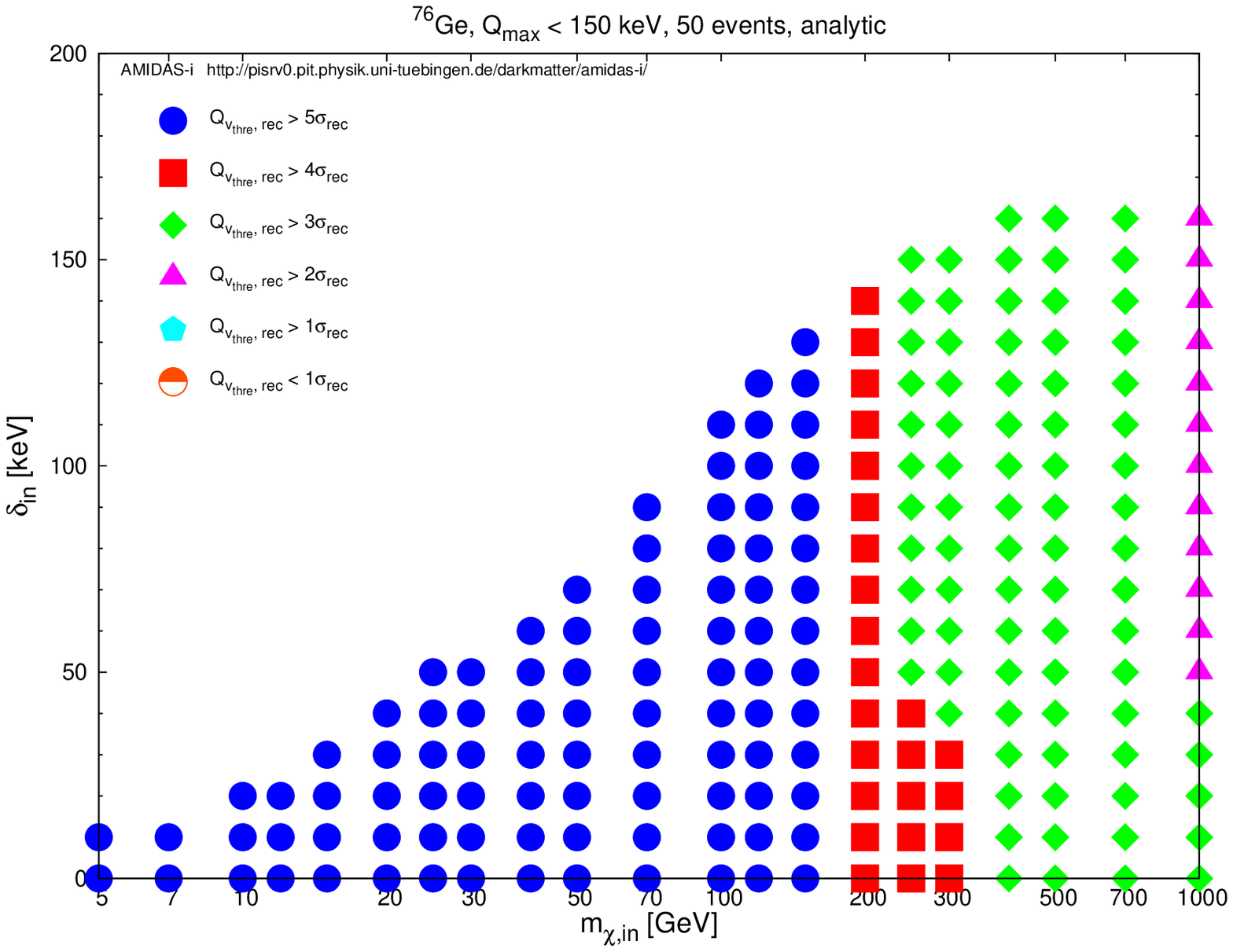}
\includegraphics[width=8.5cm]{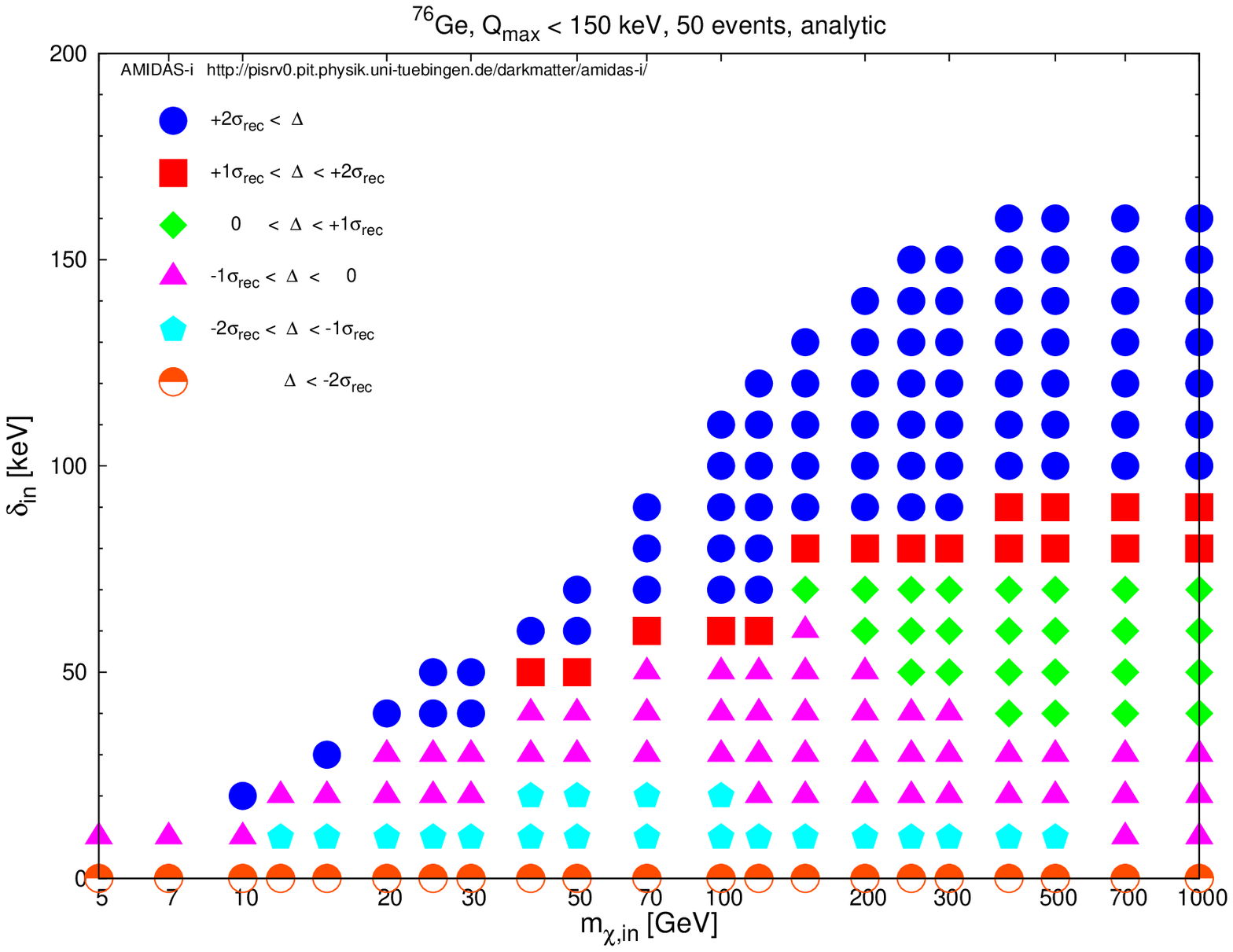} \hspace*{-1.6cm} \\
\vspace{-0.25cm}
\end{center}
\caption{
 Left:
 confidence levels (CLs) of the positivity
 of the analytically reconstructed $\Qvthre$
 (reconstructed $\Qvthre$
  in units of the estimated 1$\sigma$ lower
  statistical uncertainties
  $\sigma_{\rm lo}\abrac{\Qvthre}$):
 $> 5\sigma$ CL (blue    filled           circles),
 $> 4\sigma$ CL (red     filled           squares),
 $> 3\sigma$ CL (green   filled           diamonds),
 $> 2\sigma$ CL (magenta filled           up--triangles),
 $> 1\sigma$ CL (cyan    filled           pentagons),
 $< 1\sigma$ CL (orange  up--half--filled circles).
 Right:
 deviations of the analytically reconstructed $\Qvthre$
 from the theoretical values
 (\mbox{$\Delta \equiv Q_{v_{\rm thre}, {\rm th}} - Q_{v_{\rm thre}, {\rm rec}}$})
  in units of the estimated 1$\sigma$ lower
  statistical uncertainties
  $\sigma_{\rm lo}\abrac{\Qvthre}$:
 $ 2\sigma < \Delta           $ (blue    filled           circles),
 $ 1\sigma < \Delta <  2\sigma$ (red     filled           squares),
 $ 0       < \Delta <  1\sigma$ (green   filled           diamonds),
 $-1\sigma < \Delta <  0      $ (magenta filled           up--triangles),
 $-2\sigma < \Delta < -1\sigma$ (cyan    filled           pentagons),
 $           \Delta < -2\sigma$ (orange  up--half--filled circles).
 Here we use $\rmXA{Ge}{76}$
 as the target nucleus
 and check 21 different input WIMP masses
 between 5 GeV and 1 TeV
 with 21 different input mass splittings
 between 0 and \mbox{200 keV}.
 Other parameters are as
 in Figs.~\ref{fig:idRdQ-Ge76-100-025-050}.
 See the text for further details.
}
\label{fig:Qvthre-Ge76-ana-050}
\end{figure}
\begin{figure}[b!]
\begin{center}
\hspace*{-1.6cm}
\includegraphics[width=8.5cm]{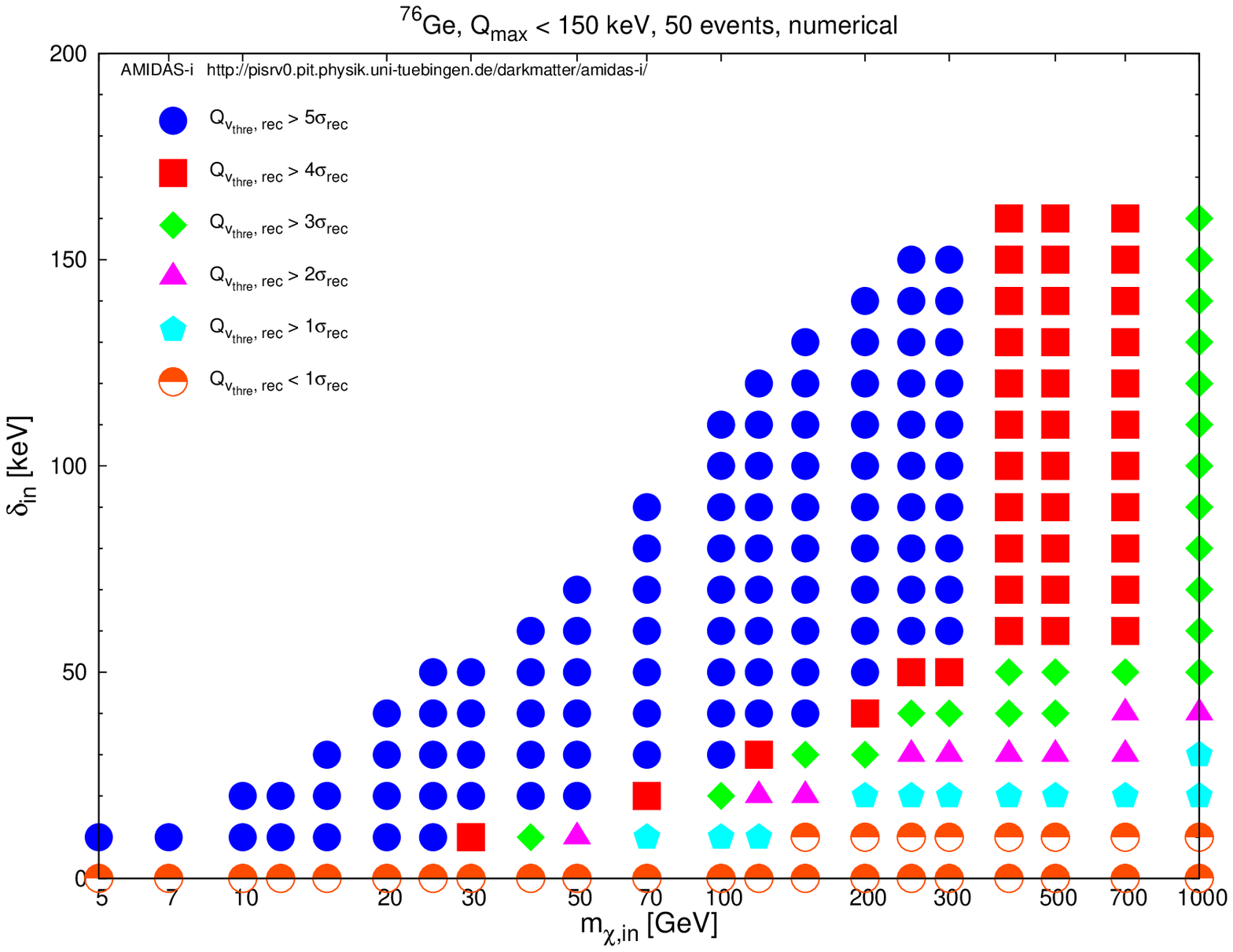}
\includegraphics[width=8.5cm]{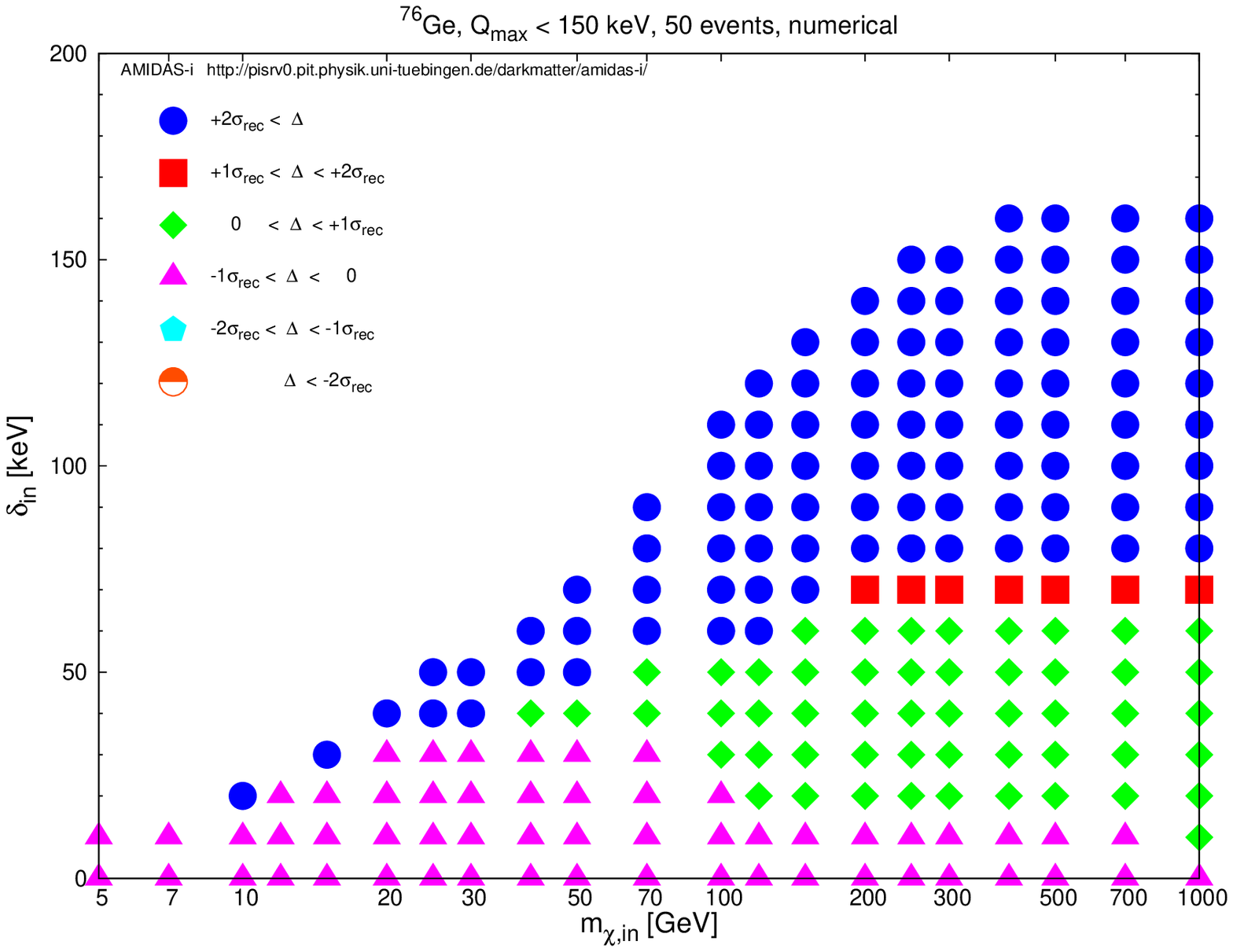} \hspace*{-1.6cm} \\
\vspace{-0.25cm}
\end{center}
\caption{
 As in Figs.~\ref{fig:Qvthre-Ge76-ana-050},
 except that
 $\Qvthre$ have been reconstructed numerically.
}
\label{fig:Qvthre-Ge76-num-050}
\end{figure}
\begin{figure}[t!]
\begin{center}
\hspace*{-1.6cm}
\includegraphics[width=8.5cm]{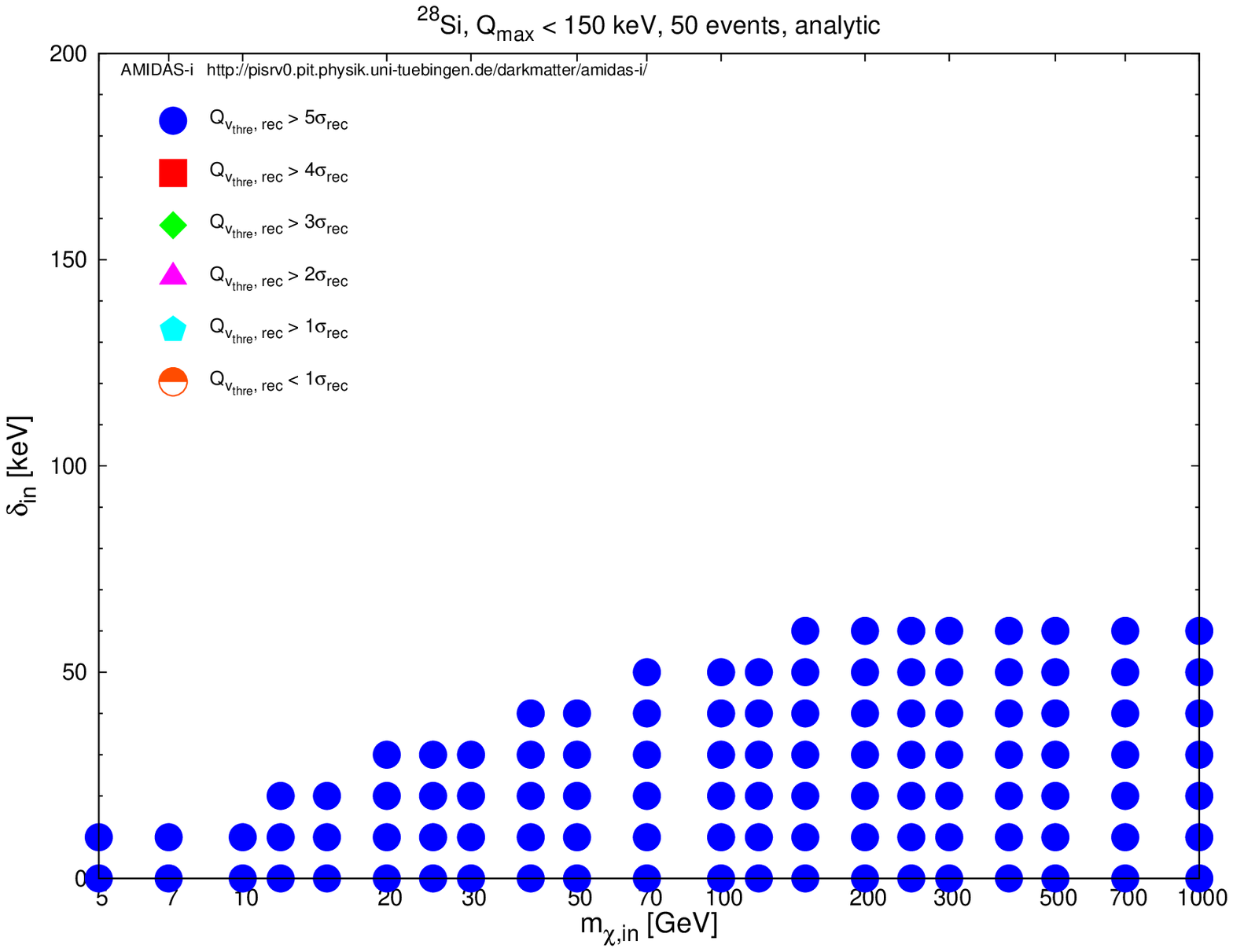}
\includegraphics[width=8.5cm]{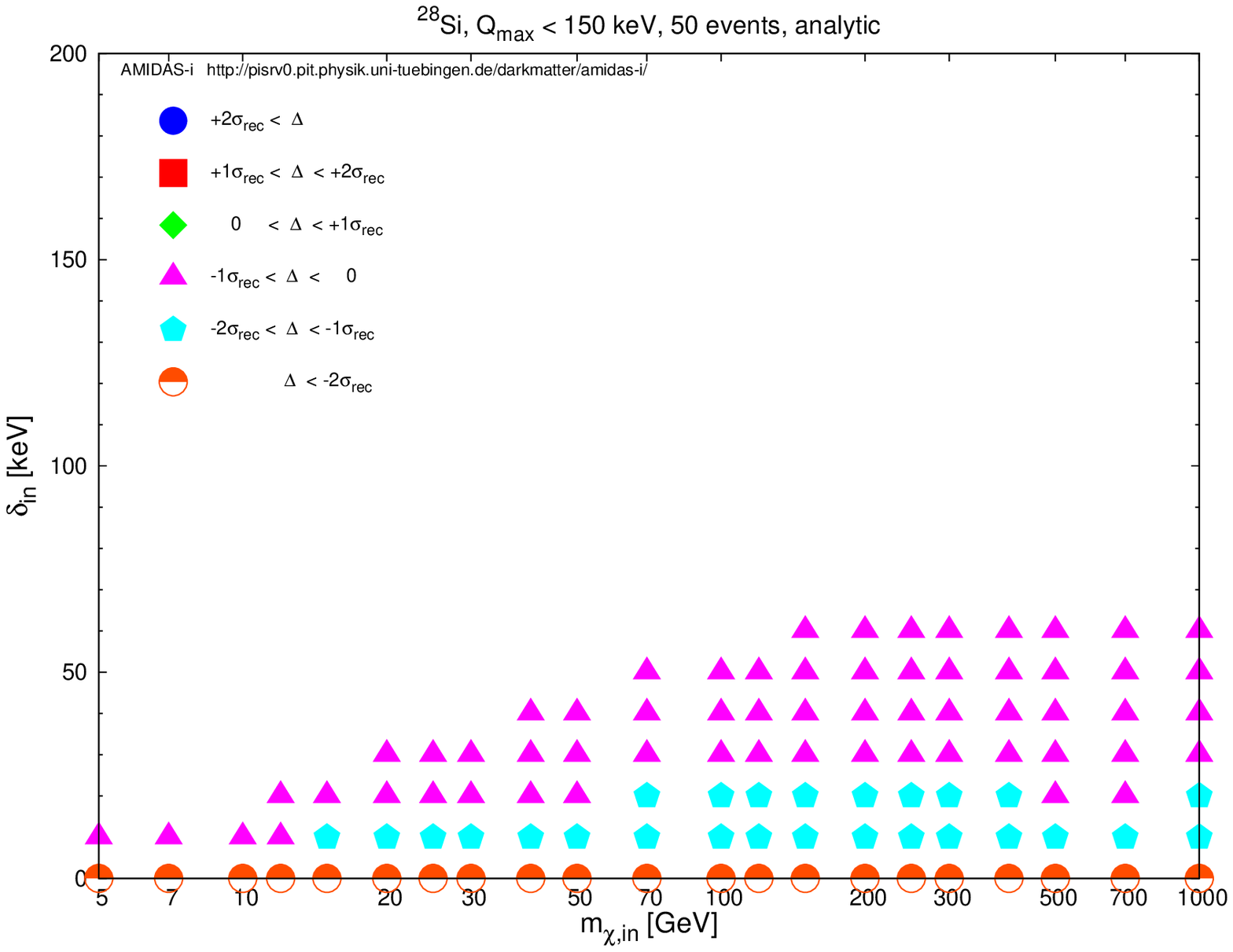} \hspace*{-1.6cm} \\
 ($\rmXA{Si}{28}$) \\
\vspace{0.75cm}
\hspace*{-1.6cm}
\includegraphics[width=8.5cm]{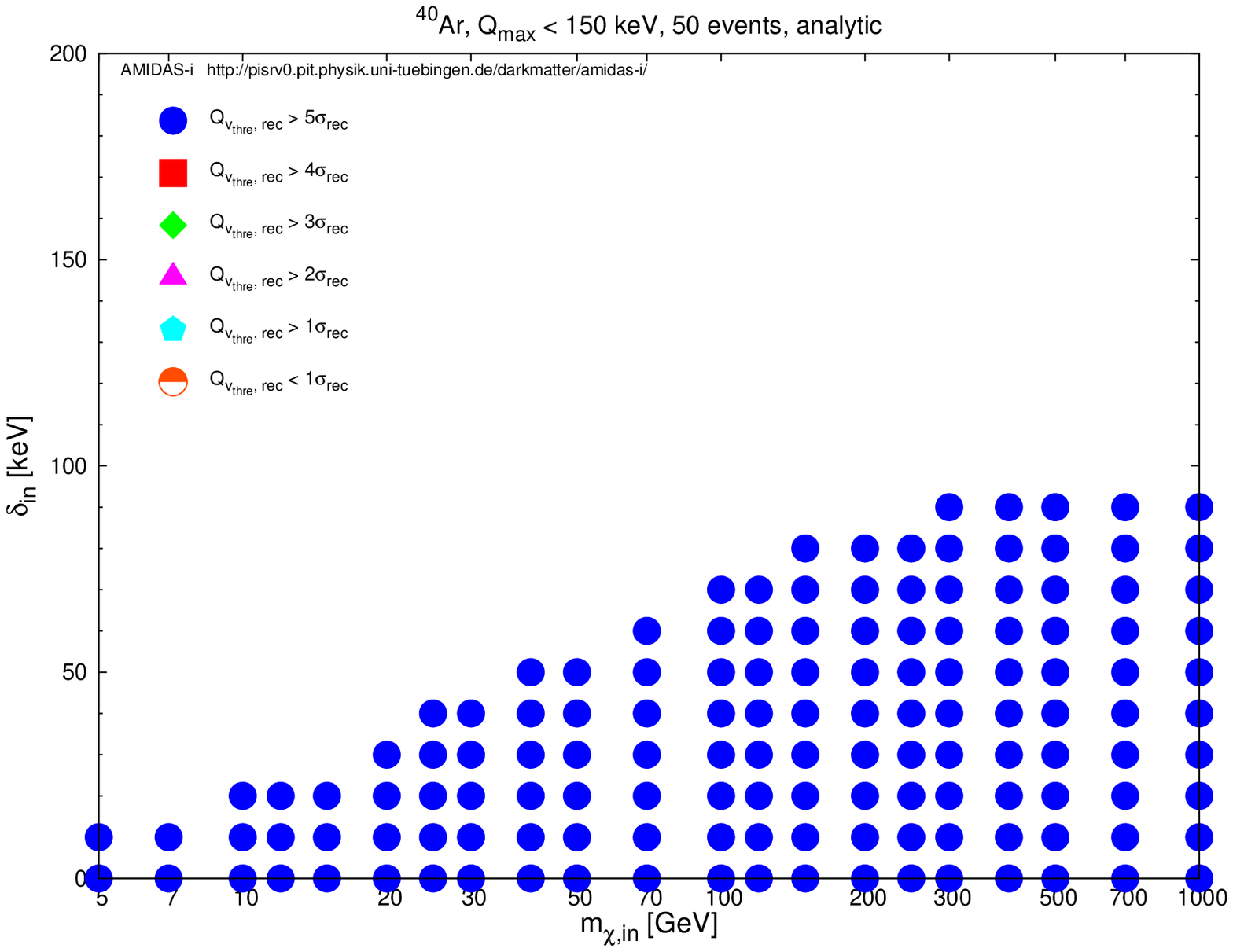}
\includegraphics[width=8.5cm]{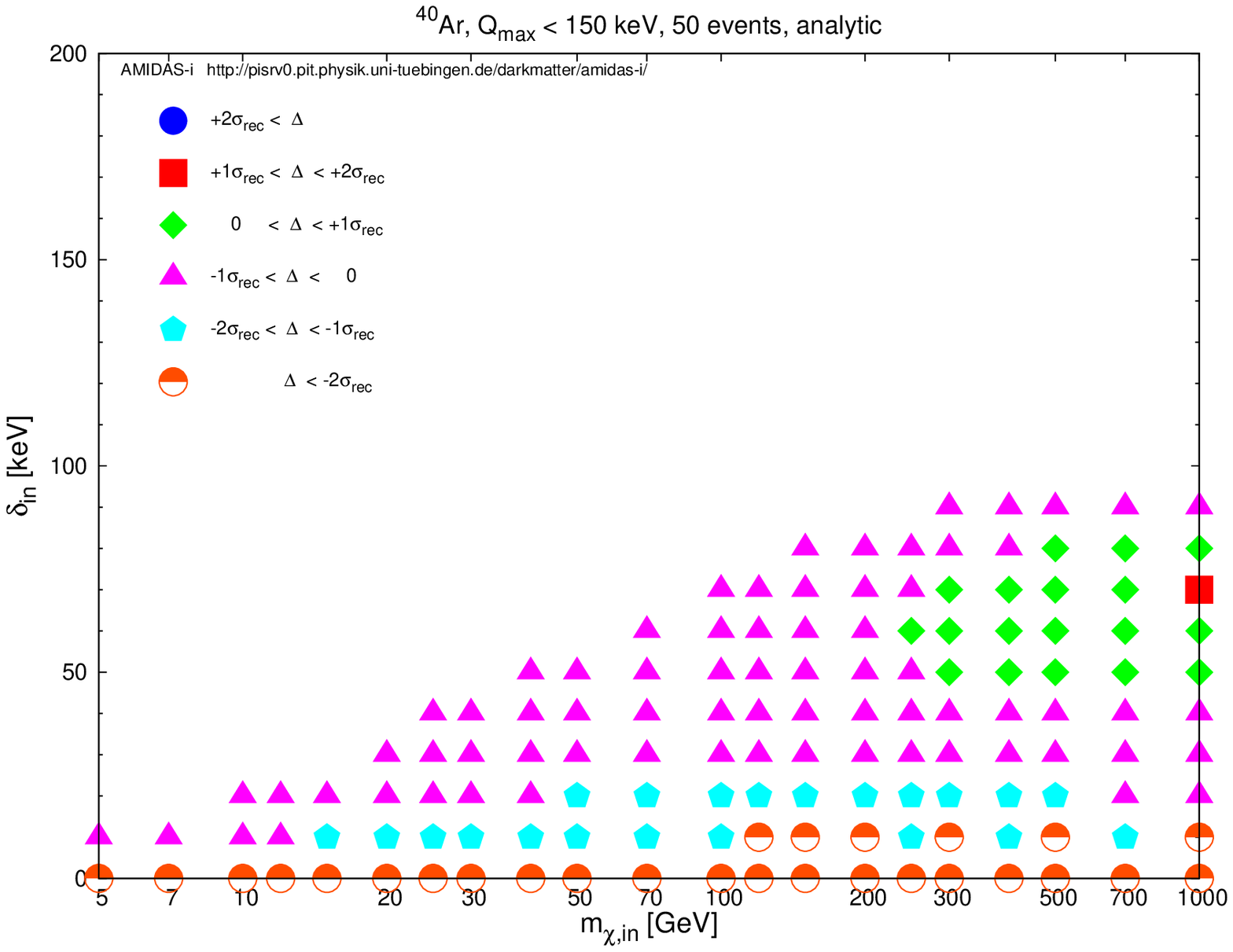} \hspace*{-1.6cm} \\
 ($\rmXA{Ar}{40}$) \\
\vspace{0.75cm}
\hspace*{-1.6cm}
\includegraphics[width=8.5cm]{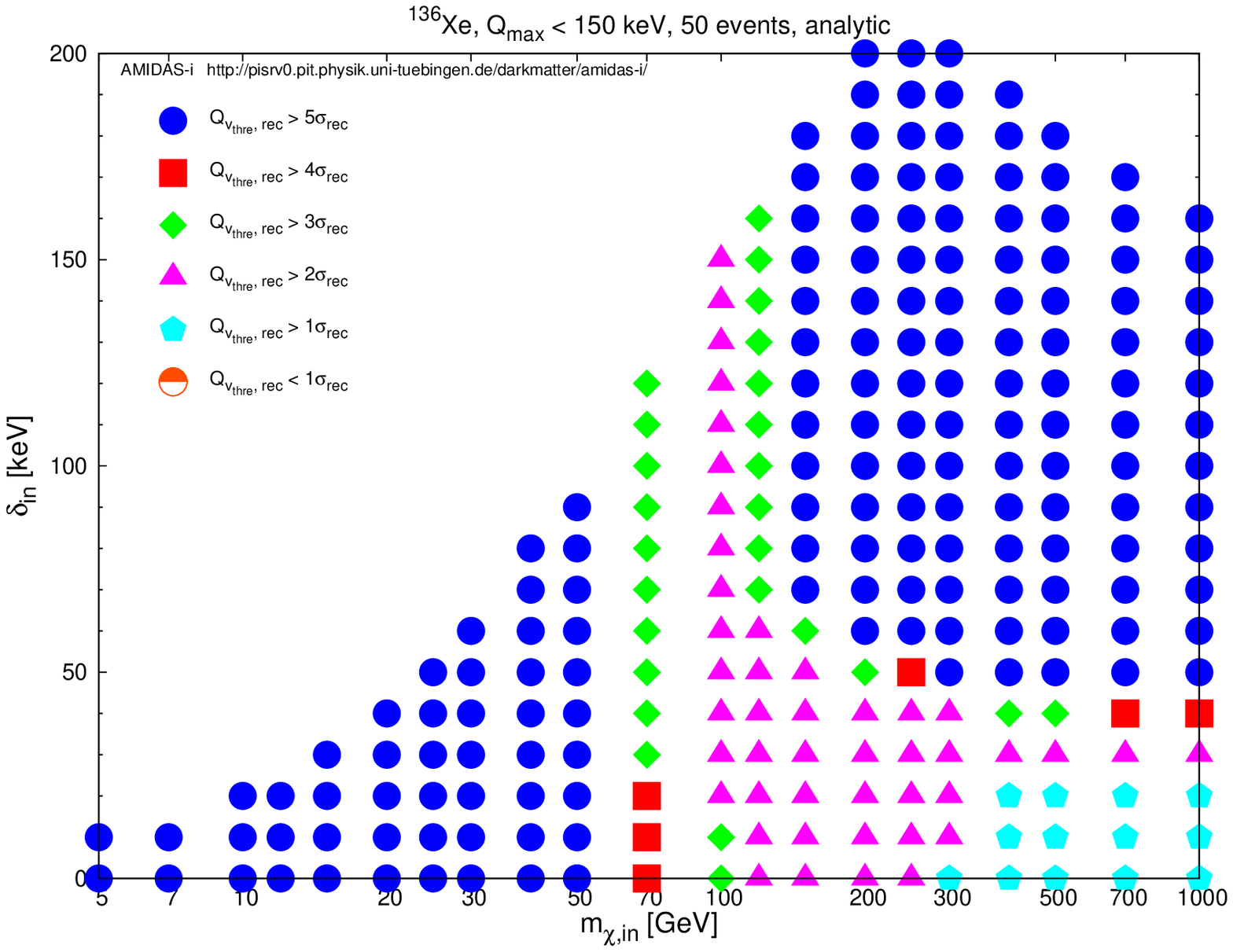}
\includegraphics[width=8.5cm]{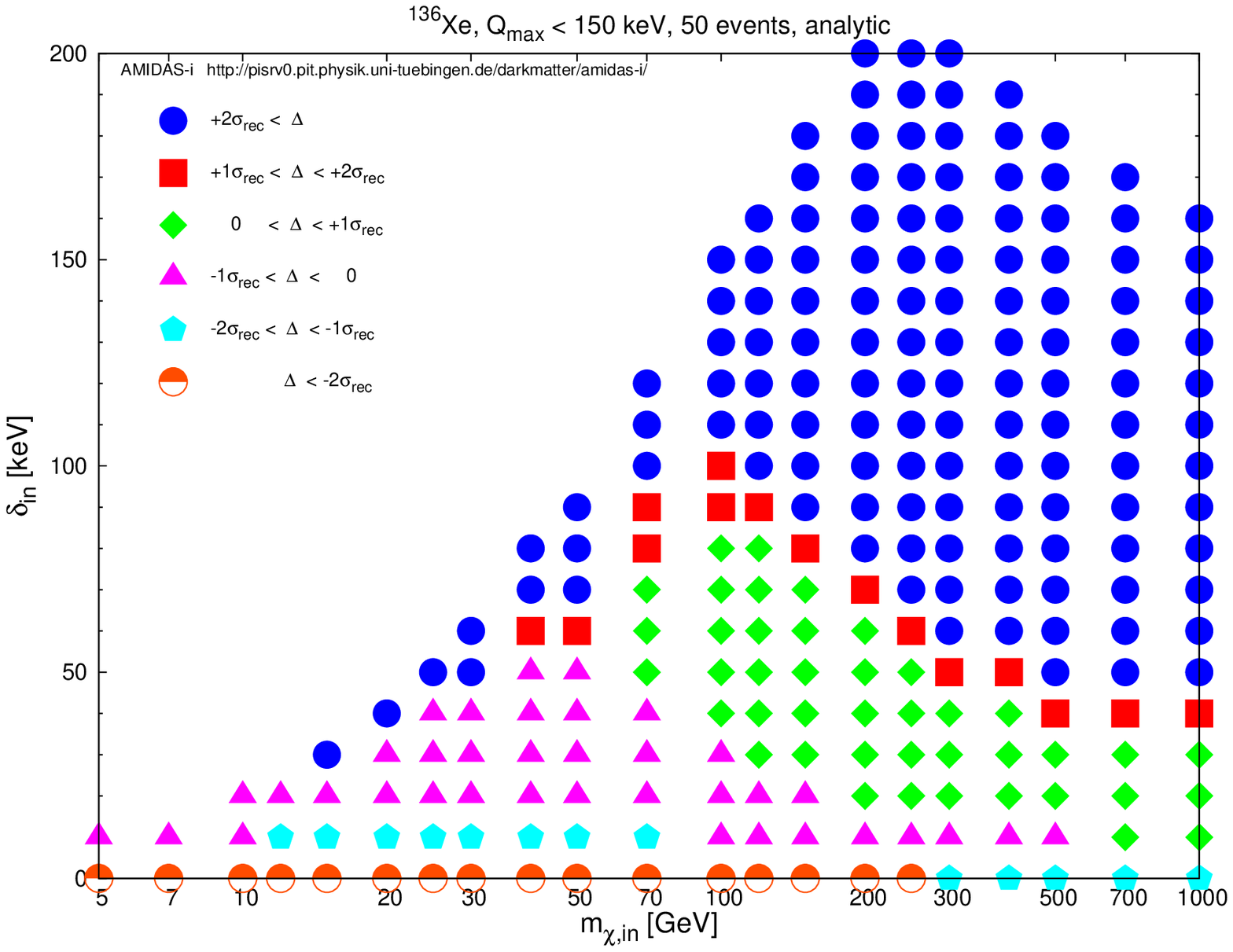} \hspace*{-1.6cm} \\
 ($\rmXA{Xe}{136}$) \\
\vspace{-0.25cm}
\end{center}
\caption{
 As in Figs.~\ref{fig:Qvthre-Ge76-ana-050}
 ({\em analytically} estimated $\Qvthre$),
 except that
 $\rmXA{Si}{28}$  (top),
 $\rmXA{Ar}{40}$  (middle) and
 $\rmXA{Xe}{136}$ (bottom)
 have been used as target nuclei.
}
\label{fig:Qvthre-ana-050}
\end{figure}

 In this subsection,
 we study the positivity of
 the reconstructed characteristic energy $\Qvthre$
 (as the most important criterion of
  the identification of inelastic WIMP--nucleus scattering)
 in details.

 In the left frame of Figs.~\ref{fig:Qvthre-Ge76-ana-050},
 we show
 the confidence levels (CLs) of the positivity
 of the analytically reconstructed $\Qvthre$
 (reconstructed $\Qvthre$
  in units of the estimated 1$\sigma$ {\em lower}
  statistical uncertainties
  $\sigma_{\rm lo}\abrac{\Qvthre}$):
 $> 5\sigma$ CL (blue    filled           circles),
 $> 4\sigma$ CL (red     filled           squares),
 $> 3\sigma$ CL (green   filled           diamonds),
 $> 2\sigma$ CL (magenta filled           up--triangles),
 $> 1\sigma$ CL (cyan    filled           pentagons),
 $< 1\sigma$ CL (orange  up--half--filled circles).
 Here we use $\rmXA{Ge}{76}$
 as the target nucleus
 and check 21 different input WIMP masses
 between 5 GeV and 1 TeV
 with 21 different input mass splittings
 between 0 and \mbox{200 keV}.
 Note that
 the empty areas on the upper part of these plots
 indicate that
 the ability for reconstructing the fitting parameters $k$ and $k'$
 and then for solving $\Qvthre$
 would be limited by
 the kinematic minimal and maximal cut--off energies
 $Q_{\rm (min, max), kin}$ given
 in Eqs.~(\ref{eqn:Qmin_kin_in}) and (\ref{eqn:Qmax_kin_in}),
 since either the WIMP mass is too small or the mass splitting is too large.
 It can be found here that,
 for WIMP masses $\lsim$ 150 GeV
 (with $\rmXA{Ge}{76}$ as the target nucleus),
 the analytically reconstructed $\Qvthre$ should in principle be
 at least 5$\sigma$ CL apart from zero;
 for larger masses 200 GeV $\lsim~\mchi~\lsim$ 1 TeV,
 $\Qvthre$ should still be $3\sigma - 4 \sigma$ CL apart from zero.

 However,
 as shown in the previous subsection,
 the (analytically) reconstructed $\Qvthre$ would be
 either {\em overestimated} (for smaller $\delta$)
 or {\em underestimated} (for larger $\delta$).
 Hence,
 in the right frame of Figs.~\ref{fig:Qvthre-Ge76-ana-050}
 we also check
 the deviations of the analytically reconstructed $\Qvthre$
 from the theoretical values
 (\mbox{$\Delta \equiv Q_{v_{\rm thre}, {\rm th}} - Q_{v_{\rm thre}, {\rm rec}}$})
  in units of the estimated 1$\sigma$ {\em lower}
  statistical uncertainties
  $\sigma_{\rm lo}\abrac{\Qvthre}$:
 $ 2\sigma < \Delta           $ (blue    filled           circles),
 $ 1\sigma < \Delta <  2\sigma$ (red     filled           squares),
 $ 0       < \Delta <  1\sigma$ (green   filled           diamonds),
 $-1\sigma < \Delta <  0      $ (magenta filled           up--triangles),
 $-2\sigma < \Delta < -1\sigma$ (cyan    filled           pentagons),
 $           \Delta < -2\sigma$ (orange  up--half--filled circles).
 It can be found that,
 firstly,
 for WIMP masses $\lsim$ 50 GeV
 (with $\rmXA{Ge}{76}$ as the target nucleus),
 the analytically reconstructed $\Qvthre$ would be {\em overestimated}
 ($\Delta < 0$),
 once the mass splitting \mbox{$\delta~\lsim$ 40 keV};
 this upper bound could be reduced to \mbox{$\sim$ 30 keV}
 for heavier WIMP masses ($\mchi~\gsim$ 300 GeV).
 Secondly,
 for cases of the non--zero mass splitting ($\delta > 0$),
 the deviations would in principle be
 maximal 2$\sigma$ or less than 1$\sigma$ {\em overestimated}
 ($-2 \sigma_{\rm lo}\abrac{\Qvthre} < \Delta$).

 Now, 
 by combining results showing
 in two frames of Figs.~\ref{fig:Qvthre-Ge76-ana-050},
 one could conclude that,
 although the analytically reconstructed $\Qvthre$
 could be $1\sigma - 2\sigma$ overestimated,
 with around 50 total events
 one could in principle already identify the positivity of $\Qvthre$
 with a high (3$\sigma$ to 5$\sigma$) confidence level,
 except of the $\delta = 0$ (elastic WIMP scattering) case.
 This observation can in turn be used for distinguishing
 the inelastic WIMP--nucleus scattering scenarios
 from the elastic one.

 As comparison,
 in Figs.~\ref{fig:Qvthre-Ge76-num-050},
 we show the confidence levels of the positivity
 as well as the deviations
 of the numerically reconstructed $\Qvthre$
 with $\rmXA{Ge}{76}$ as the target nucleus.
 As shown in Figs.~\ref{fig:idRdQ-Ge76-100-025-050}, 
 \ref{fig:idRdQ-100-010-050} and \ref{fig:idRdQ-100-050-050},
 the statistical uncertainty
 on the numerically reconstructed $\Qvthre$
 as well as $\Qvthre$ itself
 would in principle be smaller than
 the (uncertainty on the) analytically reconstructed one.
 Hence,
 firstly,
 for WIMP masses $\lsim$ 300 GeV,
 the numerically reconstructed $\Qvthre$ could be
 5$\sigma$ CL apart from zero;
 for larger masses 300 GeV $\lsim~\mchi~\lsim$ 1 TeV,
 $\Qvthre$ could still be 4$\sigma$ CL apart from zero.
 Meanwhile,
 in the right frame
 it can be seen clearly that
 the boundary line
 between over-- and underestimations of $\Qvthre$
 is reduced to \mbox{$\sim$ 30 keV} (for lighter WIMPs \mbox{$\mchi~\lsim$ 100 GeV})
 to \mbox{$\sim$ 20 keV} (for heavier WIMPs \mbox{$\mchi~\gsim$ 100 GeV}).

 Moreover,
 in Figs.~\ref{fig:Qvthre-ana-050}
 we check the confidence levels of the positivity
 as well as the deviations
 of the {\em analytically}%
\footnote{
 Since,
 as discussed in the previous subsection,
 the analytically reconstructed fitting parameters $k$ and $k'$
 as well as the further solved $\Qvthre$
 would in principle be more reliable
 for higher mass splittings.
}
 reconstructed $\Qvthre$ with
 $\rmXA{Si}{28}$  (top),
 $\rmXA{Ar}{40}$  (middle) and
 $\rmXA{Xe}{136}$ (bottom)
 as detector materials.
 Although results offered by using light target nuclei,
 e.g.~Si and Ar,
 would (almost) always be {\em overestimated},
 an at least $3\sigma$ difference
 between reconstructed $\Qvthre$ and zero
 could still be observed.
 Meanwhile,
 by using heavier nuclei,
 e.g.~Xe,
 one could not only test a wilder area on the $\delta - \mchi$ plane
 but also give results with a higher confidence level.

\subsection{Determining \boldmath $\mchi$}

 In this subsection,
 we present the simulation results of
 the reconstruction of one of the key properties of inelastic WIMPs,
 i.e.~the WIMP mass $\mchi$.%
\footnote{
 Note that
 in all simulations shown in this and the next subsection,
 we always reconstruct the parameters $\mchi$ and $\delta$
 {\em simultaneously} in each simulated experiment.
 This means that
 neither $\mchi$ nor $\delta$ has been fixed
 (as the input value)
 in our data analysis procedure.
}

\begin{figure}[t!]
\begin{center}
\hspace*{-1.6cm}
\includegraphics[width=6.2cm]{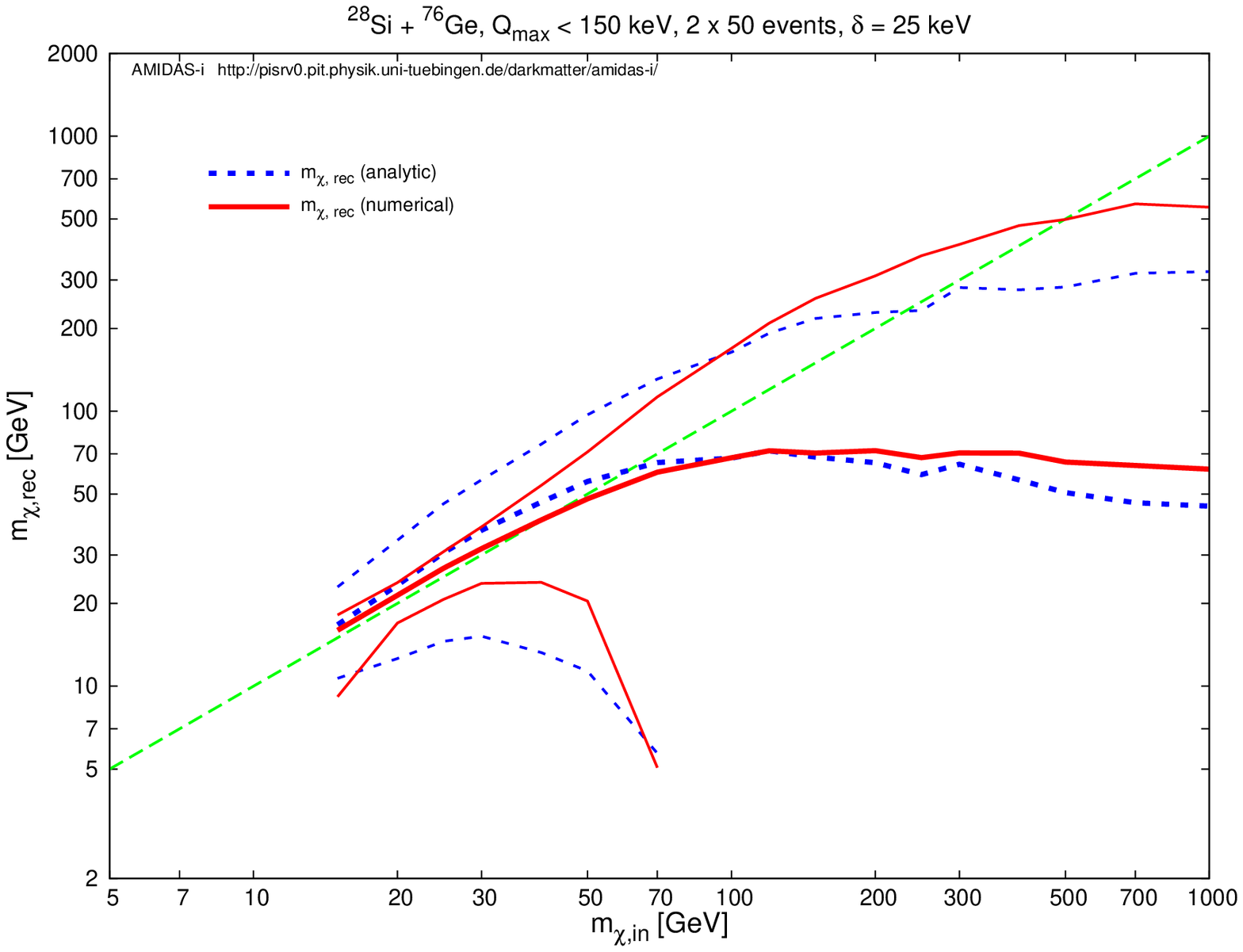}         \hspace*{-0.5cm}
\includegraphics[width=6.2cm]{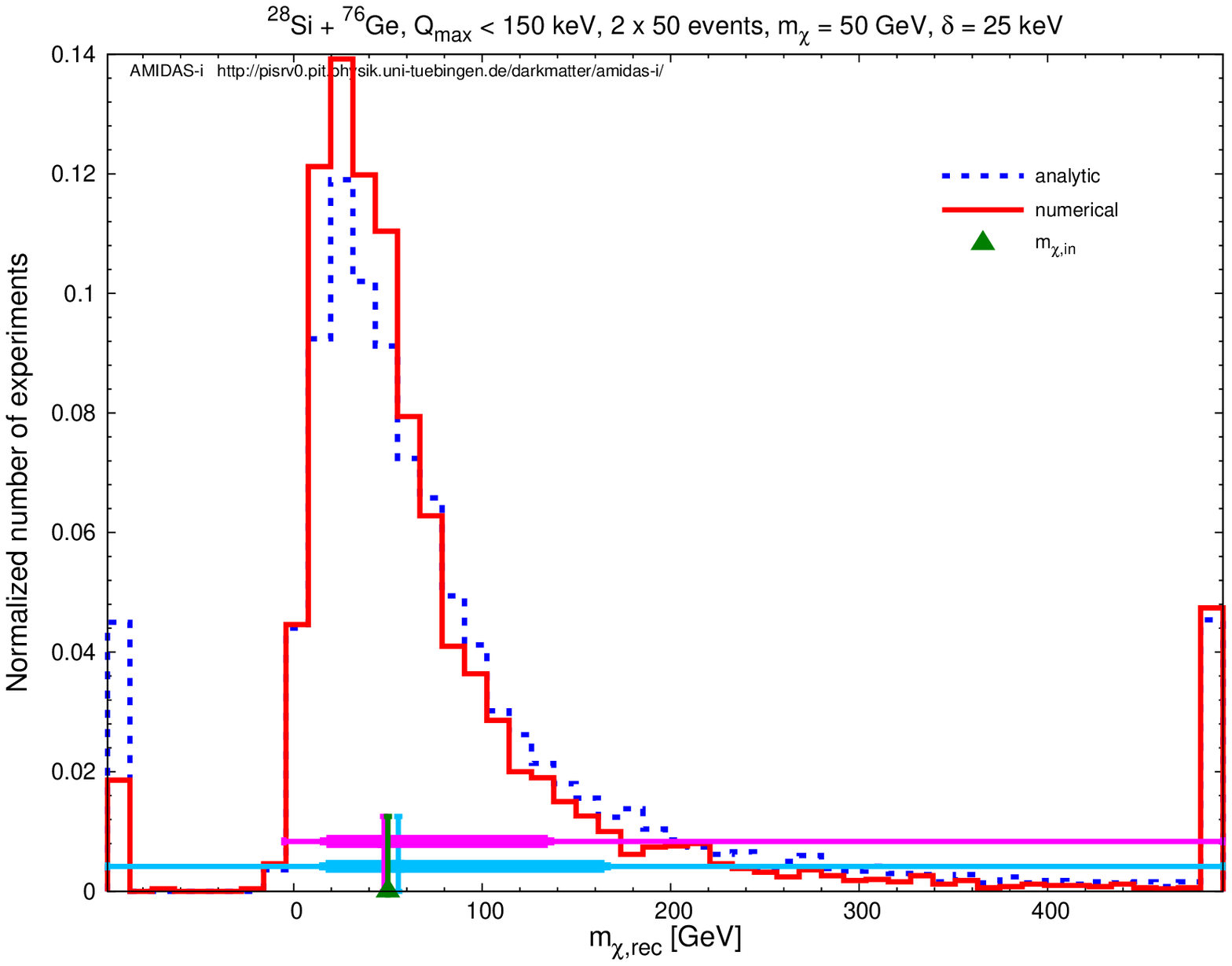} \hspace*{-0.5cm}
\includegraphics[width=6.2cm]{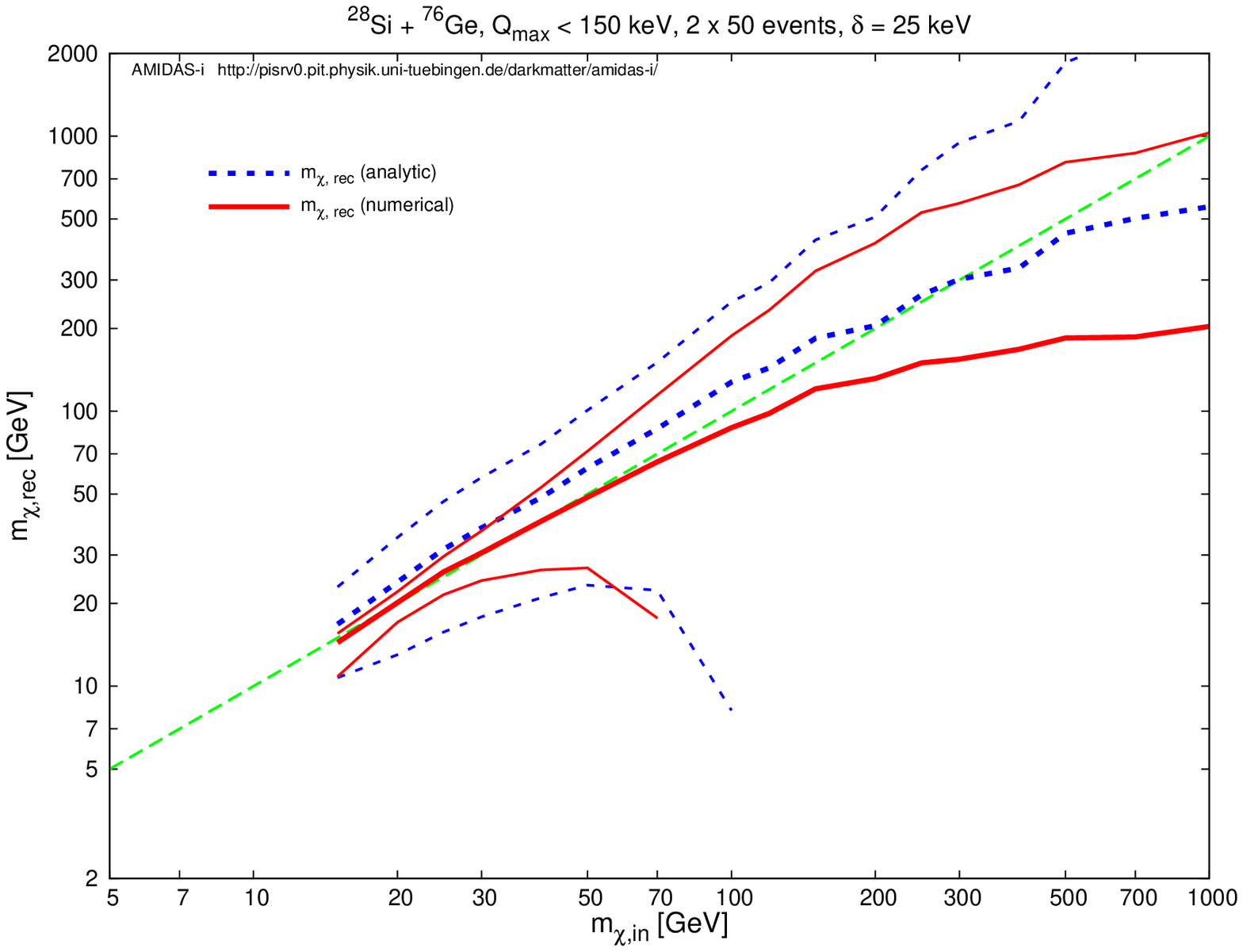}  \hspace*{-1.6cm} \\
\vspace{-0.25cm}
\end{center}
\caption{
 Left:
 reconstructed WIMP mass
 and the lower and upper bounds of
 the 1$\sigma$ statistical uncertainty
 as functions of the input WIMP mass.
 The dashed blue curves indicate the 1$\sigma$ band
 given with the parameters $k$ and $k'$ estimated analytically
 by Eqs.~(\ref{eqn:k_in_ana}) and (\ref{eqn:kp_in_ana}),
 whereas
 the solid red curves indicate the band
 given with the numerically estimated $k$ and $k'$.
 Middle:
 distributions of the reconstructed WIMP masses with
 analytically (dashed blue) and numerically (solid red)
 estimated $k$ and $k'$.
 While
 the cyan (magenta) vertical lines indicate
 the median values of the simulated results
 (corresponding to the blue and red distribution histograms)
 and
 the horizontal thick (thin) bars show the 1$\~$(2)$\~\sigma$
 (68.27\% (95.45\%)) ranges of the results,
 the green vertical line indicates
 the true (input) WIMP mass of \mbox{$\mchi = 50$ GeV}.
 Right:
 reconstructed WIMP mass
 and the 1$\sigma$ statistical uncertainty bands
 given by the median values of the reconstructed $\Qvthre$ with
 $k$ and $k'$ estimated analytically (dashed blue)
 and numerically (solid red),
 respectively.
 $\rmXA{Si}{28}$ and $\rmXA{Ge}{76}$
 have been chosen as two target nuclei.
 50 total events on average in each data set
 have been simulated.
 The input mass splitting has been set as \mbox{$\delta = 25$ keV}.
 Other parameters are as
 in Figs.~\ref{fig:idRdQ-Ge76-100-025-050}.
 See the text for further details.
}
\label{fig:mchi-SiGe-025-050}
\end{figure}

 In the left frame of Figs.~\ref{fig:mchi-SiGe-025-050},
 we show the reconstructed WIMP mass $\mchi$
 estimated by Eq.~(\ref{eqn:mchi_in})
 and the lower and upper bounds of
 the 1$\sigma$ statistical uncertainty
 estimated by Eq.~(\ref{eqn:sigma_mchi_in})
 as functions of the input WIMP mass
 for the case of $\delta = 25$ keV
 by using
 $\rmXA{Si}{28}$ and $\rmXA{Ge}{76}$
 as two target nuclei%
\footnote{
 From our results shown
 in Figs.~\ref{fig:Qvthre-Ge76-ana-050} to \ref{fig:Qvthre-ana-050},
 it seems that
 the WIMP mass (and the mass splitting)
 would be reconstructed better
 by combining two heavy target nuclei.
 However,
 our simulations show that,
 in practice,
 a combination of one light and one heavy target nucleus
 should be more suitable for reconstructing the WIMP mass.
 This can be understood as follows.
 The estimator given in Eq.~(\ref{eqn:mchi_in}) (Eq.~(\ref{eqn:delta_in}))
 of the (statistical uncertainty on the) reconstructed WIMP mass
 is inversely proportional to the (squared) difference between
 the characteristic energies of the recoil spectrum of two target nuclei
 (multiplied by the atomic mass of each nucleus).
 Hence,
 since the difference between
 the characteristic energies of the recoil spectrum of two heavy target nuclei
 is (much) smaller than
 that of one light and one heavy target nucleus,
 the statistical fluctuation caused by the use of only a few tens events
 (from one data set)
 would affect strongly our estimation of
 the (median values of the) reconstructed results,
 especially for heavier WIMP masses (\mbox{$\mchi~\gsim$ 40 GeV}).
}
 with 50 total events on average in each data set.
 The dashed blue curves indicate the 1$\sigma$ band
 given with the parameters $k$ and $k'$ estimated analytically
 by Eqs.~(\ref{eqn:k_in_ana}) and (\ref{eqn:kp_in_ana}),
 whereas
 the solid red curves indicate the band
 given with the numerically estimated $k$ and $k'$.

\begin{figure}[t!]
\begin{center}
\hspace*{-1.6cm}
\includegraphics[width=6.2cm]{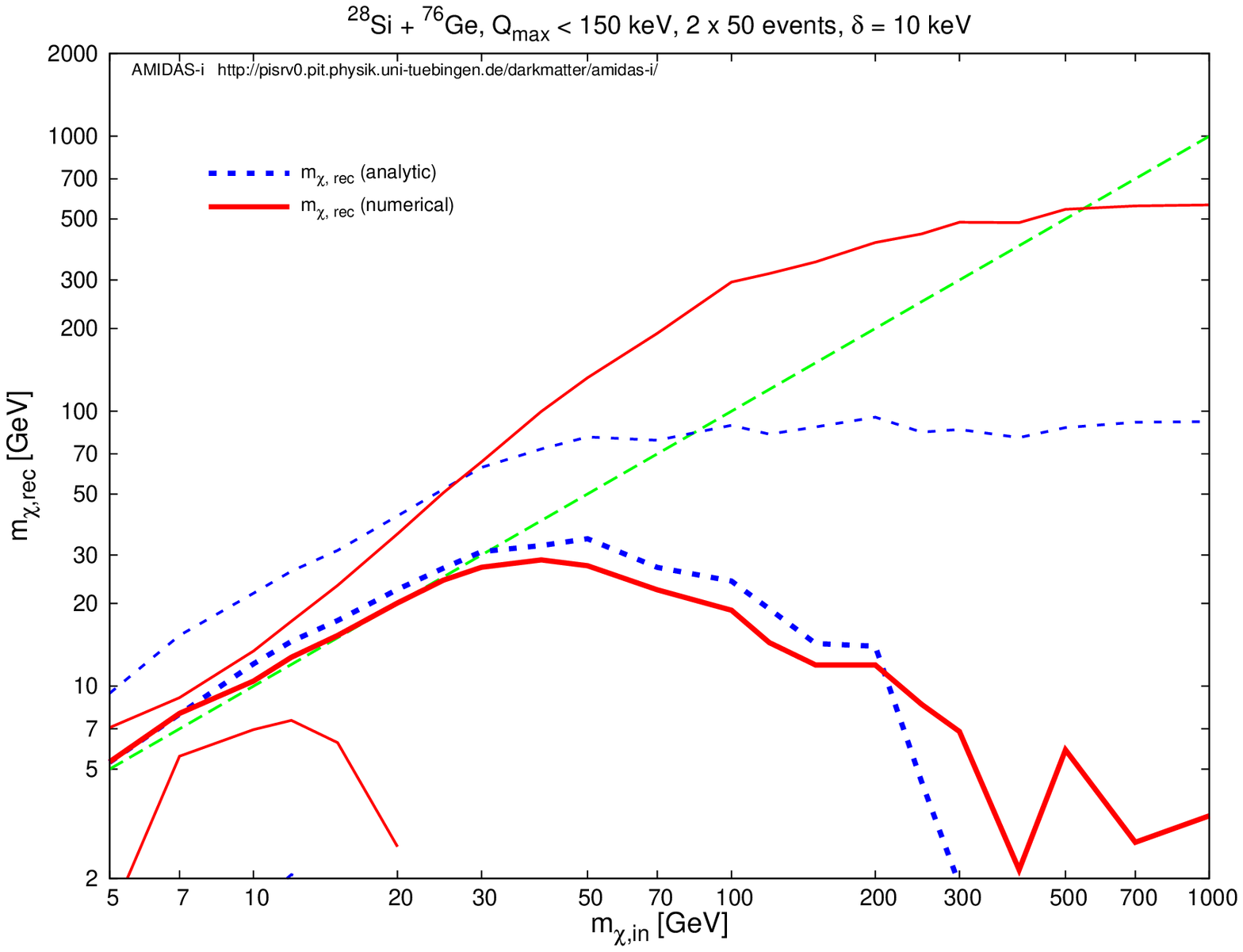}         \hspace*{-0.5cm}
\includegraphics[width=6.2cm]{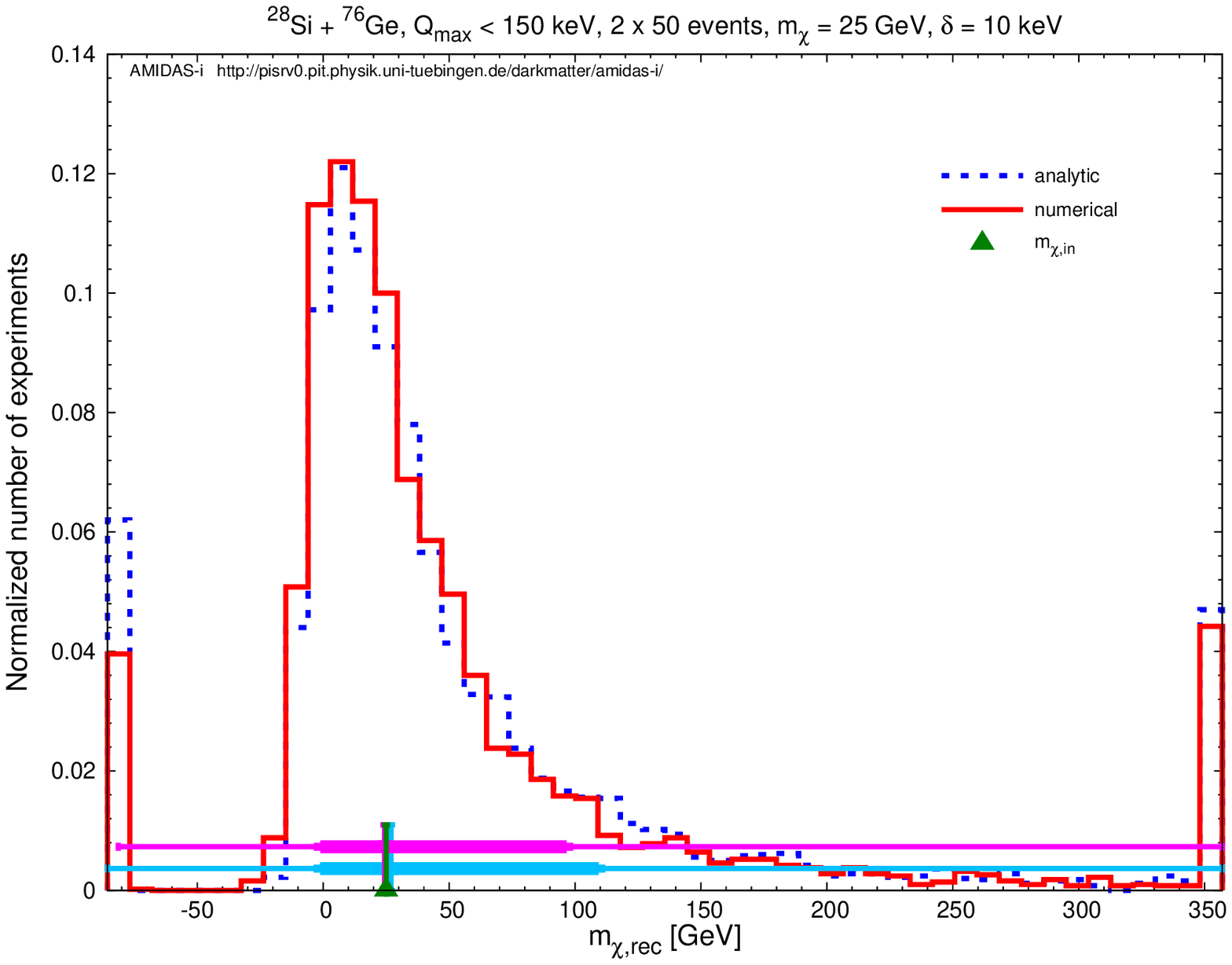} \hspace*{-0.5cm}
\includegraphics[width=6.2cm]{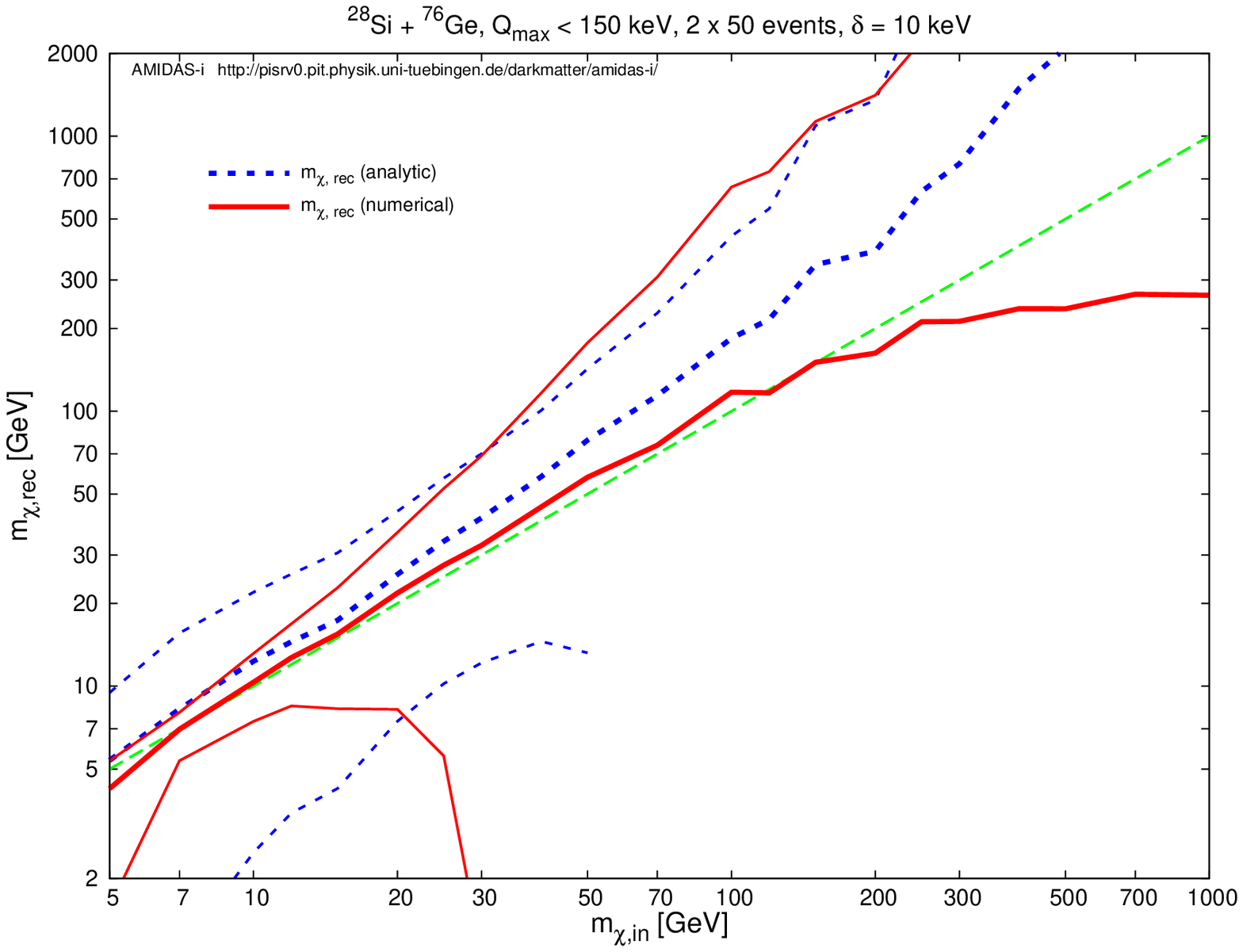}  \hspace*{-1.6cm} \\
 ($\delta = 10$ keV) \\
\vspace{0.75cm}
\hspace*{-1.6cm}
\includegraphics[width=6.2cm]{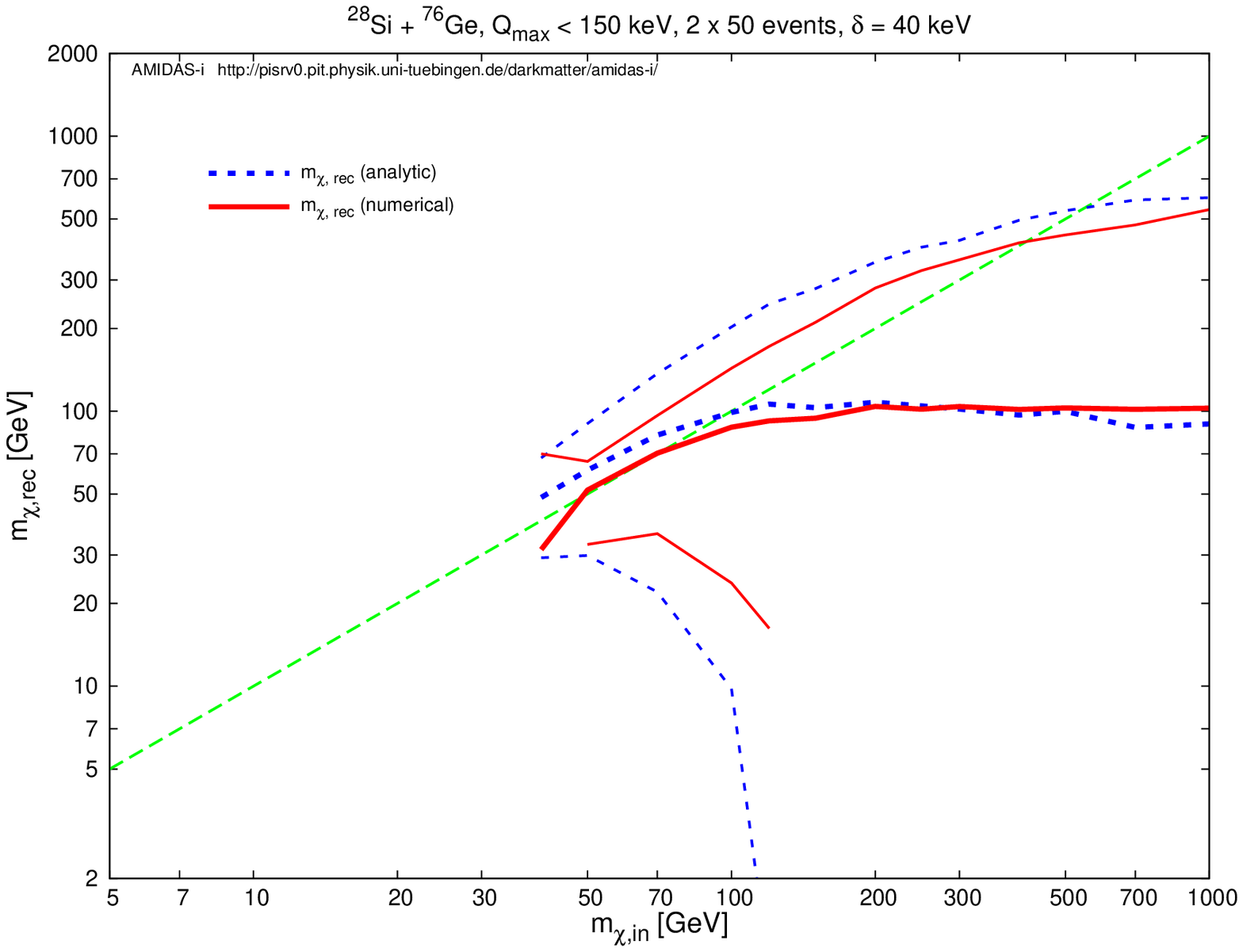}         \hspace*{-0.5cm}
\includegraphics[width=6.2cm]{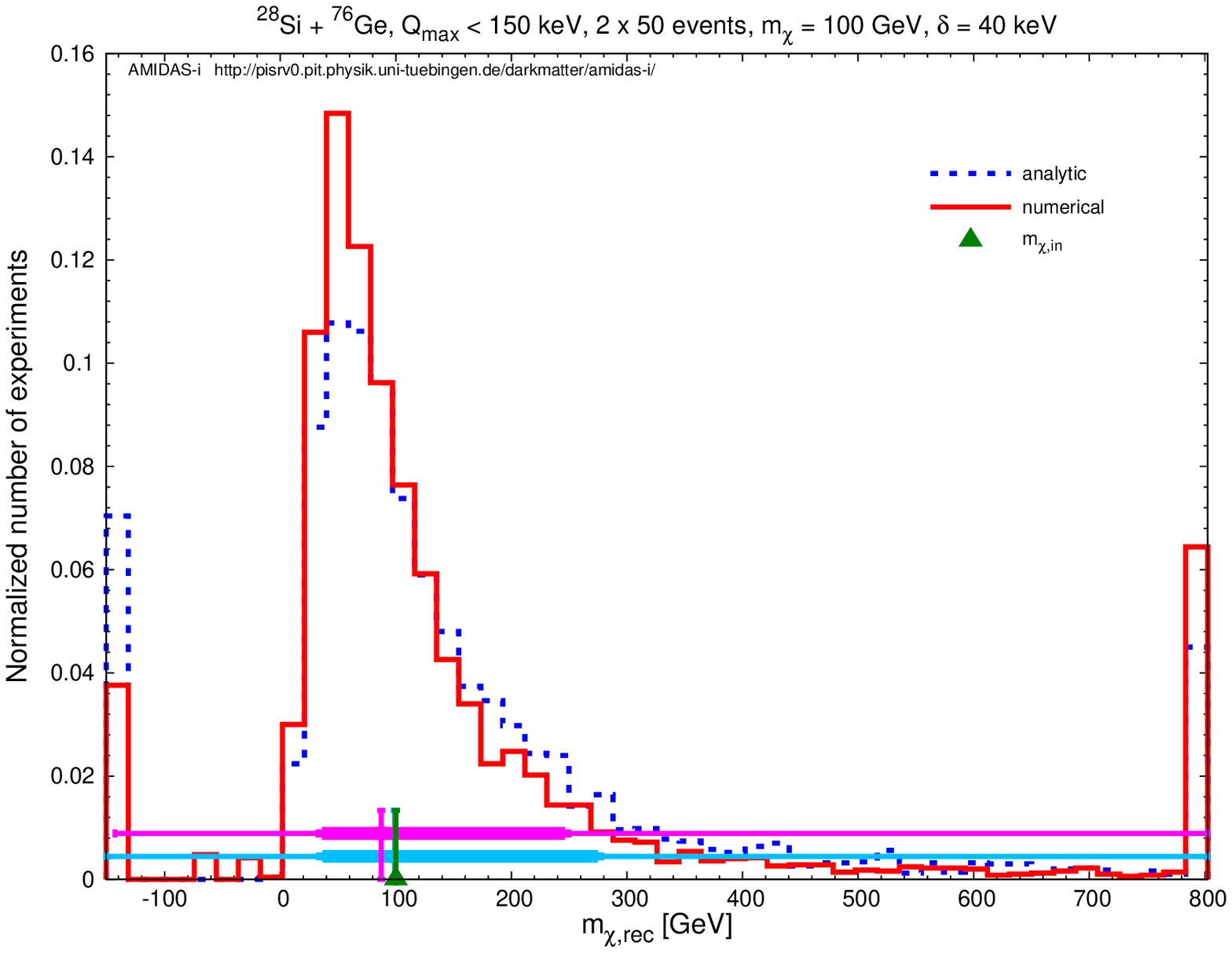} \hspace*{-0.5cm}
\includegraphics[width=6.2cm]{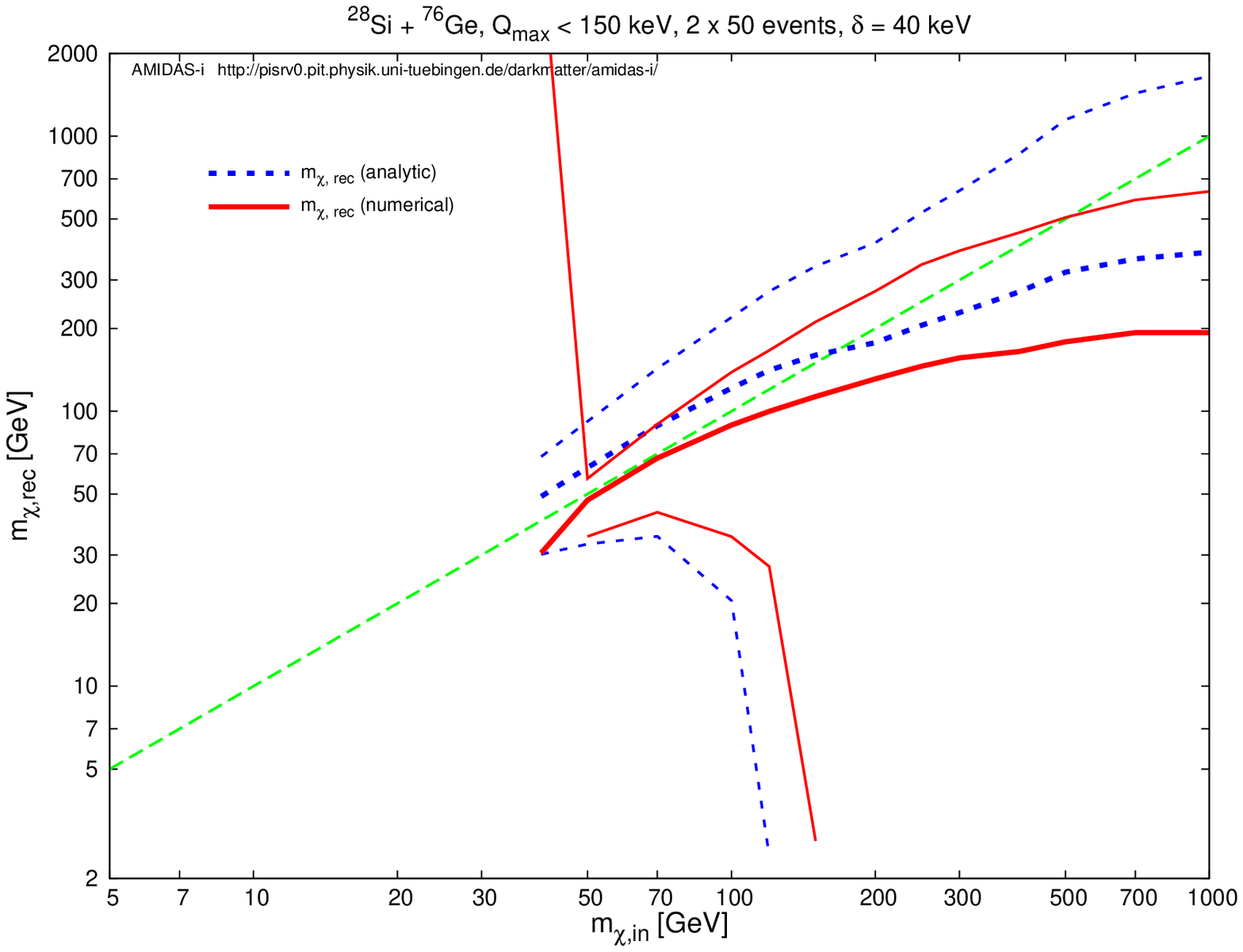}  \hspace*{-1.6cm} \\
 ($\delta = 40$ keV) \\
\vspace{-0.25cm}
\end{center}
\caption{
 As in Figs.~\ref{fig:mchi-SiGe-025-050},
 except that
 the input mass splitting has been set as
 \mbox{$\delta = 10$ keV} (upper) and
 \mbox{$\delta = 40$ keV} (lower).
 Meanwhile,
 in the middle frames
 the input WIMP masses have been set as
 \mbox{$\mchi =  25$ GeV} (upper) and
 \mbox{$\mchi = 100$ GeV} (lower).
}
\label{fig:mchi-SiGe-050}
\end{figure}

 It can be seen that,
 firstly,
 for input WIMP masses 20 GeV $\lsim~\mchi~\lsim~70$ GeV,
 while
 the analytically reconstructed WIMP mass (dashed blue)
 could be a bit {\em overestimated},
 one could in principle reconstruct $\mchi$
 by the numerical method (solid red) pretty well.
 However,
 for heavier WIMP masses 70 GeV $\lsim~\mchi~\lsim~500$ GeV,
 $\mchi$ reconstructed by both methods
 would be (strongly) {\em underestimated}.
 Nevertheless,
 the 1$\sigma$ upper bound could still cover the input (true) value
 and therefore offer at least an {\em upper} constraint
 on the mass of inelastic WIMPs.

 Meanwhile,
 in the middle frame of Figs.~\ref{fig:mchi-SiGe-025-050}
 we show the distributions of the reconstructed WIMP masses with
 analytically (dashed blue) and numerically (solid red)
 estimated $k$ and $k'$.
 While
 the cyan (magenta) vertical lines indicate
 the median values of the simulated results
 (corresponding to the blue and red distribution histograms)
 and
 the horizontal thick (thin) bars show the 1$\~$(2)$\~\sigma$
 (68.27\% (95.45\%)) ranges of the results,
 the green vertical line indicates
 the true (input) WIMP mass of \mbox{$\mchi = 50$ GeV}.
 This plot shows that,
 for input WIMP masses 20 GeV $\lsim~\mchi~\lsim~70$ GeV,
 the reconstructed WIMP masses should be concentrated
 around the true (input) value
 with tails in the high--mass range.
 However,
 it has also been found that,
 once the WIMP mass is heavy (\mbox{$\mchi~\gsim$ 100 GeV}),
 the distributions of the reconstructed $\mchi$
 could extend pretty widely
 (not only to the high--$\mchi$ range,
  but also to the unphysical, negative range).

 On the other hand,
 in contrast to a wide spread of
 the distribution of the reconstructed $\mchi$,
 the distribution of the reconstructed $\Qvthre$
 has been found to in principle be concentrated and converged
 around the theoretical value.
 Therefore,
 in the right frame of Figs.~\ref{fig:mchi-SiGe-025-050}
 we show the reconstructed WIMP mass
 and the 1$\sigma$ statistical uncertainty bands
 given by the median values of the reconstructed $\Qvthre$ with
 $k$ and $k'$ estimated analytically (dashed blue)
 and numerically (solid red),
 respectively.
 This plot shows that
 $\mchi$ estimated by Eq.~(\ref{eqn:mchi_in})
 with the median values of the reconstructed $\Qvthre$
 by using several data sets (with the same target nuclei)
 could indeed be (much) better than
 the median values of $\mchi$ reconstructed by Eq.~(\ref{eqn:mchi_in})
 with each single pair of data sets,
 especially for heavier input WIMP masses
 (\mbox{$\mchi~\gsim$ 100 GeV}).

 As comparison,
 in Figs.~\ref{fig:mchi-SiGe-050}
 we consider the case of a smaller mass splitting of
 $\delta = 10$ keV (upper)
 and that of a larger one of
 $\delta = 40$ keV (lower).
 The left frame in the upper row of Figs.~\ref{fig:mchi-SiGe-050}
 indicates that,
 while
 one could reconstruct the WIMP mass pretty well
 by both of the analytic and numerical methods
 for {\em light} WIMP masses (\mbox{$\mchi~\lsim~30$ GeV}),
 for heavier WIMP masses (\mbox{$\mchi~\gsim~50$ GeV}),
 the reconstructed $\mchi$ would be (strongly) {\em underestimated}.
 Fortunately,
 a 1$\sigma$ upper bound
 as a constraint on the mass of inelastic WIMPs
 could still be given.

 The middle frame in the upper row of Figs.~\ref{fig:mchi-SiGe-050}
 shows that,
 (only) for available WIMP mass range
 ($10^6 \delta~\lsim~\mchi~\lsim$ a few $\times$ $10^6 \delta$),
 the distribution of the reconstructed $\mchi$
 could in principle be concentrated around the true (input) value,
 with however tails in the high--
 and negative--mass ranges.
 Moreover,
 the right frame
 shows that,
 for lighter mass splitting 
 (e.g.~\mbox{$\delta = 10$ keV} shown here),
 $\mchi$ reconstructed by using the median values of $\Qvthre$
 could also be (much) better than
 the median values of $\mchi$ reconstructed
 with each single pair of data sets,
 up to even heavier WIMP mass of $\mchi~\gsim$ 1 TeV.

 On the other hand,
 the lower row of Figs.~\ref{fig:mchi-SiGe-050}
 shows that,
 once the mass splitting \mbox{$\delta~\gsim$ 40 keV},
 except of the mass range between 40 GeV and \mbox{$\sim$ 120 GeV},
 the distribution of the reconstructed WIMP mass
 could spread pretty widely
 and the reconstructed $\mchi$
 could anyway be (strongly) underestimated.

 Generally speaking,
 for the reconstruction of the WIMP mass
 by using the target combination of $\rmXA{Si}{28}$ and $\rmXA{Ge}{76}$,
 our simulations show that,
 for mass splittings \mbox{$\delta~\lsim$ 40 keV},
 one could in principle reconstruct the WIMP mass
 in the range between \mbox{$10^6 \delta$} and \mbox{$(3 - 5) \times 10^6 \delta$}
 pretty precisely with a statistical uncertainty
 of \mbox{$\sim$ 30\%} (for \mbox{$\mchi \simeq 10^6 \delta$})
 to a factor of \mbox{$\sim$ 2} (\mbox{$\mchi \simeq 5 \times 10^6 \delta$}).

\begin{figure}[t!]
\begin{center}
\hspace*{-1.6cm}
\includegraphics[width=6.2cm]{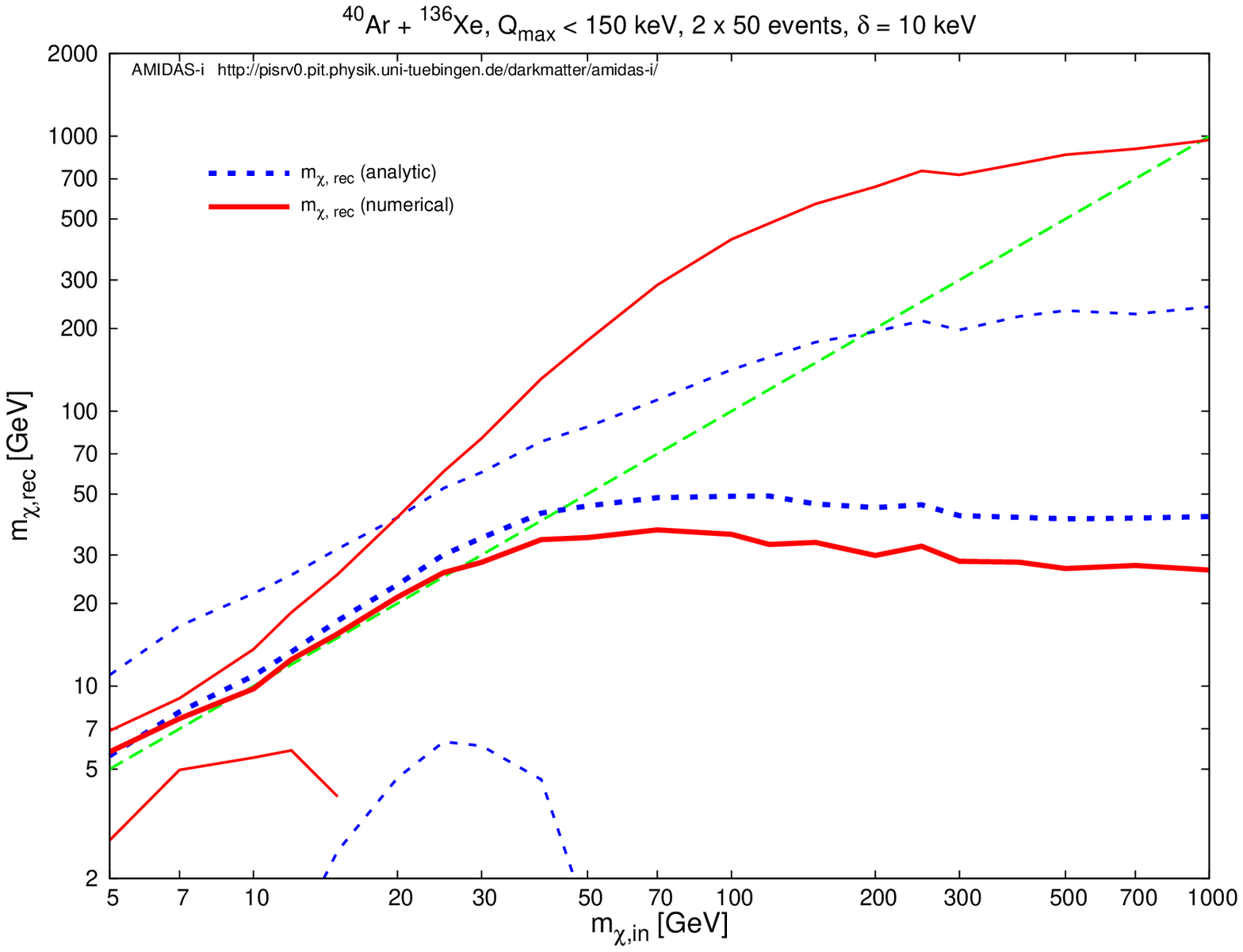}         \hspace*{-0.5cm}
\includegraphics[width=6.2cm]{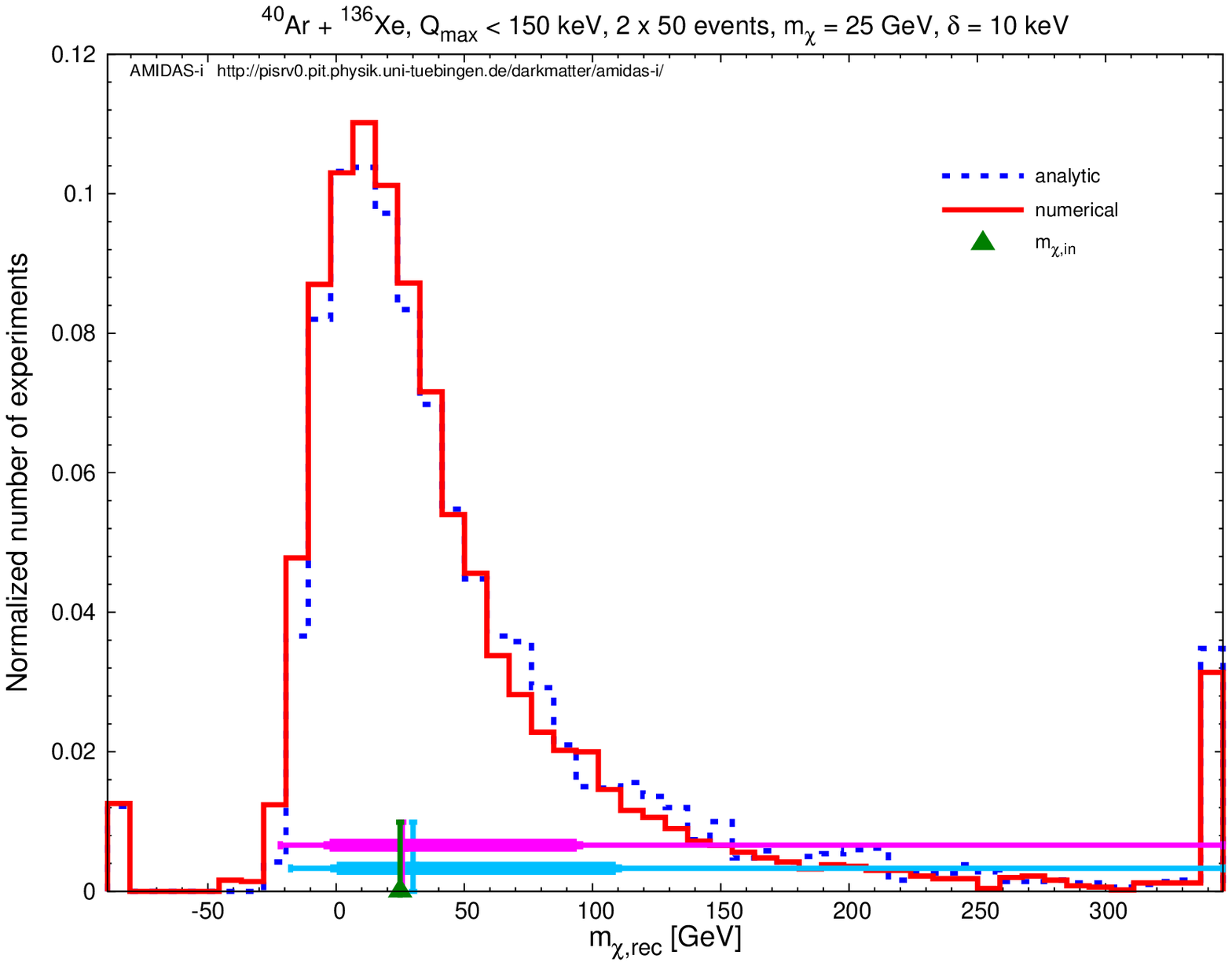} \hspace*{-0.5cm}
\includegraphics[width=6.2cm]{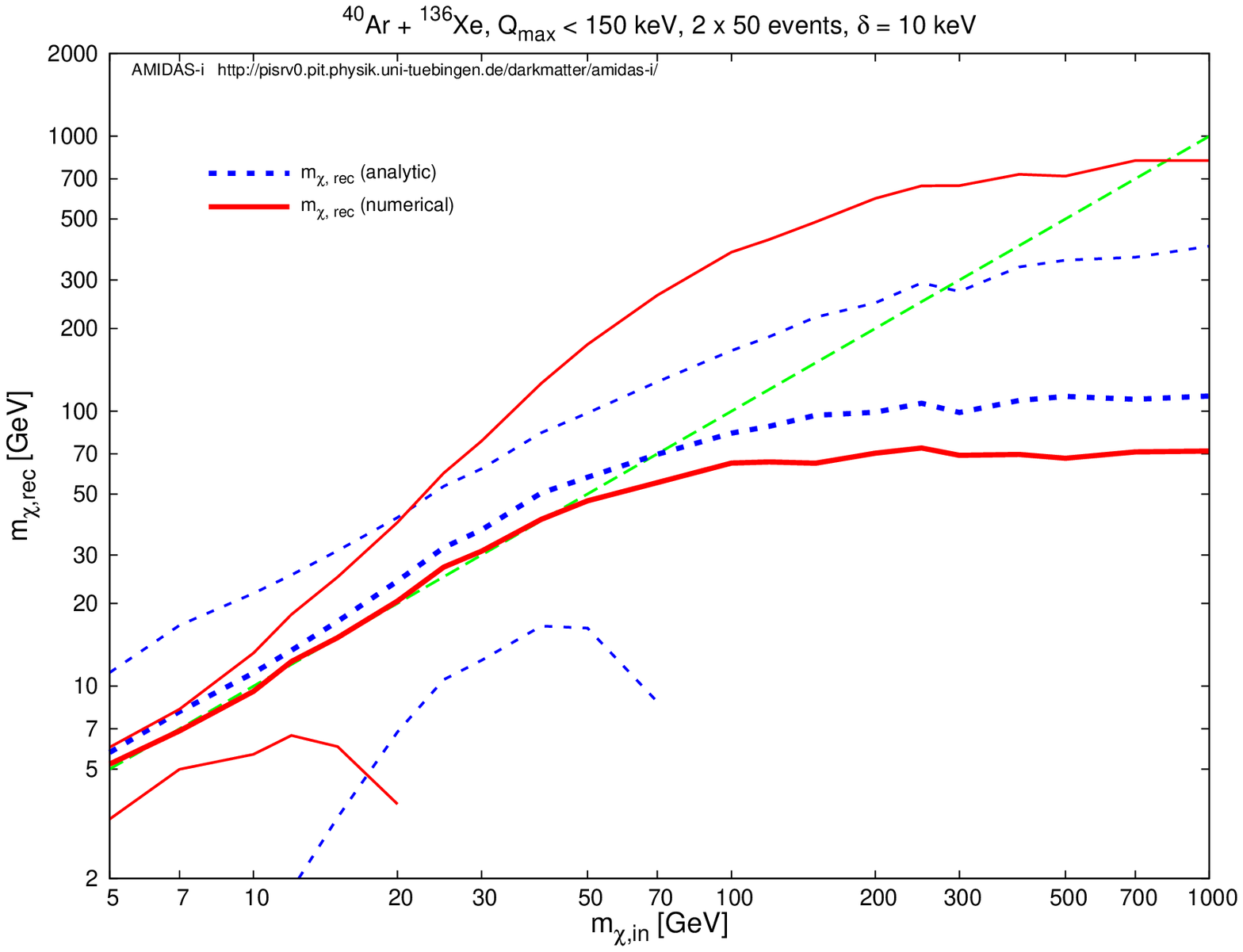}  \hspace*{-1.6cm} \\
\vspace{-0.25cm}
\end{center}
\caption{
 As in the upper frame of Figs.~\ref{fig:mchi-SiGe-050}
 (the input WIMP mass
  and the mass splitting has been set as
  \mbox{$\mchi = 25$ GeV} and \mbox{$\delta = 10$ keV},
  respectively),
 except that
 $\rmXA{Ar}{40}$ and $\rmXA{Xe}{136}$
 have been chosen as two target nuclei.
}
\label{fig:mchi-ArXe-010-050}
\end{figure}

 Finally,
 in Figs.~\ref{fig:mchi-ArXe-010-050}
 we present our simulation results of the $\mchi$ reconstruction
 by using the combination of $\rmXA{Ar}{40}$ and $\rmXA{Xe}{136}$
 as our target nuclei.
 Here we show only the case of
 a light mass splitting of \mbox{$\delta = 10$ keV}.
 It has been found
 interestingly and a bit unexpectedly
 that,
 by using the target combination of $\rmXA{Ar}{40}$ and $\rmXA{Xe}{136}$
 the strong underestimation of the reconstructed WIMP mass
 in the high--mass range
 could be alleviated
 with a much smaller statistical uncertainty
 (left)
 and the distribution of the reconstructed $\mchi$
 would also be more concentrated
 (middle).

\subsection{Determining \boldmath $\delta$}

 In this subsection,
 we present the simulation results of
 the reconstruction of the second key property of inelastic WIMPs,
 i.e.~the mass splitting $\delta$.

\begin{figure}[t!]
\begin{center}
\hspace*{-1.6cm}
\includegraphics[width=6.2cm]{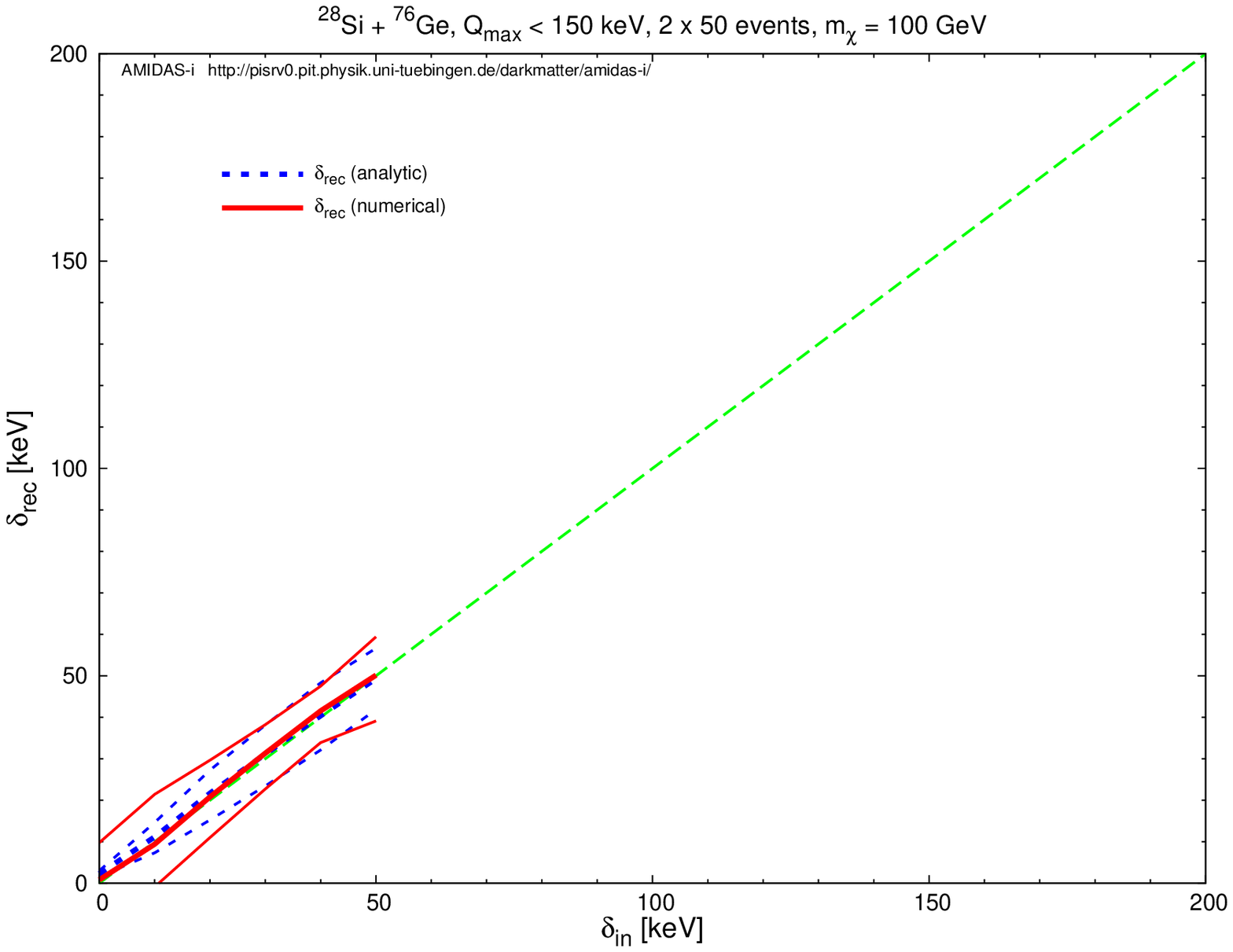}         \hspace*{-0.5cm}
\includegraphics[width=6.2cm]{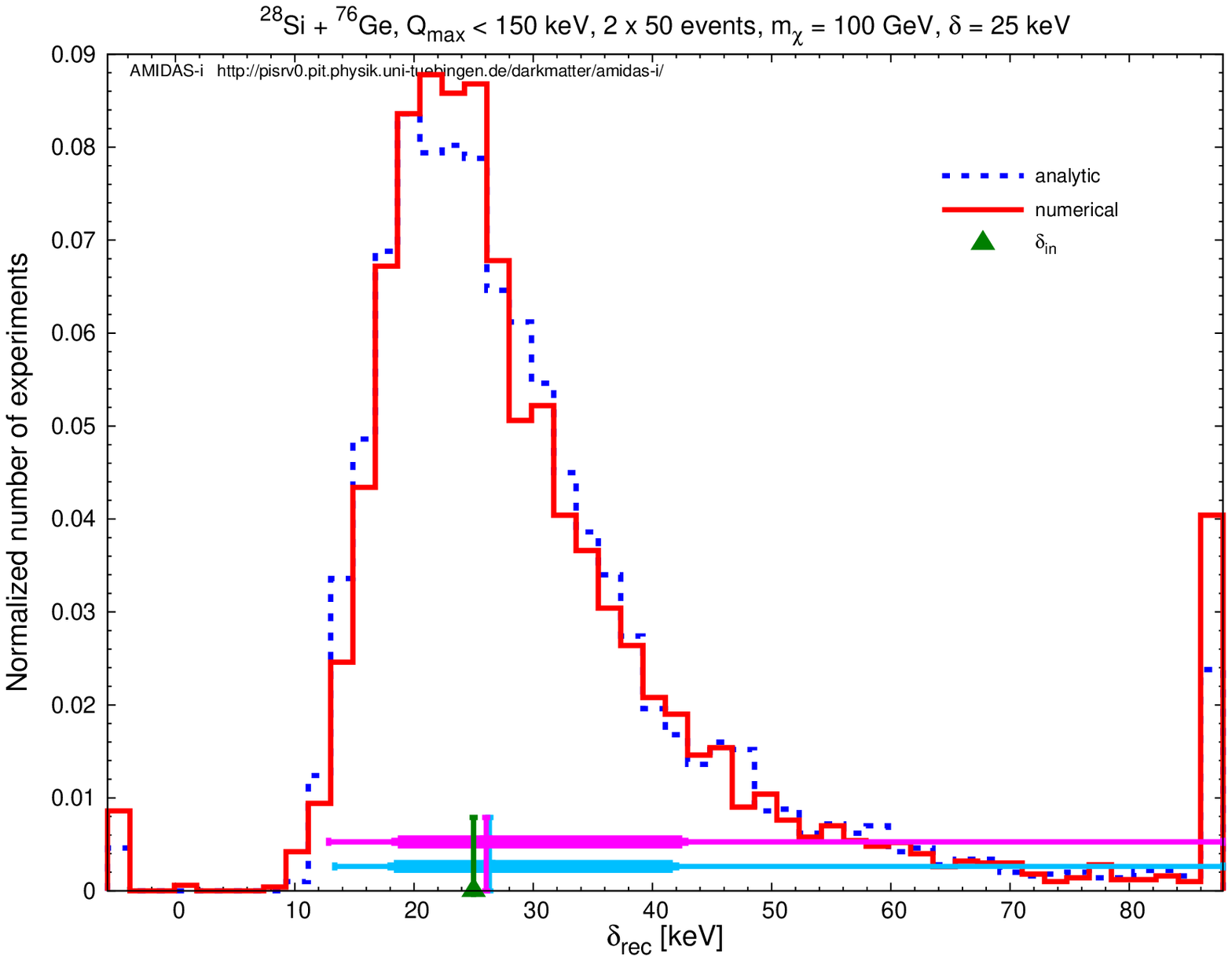} \hspace*{-0.5cm}
\includegraphics[width=6.2cm]{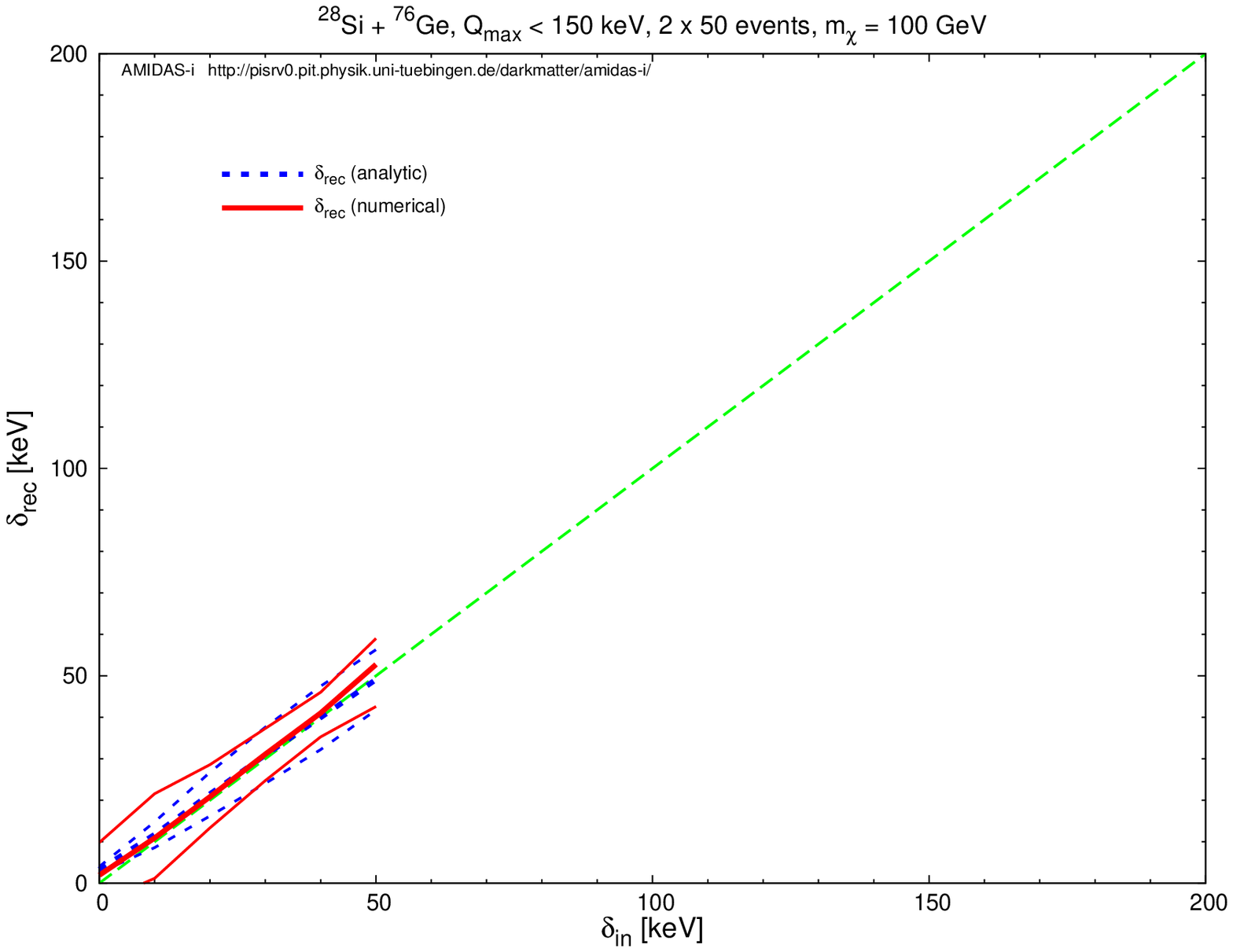}  \hspace*{-1.6cm} \\
\vspace{-0.25cm}
\end{center}
\caption{
 Left:
 reconstructed mass splitting
 and the lower and upper bounds of
 the 1$\sigma$ statistical uncertainty
 as functions of the input mass splitting.
 The dashed blue curves indicate the 1$\sigma$ band
 given with the parameters $k$ and $k'$ estimated analytically
 by Eqs.~(\ref{eqn:k_in_ana}) and (\ref{eqn:kp_in_ana}),
 whereas
 the solid red curves indicate the band
 given with the numerically estimated $k$ and $k'$.
 Middle:
 distributions of the reconstructed mass splittings with
 analytically (dashed blue) and numerically (solid red)
 estimated $k$ and $k'$.
 While
 the cyan (magenta) vertical lines indicate
 the median values of the simulated results
 (corresponding to the blue and red distribution histograms)
 and
 the horizontal thick (thin) bars show the 1$\~$(2)$\~\sigma$
 (68.27\% (95.45\%)) ranges of the results,
 the green vertical line indicates
 the true (input) mass splitting of \mbox{$\delta = 25$ keV}.
 Right:
 reconstructed mass splitting
 and the 1$\sigma$ statistical uncertainty bands
 given by the median values of the reconstructed $\Qvthre$ with
 $k$ and $k'$ estimated analytically (dashed blue)
 and numerically (solid red),
 respectively.
 $\rmXA{Si}{28}$ and $\rmXA{Ge}{76}$
 have been chosen as two target nuclei.
 50 total events on average in each data set
 have been simulated.
 The input WIMP mass has been set as \mbox{$\mchi = 100$ GeV}.
 Other parameters are as
 in Figs.~\ref{fig:idRdQ-Ge76-100-025-050}.
 See the text for further details.
}
\label{fig:delta-SiGe-100-050}
\end{figure}

 In the left frame of Figs.~\ref{fig:delta-SiGe-100-050},
 we show the reconstructed mass splitting
 estimated by Eq.~(\ref{eqn:delta_in})
 and the lower and upper bounds of
 the 1$\sigma$ statistical uncertainty
 estimated by Eq.~(\ref{eqn:sigma_delta_in})
 as functions of the input mass splitting
 for the case of $\mchi = 100$ GeV
 by using
 $\rmXA{Si}{28}$ and $\rmXA{Ge}{76}$
 as two target nuclei
 with 50 total events on average in each data set.
 The dashed blue curves indicate the 1$\sigma$ band
 given with the parameters $k$ and $k'$ estimated analytically
 by Eqs.~(\ref{eqn:k_in_ana}) and (\ref{eqn:kp_in_ana}),
 whereas
 the solid red curves indicate the band
 given with the numerically estimated $k$ and $k'$.

\begin{figure}[t!]
\begin{center}
\hspace*{-1.6cm}
\includegraphics[width=6.2cm]{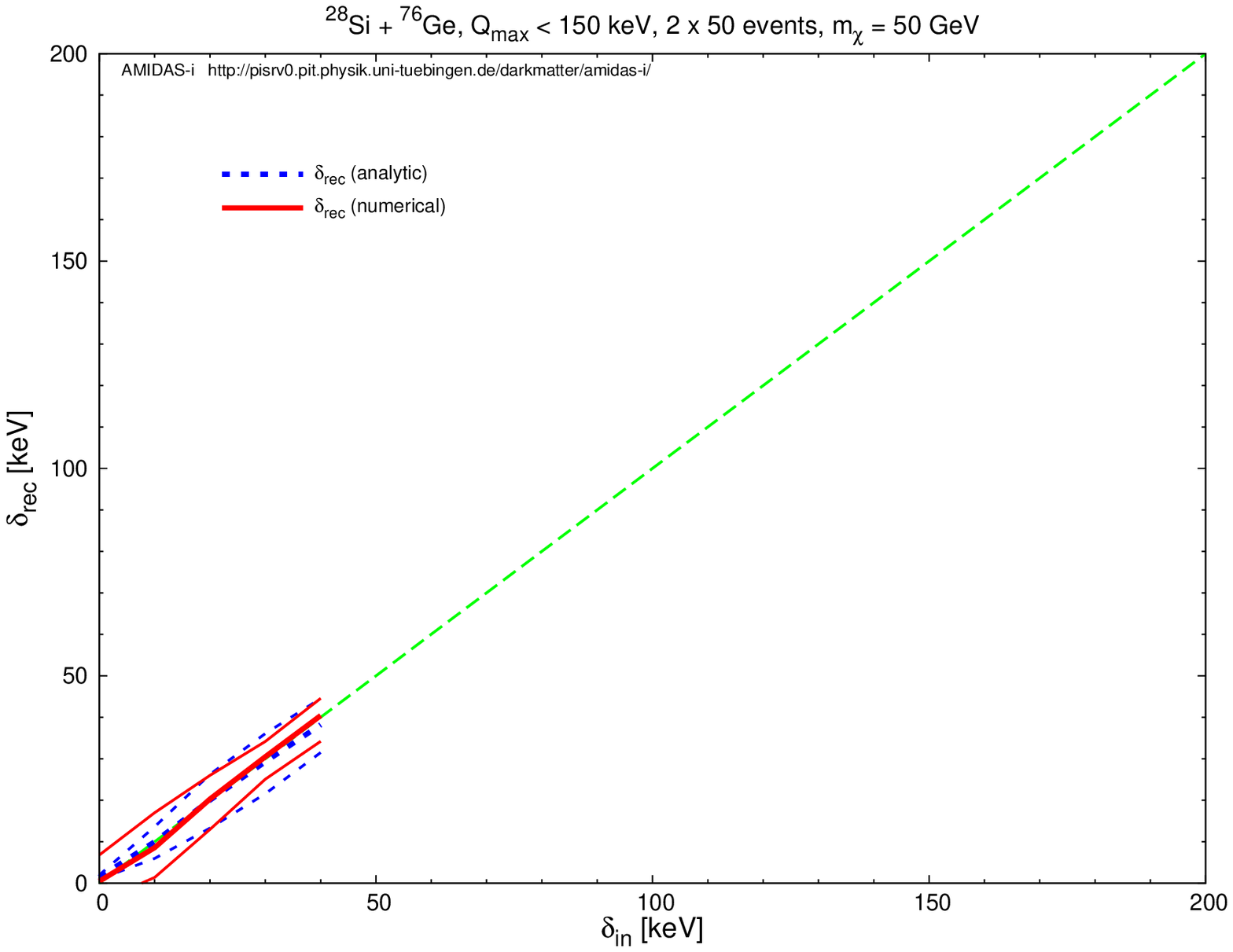}         \hspace*{-0.5cm}
\includegraphics[width=6.2cm]{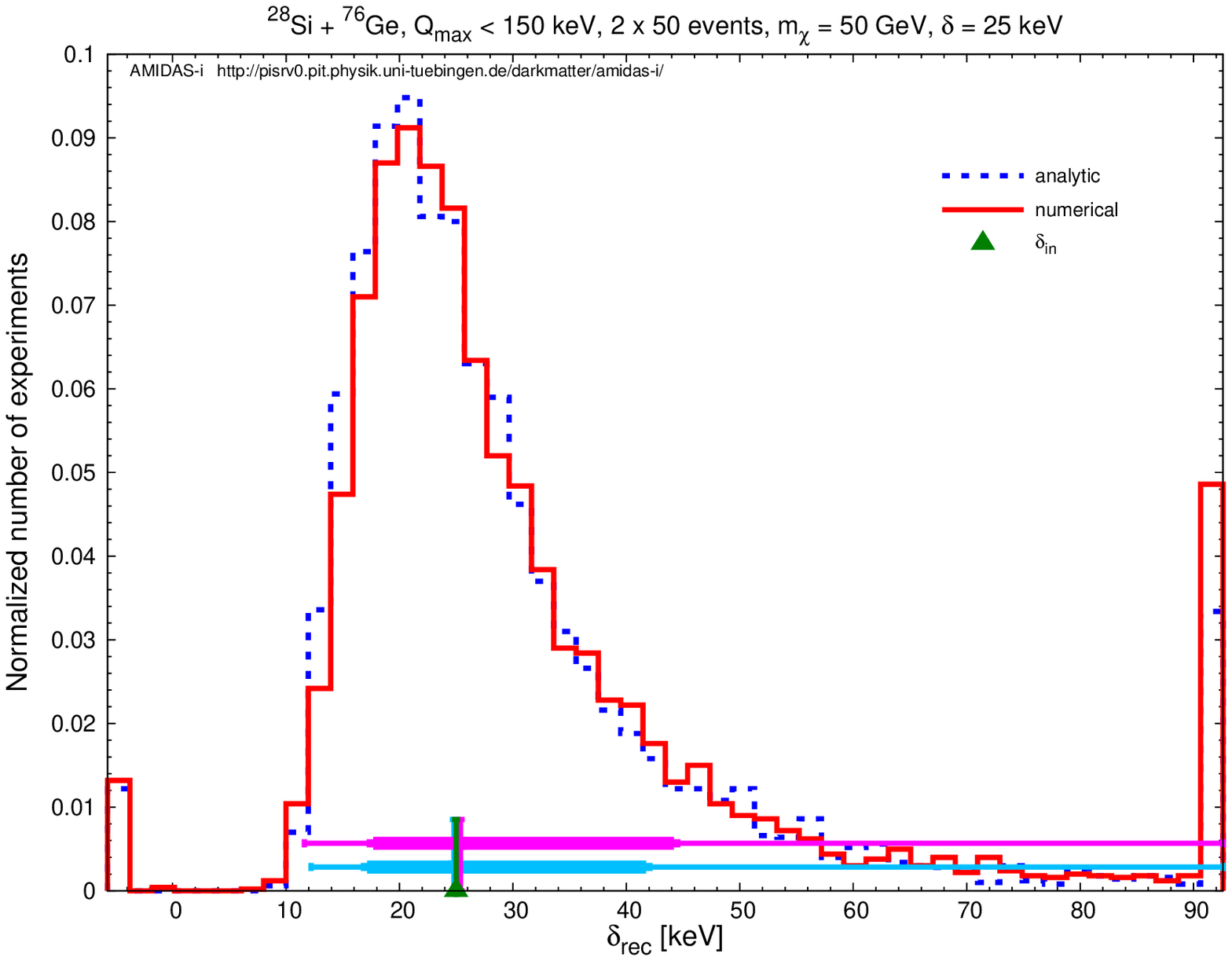} \hspace*{-0.5cm}
\includegraphics[width=6.2cm]{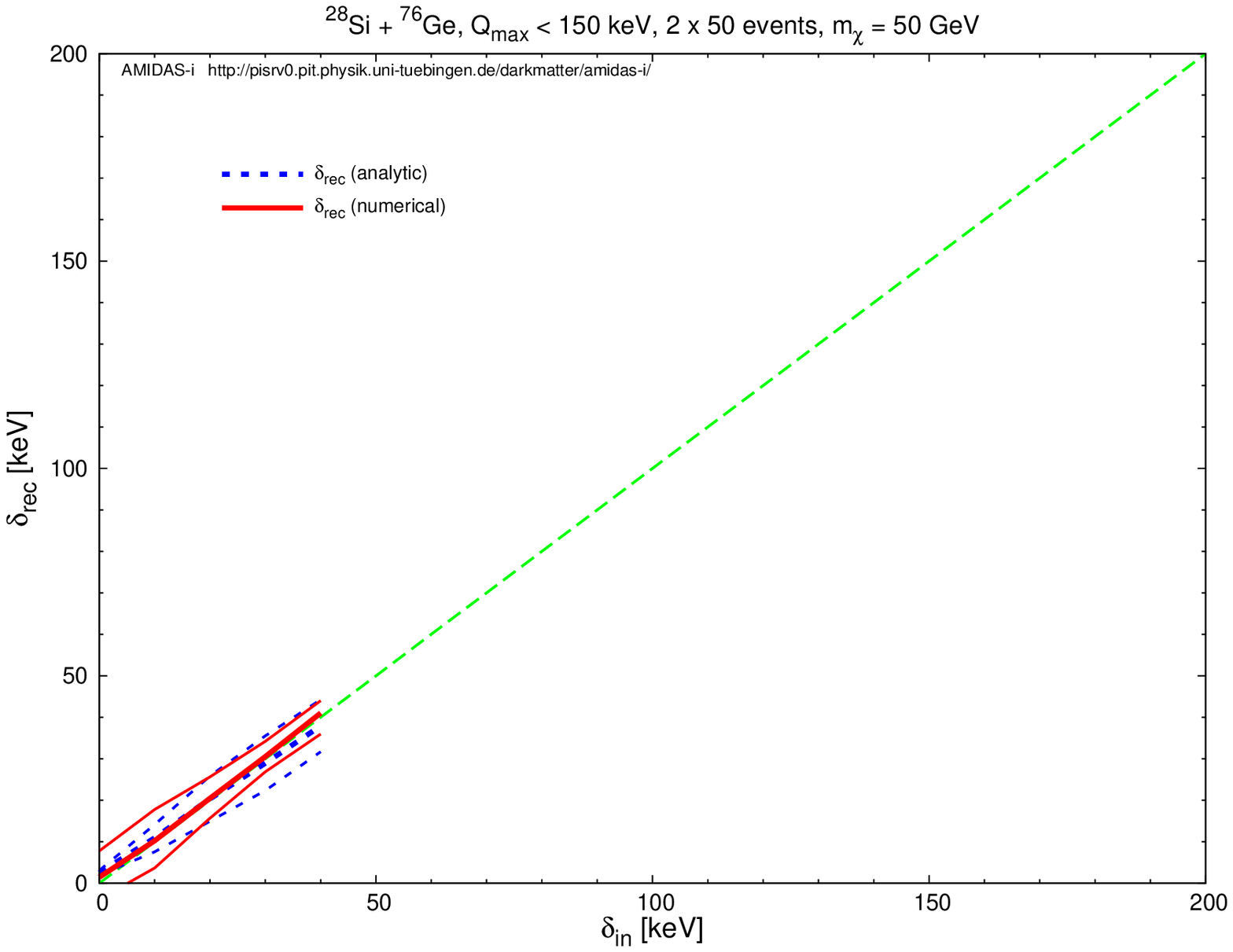}  \hspace*{-1.6cm} \\
 ($\mchi =  50$ GeV) \\
\vspace{0.75cm}
\hspace*{-1.6cm}
\includegraphics[width=6.2cm]{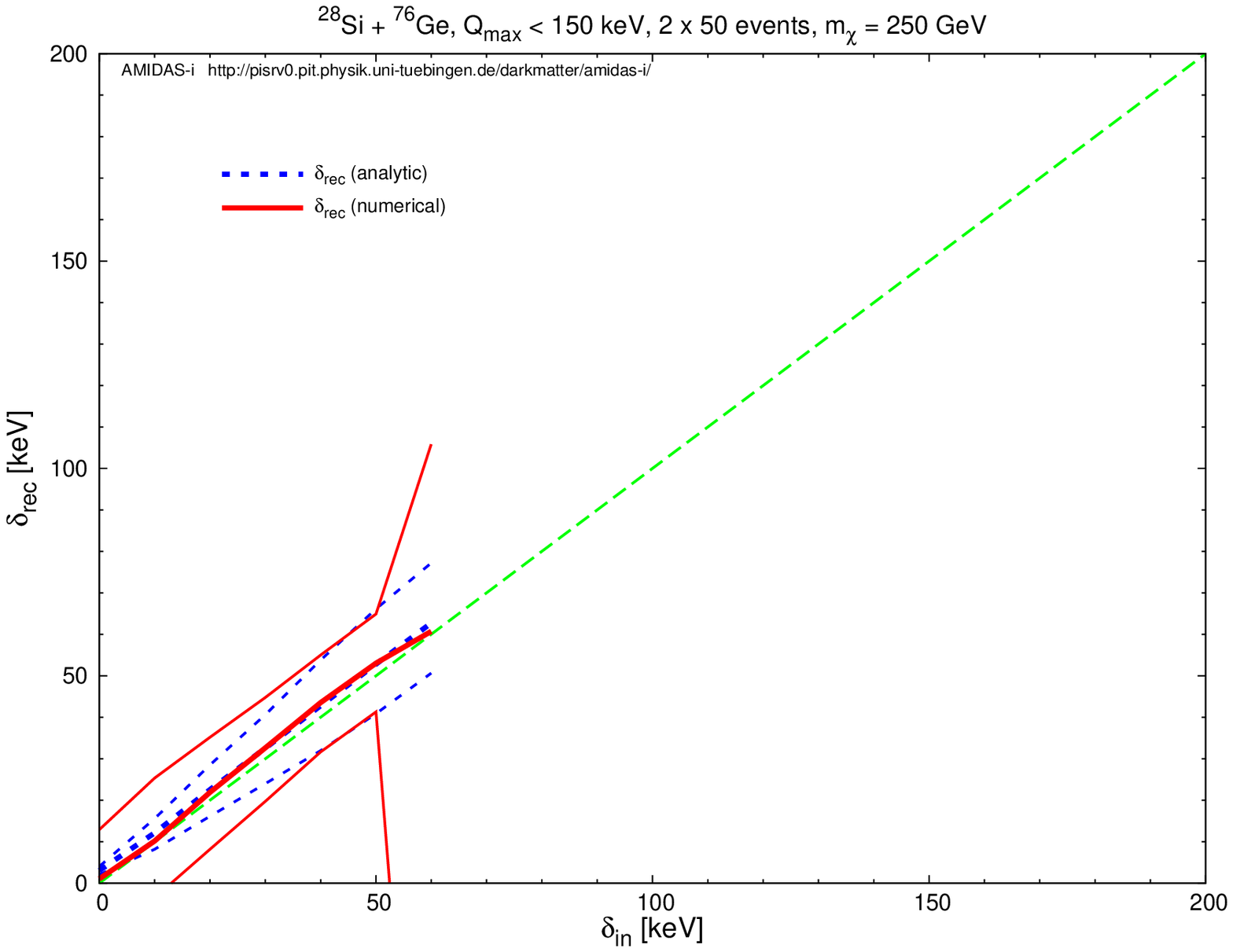}         \hspace*{-0.5cm}
\includegraphics[width=6.2cm]{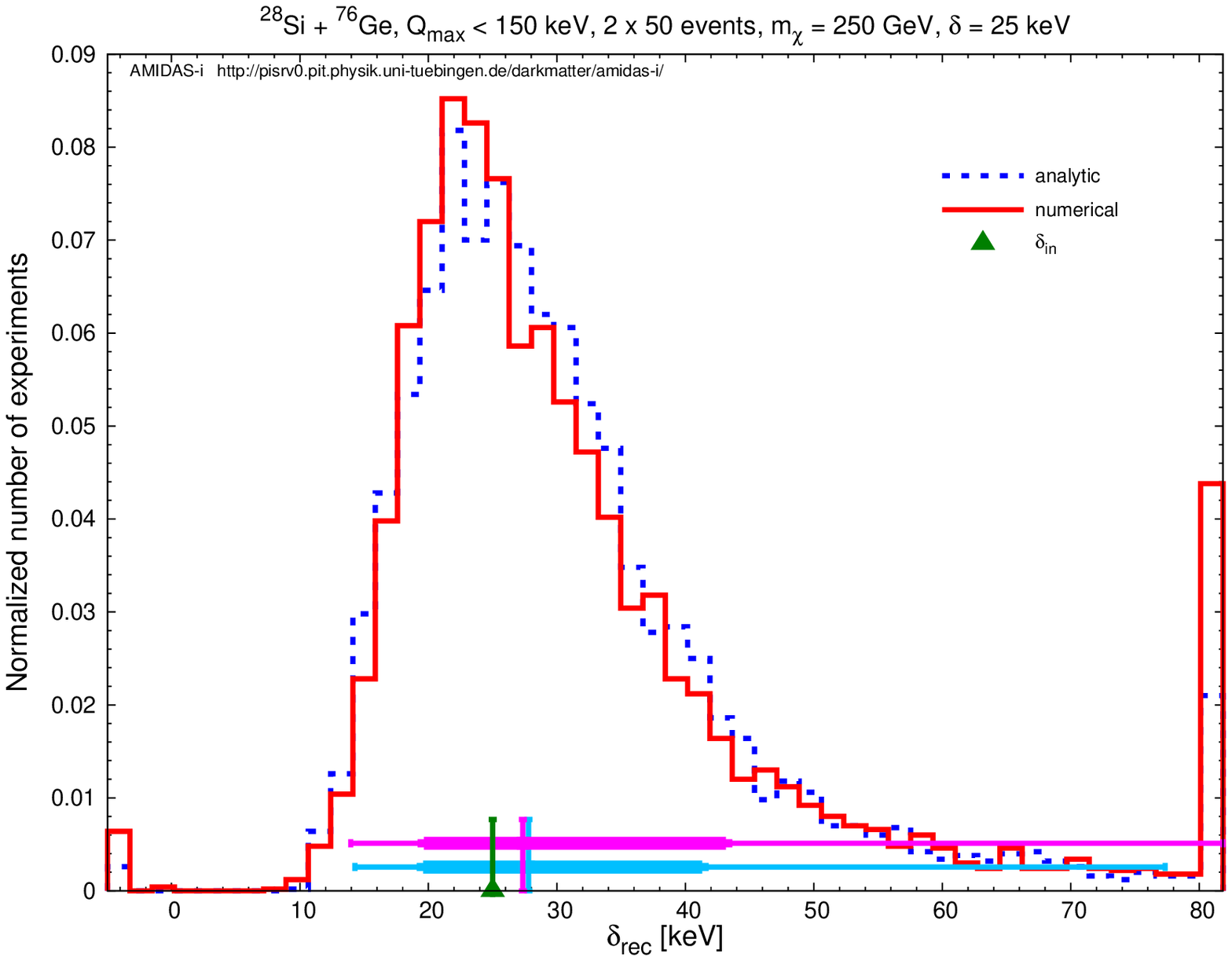} \hspace*{-0.5cm}
\includegraphics[width=6.2cm]{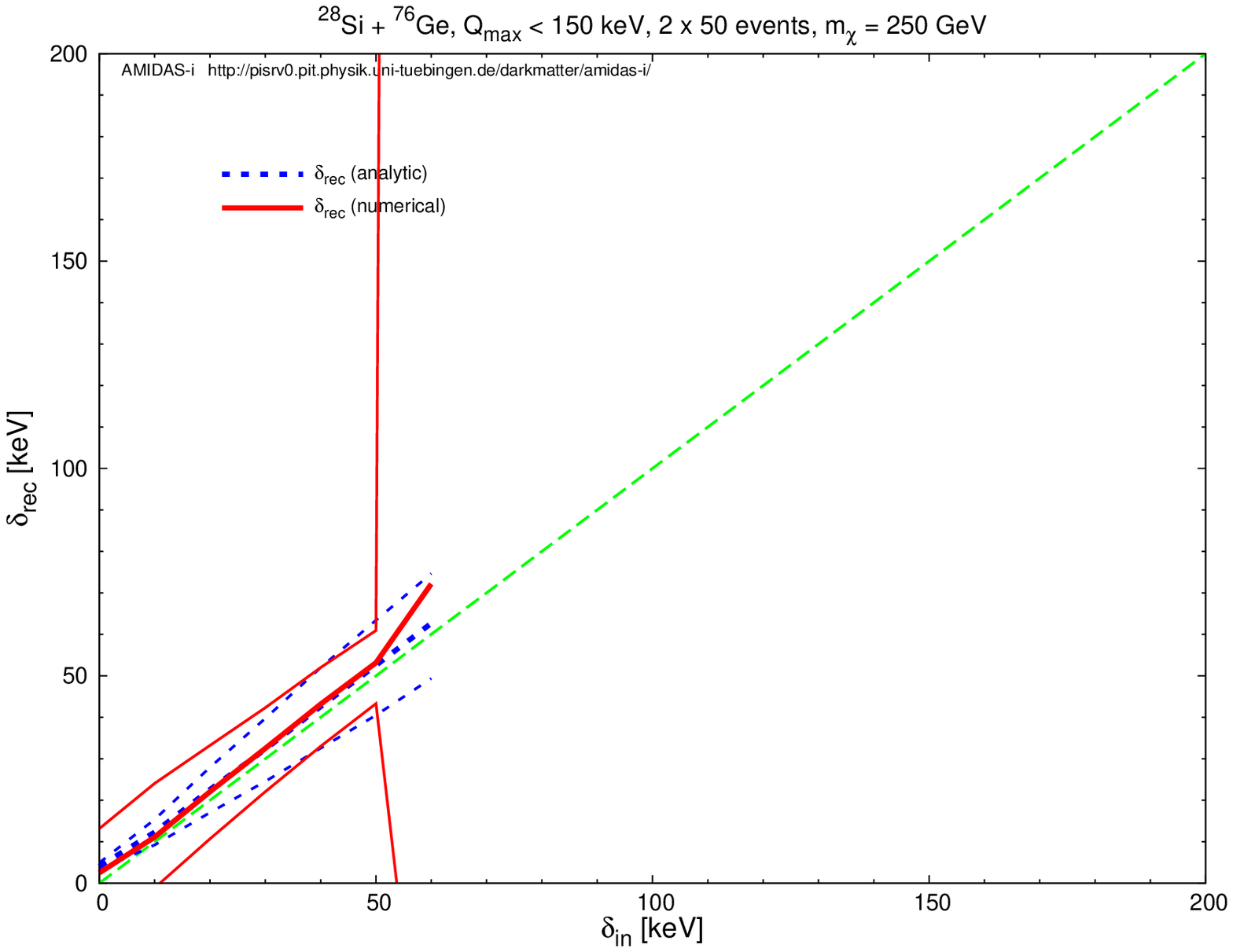}  \hspace*{-1.6cm} \\
 ($\mchi = 250$ GeV) \\
\vspace{-0.25cm}
\end{center}
\caption{
 As in Figs.~\ref{fig:delta-SiGe-100-050},
 except that
 the input WIMP mass has been set as
 \mbox{$\mchi =  50$ GeV} (upper) and
 \mbox{$\mchi = 250$ GeV} (lower).
 Meanwhile,
 in two middle frames
 the input mass splittings have been set the same as
 \mbox{$\delta = 25$ keV}.
}
\label{fig:delta-SiGe-050}
\end{figure}

 It can be seen that,
 firstly,
 due to the maximal kinematic cut--off energy $Q_{\rm max, kin}$
 in Eq.~(\ref{eqn:Qmax_kin_in}),
 one could reconstruct the mass splitting
 only up to \mbox{$\delta \simeq$ 50 keV}
 (for a WIMP mass of \mbox{$\sim$ 100 GeV}).
 However,
 the middle frame of Figs.~\ref{fig:delta-SiGe-100-050}
 shows that,
 once this reconstruction is achievable,
 the mass splitting could in principle be reconstructed pretty well:
 not only
 the median values of $\delta$
 reconstructed both analytically and numerically
 could match the true (input) one,
 but also
 the distribution of the reconstructed $\delta$
 would be pretty concentrated.
 Moreover,
 the middle frame of Figs.~\ref{fig:delta-SiGe-100-050}
 shows also that
 a 3$\sigma$ to 5$\sigma$ ($>$ 99\%) CL of
 the positivity of the reconstructed mass splitting
 (as the confirmation of the inelastic WIMP scattering)
 could in principle be identified.

 Finally,
 as in Figs.~\ref{fig:mchi-SiGe-025-050},
 in the right frame of Figs.~\ref{fig:delta-SiGe-100-050}
 we show the reconstructed mass splitting
 and the 1$\sigma$ statistical uncertainty bands
 given by the median values of the reconstructed $\Qvthre$ with
 $k$ and $k'$ estimated analytically (dashed blue)
 and numerically (solid red),
 respectively.
 Not surprisingly,
 $\delta$ estimated by Eq.~(\ref{eqn:delta_in})
 with the median values of the reconstructed $\Qvthre$
 could be so good as
 the median values of $\delta$ reconstructed by Eq.~(\ref{eqn:delta_in})
 with each single pair of data sets;
 the former could have little bit smaller statistical uncertainties.

 As comparison,
 in Figs.~\ref{fig:delta-SiGe-050}
 we consider the cases of a smaller WIMP mass of
 \mbox{$\mchi =  50$ GeV} (upper)
 and a larger one of
 \mbox{$\mchi = 250$ GeV} (lower).
 In contrast to the reconstruction of the WIMP mass
 shown in Figs.~\ref{fig:mchi-SiGe-025-050} and \ref{fig:mchi-SiGe-050},
 our simulations show here clearly that,
 for a rather small mass splitting $\delta \sim$ a few (tens) keV,
 one could always reconstruct $\delta$ pretty well,
 with only a bit larger statistical uncertainty
 for large WIMP masses.
 More importantly,
 a clear 3$\sigma$ to 5$\sigma$ (\mbox{$>$ 99\%}) CL of
 the positivity of the reconstructed mass splitting
 could always be identified.

 In fact,
 it has been found that,
 up to a WIMP mass of $\sim$ 1 TeV,
 one could in principle always reconstruct the mass splitting
 pretty precisely
 with a deviation of $\lsim$ 20\% (analytically)
 or even $\lsim$ 10\% (numerically)
 and a statistical uncertainty
 of \mbox{$\sim$ 15\%} to \mbox{$\sim$ 50\%}  (analytically)
 or \mbox{$\sim$ 10\%} to a factor of \mbox{$\sim$ 2} (numerically).
 Note that,
 as shown in Figs.~\ref{fig:delta-SiGe-100-050} and \ref{fig:delta-SiGe-050},
 the statistical uncertainty on the analytically reconstructed mass splitting
 increases with increasing the mass splitting,
 whereas
 that of the numerically reconstructed one
 is approximately the same for all values of the mass splitting.

\subsection{\boldmath $\delta = 0$ (elastic WIMP--nucleus scattering) case}
\begin{figure}[t!]
\begin{center}
\hspace*{-1.6cm}
\includegraphics[width=4.8cm]{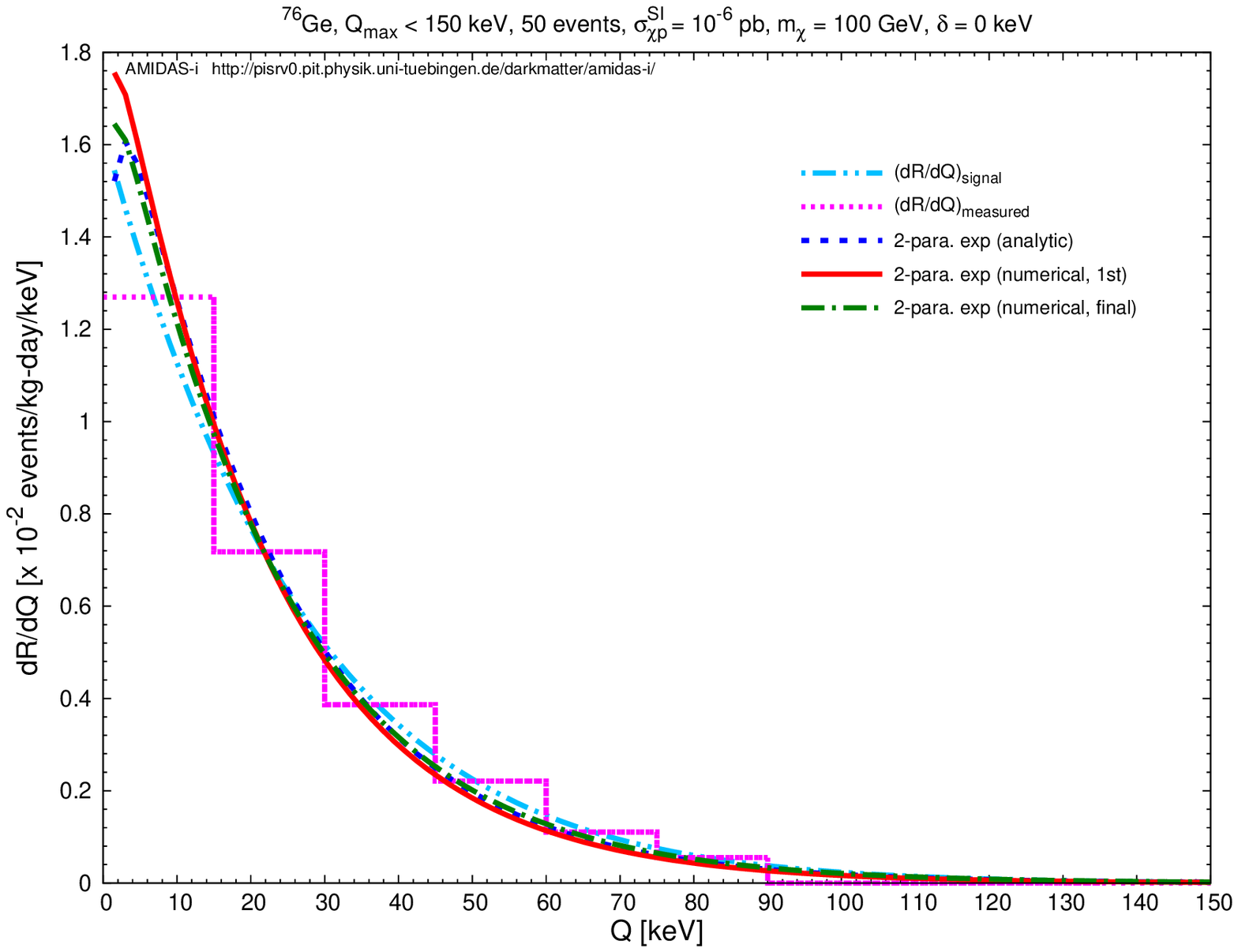}      \hspace*{-0.5cm}
\includegraphics[width=4.8cm]{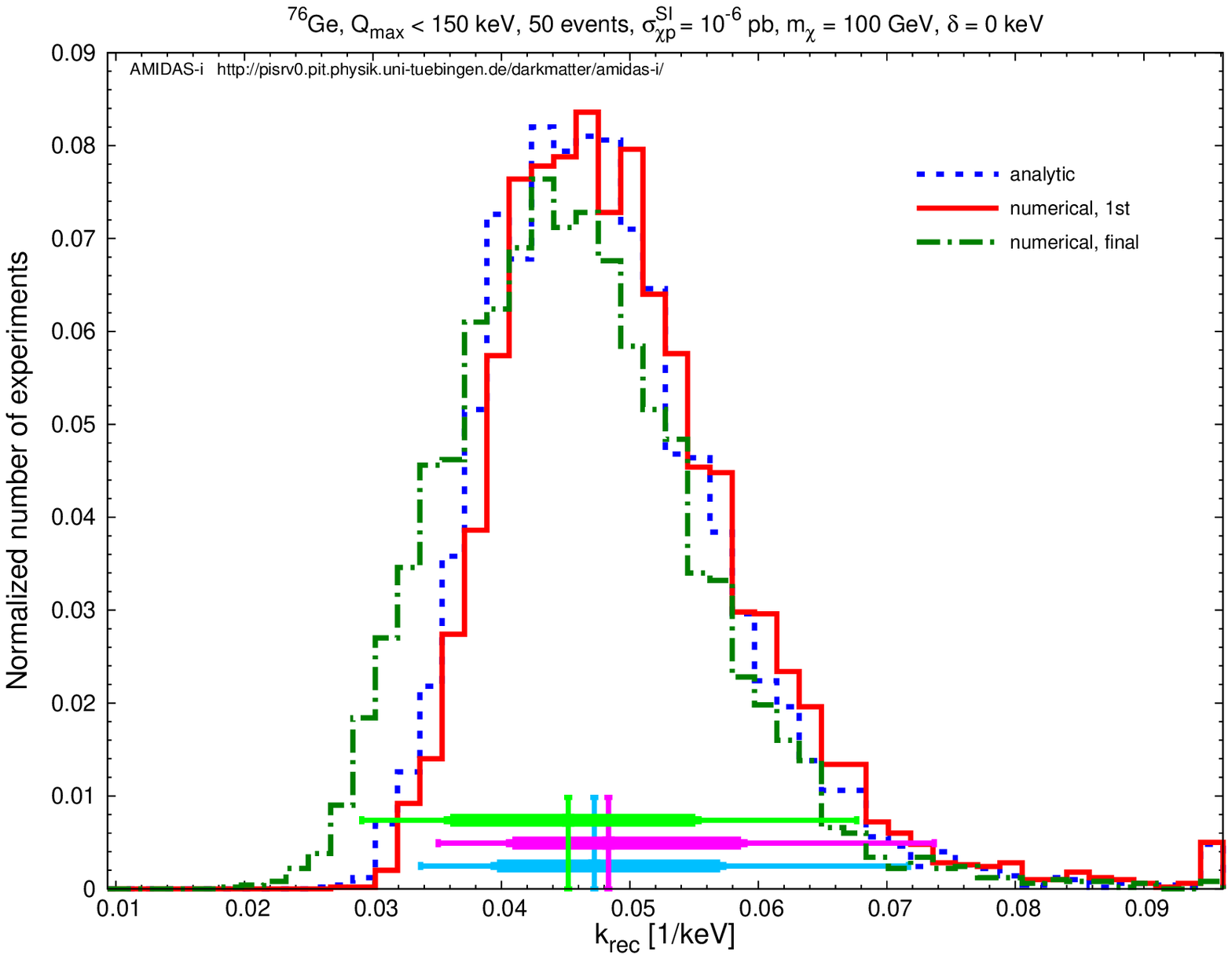}      \hspace*{-0.5cm}
\includegraphics[width=4.8cm]{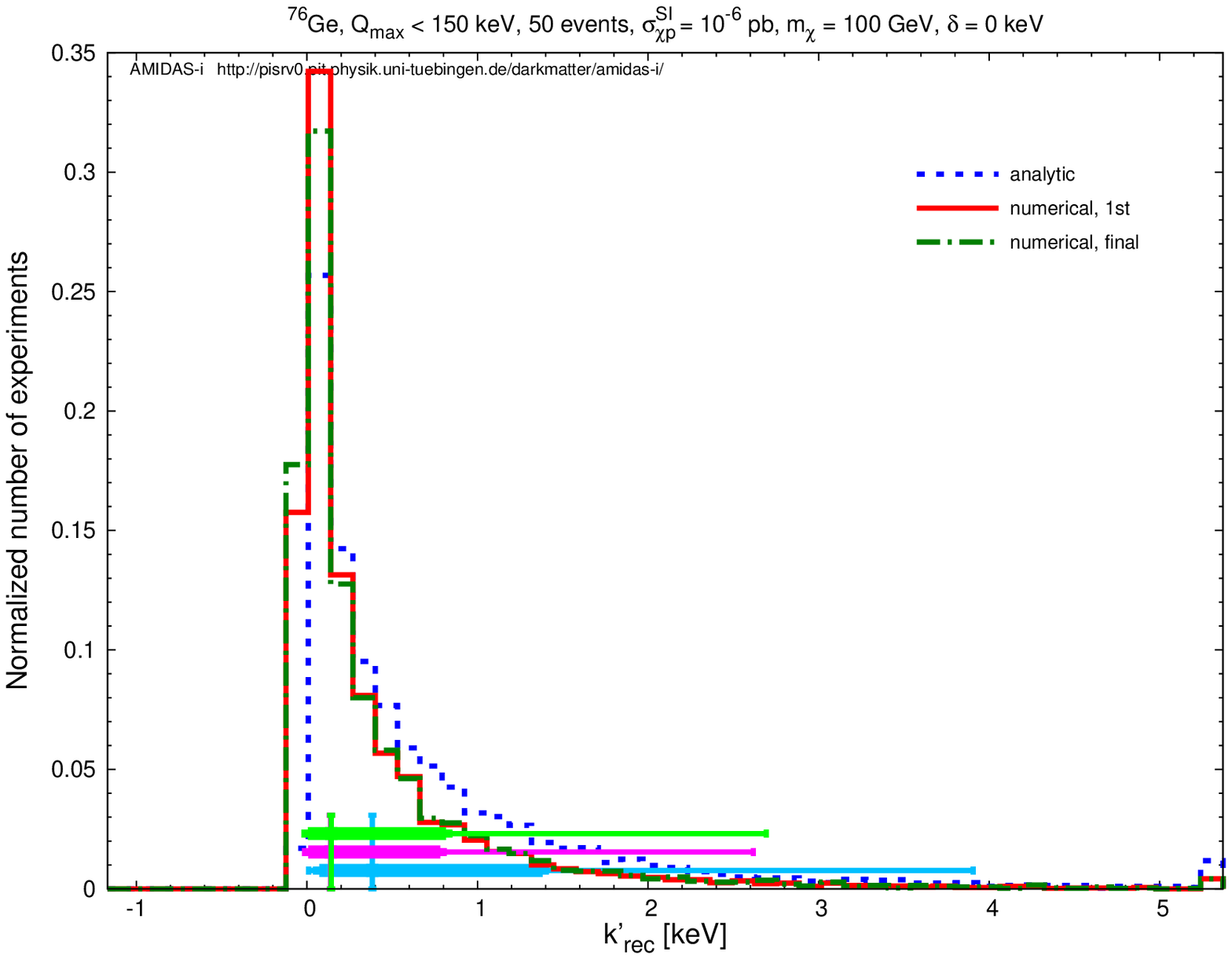}     \hspace*{-0.5cm}
\includegraphics[width=4.8cm]{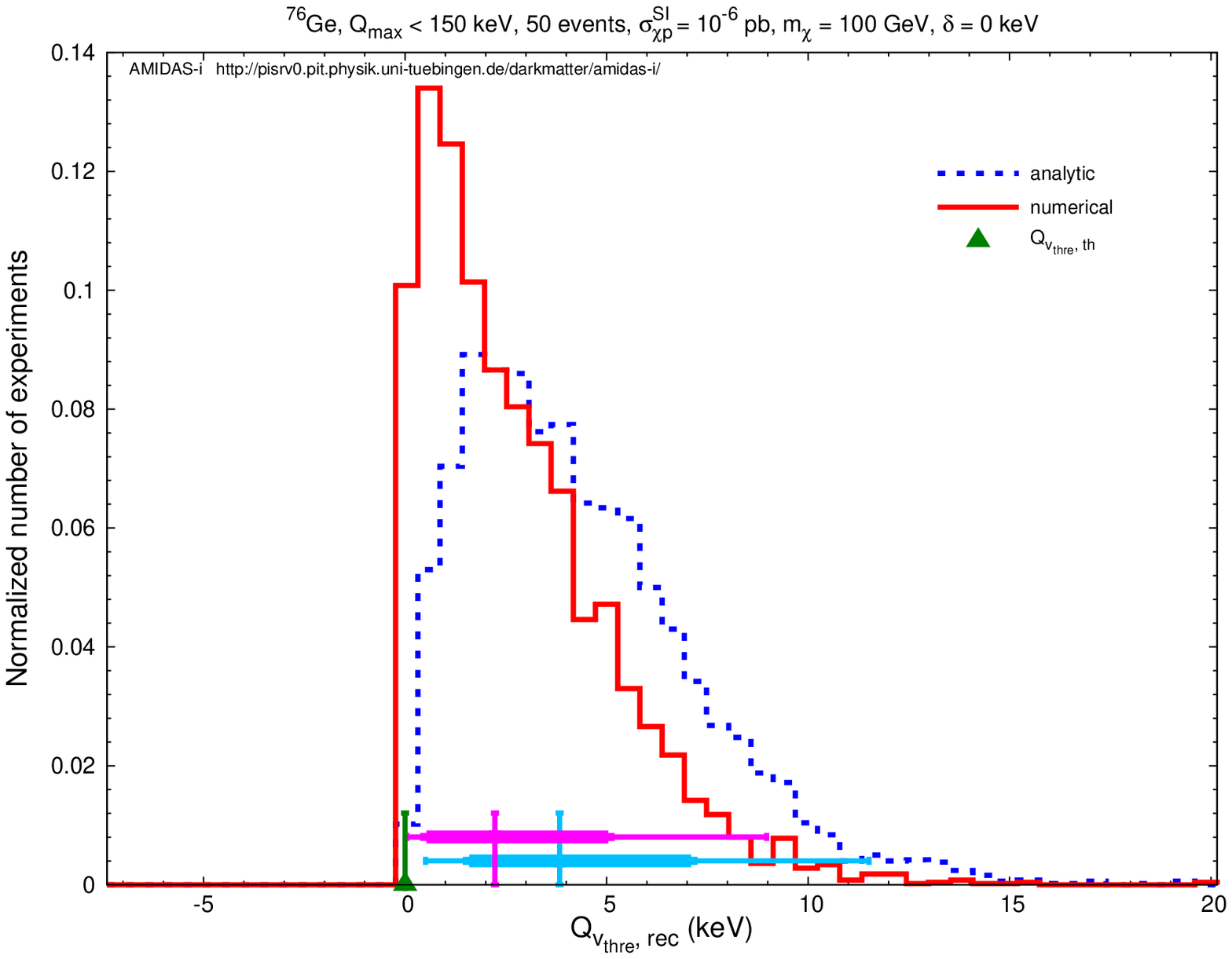} \hspace*{-1.6cm} \\
 ($\rmXA{Ge}{76}$) \\
\vspace{0.75cm}
\hspace*{-1.6cm}
\includegraphics[width=4.8cm]{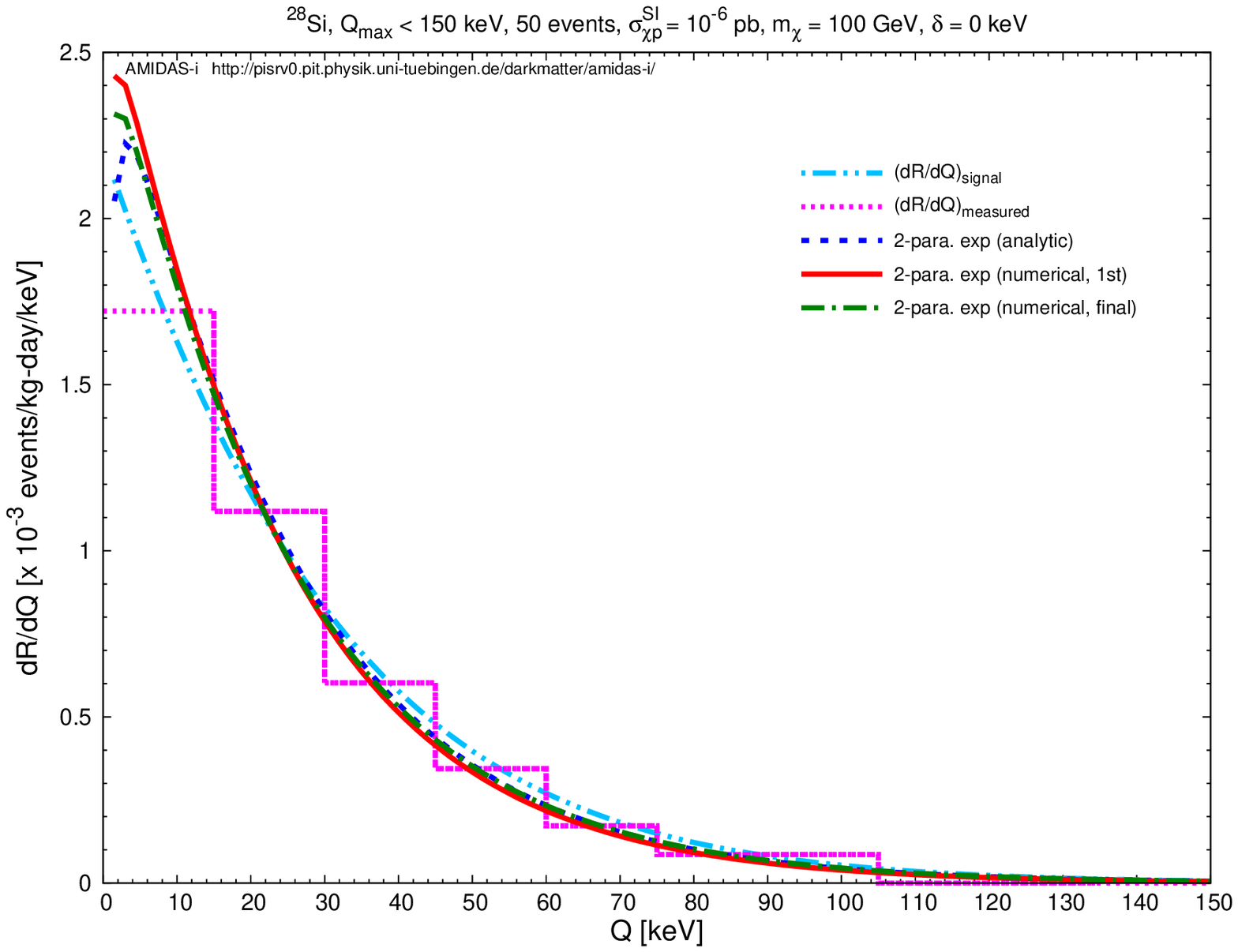}      \hspace*{-0.5cm}
\includegraphics[width=4.8cm]{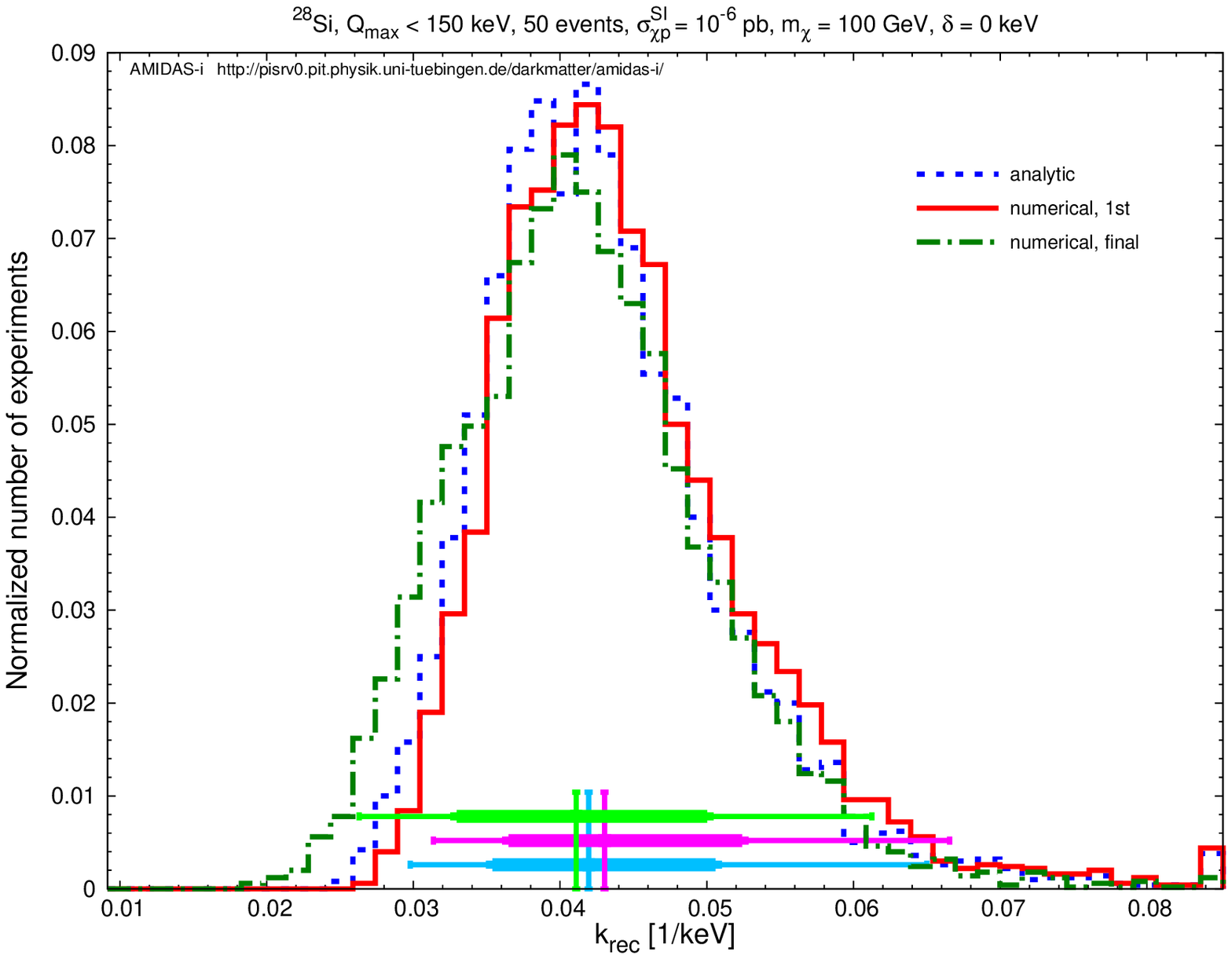}      \hspace*{-0.5cm}
\includegraphics[width=4.8cm]{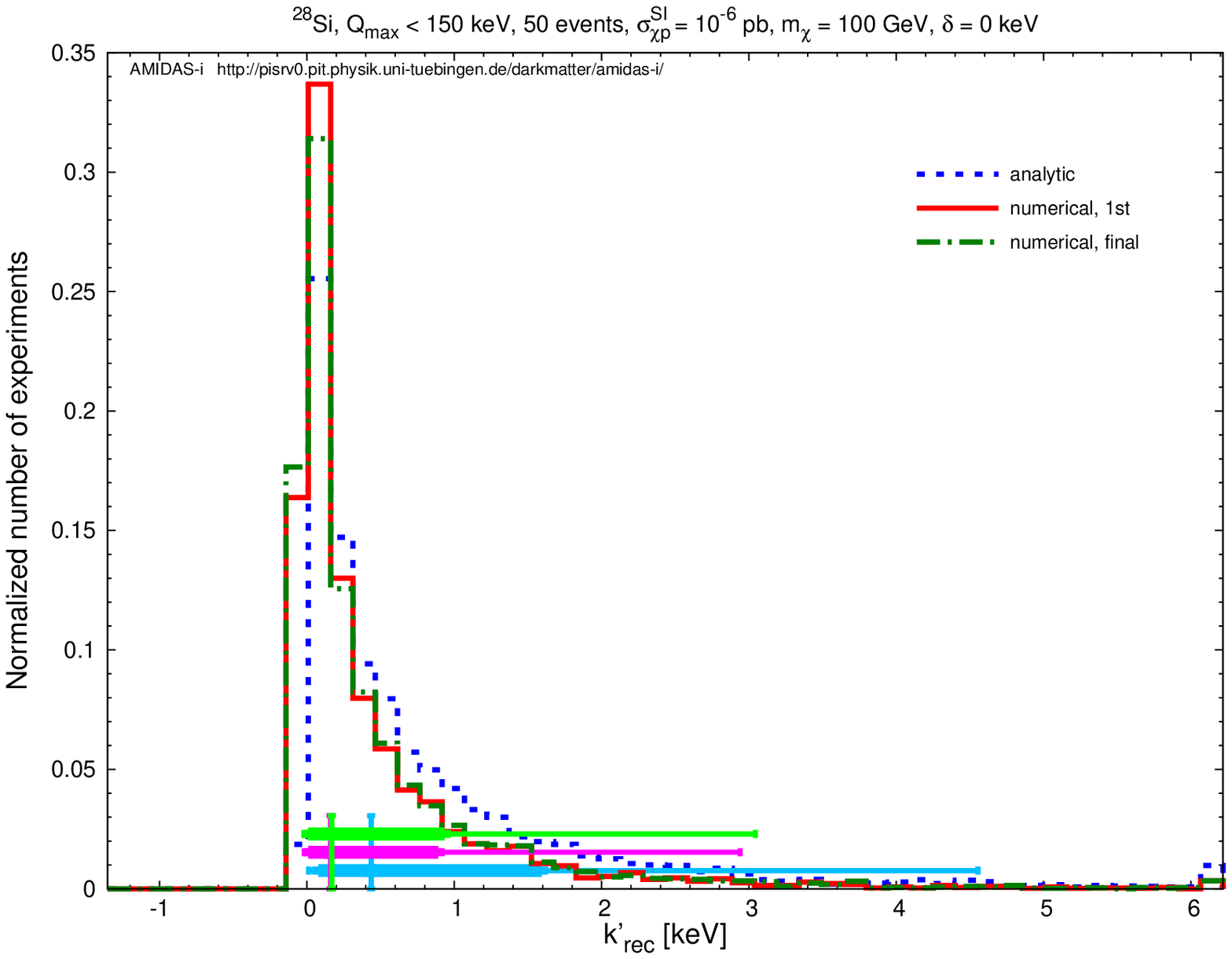}     \hspace*{-0.5cm}
\includegraphics[width=4.8cm]{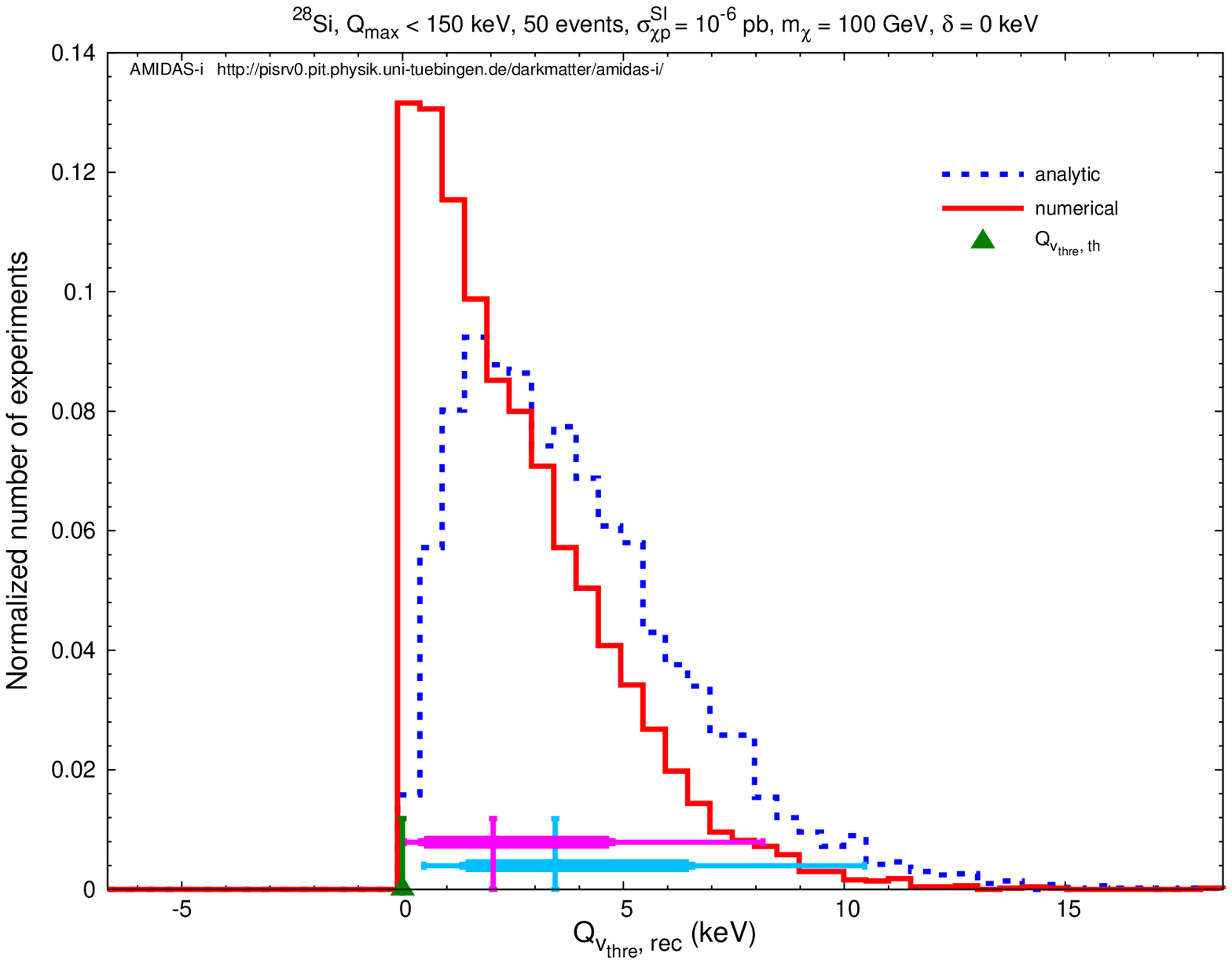} \hspace*{-1.6cm} \\
 ($\rmXA{Si}{28}$) \\
\vspace{0.75cm}
\hspace*{-1.6cm}
\includegraphics[width=4.8cm]{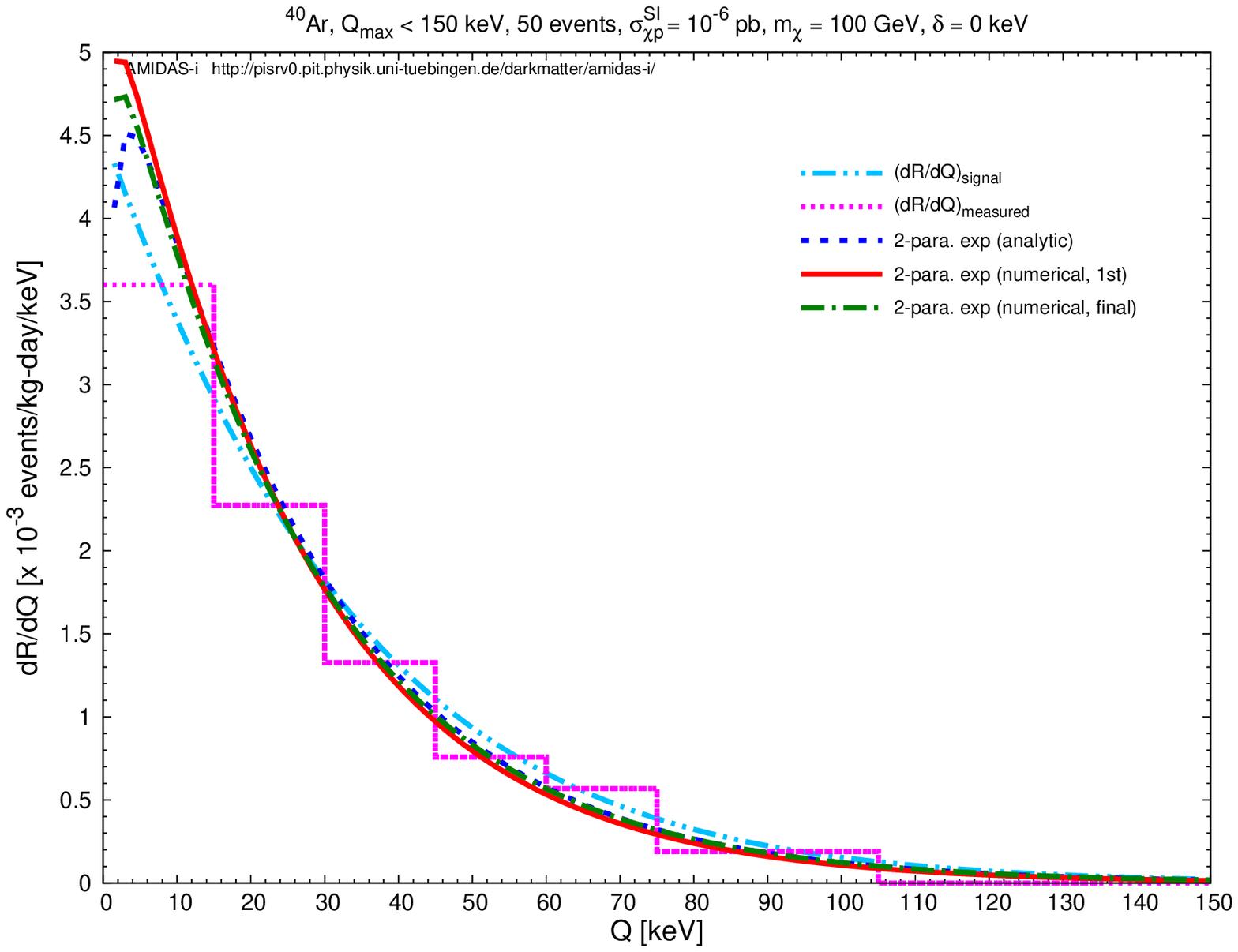}      \hspace*{-0.5cm}
\includegraphics[width=4.8cm]{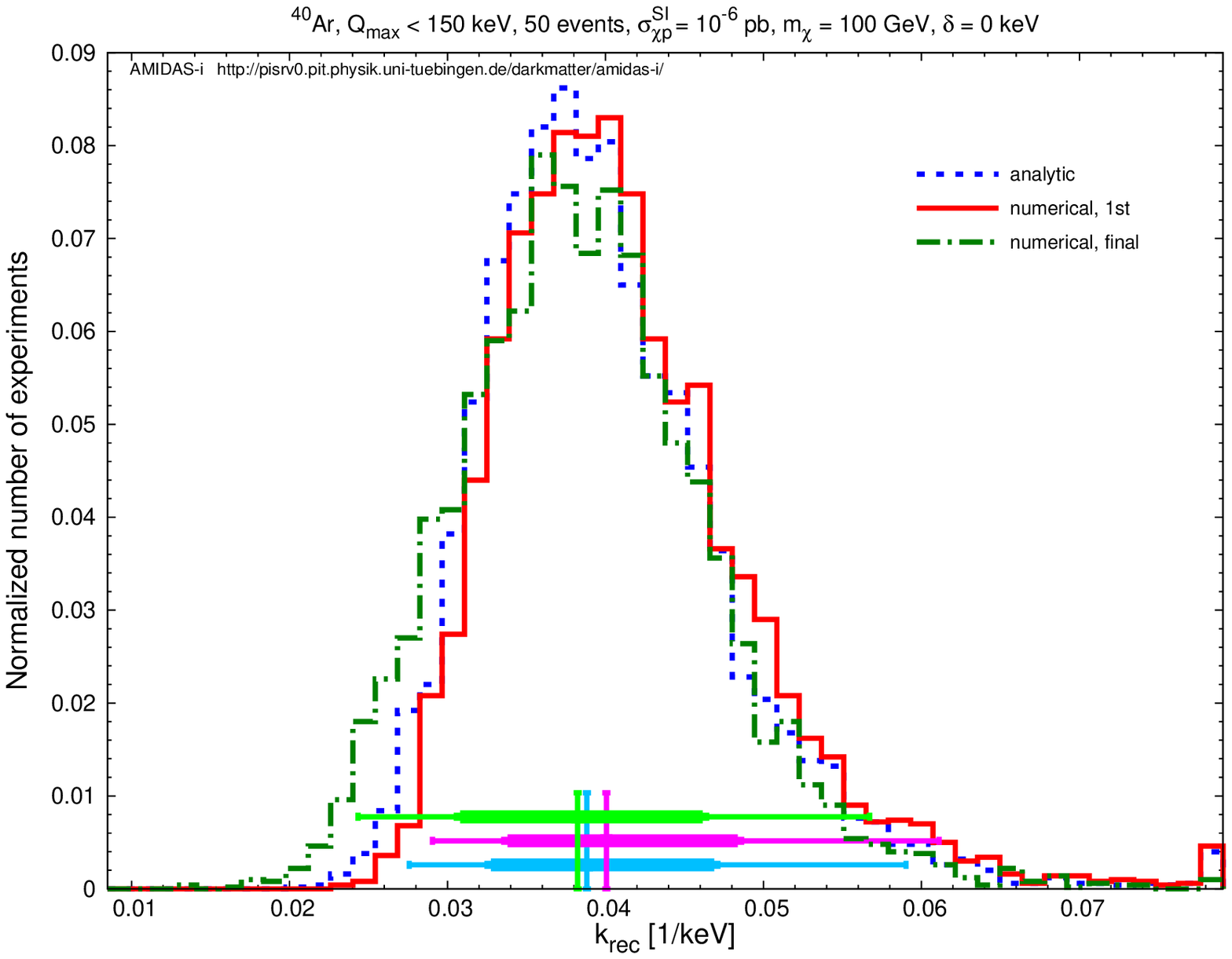}      \hspace*{-0.5cm}
\includegraphics[width=4.8cm]{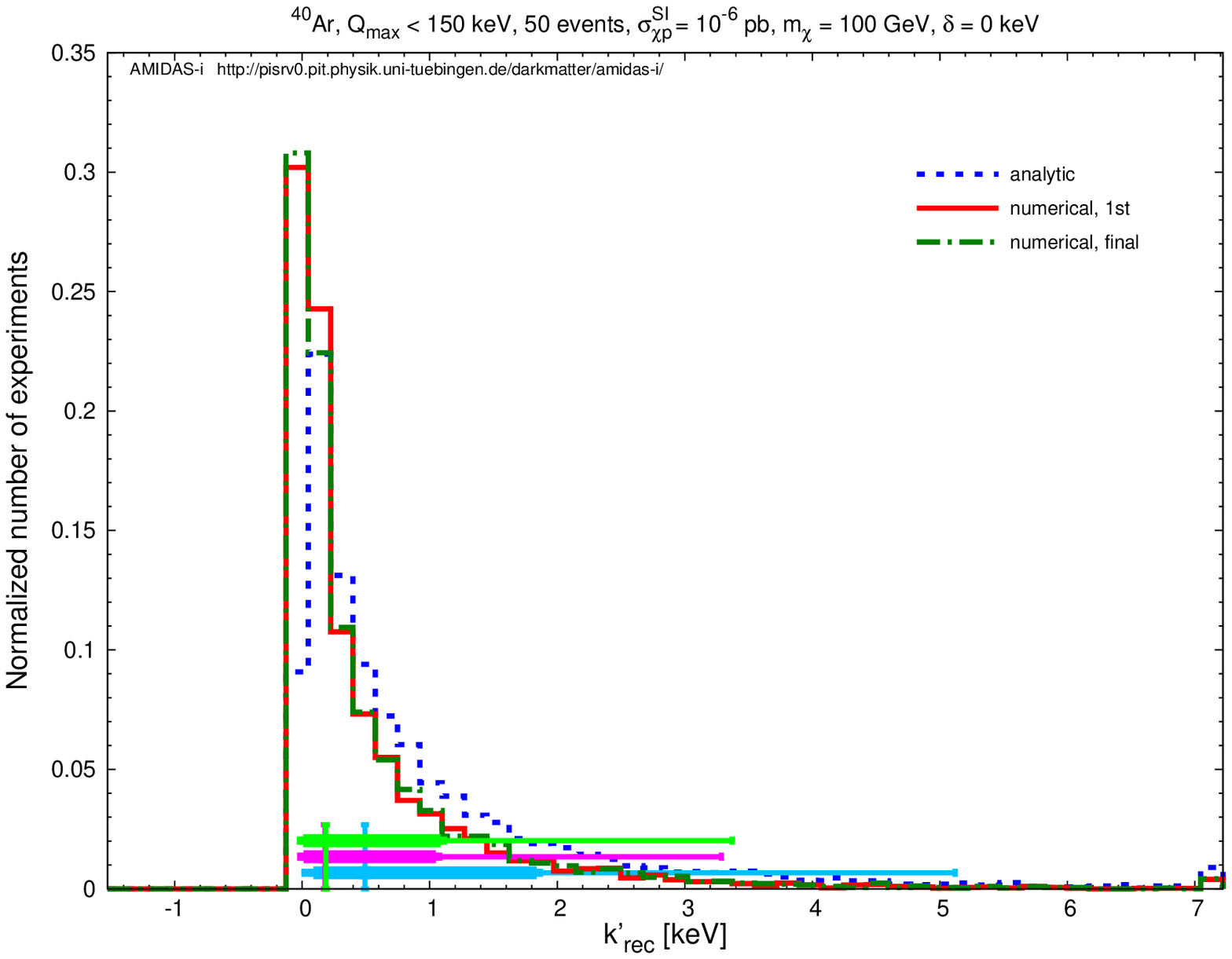}     \hspace*{-0.5cm}
\includegraphics[width=4.8cm]{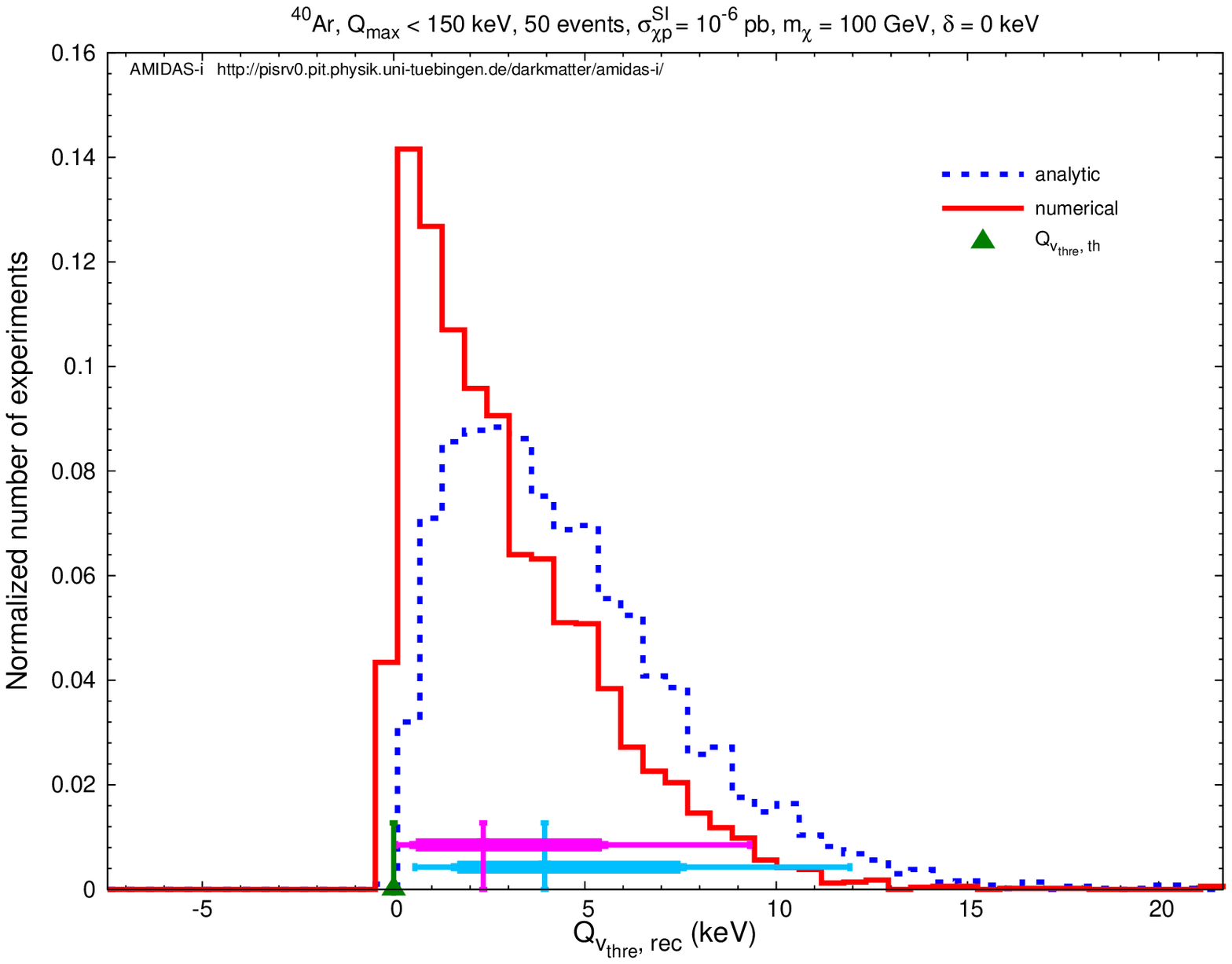} \hspace*{-1.6cm} \\
 ($\rmXA{Ar}{40}$) \\
\vspace{0.75cm}
\hspace*{-1.6cm}
\includegraphics[width=4.8cm]{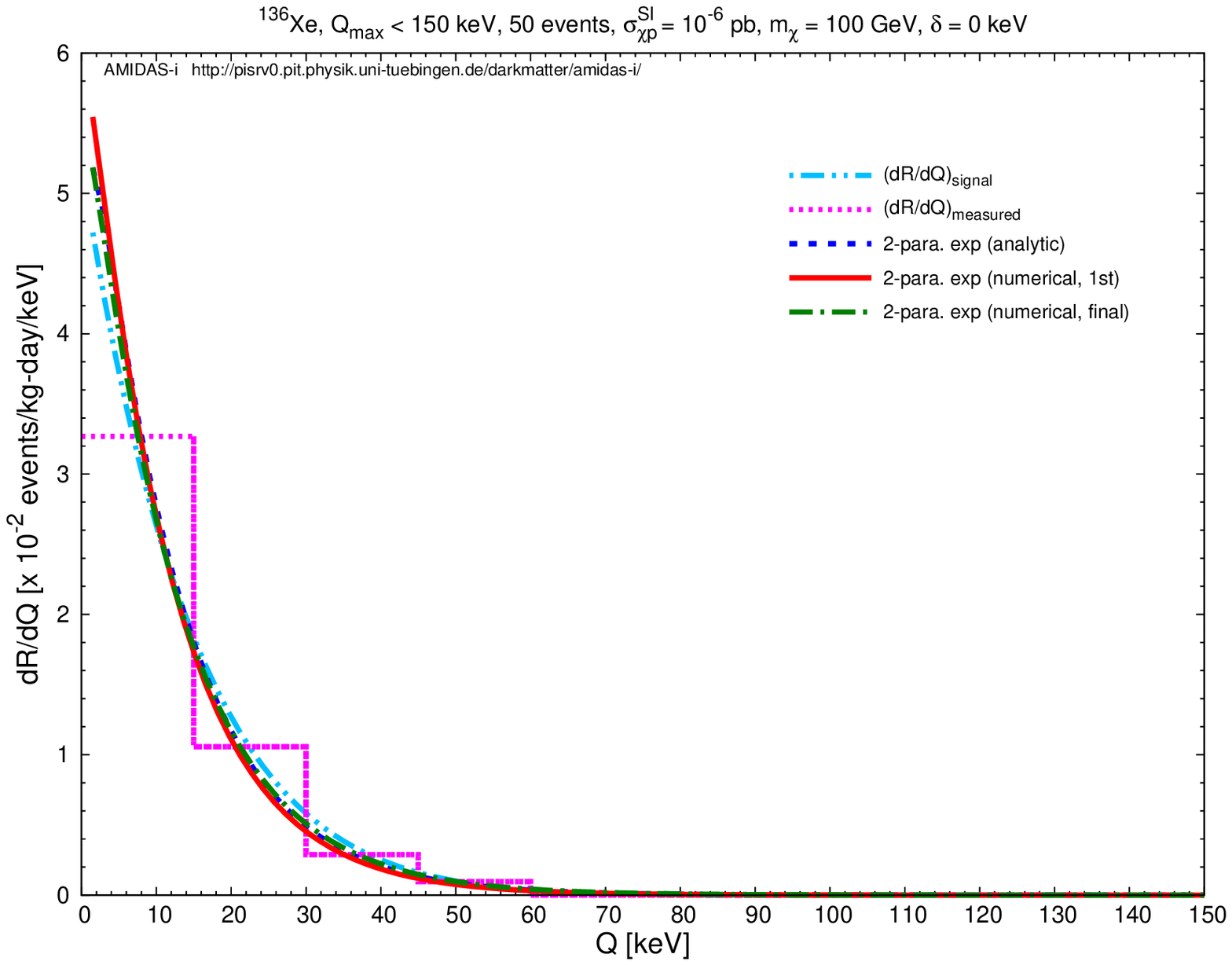}      \hspace*{-0.5cm}
\includegraphics[width=4.8cm]{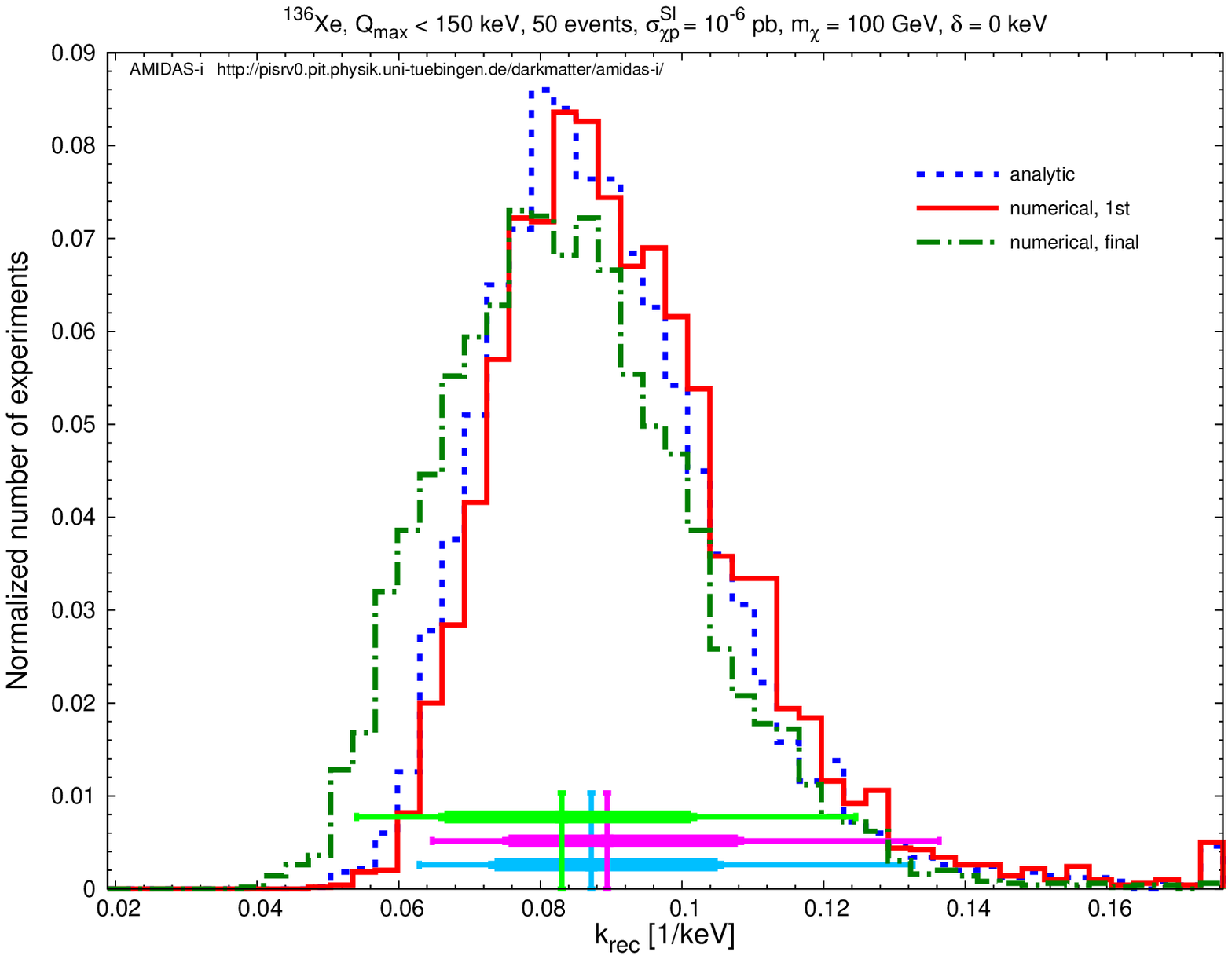}      \hspace*{-0.5cm}
\includegraphics[width=4.8cm]{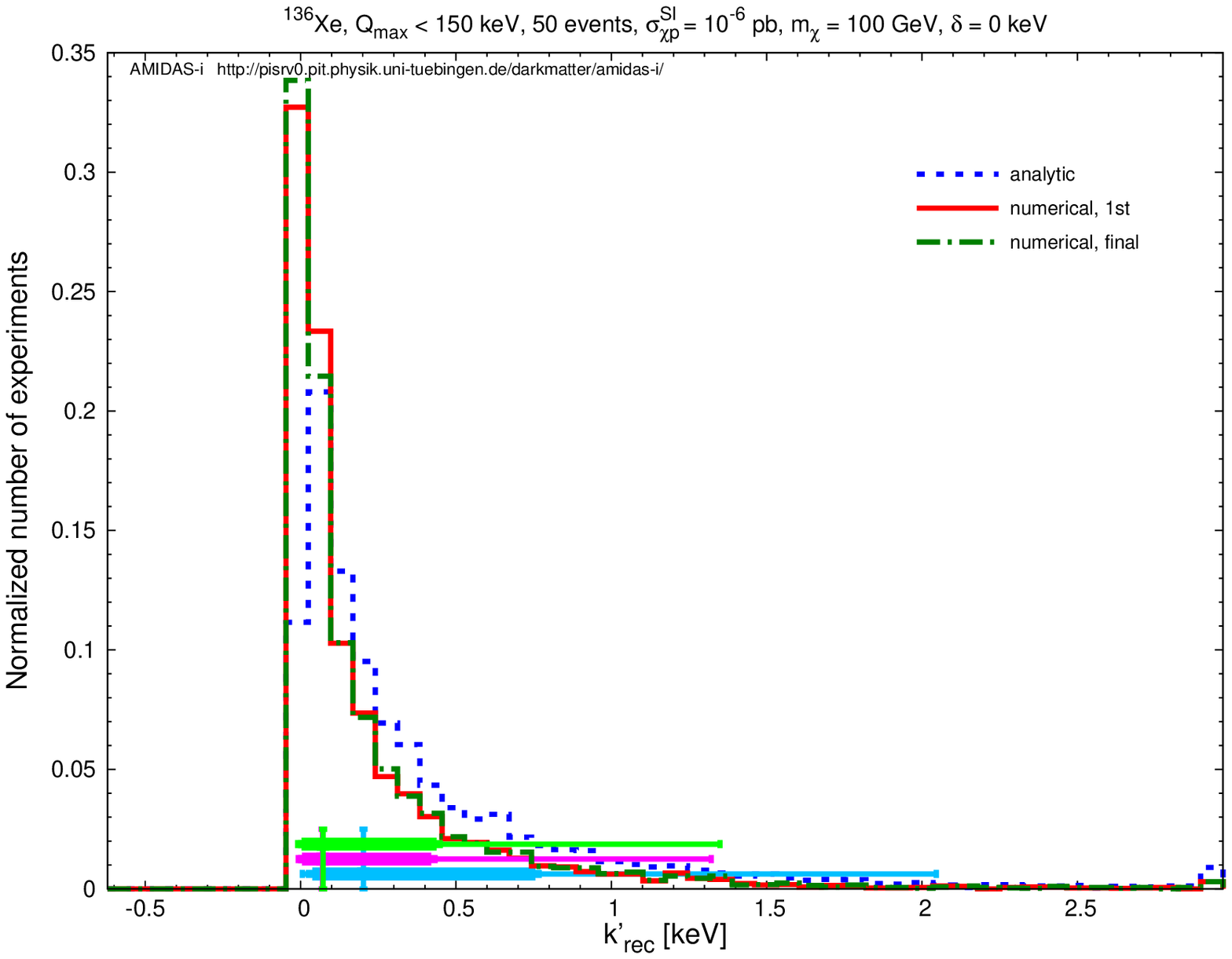}     \hspace*{-0.5cm}
\includegraphics[width=4.8cm]{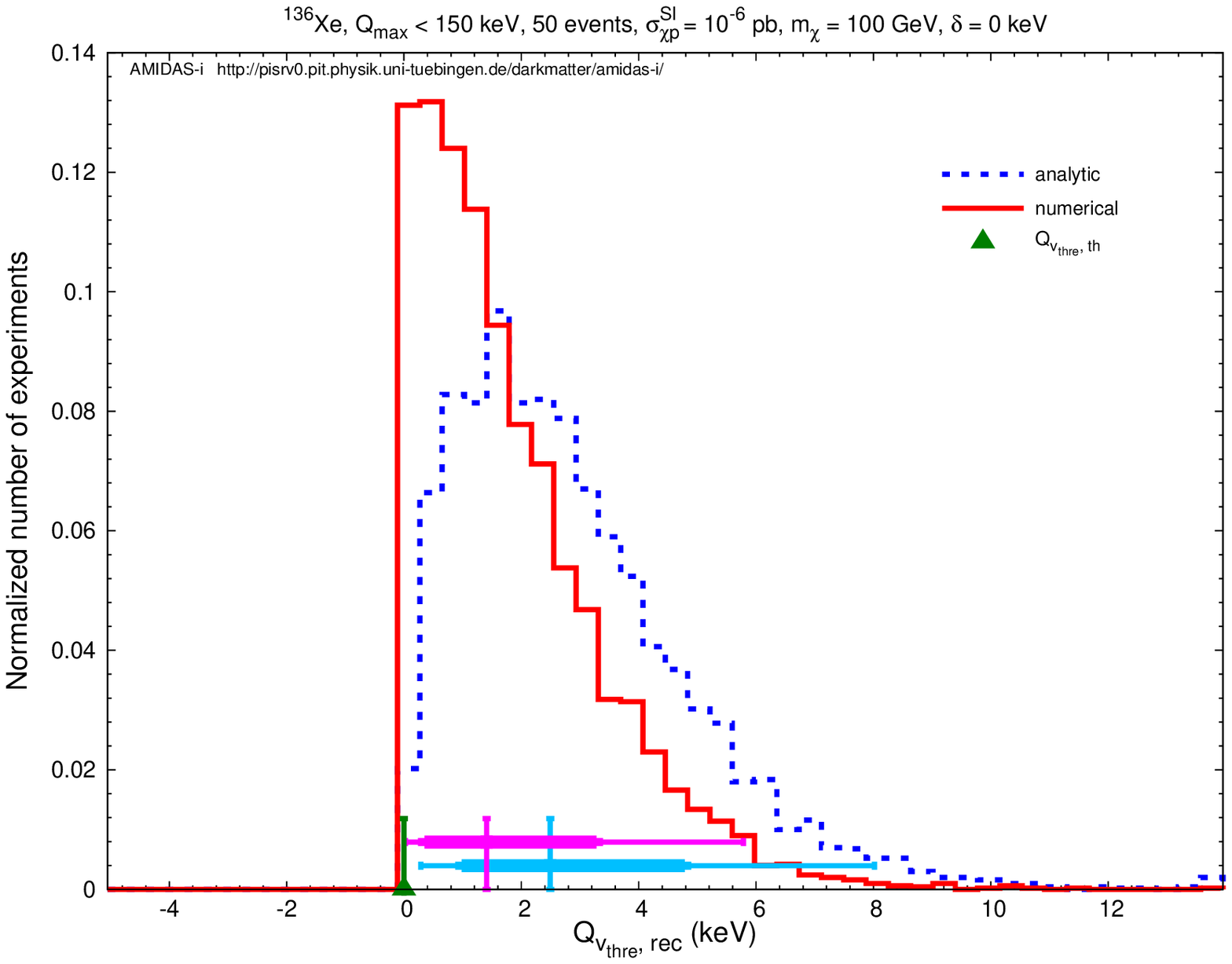} \hspace*{-1.6cm} \\
 ($\rmXA{Xe}{136}$) \\
\vspace{-0.25cm}
\end{center}
\caption{
 As in Figs.~\ref{fig:idRdQ-Ge76-100-025-050}
 and \ref{fig:idRdQ-100-025-050},
 except that
 the zero mass splitting \mbox{$\delta = 0$} has been used.
}
\label{fig:idRdQ-100-000-050}
\end{figure}

 In this subsection,
 we consider the special case of
 the zero mass splitting $\delta = 0$,
 i.e.~elastic WIMP--nucleus scattering,
 in order to demonstrate the ability
 of distinguishing the inelastic WIMP scattering scenarios
 from the elastic one.

 In the first column of Figs.~\ref{fig:idRdQ-100-000-050},
 we show the theoretical recoil spectra (dashed--double--dotted cyan),
 the measured energy spectra (dotted magenta histograms)
 as well as the analytically and numerically reconstructed energy spectra
 with four different target nuclei.
 It can be seen that,
 for this special case
 with all four simulated detector materials,
 while
 the analytically reconstructed spectra
 are still somehow peaky (with non--zero $k'$ as well as non--zero $\Qvthre$),
 the numerical iterative process
 could now indeed offer a better (and better) reconstruction of the recoil spectrum
 with much smaller $k'$ and $\Qvthre$
 (see the third and forth columns).
 Meanwhile,
 the distributions of the reconstructed $k'$ and $\Qvthre$
 indicate consistently that,
 although the {\em median} values of the (numerically)
 reconstructed $k'$ and $\Qvthre$
 would be non--zero,
 the {\em most frequently} observable values of $k'$ and $\Qvthre$
 could however be negligibly small!

 Moreover,
 it has also been found that,
 importantly,
 for this special $\delta = 0$ (elastic scattering) case,
 although the reconstructed mass splitting
 would be a little bit non--zero (positive),
 one would observe simultaneously non--physically
 a {\em negative} reconstructed WIMP mass (\mbox{$\mchi < 0$});
 the larger the true (input) WIMP mass,
 the larger the {\em absolute} value of the reconstructed one.
 This unique observation could in turn help us to confirm or rule out
 the inelastic WIMP--nucleus scattering scenarios
 down to a (very) small mass splitting (\mbox{$\delta~\lsim$ a few keV}).

 Here
 we would like to emphasize that,
 even though the measured recoil energy spectra
 (histograms shown in the first columns of Figs.~\ref{fig:idRdQ-100-010-050}
  and \ref{fig:idRdQ-100-000-050})
 look almost the same
 (and thus would be difficult to be distinguished from each other
  by conventional data analysis),
 our model--independent (analytic and numerical)
 reconstructions of the recoil spectrum
 as well as the estimations of $\Qvthre$
 could give clearly different results
 (cf.~the third and forth columns of Figs.~\ref{fig:idRdQ-100-010-050}
  and \ref{fig:idRdQ-100-000-050}).

\section{Summary and conclusions}
 In this paper
 we study direct Dark Matter detection experiments
 for the inelastic WIMP--nucleus scattering framework
 and develop model--independent methods
 for not only reconstructing the measured recoil energy spectrum,
 which can in turn be used
 for identifying the inelastic WIMP scenarios,
 but also determining the WIMP mass and the mass splitting.

 At the beginning of this paper,
 we followed our earlier work \cite{DMDDf1v}
 to derive the formula for reconstructing
 the one--dimensional velocity distribution function
 of inelastic WIMPs.
 However,
 since for inelastic WIMPs,
 there are two unknown parameters:
 the (degenerate) mass $\mchi$ and the tiny mass splitting $\delta$,
 not only our formula for reconstructing $f_1(v)$,
 but also the method for determining the WIMP mass
 introduced in Ref.~\cite{DMDDmchi},
 can not be used directly.
 Hence,
 we turned to develop a new procedure
 for determining the WIMP mass as well as the mass splitting
 model--independently.

 For this aim,
 we introduced a two--parameter exponential ansatz
 for reconstructing the measured recoil energy spectrum
 as well as determining the characteristic energy
 corresponding to the threshold (minimal required) velocity
 of incident inelastic WIMPs,
 which can produce recoil energy at all.
 In this process,
 not only
 the additional fitting parameter
 for reconstructing the recoil energy spectrum
 is approximately proportional to the squared mass splitting,
 but also the characteristic energy
 is directly proportional to the mass splitting
 (for a fixed, non--zero WIMP mass).
 Thus
 these two quantities could be good indicators
 for identifying inelastic WIMP--nucleus scattering scenarios.
 Our numerical simulations show that,
 not only for large ($\sim$ a few tens keV)
 but also for (very) small ($\sim$ a few keV) mass splittings,
 our model--independent reconstruction
 could in principle indeed identify
 the positivities of the additional fitting parameter
 and the characteristic energy
 of the recoil energy spectrum
 with a 3$\sigma$ to 5$\sigma$ confidence level
 up to a WIMP mass of $\sim$ 1 TeV
 and a mass splitting of $\sim$ 200 keV.

 Meanwhile,
 for analytically reconstructing the recoil energy spectrum,
 one has to assume that
 the minimal cut--off energy could be negligible
 and the maximal one is large enough.
 Not only because this would still be a challenge
 for most currently running and next--generation experiments,
 the kinematic minimal and maximal cut--off energies
 would also limit the available ranges of the mass (splitting),
 especially for the use of light target nuclei.
 Therefore,
 we introduced in this paper a numerical iterative procedure
 for correcting the analytic estimations of the fitting parameters
 of the measure recoil energy spectrum.
 Our simulations show that,
 basically
 this numerical correction could indeed offer
 better reconstructions
 with smaller statistical uncertainties.
 However,
 the statistical fluctuation due to
 the use of only a few tens events
 causes the divergency problem of the distribution of
 the characteristic energy.
 For larger mass splittings (\mbox{$\delta~\gsim$ 40 keV}),
 the analytic reconstructions seem to be more reliable;
 the distributions of the reconstructed spectrum fitting parameters
 and the characteristic energy
 offered by the numerical iterative procedure
 could have a (much) longer tails
 in high-- (and even low--)value ranges.
 This problem would be worse for the use of light target nuclei;
 results offered with heavy target nuclei
 could still be used as auxiliary.

 Moreover,
 we considered several different target nuclei.
 It has been found that,
 once the mass splitting is small (\mbox{$\delta~\lsim$ 40 keV}),
 all materials could identify
 inelastic WIMP scattering pretty well.
 In practice,
 light target nuclei,
 e.g.~$\rmXA{Si}{28}$ and $\rmXA{Ar}{40}$,
 could even work better,
 due to relatively higher values of the second fitting parameter
 and the characteristic energy
 of the reconstructed recoil spectrum.
 However,
 for larger mass splittings (\mbox{$\delta~\gsim$ 50 keV}),
 limited by the kinematic minimal and maximal cut--off energies
 due to the Galactic escape velocity of halo WIMPs,
 one would need heavy target nuclei,
 e.g.~$\rmXA{Ge}{76}$ and $\rmXA{Xe}{136}$,
 for the identification of inelastic WIMPs.

 Furthermore,
 for the reconstruction of the WIMP mass
 with the target combination of $\rmXA{Si}{28}$ and $\rmXA{Ge}{76}$,
 our simulations show that,
 for mass splittings \mbox{$\delta~\lsim$ 40 keV},
 one could in principle reconstruct the WIMP mass
 in the range between \mbox{$10^6 \delta$} and \mbox{$(3 - 5) \times 10^6 \delta$}
 pretty precisely with a statistical uncertainty
 of \mbox{$\sim$ 30\%} (for \mbox{$\mchi \simeq 10^6 \delta$})
 to a factor of \mbox{$\sim$ 2} (\mbox{$\mchi \simeq 5 \times 10^6 \delta$}).
 Meanwhile,
 we found also that
 the WIMP mass estimated
 with the median values of the reconstructed characteristic energy
 by using several data sets (with the same target nuclei)
 could indeed be (much) better than
 the median values of the WIMP mass reconstructed
 with each single pair of data sets,
 especially for heavier input WIMP masses
 (\mbox{$\mchi~\gsim$ 100 GeV}).
 Moreover,
 it has been found that,
 once the mass splitting is pretty light (\mbox{$\delta \sim 10$ keV}),
 by using the target combination of $\rmXA{Ar}{40}$ and $\rmXA{Xe}{136}$,
 the strongly underestimated WIMP mass in the high--mass range
 with the Si--Ge combination
 could be (strongly) alleviated
 with a much smaller statistical uncertainty;
 the distribution of the reconstructed $\mchi$
 would also be more concentrated.

 On the other hand,
 due to the limitation by the maximal kinematic cut--off energy
 and the required use of two (one heavy and one light) target nuclei,
 one could only reconstruct the mass splitting
 less than a few tens keV.
 Nevertheless,
 in the reconstructable range,
 one could in principle always reconstruct the mass splitting
 pretty precisely
 with a deviation of \mbox{$\lsim$ 20\%} (analytically)
 or even \mbox{$\lsim$ 10\%} (numerically)
 and a statistical uncertainty
 of \mbox{$\sim$ 15\%} to \mbox{$\sim$ 50\%}  (analytically)
 or \mbox{$\sim$ 10\%} to a factor of \mbox{$\sim$ 2} (numerically)
 up to a WIMP mass of \mbox{$\sim$ 1 TeV}.
 Whether and how to use
 this (pretty) precisely reconstructed mass splitting
 as a priorly determined parameter
 for further reconstructions of the mass
 as well as the one--dimensional velocity distribution
 of inelastic WIMPs
 will be investigated in the future.

 Finally,
 we consider the special case of the zero mass splitting
 (elastic WIMP--nucleus scattering).
 It has been found that,
 firstly,
 for this special case
 the numerical iterative procedure
 could offer an estimation of (almost) zero of
 the characteristic energy of the recoil spectrum.
 In addition,
 the detailed analysis of the distributions of the reconstructed results
 shows that,
 although the median values of the (numerically)
 reconstructed characteristic energies
 would be a little bit non--zero,
 the most frequently observable values
 could however be negligibly small!
 Secondly,
 although the reconstructed characteristic energy
 as well as the reconstructed mass splitting
 would be a little bit non--zero (positive),
 one would observe simultaneously non--physically
 a negative reconstructed WIMP mass;
 the larger the true (input) WIMP mass,
 the larger the absolute value of the reconstructed one.
 This unique observation could in turn help us to confirm or rule out
 the inelastic WIMP--nucleus scattering scenarios
 down to a (very) small mass splitting (\mbox{$\delta~\lsim$ a few keV}).
 Moreover,
 we would like to emphasize that
 our model--independent (analytic and numerical)
 reconstructions of the recoil spectrum
 as well as the estimations of $\Qvthre$
 could give clearly different results
 between the inelastic scattering case of a small, but non--zero mass splitting
 and the elastic one.

 In summary,
 as complementarity and extension of our earlier work
 on the development of (model--independent) methods
 for reconstructing properties of Galactic WIMPs
 by using direct DM detection data,
 we introduce in this paper new model--independent approaches
 for identifying inelastic WIMP--nucleus scattering
 as well as for reconstructing
 the mass and the mass splitting of inelastic WIMPs
 simultaneously and separately.
 Our results show that,
 with a few tens observed WIMP signals
 (from one experiment),
 one could already distinguish the inelastic WIMP scenarios
 from the elastic one.
 By using two or more data sets with positive signals,
 the WIMP mass and the mass splitting could even be reconstructed.
 As mentioned in Introduction,
 several experimental collaborations announced recently
 their positive observations of DM/WIMP signals.
 Although so far
 either the numbers of recorded events
 are too few
 (e.g.~only three candidate events were observed
  in the CDMS Si detectors
  \cite{Agnese13b})
 or the estimated background fractions are still too high
 (e.g.~$\sim$ 50\% in the CRESST-II experiment
  \cite{Angloher11})
 or even both,
 with increased numbers of cumulated (candidate) WIMP events
 and continuously (strongly) reduced background levels
 (e.g.~the LUX experiment
  \cite{Akerib13})
 we could hope that
 in the near future
 our methods presented here
 (combined probably with other approaches)
 could help our experimental colleagues to
 not only distinguish different frameworks of DM/particle physics,
 but also further constrain the parameter space
 in (various) extensions of
 the Standard Model of particle physics.

\subsubsection*{Acknowledgments}
 The authors would like to thank
 Chun-Peng Chang
 for helping to solve part of the mathematical calculations
 required in this work,
 Joakim Edsj\"o for suggesting the numerical correction,
 as well as
 the Physikalisches Institut der Universit\"at T\"ubingen
 for the technical support of the computational work
 presented in this paper.
 CLS would also like to thank
 the friendly hospitality of the
 Kavli Institute for Theoretical Physics China at the Chinese Academy of Sciences,
 Beijing,
 and the Department of Physics, Soochow University, Taipei, Taiwan,
 where part of this work was completed.
 This work
 was partially supported
 by the National Basic Research Program of China (973 Program)
 under grant no.~2010CB833000,
 the National Nature Science Foundation of China (NSFC)
 under grants no.~10975170,
 no.~10821504 and no.~10905084,
 and the Project of Knowledge Innovation Program (PKIP)
 of the Chinese Academy of Science,
 as well as
 by the National Science Council of R.O.C.~%
 under the contract no.~NSC-101-2811-M-001-033,
 and
 the LHC Physics Focus Group and
 the Focus Group on Cosmology and Particle Astrophysics,
 National Center of Theoretical Sciences, R.O.C..
\appendix
\setcounter{equation}{0}
\setcounter{figure}{0}
\renewcommand{\theequation}{A\arabic{equation}}
\renewcommand{\thefigure}{A\arabic{figure}}
%
%
%
\section{Expressions for the two--parameter exponential ansatz}
 In this section,
 we give detailed derivations and expressions needed for
 numerical estimations of different moments of
 the two--parameter exponential ansatz.

\subsection{Estimator of
            \boldmath$r_0$ given in Eq.~(\ref{eqn:r0_in_inf})}
 We start with the {\em inequality of arithmetic and geometric means}:
\beq
     \frac{1}{2} \abrac{a x + \frac{b}{x}}
 \ge \sqrt{a x \cdot \frac{b}{x}}
 =   \sqrt{a} \sqrt{b}
\~.
\eeq
 Then we can define that
\beq
        a x + \frac{b}{x}
 \equiv 2 \sqrt{a} \sqrt{b} \~ \cosh t
 \ge    2 \sqrt{a} \sqrt{b}
\~,
        ~~~~~~~~~~~~ 
        \forall~t \in [-\infty, \infty]
\~.
\label{eqn:abcosh}
\eeq
 Differentiating both sides,
 one can get
\beq
   \abrac{a - \frac{b}{x^2}} dx
 = \frac{1}{x} \abrac{a x - \frac{b}{x}} dx
 = 2 \sqrt{a} \sqrt{b} \~ \sinh t \~ dt
\~.
\label{eqn:absinh}
\eeq
 On the other hand,
 from the definition (\ref{eqn:abcosh}),
 we have
\beq
   \cosh^2 t
 = \frac{1}{4 a b} \abrac{a x + \frac{b}{x}}^2
 = \frac{1}{4 a b} \abrac{a^2 x^2 + \frac{b^2}{x^2}} + \frac{1}{2}
\~.
\eeq
 Then
\beq
   \sinh^2 t
 = \cosh^2 t - 1
 = \frac{1}{4 a b} \abrac{a^2 x^2 + \frac{b^2}{x^2}} - \frac{1}{2}
 = \frac{1}{4 a b} \abrac{a x - \frac{b}{x}}^2
\~,
\eeq
 i.e.
\beq
   2 \sqrt{a} \sqrt{b} \~ \sinh t
 = a x - \frac{b}{x}
\~.
\eeq
 Comparing the above equation with Eq.~(\ref{eqn:absinh}),
 we have
\beq
   dt
 = \frac{dx}{x}
\~.
\eeq
 Therefore,
 we can obtain that
\beqn
    \intz \afrac{1}{x} \~ e^{-a x - b / x} \~ dx
 \= \abrac{\int_0^{\sqrt{b / a}} + \int_{\sqrt{b / a}}^{\infty}}
    e^{- \abrac{a x + b / x}} \afrac{dx}{x}
    \non\\
 \= \abrac{\int_{-\infty}^0 + \int_0^{\infty}}
    e^{- 2 \sqrt{a} \sqrt{b} \cosh t} \~ dt
    \non\\
 \= 2 K_0\abrac{2 \sqrt{a} \sqrt{b}}
\~.
\label{eqn:Bessel_K_0}
\eeqn
 Here,
 firstly,
 from Eq.~(\ref{eqn:abcosh}),
 one can find that,
 once $x = \sqrt{b /a}$,
 $\cosh t = 1$,
 and thus $t = 0$.
 Secondly,
 for the last line
 we have used the integral formula
 for the modified Bessel function of the second kind:
\beq
   K_{\nu}(z)
 = \intz e^{-z \cosh t} \cosh (\nu t) \~ dt
\~.
\eeq
 Finally,
 by differentiating with respect to $a$,
 we can obtain an analytic form for the denominator of $r_0$
 in Eq.~(\ref{eqn:r0_in})
 for the case of a negligible minimal cut--off energy and
 a (very) large maximal one
 as
\beq
     \intz e^{-a x - b / x} \~ dx
  =- \pp{a} \bbigg{2 K_0\abrac{2 \sqrt{a} \sqrt{b}}}
  =  2 \~ \sfrac{b}{a} \~ K_1\abrac{2 \sqrt{a} \sqrt{b}}
\~,
\eeq
 where we have used Eq.~(\ref{eqn:Bessel_K_0})
 and the recursion relation of $ K_n(z)$:
\beq
    \dd{z} \bBig{z^{-n} K_n(z)}
 =- z^{-n} K_{n + 1}(z)
\~,
    ~~~~~~~~~~~~~~~~~~~~ 
    n
 =  0,~\pm 1,~\pm 2, \cdots
\eeq
\subsection{Moments of the two--parameter exponential ansatz}
 Since we have found that
\beq
        \Expv{1 / \sqrt{x}}_{\rm inf}
 \equiv \intz \frac{1}{\sqrt{x  }} \~ e^{-a x - b / x} \~ dx
 =      \sfrac{\pi}{a} \~ e^{- 2 \sqrt{a} \sqrt{b}}
\~,
\label{eqn:expv_x_m1_2_inf}
\eeq
 by differentiating with respect to $a$,
 one can get
\cheqnXa{A}
\beq
     \Expv{\sqrt{x}}_{\rm inf}
 \equiv
     \intz \sqrt{x  } \~ e^{-a x - b / x} \~ dx
  =  \frac{1}{2  } \sfrac{\pi}{a} \~ e^{- 2 \sqrt{a} \sqrt{b}}
     \abrac{\sfrac{b}{a} \cdot 2 + \frac{1}{a}}
\~.
\label{eqn:expv_x_p1_2_inf}
\eeq
 Similarly,
 by differentiating with respect to $b$,
 we have
\cheqnXb{A}
\beq
     \Expv{1 / \sqrt{x^3}}_{\rm inf}
 \equiv
     \intz \frac{1}{\sqrt{x^3}} \~ e^{-a x - b / x} \~ dx
  =  \sfrac{\pi}{b} \~ e^{- 2 \sqrt{a} \sqrt{b}}
\~,
\label{eqn:expv_x_m3_2_inf}
\eeq
 and
\cheqnXc{A}
\beq
     \Expv{1 / \sqrt{x^5}}_{\rm inf}
 \equiv
     \intz \frac{1}{\sqrt{x^5}} \~ e^{-a x - b / x} \~ dx
  =  \frac{1}{2  } \sfrac{\pi}{b} \~ e^{- 2 \sqrt{a} \sqrt{b}}
     \abrac{\sfrac{a}{b} \cdot 2 + \frac{1}{b}}
\~.
\label{eqn:expv_x_m5_2_inf}
\eeq
\cheqnX{A}%
 Then we can obtain that
\beq
   \frac{\Expv{1 / \sqrt{x}}_{\rm inf}}{\Expv{1 / \sqrt{x^3}}_{\rm inf}}
 = \sfrac{b}{a}
\~,
\eeq
 as well as
\cheqnXa{A}
\beq
    \frac{\Expv{\sqrt{x}}_{\rm inf}}{\Expv{1 / \sqrt{x}}_{\rm inf}}
 =  \sfrac{b}{a}
  + \frac{1}{2 a}
 =  \frac{\Expv{1 / \sqrt{x}}_{\rm inf}}{\Expv{1 / \sqrt{x^3}}_{\rm inf}}
  + \frac{1}{2 a}
\~,
\eeq
 and
\cheqnXb{A}
\beq
    \frac{\Expv{1 / \sqrt{x^5}}_{\rm inf}}{\Expv{1 / \sqrt{x^3}}_{\rm inf}}
 =  \sfrac{a}{b}
  + \frac{1}{2 b}
 =  \frac{\Expv{1 / \sqrt{x^3}}_{\rm inf}}{\Expv{1 / \sqrt{x}}_{\rm inf}}
  + \frac{1}{2 b}
\~.
\eeq
\cheqnX{A}
 These give us that
\cheqnXa{A}
\beq
   a
 = \frac{1}{2}
   \abrac{  \frac{\Expv{\sqrt{x}}_{\rm inf}}    {\Expv{1 / \sqrt{x  }}_{\rm inf}}
          - \frac{\Expv{1 / \sqrt{x}}_{\rm inf}}{\Expv{1 / \sqrt{x^3}}_{\rm inf}} }^{-1}
 = \frac{1}{2}
   \afrac{  \Expv{1 / \sqrt{x}}_{\rm inf} \Expv{1 / \sqrt{x^3}}_{\rm inf}}
         {  \Expv{\sqrt{x}}_{\rm inf}     \Expv{1 / \sqrt{x^3}}_{\rm inf}
          - \Expv{1 / \sqrt{x}}_{\rm inf}^2}
\~,
\label{eqn:a_expv_x}
\eeq
 and
\cheqnXb{A}
\beq
   b
 = \frac{1}{2}
   \abrac{  \frac{\Expv{1 / \sqrt{x^5}}_{\rm inf}}{\Expv{1 / \sqrt{x^3}}_{\rm inf}}
          - \frac{\Expv{1 / \sqrt{x^3}}_{\rm inf}}{\Expv{1 / \sqrt{x  }}_{\rm inf}} }^{-1}
 = \frac{1}{2}
   \afrac{\Expv{1 / \sqrt{x}}_{\rm inf} \Expv{1 / \sqrt{x^3}}_{\rm inf}}
         {  \Expv{1 / \sqrt{x}}_{\rm inf} \Expv{1 / \sqrt{x^5}}_{\rm inf}
          - \Expv{1 / \sqrt{x^3}}_{\rm inf}^2}
\~.
\label{eqn:b_expv_x}
\eeq
\cheqnX{A}

 More generally,
 we have found that
\beqn
        \Expv{1 / \sqrt{x}}(a, b; x)
 \eqnequiv
        \int \frac{1}{\sqrt{x}} \~ e^{-a x - b / x} \~ dx
        \non\\
 \=     \frac{1}{2} \sfrac{\pi}{a}
        \bbigg{  e^{  2 \sqrt{a} \sqrt{b}} \~
                 \erf\abrac{\T \sqrt{a} \sqrt{x} + \frac{\sqrt{b}}{\sqrt{x}} }
               + e^{- 2 \sqrt{a} \sqrt{b}} \~
                 \erf\abrac{\T \sqrt{a} \sqrt{x} - \frac{\sqrt{b}}{\sqrt{x}} } }
\~.
\label{eqn:expv_x_m1_2}
\eeqn
 Therefore,
 by differentiating with respect to $a$,
 one can get
\cheqnXa{A}
\beqn
 \conti
        \Expv{\sqrt{x}}(a, b; x)
        \non\\
 \eqnequiv
        \int \sqrt{x} \~ e^{-a x - b / x} \~ dx
     \non\\
 \=  \frac{1}{2} \sfrac{\pi}{a}
     \cBiggl{  \frac{1}{2 a}
               \bbigg{  e^{  2 \sqrt{a} \sqrt{b}} \~
                        \erf\abrac{\T \sqrt{a} \sqrt{x} + \frac{\sqrt{b}}{\sqrt{x}} }
                      + e^{- 2 \sqrt{a} \sqrt{b}} \~
                        \erf\abrac{\T \sqrt{a} \sqrt{x} - \frac{\sqrt{b}}{\sqrt{x}} } } }
     \non\\
 \conti ~~~~~~~~~~~~ 
     \cBiggr{- \sfrac{b}{a}
               \bbigg{  e^{  2 \sqrt{a} \sqrt{b}} \~
                        \erf\abrac{\T \sqrt{a} \sqrt{x} + \frac{\sqrt{b}}{\sqrt{x}} }
                      - e^{- 2 \sqrt{a} \sqrt{b}} \~
                        \erf\abrac{\T \sqrt{a} \sqrt{x} - \frac{\sqrt{b}}{\sqrt{x}} } } }
     \non\\
 \conti ~~~~
   - \frac{1}{a}
     \abigg{  \sqrt{x} \~ e^{- a x - b / x} }
\~.
\label{eqn:expv_x_p1_2}
\eeqn
 Similarly,
 with respect to $b$,
 we have
\cheqnXb{A}
\beqn
 \conti
        \Expv{1 / \sqrt{x^3}}(a, b; x)
        \non\\
 \eqnequiv
        \int \frac{1}{\sqrt{x^3}} \~ e^{-a x - b / x} \~ dx
     \non\\
 \=- \frac{1}{2} \sfrac{\pi}{b}
     \bbigg{  e^{  2 \sqrt{a} \sqrt{b}} \~
              \erf\abrac{\T \sqrt{a} \sqrt{x} + \frac{\sqrt{b}}{\sqrt{x}} }
            - e^{- 2 \sqrt{a} \sqrt{b}} \~
              \erf\abrac{\T \sqrt{a} \sqrt{x} - \frac{\sqrt{b}}{\sqrt{x}} } }
\~,
\label{eqn:expv_x_m3_2}
\eeqn
 and
\cheqnXc{A}
\beqn
 \conti
        \Expv{1 / \sqrt{x^5}}(a, b; x)
        \non\\
 \eqnequiv
        \int \frac{1}{\sqrt{x^5}} \~ e^{-a x - b / x} \~ dx
     \non\\
 \=- \frac{1}{2} \sfrac{\pi}{b}
     \cBiggl{  \frac{1}{2 b}
               \bbigg{  e^{  2 \sqrt{a} \sqrt{b}} \~
                        \erf\abrac{\T \sqrt{a} \sqrt{x} + \frac{\sqrt{b}}{\sqrt{x}} }
                      - e^{- 2 \sqrt{a} \sqrt{b}} \~
                        \erf\abrac{\T \sqrt{a} \sqrt{x} - \frac{\sqrt{b}}{\sqrt{x}} } } }
     \non\\
 \conti ~~~~~~~~~~~~~~ 
     \cBiggr{- \sfrac{a}{b}
               \bbigg{  e^{  2 \sqrt{a} \sqrt{b}} \~
                        \erf\abrac{\T \sqrt{a} \sqrt{x} + \frac{\sqrt{b}}{\sqrt{x}} }
                      + e^{- 2 \sqrt{a} \sqrt{b}} \~
                        \erf\abrac{\T \sqrt{a} \sqrt{x} - \frac{\sqrt{b}}{\sqrt{x}} } } }
        \non\\
 \conti ~~~~~~ 
   + \frac{1}{b}
     \abigg{  \frac{1}{\sqrt{x}} \~ e^{- a x - b / x} }
\~.
\label{eqn:expv_x_m5_2}
\eeqn
\cheqnX{A}
 Here we have used that
\beqN
   \dd{x}\bBig{\erf(x)}
 = \frac{2}{\sqrt{\pi}} \bbrac{\dd{x} \int_0^{x} e^{-t^2} \~ dt}
 = \frac{2}{\sqrt{\pi}}~e^{-x^2}
\~.
\eeqN

 On the other hand,
 by setting $x = 1 / y$,
 one can find that
\beqn
    \int_{x_1}^{x_2} x^{-1 - \lambda} \~ e^{-a x - b / x} \~ dx
 =- \int_{1 / x_1}^{1 / x_2} y^{-1 + \lambda} \~ e^{-b y - a / y} \~ dy
\~,
\label{eqn:expv_x_symmetry}
\eeqn
 and,
 as a special case,
\beqn
    \intz x^{-1 - \lambda} \~ e^{-a x - b / x} \~ dx
 =  \intz y^{-1 + \lambda} \~ e^{-b y - a / y} \~ dy
\~,
\label{eqn:expv_x_symmetry_inf}
\eeqn
 for $\lambda = 1/2,~1,~3/2,~2,~\cdots$.
 Hence,
 one can obtain Eqs.~(\ref{eqn:expv_x_m3_2_inf}), (\ref{eqn:expv_x_m5_2_inf}),
 (\ref{eqn:expv_x_m3_2}) and (\ref{eqn:expv_x_m5_2})
 from Eqs.~(\ref{eqn:expv_x_m1_2_inf}) and (\ref{eqn:expv_x_p1_2_inf})
 and Eqs.~(\ref{eqn:expv_x_m1_2}) and (\ref{eqn:expv_x_p1_2})
 by exchanging $a \getsto b$ and $x \getsto 1 / x$
 and using
\beqN
    \erf(-x)
 = -\erf(x)
\~.
\eeqN
%

%
%
%
\section{Solving the fitting parameters \boldmath$k$ and $k'$ numerically}
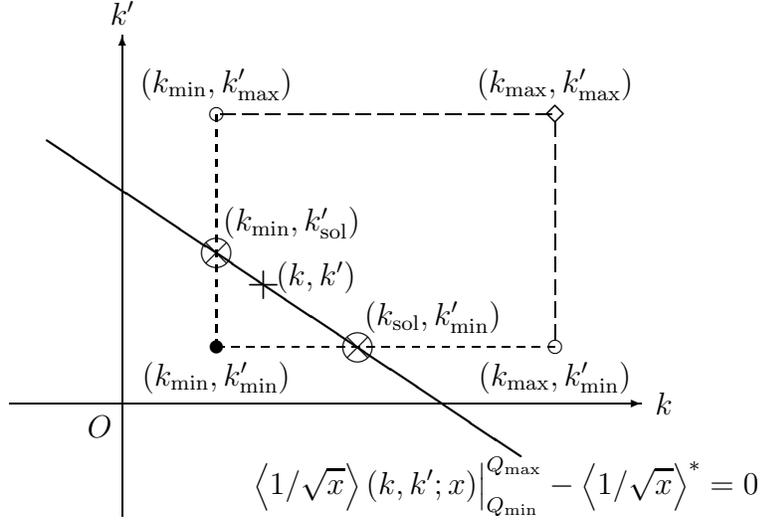
\begin{figure}[t!]
\begin{center}
\begin{picture}(10  , 7  )
\put(0    , 0    ){\makebox(10  , 7   ){}}
%
\put(0    , 1.5  ){\vector (1   , 0   ){8.4}}    
\put(1.5  , 0    ){\vector (0   , 1   ){6.4}}    
\put(8.4  , 1.2  ){\makebox(0.6 , 0.6 ){$k$}}    
\put(1.2  , 6.4  ){\makebox(0.6 , 0.6 ){$k'$}}   
\put(0.9  , 0.9  ){\makebox(0.6 , 0.6 ){$O $}}   
%
\thicklines
\put(0.5  , 5    ){\line   (3   ,-2   ){6.3}}
\put(3.1  , 0.1  ){\makebox(7 , 0.6 )
{$\D \Expv{1 / \sqrt{x}}(k, k'; x)\Big|_{\Qmin}^{\Qmax} - \Expv{1 / \sqrt{x}}^{\ast} = 0$}}
\thinlines
%
\put(2.75 , 2.25 ){\circle*{0.18}}
\put(1.75 , 1.65 ){\makebox(2   , 0.4 ){$(k_{\rm min}, k'_{\rm min})$}}
%
\multiput(2.75 , 2.25 )(0.2 , 0   ){23}{\line(1, 0){0.1}}
\multiput(2.75 , 2.25 )(0   , 0.2 ){16}{\line(0, 1){0.1}}
\put(7.25 , 2.25 ){\circle {0.18}}
\put(6.25 , 1.65 ){\makebox(2   , 0.4 ){$(k_{\rm max}, k'_{\rm min})$}}
\put(2.75 , 5.35 ){\circle {0.18}}
\put(1.75 , 5.55 ){\makebox(2   , 0.4 ){$(k_{\rm min}, k'_{\rm max})$}}
%
\multiput(2.8  , 5.35 )(0.3 , 0   ){15}{\line(1, 0){0.2}}
\multiput(7.25 , 2.35 )(0   , 0.3 ){10}{\line(0, 1){0.2}}
%
\put(7    , 5.1  ){\makebox(0.5 , 0.5 ){\large$\diamond$}}
\put(6.25 , 5.55 ){\makebox(2   , 0.4 ){$(k_{\rm max}, k'_{\rm max})$}}
%
\put(4.375, 2    ){\makebox(0.5 , 0.5 ){\Large$\otimes$}}
\put(4.625, 2.45 ){\makebox(2   , 0.4 ){$(k_{\rm sol}, k'_{\rm min})$}}
\put(2.5  , 3.25 ){\makebox(0.5 , 0.5 ){\Large$\otimes$}}
\put(2.75 , 3.7 ){\makebox(2   , 0.4 ){$(k_{\rm min}, k'_{\rm sol})$}}
%
\put(3.125, 2.833){\makebox(0.5 , 0.5 ){\Large$+$}}
\put(3.525, 2.983){\makebox(1.1 , 0.4 ){$(k, k')$}}
\end{picture}
\vspace{-0.25cm}
\end{center}
\caption{
 Sketch of the numerical procedure
 for giving a linear equation of $k$ and $k'$
 by using the minus--one--half moment of
 the two--parameter exponential ansatz.
}
\label{fig:k_kp_in_sol}
\end{figure}

 In this section,
 we describe our numerical iterative procedure
 for solving the fitting parameters of the measured recoil energy spectrum
 based on the analytically estimation of these two parameters.

 We start with the point $(k_{\rm ana}, k'_{\rm ana})$,
 which has been estimated by Eqs.~(\ref{eqn:k_in_ana}) and (\ref{eqn:kp_in_ana}).
 Check at first whether
\(
\D
        \Expv{1 / \sqrt{x}}(k_{\rm ana}, k'_{\rm ana}; x)\Big|_{\Qmin}^{\Qmax}
 -      \Expv{1 / \sqrt{x}}^{\ast}
 \equiv F_{-1/2}(k = k_{\rm ana}, k' = k'_{\rm ana})
 >      0
\).
 Here $\D \Expv{1 / \sqrt{x}}(k, k'; x)$ is the function of $k$ and $k'$
 given by Eq.~(\ref{eqn:expv_x_m1_2})
 and
\beq
        \Expv{1 / \sqrt{x}}^{\ast}
 \equiv \abrac{\int_{\Qmin}^{\Qmax} e^{-k^{\ast} x - k'^{\ast} / x} \~ dx}
        \abrac{\frac{1}{N_{\rm tot}} \sum_{a} Q_a^{\lambda}}
\label{eqn:expv_x_m1_2_ast}
\eeq
 can be estimated from measured recoil energies directly.
 Note that
 the integral on the right--hand side of Eq.~(\ref{eqn:expv_x_m1_2_ast})
 has an upper (lower) bound of $\Qmax$ ($\Qmin$)
 and can be estimated numerically
 by setting,
 e.g.~$k^{(\prime) \ast} = k^{(\prime)}_{\rm ana}$.

 Without losing the generality,
 we assume that
 $F_{-1/2}(k_{\rm ana}, k'_{\rm ana}) > 0$
 and set $(k_{\rm ana}, k'_{\rm ana})$ as $(k_{\rm min}, k'_{\rm min})$.
 Note that
 the function $\D \Expv{1 / \sqrt{x}}(k, k'; x)\Big|_{\Qmin}^{\Qmax}$
 should decrease monotonically
 as $k$ and/or $k'$ increases.
 As sketched in Fig.~\ref{fig:k_kp_in_sol},
 by,
 e.g.~fixing $k' = k'_{\rm min}$ and then
 checking whether $F_{-1/2}(k, k') > 0$
 by setting $k = k_{\rm min} + n \~ (k_{\rm min} / 10)$,
 $n = 1,~2,~3,~\cdots$,
 one should be able to find two auxiliary points
 $(k_{\rm max}, k'_{\rm min})$ and $(k_{\rm min}, k'_{\rm max})$
 at which the function $F_{-1/2}(k, k')$ has a different sign
 ($< 0$ corresponding to our current assumption).
 Now solve
 $k _{\rm sol} \in [k _{\rm min}, k _{\rm max}]$ and
 $k'_{\rm sol} \in [k'_{\rm min}, k'_{\rm max}]$,
 which satisfy
 $\vbrac{F_{-1/2}(k_{\rm sol}, k'_{\rm min})} < 10^{-5} \Expv{1 / \sqrt{x}}^{\ast}$ and
 $\vbrac{F_{-1/2}(k_{\rm min}, k'_{\rm sol})} < 10^{-5} \Expv{1 / \sqrt{x}}^{\ast}$.

 Since the intersection boundary of $F_{-1/2}(k, k') = 0$ on the $k-k'$ plane
 is almost a straight line
 and the analytic estimates $(k_{\rm ana}, k'_{\rm ana})$
 should in principle be pretty close to
 the numerical solution $(k_{\rm num}, k'_{\rm num})$,
 we approximate the equation of the boundary to a linear equation by
\beq
   \frac{k  - k _{\rm sol, -1/2}}{k _{\rm min, -1/2} - k _{\rm sol, -1/2}}
 = \frac{k' - k'_{\rm min, -1/2}}{k'_{\rm sol, -1/2} - k'_{\rm min, -1/2}}
\~,
\eeq
 namely,
\beqn
 \conti
     \abrac{k'_{\rm sol, -1/2} - k'_{\rm min, -1/2}} k
   + \abrac{k _{\rm sol, -1/2} - k _{\rm min, -1/2}} k'
     \non\\
 \=  k_{\rm sol, -1/2} k'_{\rm sol, -1/2}
   - k_{\rm min, -1/2} k'_{\rm min, -1/2}
\~.
\label{eqn:k_kp_in_num_sol_m1_2}
\eeqn

 Similarly,
 we define
\(
\D
        \Expv{1 / \sqrt{x^3}}(k, k'; x)\Big|_{\Qmin}^{\Qmax}
 -      \Expv{1 / \sqrt{x^3}}^{\ast}
 \equiv F_{-3/2}(k, k')
\)
 and asssum that
 $F_{-3/2}(k_{\rm ana}, k'_{\rm ana}) > 0$.
 Here $\D \Expv{1 / \sqrt{x^3}}(k, k'; x)$ is the function of $k$ and $k'$
 given by Eq.~(\ref{eqn:expv_x_m3_2}).
 Then,
 by repeating the above process,
 one have
\beqn
 \conti
     \abrac{k'_{\rm sol, -3/2} - k'_{\rm min, -3/2}} k
   + \abrac{k _{\rm sol, -3/2} - k _{\rm min, -3/2}} k'
     \non\\
 \=  k_{\rm sol, -3/2} k'_{\rm sol, -3/2}
   - k_{\rm min, -3/2} k'_{\rm min, -3/2}
\~.
\label{eqn:k_kp_in_num_sol_m3_2}
\eeqn
 Finally,
 the numerical solution of $(k, k')$ can be given by
\beq
   k_{\rm num, 1}
 = \frac{\Delta_{k }}{\Delta}
\~,
   ~~~~~~~~~~~~~~~~~~~~ 
   k'_{\rm num, 1}
 = \frac{\Delta_{k'}}{\Delta}
\~.
\eeq
 Here we define
\beqn
            \Delta
 \eqnequiv  \abrac{k'_{\rm sol, -1/2} - k'_{\rm min, -1/2}}
            \abrac{k _{\rm sol, -3/2} - k _{\rm min, -3/2}}
            \non\\
 \conti     ~~~~~~~~ 
          - \abrac{k _{\rm sol, -1/2} - k _{\rm min, -1/2}}
            \abrac{k'_{\rm sol, -3/2} - k'_{\rm min, -3/2}}
\~,
\eeqn
 and
\cheqna
\beqn
            \Delta_{k}
 \eqnequiv  \abrac{  k_{\rm sol, -1/2} k'_{\rm sol, -1/2}
                   - k_{\rm min, -1/2} k'_{\rm min, -1/2}}
            \abrac{k _{\rm sol, -3/2} - k _{\rm min, -3/2}}
            \non\\
 \conti     ~~~~~~ 
          - \abrac{  k_{\rm sol, -3/2} k'_{\rm sol, -3/2}
                   - k_{\rm min, -3/2} k'_{\rm min, -3/2}}
            \abrac{k _{\rm sol, -1/2} - k _{\rm min, -1/2}}
\~,
\eeqn
\cheqnb
\beqn
            \Delta_{k'}
 \eqnequiv  \abrac{  k_{\rm sol, -3/2} k'_{\rm sol, -3/2}
                   - k_{\rm min, -3/2} k'_{\rm min, -3/2}}
            \abrac{k'_{\rm sol, -1/2} - k'_{\rm min, -1/2}}
            \non\\
 \conti     ~~~~~~ 
          - \abrac{  k_{\rm sol, -1/2} k'_{\rm sol, -1/2}
                   - k_{\rm min, -1/2} k'_{\rm min, -1/2}}
            \abrac{k'_{\rm sol, -3/2} - k'_{\rm min, -3/2}}
\~.
\eeqn
\cheqn
 Here $k^{(\prime)}_{\rm num, 1}$ indicate
 the numerical estimates of the fitting parameters
 in the first round.
 One could process the whole numerical procedure iteratively.
 Remind however that
 statistical fluctuation could cause a divergency problem
 and the results from the later rounds might be worse
 than that from the first or second round.

%
%
%
\section{Expressions of the derivatives of \boldmath $\Qvthre$, $k$ and $k'$}
 Firstly,
 differentiating the expression (\ref{eqn:Qvthre_sol})
 for solving the characteristic energy $\Qvthre$
 with respect to $k$,
 one find that
\cheqnXa{A}
\beq
          1
        + \frac{2 k'}{\Qvthre^3} \aPp{\Qvthre}{k}
        + \dd{\Qvthre}\bbrac{\frac{2}{F(\Qvthre)} \aDd{F}{Q}_{Q = \Qvthre}} \aPp{\Qvthre}{k}
  =       0
\~.
\eeq
 Then we can get 
\beq
    \Pp{\Qvthre}{k}
 =- \frac{1}{2}
    \cbrac{  \frac{k'}{Q^3} 
           + \dd{Q}\bbrac{\frac{1}{F(Q)} \aDd{F}{Q}} }_{Q = \Qvthre}^{-1}
\~.
\label{eqn:dQvthre_dk}
\eeq
 Similarly,
 differentiating the expression (\ref{eqn:Qvthre_sol})
 with respect to $k'$,
 one has
\cheqnXNx{A}{-2}{b}
\beq
        - \frac{1}{\Qvthre^2}
        + \frac{2 k'}{\Qvthre^3} \aPp{\Qvthre}{k'}
        + \dd{\Qvthre}\bbrac{\frac{2}{F(\Qvthre)} \aDd{F}{Q}_{Q = \Qvthre}} \aPp{\Qvthre}{k'}
          \non\\
  =       0
\~.
\eeq
 Thus,
 it can be found that
\beq
    \Pp{\Qvthre}{k'}
 =  \frac{1}{\Qvthre^2}
    \cbrac{  \frac{2 k'}{Q^3}
           + \dd{Q}\bbrac{\frac{2}{F(Q)} \aDd{F}{Q}} }_{Q = \Qvthre}^{-1}
 =- \frac{1}{\Qvthre^2}
    \aPp{\Qvthre}{k}
\~.
\label{eqn:dQvthre_dkp}
\eeq
\cheqnX{A}
 Note that,
 as Eqs.~(\ref{eqn:f1v_in_rec}) and (\ref{eqn:Qvthre_sol}),
 Eqs.~(\ref{eqn:dQvthre_dk}) and (\ref{eqn:dQvthre_dkp})
 can be used for both of
 the analytically and numerically estimated
 $k$ and $k'$.

\subsection{For the analytic estimates}

 For the analytic estimate of $k$
 given by Eq.~(\ref{eqn:k_in_ana}),
 we have
\cheqnXa{A}
\beq
    \Pp{k_{\rm ana}}{\D \Expv{Q^{1 / 2}}_{\rm inf}}
 =- \bfrac{\D  \Expv{Q^{-3 / 2}}_{\rm inf}}
          {\D  \Expv{Q^{ 1 / 2}}_{\rm inf} \Expv{Q^{-3 / 2}}_{\rm inf}
             - \Expv{Q^{-1 / 2}}_{\rm inf}^2}
    k_{\rm ana}
\~,
\label{eqn:dk_dexpv_Q_p1_2_inf}
\eeq
\cheqnXb{A}
\beq
    \Pp{k_{\rm ana}}{\D \Expv{Q^{-1 / 2}}_{\rm inf}}
 =  \frac{1}{2}
    \bfrac{\D \Expv{Q^{-3 / 2}}_{\rm inf}
              \abrac{  \Expv{Q^{ 1 / 2}}_{\rm inf} \Expv{Q^{-3 / 2}}_{\rm inf} 
                     + \Expv{Q^{-1 / 2}}_{\rm inf}^2  }  }
          {\D \abrac{  \Expv{Q^{ 1 / 2}}_{\rm inf} \Expv{Q^{-3 / 2}}_{\rm inf}
                     - \Expv{Q^{-1 / 2}}_{\rm inf}^2  }^2}
\~,
\label{eqn:dk_dexpv_Q_m1_2_inf}
\eeq
\cheqnXc{A}
\beq
    \Pp{k_{\rm ana}}{\D \Expv{Q^{-3 / 2}}_{\rm inf}}
 =- \frac{1}{2}
    \bfrac{\D \Expv{Q^{-1 / 2}}_{\rm inf}^3}
          {\D  \abrac{  \Expv{Q^{ 1 / 2}}_{\rm inf} \Expv{Q^{-3 / 2}}_{\rm inf}
                      - \Expv{Q^{-1 / 2}}_{\rm inf}^2  }^2}
\~,
\label{eqn:dk_dexpv_Q_m3_2_inf}
\eeq
 and
\cheqnXNx{A}{-1}{d}
\beq
    \Pp{k_{\rm ana}}{\D \Expv{Q^{-5 / 2}}_{\rm inf}}
 =  0
\~.
\label{eqn:dk_dexpv_Q_m5_2_inf}
\eeq
\cheqnX{A}
 Similarly,
 from the expression (\ref{eqn:kp_in_ana}) of $k'$,
 we have
\cheqnXa{A}
\beq
    \Pp{k'_{\rm ana}}{\D \Expv{Q^{1 / 2}}_{\rm inf}}
 =  0
\~,
\label{eqn:dkp_dexpv_Q_p1_2_inf}
\eeq
\cheqnXb{A}
\beq
    \Pp{k'_{\rm ana}}{\D \Expv{Q^{-1 / 2}}_{\rm inf}}
 =- \frac{1}{2}
    \bfrac{\D \Expv{Q^{-3 / 2}}_{\rm inf}^3}
          {\D \abrac{  \Expv{Q^{-1 / 2}}_{\rm inf} \Expv{Q^{-5 / 2}}_{\rm inf}
                     - \Expv{Q^{-3 / 2}}_{\rm inf}^2  }^2}
\~,
\label{eqn:dkp_dexpv_Q_m1_2_inf}
\eeq
\cheqnXc{A}
\beq
    \Pp{k'_{\rm ana}}{\D \Expv{Q^{-3 / 2}}_{\rm inf}}
 =  \frac{1}{2}
    \bfrac{\D \Expv{Q^{-1 / 2}}_{\rm inf}
              \abrac{  \Expv{Q^{-1 / 2}}_{\rm inf} \Expv{Q^{-5 / 2}}_{\rm inf}
                     + \Expv{Q^{-3 / 2}}_{\rm inf}^2  } }
          {\D \abrac{  \Expv{Q^{-1 / 2}}_{\rm inf} \Expv{Q^{-5 / 2}}_{\rm inf}
                     - \Expv{Q^{-3 / 2}}_{\rm inf}^2  }^2}
\~,
\label{eqn:dkp_dexpv_Q_m3_2_inf}
\eeq
 and
\cheqnXNx{A}{-1}{d}
\beq
    \Pp{k'_{\rm ana}}{\D \Expv{Q^{-5 / 2}}_{\rm inf}}
 =- \bfrac{\D  \Expv{Q^{-1 / 2}}_{\rm inf}}
          {\D  \Expv{Q^{-1 / 2}}_{\rm inf} \Expv{Q^{-5 / 2}}_{\rm inf}
             - \Expv{Q^{-3 / 2}}_{\rm inf}^2}
    k'_{\rm ana}
\~.
\label{eqn:dkp_dexpv_Q_m5_2_inf}
\eeq
\cheqnX{A}
\subsection{For the numerical solutions}
 Firstly,
 from Eq.~(\ref{eqn:expv_Qlambda_k_kp}),
 we have
\beqn
    \Pp{\D \Expv{Q^{\lambda}}}{\D k^{(\prime)}}
 \= \frac{1}{\D \int_{\Qmin}^{\Qmax} e^{-k^{\ast} x - k'^{\ast} / x} \~ dx}
    \bbrac{  \Pp{\D \Expv{x^{\lambda}}(k, k'; x = \Qmax)}{\D k^{(\prime)}}
           - \Pp{\D \Expv{x^{\lambda}}(k, k'; x = \Qmin)}{\D k^{(\prime)}}}
\~,
    \non\\
\eeqn
 namely,
\beq
   \Pp{\D k^{(\prime)}}{\D \Expv{Q^{\lambda}}}
 = \abrac{\int_{\Qmin}^{\Qmax} e^{-k^{\ast} x - k'^{\ast} / x} \~ dx}
   \bbrac{\vright{\Pp{\D \Expv{x^{\lambda}}(k, k'; x)}{\D k^{(\prime)}}}
                  _{x = \Qmin}^{x = \Qmax}}^{-1}
\~.
\eeq
 Therefore,
 by using definitions (\ref{eqn:expv_x_m1_2}),
 (\ref{eqn:expv_x_p1_2}) and (\ref{eqn:expv_x_m3_2}),
 one can get that
\cheqnXa{A}
\beqn
     \Pp{\D k_{\rm num}}{\D \Expv{Q^{-1 / 2}}}
 \=  \abrac{\int_{\Qmin}^{\Qmax} e^{-k^{\ast} x - k'^{\ast} / x} \~ dx}
     \bbrac{\vright{\Pp{\D \Expv{1 / \sqrt{x}}(k, k'; x)}{\D k}}
                    _{x = \Qmin}^{x = \Qmax}}^{-1}
     \non\\
 \=  \abrac{\int_{\Qmin}^{\Qmax} e^{-k^{\ast} x - k'^{\ast} / x} \~ dx}
     \bbrac{\vbiggr{- \Expv{\sqrt{x}}(k, k'; x)}
                    _{x = \Qmin}^{x = \Qmax}}^{-1}
     \non\\
 \=- \frac{1}{\D \Expv{Q^{1 / 2}}}
\~,
\label{eqn:dk_dexpv_Q_m1_2}
\eeqn
 and
\cheqnXb{A}
\beq
     \Pp{\D k_{\rm num}}{\D \Expv{Q^{-3 / 2}}}
  =- \frac{1}{\D \Expv{Q^{-1 / 2}}}
\~.
\label{eqn:dk_dexpv_Q_m3_2}
\eeq
\cheqnX{A}
 Similarly,
 for $k'$,
 one can get
\cheqnXa{A}
\beq
     \Pp{\D k'_{\rm num}}{\D \Expv{Q^{-1 / 2}}}
  =- \frac{1}{\D \Expv{Q^{-3 / 2}}}
\~,
\label{eqn:dkp_dexpv_Q_m1_2}
\eeq
 and
\cheqnXb{A}
\beq
     \Pp{\D k'_{\rm num}}{\D \Expv{Q^{-3 / 2}}}
  =- \frac{1}{\D \Expv{Q^{-5 / 2}}}
\~.
\label{eqn:dkp_dexpv_Q_m3_2}
\eeq
\cheqnX{A}
\end{document}